\documentclass[twocolumn,aps,amssymb,noshowpacs,superscriptaddress,nofootinbib,longbibliography]{revtex4-1}

\usepackage{amsmath,amssymb}
\usepackage{graphicx}
\usepackage{verbatim}
\usepackage{amsfonts}
\usepackage{dcolumn}
\usepackage{bm}
\usepackage[section]{placeins}
\usepackage{color}
\usepackage{mathrsfs}
\usepackage[colorlinks,bookmarks=true,citecolor=blue,linkcolor=red,urlcolor=blue]{hyperref}
\usepackage{afterpage}
\usepackage{ulem}
\usepackage{url} 
\usepackage[T1]{fontenc}
\usepackage{lmodern}
\usepackage[utf8]{inputenc} 

\usepackage{comment}
\usepackage{cleveref}
\usepackage{siunitx}
\crefname{equation}{Eq.}{Eqs.}
\crefname{figure}{Fig.}{Figs.}

\usepackage{longtable}
\usepackage{multirow}

\newcommand{\beq}{\begin{eqnarray} }
\newcommand{\eeq}{\end{eqnarray} }
\newcommand{\Beq}{\begin{eqnarray*} }
\newcommand{\Eeq}{\end{eqnarray*} }

\newcommand{\RNum}[1]{\uppercase\expandafter{\romannumeral #1\relax}}

\newcommand{\R}{\color{red}}

\newcommand{\beginsupplement}{%
        \setcounter{table}{0}
        \renewcommand{\thetable}{S\arabic{table}}%
        \setcounter{figure}{0}
        \renewcommand{\thefigure}{S\arabic{figure}}%
        \setcounter{section}{0}
        \renewcommand{\thesection}{S\arabic{section}}%
        \setcounter{equation}{0}
        \renewcommand{\theequation}{S\arabic{equation}}%
     }

\begin{document}
\title{Symmetry invariants and classes of quasiparticles in magnetically ordered systems having weak spin-orbit coupling}
\author{Jian Yang}
\affiliation{Beijing National Laboratory for Condensed Matter Physics and Institute of Physics, Chinese Academy of Sciences, Beijing 100190, China}
\author{Zheng-Xin Liu}
\email{liuzxphys@ruc.edu.cn}
\affiliation{School of Physics and Beijing Key Laboratory of Opto-electronic Functional Materials and Micro-nano Devices, Renmin University of China, Beijing, 100872, China}
\affiliation{Key Laboratory of Quantum State Construction and Manipulation (Ministry of Education), Renmin University of China, Beijing, 100872, China}
\author{Chen Fang}
\email{cfang@iphy.ac.cn}
\affiliation{Beijing National Laboratory for Condensed Matter Physics and Institute of Physics, Chinese Academy of Sciences, Beijing 100190, China}
\affiliation{Kavli Institute for Theoretical Sciences, Chinese Academy of Sciences, Beijing 100190, China}

\begin{abstract}
Symmetry invariants of a group specify the classes of quasiparticles, namely the classes of projective irreducible co-representations in systems having that symmetry.
More symmetry invariants exist in discrete point groups than the full rotation group $\mathrm{O(3)}$, leading to new quasiparticles restricted to lattices that do not have any counterpart in a vacuum.
We focus on the fermionic quasiparticle excitations under ``spin-space group'' symmetries, applicable to materials where long-range magnetic order and itinerant electrons coexist.
We provide a list of 218 classes of new quasiparticles that can only be realized in the spin-space groups.
These quasiparticles have at least one of the following properties that are qualitatively distinct from those discovered in magnetic space group(MSG)s, and distinct from each other:(i) degree of degeneracy,(ii) dispersion as function of momentum, and(iii) rules of coupling to external probe fields.
We rigorously prove this result as a theorem that directly relates these properties to the symmetry invariants, and then illustrate this theorem with a concrete example, by comparing three 12-fold fermions having different sets of symmetry invariants including one discovered in MSG.
Our approach can be generalized to realize more quasiparticles whose little co-groups are beyond those considered in our work.
\end{abstract}
\maketitle

\section{Introduction}\label{intro}

\begin{figure}
\centering
\includegraphics[width=.5\textwidth]{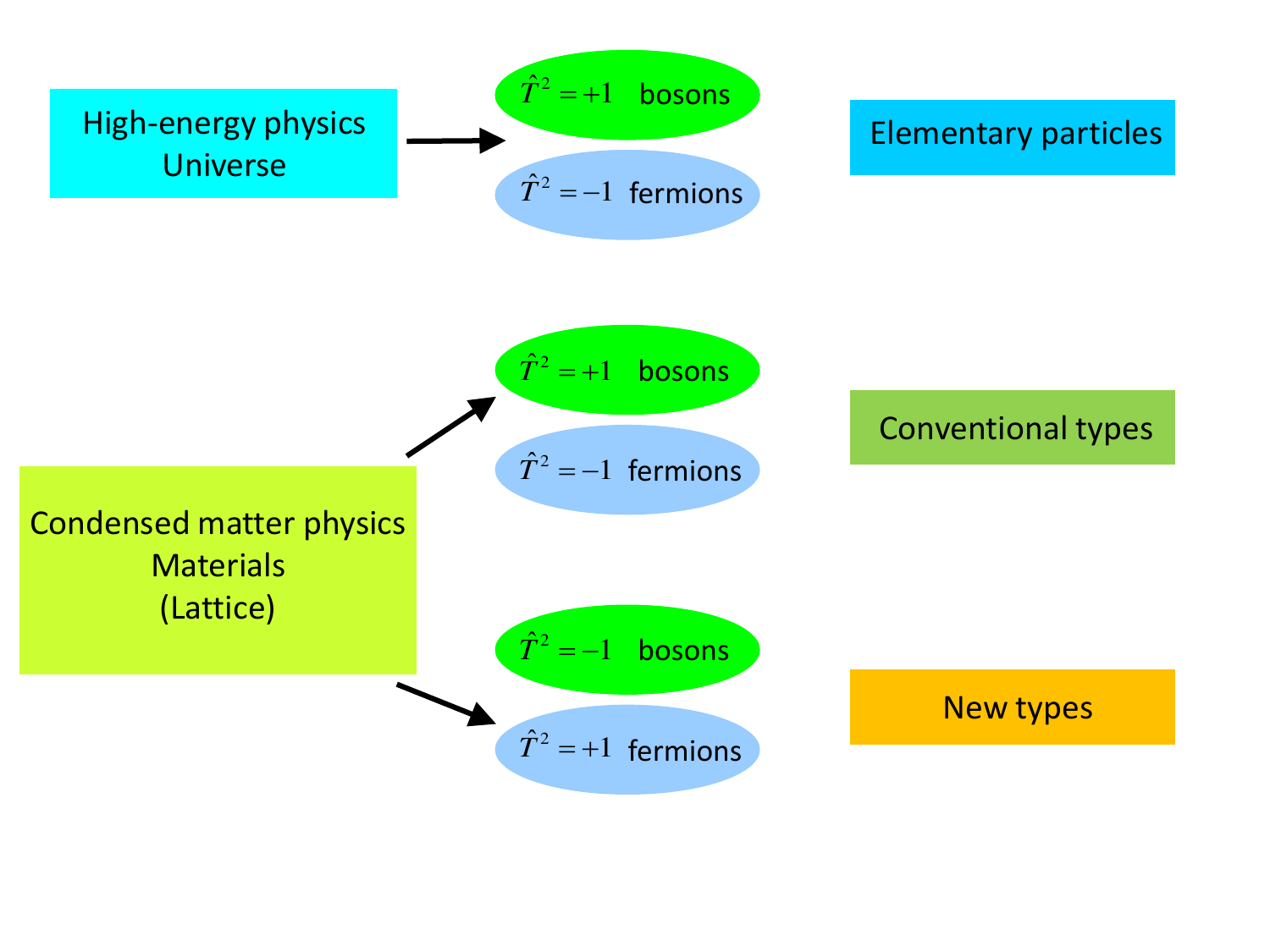}
\caption{Types of (quasi)particles in high-energy physics and condensed matter physics.
$\hat{T}^2=+1$ for bosons and $\hat{T}^2=-1$ for fermions are allowed in high-energy physics and condensed matter physics. But the new types, $\hat{T}^2=-1$ for bosons and $\hat{T}^2=+1$ for fermions, are only allowed in condensed matter physics.}
\label{fig:1}
\end{figure}

In high-energy physics, elementary particles are classified into bosons and fermions, characterized by their different statistics, $(-1)^{2s}$, determined by their spin quantum number $s$, where $s(s+1)\hbar^2 = \mathbf{S}^2$.
A similar distinction exists in their behavior under the anti-unitary time-reversal operation $T$, specifically $\hat{T}^2 = (-1)^{2s}$, with $\hat{T} = e^{i\hat{S}_{y}\pi/\hbar}K$ \cite{Sakurai2017}, where $\hat{S}_{y}$ is the $y$-component of the spin operator $\hat{\mathbf{S}}$, and $K$ is the complex-conjugate operator.
The hatted quantities are the representations of operators acting on the single-particle Hilbert space.
Therefore, the quantity $\hat{T}^2$, referred to as a symmetry invariant in later discussions, distinguishes fermions from bosons: $\hat{T}^2 = +1$ holds for particles with integer spin (bosons), while $\hat{T}^2 = -1$ is valid for those with half-odd-integer spin (fermions).
About 60 years ago, Eugene Wigner first noted that the time-reversal invariant $\eta_T \equiv \hat{T}^2$ of elementary particles, as an independent quantity, should not necessarily be related to $(-1)^{2s}$.
This principle led to the proposal of additional types of elementary particles.
Unfortunately, all particles discovered in high-energy physics satisfy the constraint $\hat{T}^2 = (-1)^{2s}$ \cite{Weinberg1995} (see Fig.\ref{fig:1}), and Wigner's idea did not attract much attention.

In condensed matter physics, massless elementary particles like Dirac \cite{Dirac_Kane14} or Weyl \cite{huang2015weyl} fermions can emerge as quasiparticle excitations in the low-energy limit.
Notably, since spatial symmetries (such as lattice translations and rotations) are discretized, symmetry invariants like $\eta_T = \hat{T}^2 = \pm 1$ and $\eta_{x,y}^{S} = (-1)^{2s} = \pm 1$ can take independent values, leading to the realization of new classes of particles proposed by Wigner.
For instance, massless quasiparticles with $\hat{T}^2 \neq \eta_{x,y}^S$ can arise in electronic energy bands of systems with anti-ferromagnetic order, where the nonsymmorphic time-reversal operation is associated with a fractional translation.
New types of quasiparticles, illustrated in Fig.\ref{fig:1}, exist in condensed matter physics, featuring $\hat{T}^2 = -1$ for bosons and $\hat{T}^2 = +1$ for fermions, without counterparts in high-energy physics.
These emergent quasiparticles in materials are explored using the representation theory of groups.
The symmetry operations of a system form a group, and point-like quasiparticle excitations are described by their projective irreducible co-representations (irReps).
Specifically, quasiparticles and their corresponding projective irReps can be grouped into several projective classes specified by a set of discrete values of the symmetry invariants \cite{ChenGuWen, ChenGuLiuWen, YangLiu, Yang2020, Yang2021, QiYang, chen2023classification}.
For example, time-like elementary particles in the standard model belong to two different projective classes (labeled by the symmetry invariant $\eta_{x,y}^{S} = \eta_{T} = \pm 1$) of the little group $\mathrm{O}(3) \times Z_2^T$ for massive particles.
In solids, the symmetry of a particle is reduced from continuous rotation O(3) to 32 subgroups known as crystallographic point groups.
Thus, there is interest in the projective classes of $P \times Z_2^T$, where $P$ is one of the 32 crystallographic point groups, and $Z_2^T$ is the time-reversal group.
To date, researchers have enumerated all quasiparticles protected by type-II \cite{rslager12,CLKane2012,BJYNagaosa14,cf15,watan16,Bradlyn2016,rslager17,bernevig17,Armitage2018,cf18,zhang2019catalogue,bernevig19,xgw19,cano21,tang2023wpprb} and type-IV \cite{Yang2020,gangxu16,Liang2016,Watanabe2018,Hua2018,mfvjur19,Cano2019,bernevig20,rslager21,Yang2021,wanxg2021,YiJiang2021,bernevigmtqc21,liu2021encyclopedia,zhang2021encyclopedia,lenggenhager2022triple,tang2022complete,bernevig2022progress,wanxgtangf2022prm,lin2022spin} Shubnikov's magnetic space groups (MSGs) \cite{Bradley1968,Lifshitz2004,Bradley2010,Litvin2013MagneticGT,gallego2016magndataI,gonzalez2021extension,LiuYaomsgcorep2023}.
Two questions naturally arise: (i) does this list exhaust the projective representations for symmetries of the form $P \times Z_2^T$? (ii) If not, where can we find the remaining quasiparticles?
For the second question, recent advances in spin-space groups (SSGs) provide a clue, as they offer a complete description of the symmetries of magnetic orders beyond MSGs.
SSGs also describe the symmetry of itinerant electrons hopping on the magnetic lattice, provided the spin-orbit coupling (SOC) of the electrons is much smaller than the Zeeman splitting caused by the local magnetic moment \cite{Liu2022prx,liu2022chiral,Corticelli2021,smejkal2021altermagnetism}.
Under weak SOC, unusual electronic structures and spin splittings emerge \cite{hayami2019momentum,hayami2020prb,ruo2aheconbs6,mazin2021prediction}.
An increasing number of experimental and theoretical studies on magnetic materials with weak SOC, including \textit{altermagnetism} with SSGs as approximate symmetries \cite{smejkal2021altermagnetism,TJungwirth2022,mazin2022altermagnetism,hariki2023xray,papaj2023andreev,ghorashi2023altermagnetic,steward2023dynamic,fernandes2023topological,sayed2023quantum}, have been conducted.
Candidate materials include $\mathrm{Mn}_{5}\mathrm{Si}_{3}$ \cite{reichlova2020macroscopic}, $\mathrm{RuO}_{2}$ \cite{feng2022anomalous,ruo2aheconbs6}, $\mathrm{MnTe}$ \cite{GBetancourtmnte2023}, $\mathrm{MnTe}_{2}$ \cite{liuchangmnte22024}, and $\mathrm{CoNb}_3\mathrm{S}_6$ \cite{liu2022chiral,Liuqhconb3s62023}.
A few examples of new quasiparticles with spin point group symmetry have been theoretically predicted \cite{pjguo2021prl,pjguo2022,Liu2022prx,liu2022chiral,Liuqhconb3s62023,Corticelli2021}.
Additionally, there are works discussing the representation theory and new quasiparticles of fermion and magnon bands within the framework of SSGs \cite{xiao2023spin,ren2023enumeration,chen2023spin}.
However, a comprehensive classification of these quasiparticles has yet to be accomplished.

\begin{figure}
\centering
\includegraphics[width=.5\textwidth]{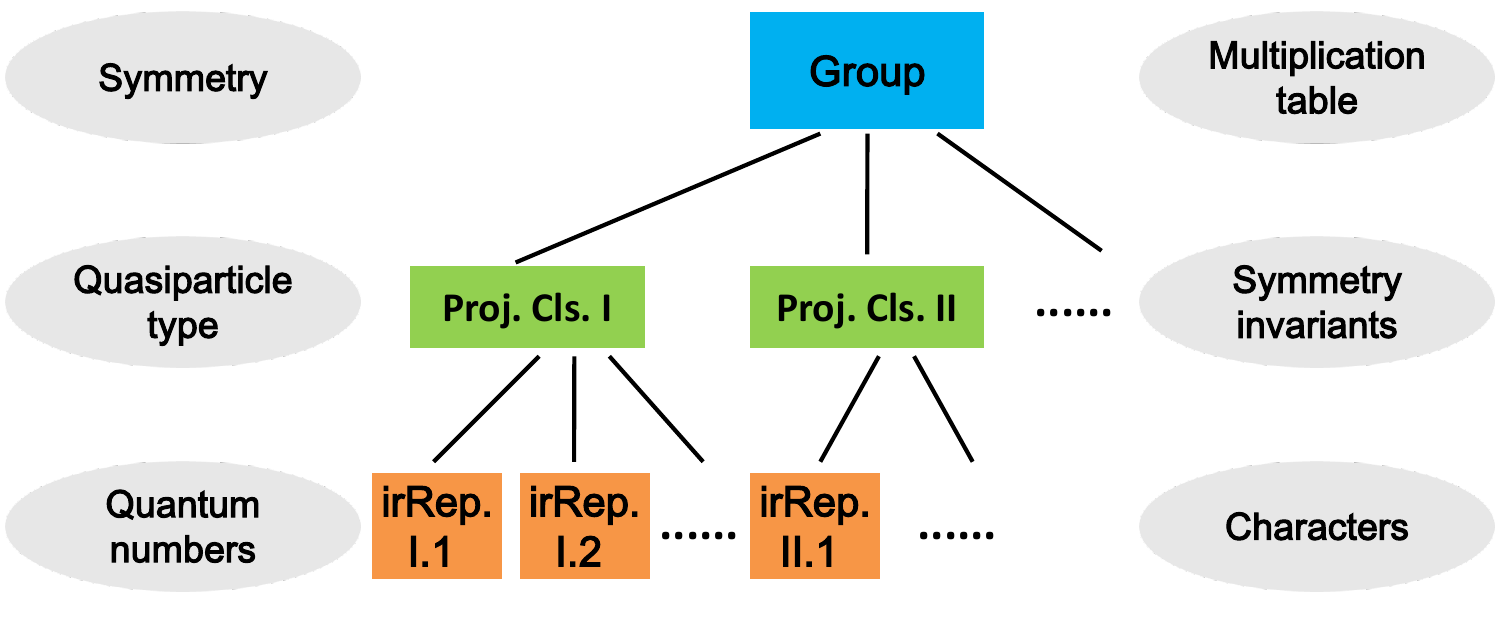}
\caption{Group, projective class and irRep.
The projective classes as intermediate between the group and the irRep, with their corresponding physical concepts to the left, and the quantities that specify them on the right.}\label{fig:2}
\end{figure}

In this paper, our results fully address the above open questions, using symmetry invariants as the primary tool.
Symmetry invariants, a set of variables formed by factor systems, generate the second group-cohomology group and can be used to classify and identify the various projective classes for the same symmetry group.
The projective classes form an intermediate layer between the groups and their (co)representations (Reps for short) and are less familiar to physicists.
In Fig.\ref{fig:2}, we summarize the relations among the groups, the projective classes, and the Reps, and also illustrate the correspondence between the physical entities (on the left) and the mathematical quantities (on the right).
Generally, a symmetry invariant takes the value of a certain root of 1, but in the present work, it is always $\pm 1$ valued (see Methods section for more details).
For any given point group $P$, there are $m_P$ number of $\mathbb{Z}_2$-valued symmetry invariants, resulting in $2^{m_P}$ projective classes and, consequently, $2^{m_P}$ quasiparticle types.
We identify all these invariants and find the rigorous limit of 680 projective classes, or quasiparticle types, for all $P \times {Z}_2^T$ groups.
Using the complete set of invariants, we show that all known quasiparticles protected by MSGs fall into 386 projective classes.
To demonstrate that symmetry invariants provide not only a mathematical classification but also a physical classification of quasiparticles, we prove a theorem that any two quasiparticles with different values for symmetry invariants must differ in at least one of the following aspects: degeneracy, dispersion, or linear couplings to external fields.
We then search for the remaining 294 quasiparticles unrealizable in MSGs and find that 218 can be realized in SSGs.
Two of these are illustrated in detail to show their uniqueness: a spin-1/2 fermion with 12-fold degeneracy, as shown in Fig.\ref{fig:3}, and another spin-1/2 fermion with 4-fold degeneracy, connected with 13 nodal lines, as shown in Fig.\ref{fig:4}.
We then attempt to scan magnetic materials with known magnetic structures for candidates hosting any of the 218 quasiparticles, and the search yields one candidate, Ce$_3$NIn.

\section{Symmetry invariants of quasiparticles}

We focus on quasiparticles with crystal momenta near a high-symmetry point in the Brillouin zone (BZ).
Due to their high symmetry, these momenta are usually the centers of carrier pockets, around which effective models are constructed.
In the long-wavelength limit, the quasiparticles move around a high-symmetry momentum and are thus subject to the little co-group at that momentum.

Assume the full symmetry of a Hamiltonian on a 3-dimensional lattice to be $\mathcal{G}$.
The three lattice vectors define a translation group $\mathcal{T}$, which is a normal subgroup of $\mathcal{G}$.
At $\mathbf{K} \in {\rm BZ}$, the little group $\mathcal{G}_\mathbf{K}$ is the subgroup of $\mathcal{G}$ that preserves $\mathbf{K}$ up to a reciprocal lattice vector, and we define the little co-group $G_\mathbf{K} \equiv \mathcal{G}_\mathbf{K} / \mathcal{T}$.
With the translation symmetry "modded out", $G_\mathbf{K}$ describes the effective symmetries for quasiparticles at $\mathbf{K}$.
Physically, a quasiparticle at $\mathbf{K}$ in this magnetic state is covariant under $G_\mathbf{K}$, and all observables are $G_\mathbf{K}$ symmetric.

The tricky part comes when the single-particle wavefunction, not an observable by itself, is considered.
The Bloch wavefunction transforms as a projective Rep $\rho(g),\ g\in{G}_\mathbf{K}$ \cite{Bradley1968, Birman74,Bradley2010,hamermesh1989group}, such that
\begin{equation}\label{eq:2}
\rho(g_1)\rho(g_2)=\omega_2(g_1,g_2)\rho(g_1g_2),
\end{equation}
for $g_1, g_2$ being unitary.
Throughout the paper, we use hatted operators $\hat{g}$ for the Rep of $g\in\mathcal{G}$ in the entire one-particle Hilbert space, and $\rho(g)$ for the Rep of $g\in{G}_\mathbf{K}$ in the subspace spanned by Bloch wavefunctions at $\mathbf{K}$.
Compared with a Rep where $\rho(g_1)\rho(g_2)=\rho(g_1g_2)$, Eq.(\ref{eq:2}) has a factor system $\omega_{2}(g_1,g_2)\in\mathrm{U(1)}$.
It is this factor system that gives rise to all types of the "new fermions" \cite{Bradlyn2016,Cano2019,Yang2020} and more quasiparticles to be discussed later.
For crystallographic point groups, symmetry invariants are $\mathbb{Z}_2$-numbers that classify all factor systems for a given symmetry.
For anti-unitary symmetries, the definition of projective Rep slightly differs, and so does the form of symmetry invariants [See Methods section].

The most well-known example of projective Reps in physics is the half-odd-integer spin.
The point group of a Galilean vacuum is $\mathrm{O(3)}\times{Z}^T_{2}$, and if one takes two twofold axes $2_x$ and $2_y$, then
\begin{equation}\label{eq:3}
\hat{2}_{x}\hat{2}_{y}=(-1)^{2s}\hat{2}_{y}\hat{2}_{x}.
\end{equation}
The symmetry invariant $\eta^S_{x,y}=(-1)^{2s}=-1$ represents the anti-commuting relation of $\hat{2}_{x}$ and $\hat{2}_{y}$ for all particles of half-odd-integer spin, like electrons, where the superscript $S$ stands for the spin contribution.
The time-reversal symmetry $T$ introduces an additional invariant $\eta_T^S=\hat T^2 =-1$, which leads to the well-known Kramers degeneracy.

For a less known, but important, example, we look at $\mathcal{G}$ being $P2_12_12_1\times{Z}_2^T$ with three twofold screw axes $C_{2x}=\{2_x|\frac{1}{2}\frac{1}{2}0\}, C_{2y}=\{2_y|0\frac{1}{2}\frac{1}{2}\}, C_{2z}=\{2_z|\frac{1}{2}0\frac{1}{2}\}$.
One easily checks that $\hat{C}_{2i} \hat{C}_{2j}= \hat{t}_{x+y+z} \hat{C}_{2j} \hat{C}_{2i}$, for $i,j=x,y,z$, $i\neq{j}$ and $t_{x+y+z}\equiv\{E|111\}$.
Now we look at phonons (bosons) near $\mathbf{K}=(\pi,\pi,\pi)$, where $G_\mathbf{K}=D_2\times{Z}_2^T$.
At $\mathbf{K}$, the translation operator $\hat{t}_{x+y+z}$ becomes $-1$, and we have
\begin{equation}\label{eq:5}
\rho(C_{2i})\rho(C_{2j})=-\rho(C_{2j})\rho(C_{2i})
\end{equation}
for $i\neq{j}$, yielding $\eta_{x,y}^{L}=-1$, where the superscript $L$ stands for the lattice contribution.
At the same time, $T$ is simply time reversal and $\eta_T^L\equiv[\rho(T){K}]^2=+1$, with $K$ being the complex conjugation.
Comparing Eq.(\ref{eq:3}) and Eq.(\ref{eq:5}), one immediately finds that the phonons at $\mathbf{K}$ have $(\eta_{x,y}^{L}=-1,\eta_T^L=+1)$.
On the other hand, for electrons at the same momentum point $\mathbf{K}$, the total invariants include the contribution from both the lattice and the spin and take the values $(\eta_{x,y}=\eta_{x,y}^{L}\eta_{x,y}^{S} = 1,\eta_T=\eta_T^L\eta_T^S=-1)$, with $\eta_{x,y}^{S} =-1$ and $\eta_{T}^S =-1$.
This combination of $(\eta_{x,y},\eta_T)$ is what leads to the multi-chiral fermions known in transition-metal monosilicides\cite{Zhang2018,Rao2019,Schroter2019,Sanchez2019}.
From the two examples, we see that the symmetry invariants of $G_\mathbf{K}$ depend, and only depend, on (i) the fundamental spin, (ii) the symmetry group $\mathcal{G}$, and (iii) the crystal momentum $\mathbf{K}$, but not on any specifics of the band structure, such as the ordering of bands or the Fermi level.
This property motivates one to classify quasiparticles using symmetry invariants.
In Supplementary Table~\ref{tab:invrts}, the definitions of symmetry invariants in terms of the factor system are given, resulting in 680 quasiparticle types.

Where can we find materials that realize all these values for symmetry invariants?
To be more precise, for a given point group $P$ and specified values of invariants $\eta$, is there a lattice with symmetry $\mathcal{G}$ and a crystal momentum $\mathbf{K}$ such that $G_\mathbf{K} \cong P \times Z_{2}^{T}$ and the invariants correspond to those specified in $\eta$?
For $\mathcal{G}$ being a type II MSG (that is, a space group (SG) times time reversal), 224 sets are realized; and for $\mathcal{G}$ being a type IV MSG (applicable to, e.g., antiferromagnets with doubled unit cells), an additional 162 sets are realized.
In fact, for $P = C_{1,s,i,2,3,4,6,2v,3v,4v,6v,2h,3i}, S_{4}, D_{2,2d,2h}, T, T_{d,h}$, all sets of invariants can be realized in MSG.
However, for $P = C_{3h,4h,6h}, D_{3,4,6,3d,3h,4h,6h}, O, O_h$, there are 294 sets of invariants that cannot be realized in MSG.

In MSG, spin-orbit coupling (SOC) requires every point-group operation $p\in\mathrm{O}(3)$ to be associated with a spin rotation $\det[p]p\in\mathrm{SO(3)}$.
This means that the spin part of the operation is fixed by the spatial or lattice part of the operation.
However, in the limit of negligible SOC, this locking between spin and space operations is released, which gives rise to more possibilities for the symmetry group.
These new symmetries with unlocked spin and lattice operations, termed SSG \cite{Brinkman1966,Litvin1974}, provide a way to realize more quasiparticles.

\section{Structure of Spin-space groups}

Now we investigate the structure of SSG, which provides important information for obtaining the invariants of quasiparticles.
We consider the Hamiltonian for itinerant electrons in a magnetically ordered lattice
\begin{equation}\label{eq:1}
\hat{H}=\frac{\hat{\mathbf{p}}^2}{2m}+V({\mathbf{r}})+\mathbf{M}({\mathbf{r}})\cdot\hat{\mathbf{s}},
\end{equation}
where $\hat{\mathbf{s}}$ is the electron-spin operator, $V({\mathbf{r}})$ the lattice potential, and $\mathbf{M}({\mathbf{r}})$ the Zeeman/exchange field generated by the ordered moments.
The spin symmetry rotations are not necessarily always locked with lattice rotations due to the absence of the SOC in Eq.(\ref{eq:1}) \cite{Liu2022prx}.
For example, in the magnetic material $\mathrm{RuO}_{2}$ \cite{feng2022anomalous,ruo2aheconbs6} with collinear magnetic order along the $x$-axis, the $C_{4z}^{+}$ lattice rotation should be associated with $C_{2z}$ spin rotation, characterized by an SSG symmetry $(C_{2z}||C_{4z}^{+}|\frac{1}{2}\frac{1}{2}\frac{1}{2})$.
Such kinds of symmetry operations are called SSG operations, which form the spin-space groups.
The absence of SOC in Eq.(\ref{eq:1}) unlocks the spatial and spin degrees of freedom of electrons from each other, enabling the realization of 218 new types of quasiparticles in SSG with little co-group $P\times{Z}_2^T$.

Generally, an SSG operation $g$ has a spin part $\varphi_g \in \mathrm{SO(3)}\times Z_2^T$ and a lattice part $l_g$, with $g=(\varphi_g || l_g)$.
For later convenience, we define a unitarity indicator $\zeta_g$ such that $\zeta_g=0$ if $g$ is unitary and $\zeta_g=1$ if $g$ is anti-unitary.
The lattice part can be represented as $l_g=\{p_g|\mathbf{t}_g\}T^{\zeta_g}$, which contains a point-group part $p_g \in \mathrm{O(3)}$ and a translation vector $\mathbf{t}_g$.
The spin rotation is decoupled from $p_g$, under the constraint that $\{(\varphi_g || l_g)\}$ forms a group under multiplication, which requires $\varphi$ to be a homomorphism from the lattice part $l_g$ to the spin part $\varphi_g$ that preserves unitarity or anti-unitarity.
A pseudo-vector field $\mathbf{M}(\mathbf{r})$ is said to be invariant under $g$ if and only if $\mathbf{M}(\mathbf{r}) = s(\varphi_g)\mathbf{M}(l_g^{-1} \mathbf{r})$.
Here, $s$ is an "isomorphism" from $\mathrm{SO(3)}\times{Z}_2^T$ to $\mathrm{O(3)}$, i.e., a generalized vector Rep with $s(T) = -1_{3\times3}$, $s(TR) = s(RT) = -s(R)$, and $s(R)$ being the usual vector Rep of $R$ for $R\in\mathrm{SO(3)}$.

Here we mention four key groups associated with the SSG $\mathcal{G}$ \cite{Litvin1974}:
(i) the group formed by the lattice part $\mathcal{L}=\{l_g |g\in{\mathcal{G}}\}$,
(ii) the group formed by the spin part $S=\{\varphi_g|g\in\mathcal{G}\}$,
(iii) all SSG operations that have a trivial spin part $\mathcal{L}_0=\{l_g|g\in\mathcal{G},\varphi_g=E\}$, and
(iv) all SSG operations that have a trivial lattice part $S_0=\{\varphi_g|g\in\mathcal{G},l_g=\{E|\mathbf{0}\}\}$.
From these definitions, it immediately follows that:
(a) $\mathcal{L}$ is also the lattice part of an MSG $\mathcal M$ with $\mathcal L\cong \mathcal M$;
(b) $\mathcal{L}_0$ is a normal subgroup of $\mathcal{L}$;
(c) $S$ is a subgroup of $\mathrm{SO}(3)\times Z_{2}^{T}$, and $S_0$ is a normal subgroup of $S$.
Symbolically, they are summarized in \cite{Litvin1974} as:
\begin{equation}\label{eq:8}
\mathcal{L}_0\lhd\mathcal{L},S_0\lhd{S}, \frac{\mathcal{L}}{\mathcal{L}_0}\cong\frac{S}{S_0} \cong {\mathcal G/ (\mathcal L_0\times S_0 ) }.
\end{equation}
Very recently, the SSGs have been completely enumerated with truncated sizes of magnetic unit cells \cite{jiang2023enumeration, ssg_website, xiao2023spin,ren2023enumeration, watanabe2023symmetry,shinohara2024algorithm}.
In the present work, we mainly discuss cases where $S_0=\{E\}$ is a trivial group.
Under this assumption, $\mathcal{G}\cong\mathcal{M}$: an SSG with a trivial $S_0$ is isomorphic to an MSG.
In the case of nontrivial $S_0$, our discussion also applies thanks to the relation $\mathcal G/S_0 \cong \mathcal L \cong \mathcal M $.
We further restrict the discussion to cases where the magnetic unit cell is twice as large as the crystal unit cell, specifically considering SSGs that are:
(i) isomorphic to type IV MSGs, and
(ii) where nontrivial spin rotations are only associated with group elements that have nontrivial lattice point operations.

The difference between SSG $\mathcal{G}$ and the isomorphic MSG $\mathcal{M}$ then lies in the spin rotation $\varphi_g$ for each $g\in\mathcal{L} \cong \mathcal{M}$, which is a homomorphism from $\mathcal{L}$ to $\mathrm{SO(3)}\times{Z}_2^T$.
First, we observe that $\mathcal{L}_0$ is the kernel of this homomorphism that maps to $1_{3\times3}$.
Then, we build a surjective homomorphism from $\mathcal{L}$ to the quotient group $\mathcal{L}/\mathcal{L}_0$ and then an injective homomorphism to $\mathrm{O(3)}$:
\begin{equation}
\mathcal{L}\xrightarrow{\varphi}\frac{\mathcal{L}}{\mathcal{L}_0}\xrightarrow{s}\mathrm{O(3)}, g\mapsto{\varphi}_g\mapsto{s}(\varphi_g),
\end{equation}
under the constraint $\det[s(\varphi_g)]=+1(-1)$ if $g$ is unitary (anti-unitary).
For each $\mathcal{L}$ and its normal subgroups $\mathcal{L}_0$, the quotient group $\mathcal{L}/\mathcal{L}_0$ is finite, thus there are a finite number of homomorphisms $s\cdot\varphi$.
A similar structure also appears in the later discussion of the little co-group of quasiparticles.

Now we relate SSG to operators that act on electrons subject to $\mathbf{M}(\mathbf{r})$ as in Eq.(\ref{eq:1}).
Under an SSG operation $g=(\varphi_g||l_g)$, an electron operator $c_\alpha(\mathbf{r})$, $\alpha,\beta$ being spin index, transforms as
\begin{equation}\label{eq:9}
c^{\dag}_\alpha(\mathbf{r})\xrightarrow{g} \sum_{\beta}  d_{\beta\alpha} (\varphi_g) K^{\zeta_g} c^{\dag}_{\beta}(l_g^{-1}\mathbf{r}),
\end{equation}
where $d(\varphi_g)K^{\zeta_{g}}=u\Big(s(\varphi_g)\Big)$
for unitary $g$ and
$d(\varphi_g)K^{\zeta_{g}}=u\Big(s(T\varphi_g)\Big)(i\sigma_y K)$ for anti-unitary $g$, $u:\mathrm{SO}(3)\rightarrow\mathrm{SU}(2)$ is the natural embedding from $\mathrm{SO(3)}$ to $\mathrm{SU(2)}$.

\section{Symmetry invariants of spin-space groups}

An SSG $\mathcal{G}$ has a normal translation subgroup $\mathcal{T}_0\lhd\mathcal L_0$.
Using $\mathcal{T}_0$, the BZ and the crystal momentum $\mathbf{K}$ are defined.
The little group $\mathcal{G}_\mathbf{K}$ is the subgroup of $\mathcal{G}$ preserving $\mathbf{K}$, and the little co-group $G_\mathbf{K}=\mathcal{G}_\mathbf{K}/\mathcal{T}_0$ is essentially a spin point group \cite{Litvin1977,schiff2023spin} whose elements are given by
$(\varphi_p|| p)$.
The lattice parts $p$ of $(\varphi_p||p)$ form $L_\mathbf{K}$, the little co-group of $\mathcal{L}$. As $\mathcal{G}\cong\mathcal{M}\cong \mathcal L$, we have $G_\mathbf{K}\cong{M}_\mathbf{K}\cong{L}_\mathbf{K}$
with ${M}_\mathbf{K}$ the little co-group of $\mathcal{M}$,
recalling that the SSG $\mathcal{G}$ and the MSG $\mathcal{M}$ have the same pure translation part of $\mathcal{L}$.

The symmetry invariants may come from the spin part, as in Eq.(\ref{eq:3}), and the lattice part, as in Eq.(\ref{eq:5}).
Each invariant of $G_\mathbf{K}$ hence factors into the lattice and the spin invariant $\eta_i=\eta^L_i\eta^S_i$.
The lattice part of $G_\mathbf{K}$ is nothing but $M_\mathbf{K}$, the invariants of which can be computed as \begin{eqnarray}\label{eq:chiLp1p2}
\eta^L_{p_1,p_2}=\exp[-i(\mathbf{K}_{p_{1}}\cdot\mathbf{t}_{p_2} - \mathbf{K}_{p_{2}}\cdot\mathbf{t}_{p_1} )]
\end{eqnarray}
if $p_1, p_2$ are unitary with $p_{1}p_{2}=p_{2}p_{1}$ and
\begin{eqnarray}\label{eq:chiLpt}
\eta^L_{p}=\exp(-i\mathbf{K}_{p}\cdot\mathbf{t}_{p}),
\end{eqnarray}
if $p$ is anti-unitary with $p^2=E$.
Here the reciprocal lattice vector $\mathbf K_p$ is defined as $\mathbf K_{p} =(-1)^{\zeta_p}(p^{-1}\mathbf K -\mathbf K)$
with $\zeta_{p}=0(1)$ if $p$ is unitary (anti-unitary)
{\cite{Bradley2010,Chen1985,chen2002group}},
$\mathbf{t}_{p}$ is the translation vector of $p$.

The spin part of $G_\mathbf{K}$, which is essentially a homomorphism from $L_\mathbf{K}$ to $\mathrm{SO(3)}\times{Z}^T_2$ then to $\mathrm{O(3)}$, can be classified and enumerated in the following way.
First, $L_{0\mathbf{K}}$, the little co-group of $\mathcal{L}_0$ is the kernel of this mapping.
For a given $L_\mathbf{K}$, there are a finite number of $L_{0\mathbf{K}}$, and for each given $L_{0\mathbf{K}}$, one can enumerate the mappings from the quotient group $L_\mathbf{K}/L_{0\mathbf{K}}$ to $\mathrm{O(3)}$.
Such a mapping $s\cdot\varphi: L_\mathbf{K}\to\mathrm{O(3)}$ is a 3-dimensional real Rep of $L_\mathbf{K}$.
With $\varphi$ specified, the spin part of the invariants is
\begin{eqnarray}\label{eq:chiSp1p2}
\eta^S_{p_1,p_2}{\cdot1_{2\times 2}} = d({\varphi_{p_1}})d({\varphi_{p_2}})d^{-1}({\varphi_{p_1}})d^{-1}({\varphi_{p_2}}),
\end{eqnarray}
for unitary elements $p_{1},p_{2}$ with $p_{1}p_{2}=p_{2}p_{1}$ and
\begin{eqnarray}\label{eq:chiSpt}
 \eta^S_{p}{\cdot1_{2\times 2}}&=&[d(\varphi_p) {K}]^2,
\end{eqnarray}
for anti-unitary element $p$ with $p^{2}=E$.

We focus on those SSGs and the momentum points $\mathbf{K}$ which have $G_\mathbf{K}\cong{P}\times{Z}_{2}^T$.
Remembering that $\mathcal{G}\cong\mathcal{M}\cong \mathcal L$, $G_\mathbf{K}\cong{M}_\mathbf{K}\cong{L}_\mathbf{K}$, and noticing that the MSG $\mathcal{M}$ and magnetic point group $M_\mathbf{K}$ can be easily identified from the standard crystallographic table, for convenience we denote $M_\mathbf{K}=P\times{Z}_2^T$ (this in turn implies that $\mathcal{M}$ is a type IV MSG).
In Supplementary Table~\ref{Pdrztlat}, we have computed $P=C_{3h,4h,6h},D_{3,4,6,3d,3h,4h,6h},O,O_h$, the lattice-part invariants $\eta^L$.
For the spin part, in Supplementary Tables~\ref{C4hdrztspin}-\ref{Ohdrztspin}, we have listed the choice of $L_{0\mathbf{K}}$, the choice of homomorphism $\varphi$, and the resultant invariants {$\eta^{S}$}.
For easy reference, each unique set of $\eta^{L/S}$ is labeled by a Boolean vector $(1{-}\eta^{L/S})/2$, and the $\eta^{S}$ leading to invariants only realizable in SSG are in RED.
Finally, we obtain the full symmetry invariants for $G_\mathbf{K}$ using $\eta_i=\eta^L_i\eta^S_i$, listed in 
Supplementary Tables~\ref{C4hdrztinv}-\ref{Ohdrztinv}.
The $\eta$'s that can be realized in type II and type IV MSG are colored in GREEN and BLUE, respectively, and those only realizable in SSG, counting 218, are in RED.
The BLACK sets of invariants cannot be realized in MSG, or even SSG.
For each RED set, we have listed all possible pairs $(\eta^L,\eta^S)$.

\section{Physical properties determined by invariants}

Now we show that, for a given $P\times{Z}_2^T$, two quasiparticles from different projective classes have distinct physical properties.
They can differ from each other in the degrees of degeneracy.
Otherwise, they are distinguished by the dispersion or the coupling to an external probe field, such as the displacement (phonon) field and the Zeeman field.
First, we observe that any dispersion $\pmb\phi$ as a function of momentum, or any external field $\pmb\phi$, corresponds to a linear irRep of $P\times{Z}_2^T$, so its components $\phi^{\mu}_i$ carry a linear irRep of $P\times{Z}_2^T$, where $\mu(P\times{Z}_2^T)$ is a linear irRep, and $i=1,\cdots,\mathrm{dim}[\mu]$ labels the component within the irRep.

Then we notice that as the fermion operator $c^{\dag}_m$ transforms as a projective irRep $\rho$ with $m=1,...,\mathrm{dim}[\rho]$, the bilinear $c^\dag_mc_n$ transforms as a linear Rep $\rho\otimes{\rho^{\ast}}$\cite{Yang2021}, which reduces into direct sums of linear irRep $\mu$: $\rho\otimes{\rho^{\ast}}=\sum_{\mu}\oplus{N}_\rho(\mu)\mu$,
where the multiplicity
\cite{YangYFL_CG2021}
\beq
N_\rho(\mu) &=& {1\over 2 |P|} \sum_{g\in P} \Big[|\chi^{(\rho)}(g)|^2\chi^{(\mu)}(g) +
\nonumber\\
&&\omega_2(gT, gT)\chi^{(\rho)}\big( (gT)^2\big)\chi^{(\mu)}(gT) \Big]
\eeq
is the number of times the linear irRep $\mu$ occurs in the direct product Rep $\rho\otimes{\rho^{\ast}}$.
Here, $\chi^{(\rho)}(g)$ stands for the character of $g\in P$ in the projective irRep $\rho$, and $\chi^{(\mu)}(g)$ is the character of $g$ in the linear irRep $\mu$. $|P|$ is the number of elements in the point group $P$.
The elements of the $\mathrm{dim}[\rho]\times\mathrm{dim}[\rho]$ matrix $\Gamma^{(\mu)\tau_{\mu}i}$ are the Clebsch-Gordon coefficients combining the bases of the direct product Rep $\rho\otimes{\rho^{\ast}}$ into the $i$-th base of the linear irRep $\mu$, where $\tau_{\mu}=1,...,N_\rho(\mu)$.

In terms of $\phi_i^\mu$ and $\Gamma^{(\mu)\tau_{\mu}i}_{}$, the general symmetric coupling takes the form
\begin{equation}\label{eq:20}
\hat{H}=\sum_{\mu,\tau_{\mu},i}\lambda^{(\mu)}_{\tau_{\mu}}\phi^\mu_i
\Gamma^{(\mu)\tau_{\mu}i},
\end{equation}
where coupling parameter $\lambda^{(\mu)}_{\tau_{\mu}}\in\mathbb{R}$.
From Eq.(\ref{eq:20}), we see that the physical response of a quasiparticle (denoted by the corresponding irRep $\rho$) entirely depends on $N_\rho(\mu)$.
For example, letting $P=O_h$ and identifying $\pmb \phi$ as momentum $\mathbf k$ which transforms as
$T^-_{1u}$ ($-$ means odd under time reversal), if $N_\rho(T^-_{1u})=0$,
then there is no linear dispersion for this quasiparticle.
On the other hand, if $N_\rho(T^-_{1u})=2$, then there are two independent coupling parameters denoted as
$\lambda^{(\mu)}_{1}$ and $\lambda^{(\mu)}_{2}$ ,
hence two Fermi velocities for this quasiparticle.
The same procedure applies if $\pmb \phi$ stands for probing Zeeman field ($T^-_{1g}$ Rep), lattice displacement field ($T^+_{1u}$ Rep), strain tensor field ($T^+_{2g}$ Rep or $E^+_{g}$ Rep) and so on.

In Supplementary Note \ref{sec:prffusiontheory}, we prove a theorem stating that if projective irReps $\rho_1$ and $\rho_2$ for $P\times{Z}_2^T$ belong to different projective classes (namely, $\rho_1$ and $\rho_2$ have different values of symmetry invariants), then there must exist at least one linear Rep $\mu$ such that $N_{\rho_{1}}(\mu)\neq N_{\rho_{2}}(\mu)$.
Hence, when dispersion $\pmb\phi$ or external probe field $\pmb\phi$ carries a linear Rep $\mu$, the quasiparticle carrying irRep $\rho_1$ must behave differently compared to the quasiparticle carrying irRep $\rho_2$.
This seals the conclusion that any one of the 218 new quasiparticle types has distinct physical properties from the 386 types realizable in MSG, and the 218 classes of new quasiparticles also have distinct physical properties from each other.
This is one of the central conclusions of the present work.
In the next section, we will illustrate this result with concrete examples, namely, by comparing three 12-fold quasiparticles having different sets of symmetry invariants.

In Supplementary Tables~\ref{C4hdrztinv}-\ref{Ohdrztinv}, we have summarized all band degeneracies for every set of symmetry invariants that are only realizable in SSG colored in RED, if the minimal dimension of (co)Reps is FOUR or higher, including the degree of degeneracy, the lowest-order dispersion, and the direction of the nodal lines (if any) meeting at this point.

\section{Example: 12-fold fermions and 13 Dirac nodal lines}

\begin{figure*}
\begin{centering}
\includegraphics[width=0.9\textwidth]
{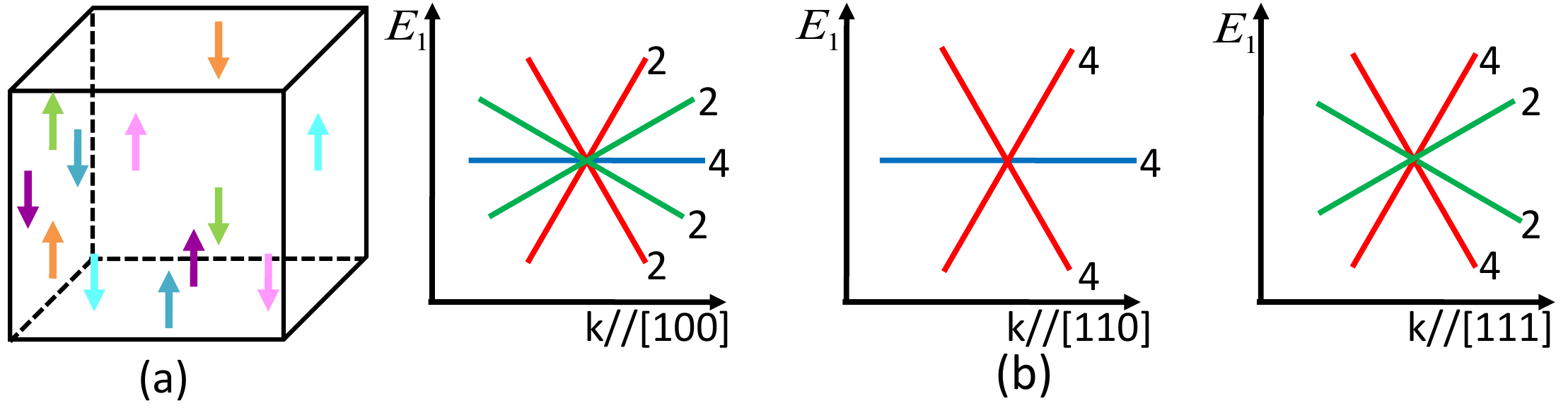}
\par\end{centering}
\protect
\caption{\label{fig:3}12-fold fermions in spin-space group 229.2.1.9 \cite{ssg_website}.
(a) A magnetic structure that is compatible with this SSG with 12 spins in one unit cell.
The 12 spins are colored in
orange, green, blue, purple, pink and sky blue, each of which
has upward and downward magnetic moments.
(b) The schematic dispersion of the 12-fold fermion along three high-symmetry lines $[100]$, $[110]$, $[111]$.
Along $[100]$, the 12-fold fermion splits into one 4-fold band colored in blue,
two 2-fold bands colored in red and two 2-fold bands colored in green.
Along $[110]$, the 12-fold fermion splits into one 4-fold band colored in blue and
two 4-fold bands colored in red.
Along $[111]$, the 12-fold fermion splits into two 4-fold bands colored in red and
two 2-fold bands colored in green.}
\end{figure*}

According to Supplementary Table~\ref{Ohdrztinv}, for $G_{\mathbf{K}}=O_h\times Z_2^T$, ${\rm Invariants}\equiv$($\eta_{C_{2x},C_{2y}}$, $\eta_T$, $\eta_{IT}$, $\eta_{TC_{2a}}$, $\eta_{I,C_{2a}}$), when the five invariants take the values $\vec{\eta}=(+1,+1,-1,-1,-1)$, the corresponding projective class is only realizable in spin-space groups.
One such example is the non-coplanar SSG \href{https://cmpdc.iphy.ac.cn/ssg/ssgs/229.2.1.9}{229.2.1.9} in the database \cite{ssg_website}, which is isomorphic to type IV MSG $\mathcal{M}=223.109$.
The elements of the SG $\mathcal{L}_0\equiv Pm\bar3n (223)$ are pure lattice operations, with quotient group $229.2.1.9/223 \cong Z_2^T$.
The little co-group at the high-symmetry point $R=(\pi,\pi,\pi)$ is $G_{\mathbf{K}}=O_{h}\times Z_{2}^{T}$, which is generated by $(E||C_{2x,2y,2z},C_{2\mathbf{n}},I)$ and $(T||TC_{2x,2y,2z},TC_{2\mathbf{n}},TI)$ as listed in Supplementary Table~\ref{Ohdrztspin}, where $I$ is spacial inversion, and $\mathbf{n}$ is a unit vector in the $\hat x \pm\hat y,\hat x \pm\hat z,\hat y \pm\hat z$ directions.

To obtain the magnetic configuration, we first illustrate a collinear configuration with 12 spins at the 12d Wyckoff positions in each unit cell, as shown in Fig.\ref{fig:3}(a).
This configuration has a collinear SSG symmetry \href{https://cmpdc.iphy.ac.cn/ssg/ssgs/229.2.1.3.L}{229.2.1.3.L}.
The complete non-coplanar spin configuration actually contains three or more sets of collinear configurations (for instance, another two sets may locate at two inequivalent 24g Wyckoff positions) with the condition that the directions of different sets of spins are mutually unparallel.
Hence, the final configuration is a non-coplanar magnetic order described by the SSG 229.2.1.9.
Similarly, the configuration of the coplanar SSG 229.2.1.6.P can be obtained if the number of collinear sets is two.
The three groups, 229.2.1.3.L, 229.2.1.6.P and 229.2.1.9 have the same lattice operations and only differ by their spin-only groups with $S_0 = \mathrm{SO}(2)\rtimes Z_2^{ C_{2x}T}, Z_2^{C_{2z}T}, \{E\}$ respectively.
Furthermore, for the common little co-group symmetry $O_h\times Z_2^T$ at the R point, the three groups 229.2.1.3.L, 229.2.1.6.P, 229.2.1.9 have the same values of symmetry invariants $(+1,+1,-1,-1,-1)$.

This set of invariants $\vec{\eta}=(+1,+1,-1,-1,-1)$ for $O_h\times Z_2^T$ allow a 12-dimensional irRep given by
\beq\label{12dfermion1}
&&\rho_{1}(C_{2i})=\tau_{0}\sigma_{0}e^{-iL_{i}\pi}, \ \
\rho_{1}(C_{2\mathbf{n}})=i\tau_{z}\sigma_{0} e^{-i\mathbf{L}\cdot\mathbf{n}\pi},\nonumber\\
&&\rho_{1}(I)=i\tau_{x}\sigma_{0}1_{3\times 3},\ \
\rho_{1}(T)K=\tau_{y}\sigma_{y}1_{3\times 3}K,
\eeq
whose dimensionality is higher than any known (co)Reps appearing in MSGs\cite{Yang2021}. Here $L_{x,y,z}$ are the three components of angular momentum for $l=1$,
$\tau_{0}(\sigma_{0})=1_{2\times 2}$,
$\tau_{x,y,z} (\sigma_{x,y,z})$ are three Pauli matrices acting on the sublattice (spin) sector, $\mathbf{n}$ is a unit vector in $\hat x \pm\hat y,\hat x \pm\hat z,\hat y \pm\hat z$ directions.
$\rho_{1}\otimes \rho_{1}^{\ast}$
can be decomposed into linear irReps of
$O_{h}\times Z_{2}^T$:
\beq\label{12dcg1}
\rho_{1}\otimes \rho^{\ast}_{1}=\cdots \oplus 4T_{1u}^{-}\oplus \cdots ,
\eeq
the linear irRep $\mu=T_{1u}^-$ corresponds to linear dispersion, the components $\phi^{\mu}_{1}=k_{x}$, $\phi^{\mu}_{2}=k_{y}$,
$\phi^{\mu}_{3}=k_{z}$.
Since multiplicity $N_{\rho_{1}}(T_{1u}^-)=4$, there are 4 coupling parameters
$\lambda^{(\mu)}_{1}=v_{1x}$,
$\lambda^{(\mu)}_{2}=v_{1y}$,
$\lambda^{(\mu)}_{3}=v_{1z}$,
$\lambda^{(\mu)}_{4}=v'_{1}$.
According to Eq.(\ref{eq:20}),
the $\mathbf{k}\cdot\mathbf{p}$ model to the lowest order in $\mathbf{k}$, the crystal momentum relative to $\mathbf{K}$, takes the form
\begin{equation}
\hat{H}_{1}(\mathbf{k})=
(
v_{1x}\tau_{z}\sigma_{x}+
v_{1y}\tau_{z}\sigma_{y}+
v_{1z}\tau_{z}\sigma_{z}
)
\mathbf{k}\cdot\mathbf{L}+v'_{1}\tau_{y}\sigma_{0}\mathbf{k}\cdot\mathbf{L}',
\end{equation}
where $L'_i\equiv|\epsilon_{ijk}|\{L_j,L_k\}$, and $v_{1x,1y,1z},v'_{1}$ material dependent Fermi velocity.
This 12-fold degeneracy is a superposition of four spin-1 double-Weyl points, two with Chern number $+2$, the other two $-2$.
The linear dispersion and band splitting of this degeneracy along different high-symmetry lines are shown in Fig.\ref{fig:3}(b).
Since there are no 12-dimensional irReps of $O_{h}\times Z_{2}^{T}$ in MSGs, the 12-fold fermion \eqref{12dfermion1} is qualitatively different from all quasiparticles with symmetry group $O_{h}\times Z_{2}^{T}$ in MSGs because of the degree of degeneracy.

This 12-dimensional irRep appears at the R point of the three SSGs 229.2.1.3.L, 229.2.1.6.P, and 229.2.1.9, even though their $S_0$ are different (the difference in $S_0$ indeed has a consequence in their irReps.
For instance, both 229.2.1.3.L and 229.2.1.6.P have an 8-dimensional irRep at the R point but the non-coplanar group 229.2.1.9 does not).

\begin{figure*}
\begin{centering}
\includegraphics[width=0.9\textwidth]{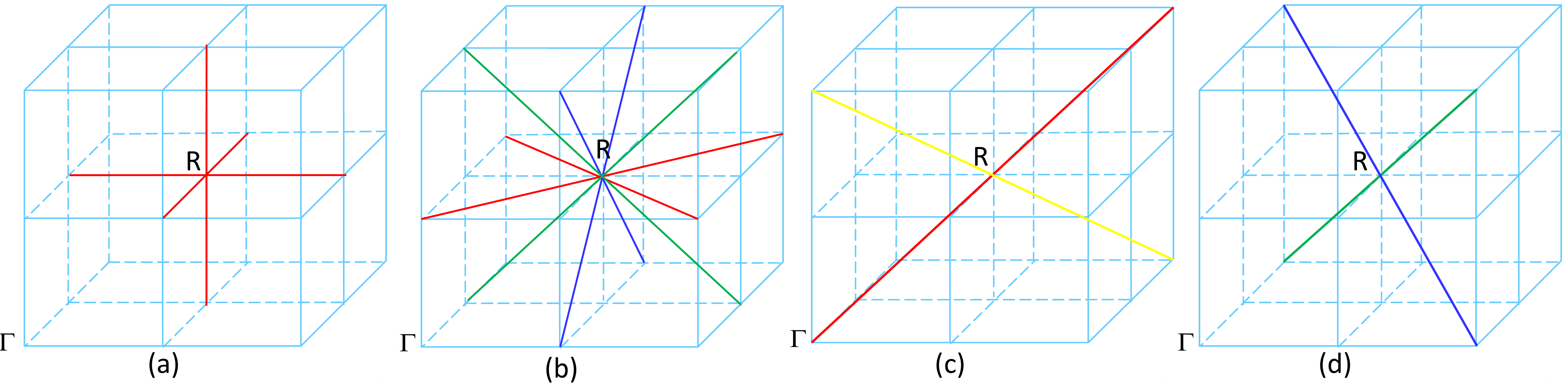}
\par\end{centering}
\protect
\caption{\label{fig:4}13-nodal-line nexus in spin-space group 229.2.1.9 \cite{ssg_website}.
The nodal lines meeting at $\mathrm{R}$ in the magnetic BZ for the 13-nodal-line nexus. The nodal line degeneracies are exact
because the irRep of $\mathrm{R}$ remains irreducible along the nodal lines. (a) The three red lines are nodal lines along $k_{x}$,$k_{y}$ and $k_{z}$ directions. (b) The two red lines are nodal lines along $k_{x}\pm k_{y}$ directions, the two green lines are nodal lines along $k_{x}\pm k_{z}$ directions, the two blue lines are nodal lines along $k_{y}\pm k_{z}$ directions. (c) The red line is nodal line along $k_{x}+k_{y}+k_{z}$ direction, the yellow line is nodal line along $k_{x}+k_{y}-k_{z}$ direction. (d) The green line is nodal line along $k_{x}-k_{y}+k_{z}$ direction, the blue line is nodal line along $k_{x}-k_{y}-k_{z}$ direction.}
\end{figure*}

With the same invariants $\vec{\eta}=(+1,+1,-1,-1,-1)$, as shown in Supplementary Table~\ref{Ohdrztinv},
there is a 4-dimensional irRep \cite{Yang2021}
\beq\label{4dfermion1}
&&\rho_{0}(C_{2i})=\tau_0\sigma_0,\ \
\rho_{0}(C_{2,01\bar1})=i\tau_0\sigma_z,
\nonumber\\
&&\rho_{0}(C_{2,110})=(i/2)\tau_0\sigma_z+(\sqrt{3}i/2)\tau_{z}\sigma_{x},
\nonumber\\
&&\rho_{0}(I)=i\tau_z\sigma_y,\ \
\rho_{0}(T)K=\tau_x\sigma_x K.
\eeq
What is special about this fourfold degeneracy is that the band splitting around this point is proportional to $k^{5}$, unprecedented in previously studied symmetry-protected band nodes:
\begin{equation}\label{quinticdisp}
{E}(\mathbf{{\mathbf{k}}})=|k_xk_yk_z|\sqrt{3(k_x^2-k_y^2)^2+(2k_z^2-k_x^2-k_y^2)^2}+O(k^6).
\end{equation}
Furthermore, as shown in Fig.\ref{fig:4}, all high-symmetry directions around $\mathrm{R}$ are Dirac nodal lines, forming a 13-nodal-line-nexus where all lines remain fourfold degenerate.
Since there are no 4-fold fermions with symmetry group $O_{h}\times Z_{2}^{T}$ having quintic band splitting dispersion in MSGs, our 4-fold fermion \eqref{4dfermion1} with 13-nodal-line-nexus only realizable in SSGs is qualitatively different from 4-fold fermions with symmetry group $O_{h}\times Z_{2}^{T}$ realized in MSGs.

\begin{figure*}
\begin{centering}
\includegraphics[width=0.7\textwidth]{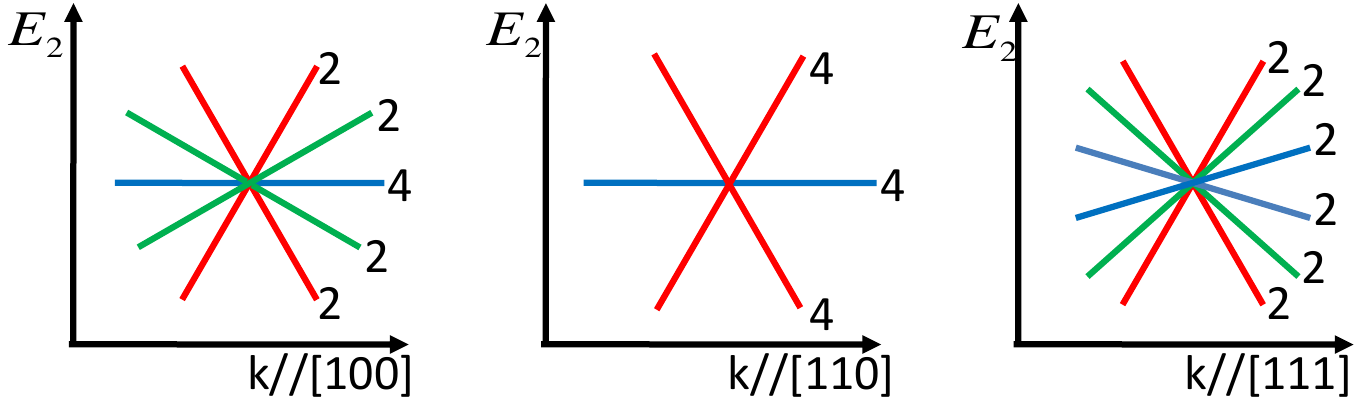}
\par\end{centering}
\protect
\caption{\label{fig:5}12-fold fermion in spin-space group 229.2.2.36 \cite{ssg_website}.
The schematic dispersion of the 12-fold fermion along three high-symmetry lines $[100]$, $[110]$, $[111]$.
Along $[100]$, the 12-fold fermion splits into one 4-fold band colored in blue,
two 2-fold bands colored in red and two 2-fold bands colored in green.
Along $[110]$, the 12-fold fermion splits into one 4-fold band colored in blue and
two 4-fold bands colored in red.
Along $[111]$, the 12-fold fermion splits into two 2-fold bands colored in red,
two 2-fold bands colored in green and two 2-fold bands colored in blue.}
\end{figure*}

Aside from the 12-fold fermion \eqref{12dfermion1}, another 12-fold fermion
can also only appear in SSGs with different values of invariants.
For instance, at the R point,
the SSG
\href{https://cmpdc.iphy.ac.cn/ssg/ssgs/229.2.2.36}{229.2.2.36}
has the invariants $(+1,-1,-1,+1,-1)$
for the little co-group $O_{h}\times Z_{2}^{T}$,
which differ from the invariants of (\ref{12dfermion1})
only by $\eta_T$ and $\eta_{TC_{2a}}$.
The corresponding 12-dimensional irRep reads,
\beq\label{12dfermion2}
&&\rho_{2}(C_{2i})=\tau_{0}\sigma_0e^{-iL_{i}\pi},\ \
\rho_{2}(C_{2\mathbf{n}})=i\tau_{0}\sigma_y e^{-i\mathbf{L}\cdot\mathbf{n}\pi}, \nonumber\\
&&\rho_{2}(I) = \tau_{0}\sigma_z 1_{3\times 3},\ \
\rho_{2}(T)K=i\tau_{y}\sigma_0 1_{3\times 3}K.
\eeq
$\rho_{2}\otimes \rho_{2}^{\ast}$
can be decomposed into linear irReps of
$O_{h}\times Z_{2}^T$:
\beq\label{12dcg2}
\rho_{2}\otimes \rho^{\ast}_{2}
&=&
\cdots \oplus 6T_{1u}^{-}\oplus \cdots.
\eeq
The 12-fold fermion also has linear dispersion $\mu=T_{1u}^-$ with the components $\phi^{\mu}_{1}=k_{x}$, $\phi^{\mu}_{2}=k_{y}$,
$\phi^{\mu}_{3}=k_{z}$.
Since multiplicity $N_{\rho_{2}}(T_{1u}^-)=6$, there are 6 coupling parameters
$\lambda^{(\mu)}_{1}=v_{2x}$,
$\lambda^{(\mu)}_{2}=v_{2y}$,
$\lambda^{(\mu)}_{3}=v_{2z}$,
$\lambda^{(\mu)}_{4}=v'_{2x}$,
$\lambda^{(\mu)}_{5}=v'_{2y}$,
$\lambda^{(\mu)}_{6}=v'_{2z}$.
According to Eq.(\ref{eq:20}),
the linear dispersion of the 12-fold fermion takes the form
\beq
\hat{H}_{2}(\mathbf{k})&=&
(
v_{2x}\tau_{x}\sigma_{y}+
v_{2y}\tau_{y}\sigma_{y}+
v_{2z}\tau_{z}\sigma_{y}
)
\mathbf{k}\cdot\mathbf{L}+
\nonumber
\\
&&
(
v'_{2x}\tau_{x}\sigma_{x}+
v'_{2y}\tau_{y}\sigma_{x}+
v'_{2z}\tau_{z}\sigma_{x}
)
\mathbf{k}\cdot\mathbf{L}',
\eeq
where $L'_i\equiv|\epsilon_{ijk}|\{L_j,L_k\}$.
The linear dispersion of this 12-fold fermion along high-symmetry lines is given in Fig.\ref{fig:5}.
Compared with Fig.\ref{fig:3}(b), the two 12-fold fermions have different responses to linear dispersion along the high-symmetry line $[111]$, which result from $N_{\rho_{1}}(T_{1u}^{-})\neq N_{\rho_{2}}(T_{1u}^{-})$ in Eqs.\eqref{12dcg1} and \eqref{12dcg2}.
These qualitative differences distinguish the two 12-fold fermions \eqref{12dfermion1} and \eqref{12dfermion2} coming from two different projective classes of $O_{h}\times Z_{2}^{T}$ with different sets of symmetry invariants.

Actually, if the little co-group is $G_\mathbf{K}=O_{h}\times{Z}_2^T\times\mathrm{SU(2)}$, instead of $O_{h}\times{Z}_2^T$, a 12-dimensional irRep may also appear in type-II MSGs without SOC. The 12-fold fermion can be realized at R point of type II MSGs 222.99, 223.105 (or H point of 230.146)\cite{tang2022complete}. In fact, there is a 6-dim irRep of $O_{h}\times Z_{2}^{T}$ in the
projective class denoted by
invariants $\vec{\eta}=(+1,+1,+1,+1,-1)$
\beq\label{12dfermion3}
\rho_{3}(C_{2i})&=&\tau_{0}\exp(-iL_{i}\pi),
\rho_{3}(C_{2\mathbf{n}})=\tau_{z}
\exp(-i\mathbf{L}\cdot\mathbf{n}\pi),
\nonumber\\
\rho_{3}(I)&=&\tau_{x}1_{3\times 3},
\rho_{3}(T)K=\tau_{0}1_{3\times 3}K.
\eeq
With spin degrees of freedom, symmetry group is $O_{h}\times Z_{2}^{T}\times\mathrm{SU(2)}$, the band degeneracies at high-symmetry points are 12-fold.

\begin{figure}
\centering
\includegraphics[width=0.9\linewidth]{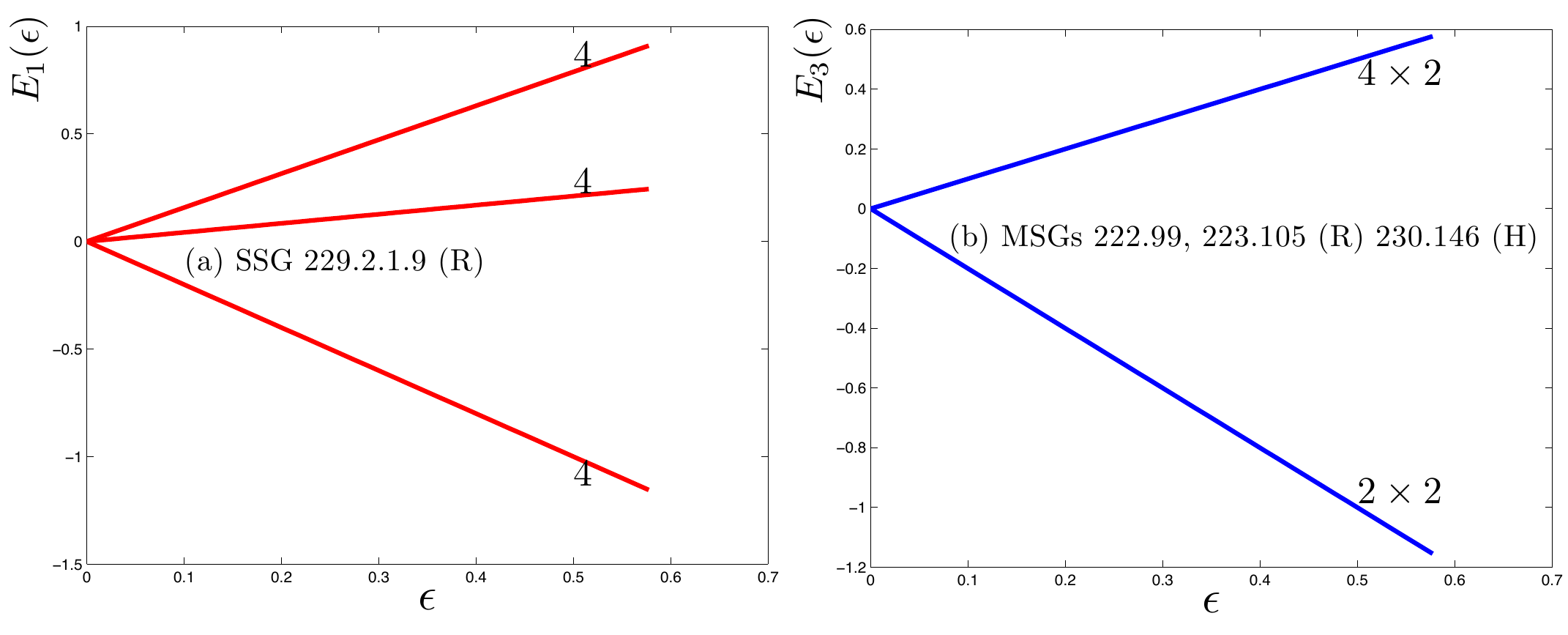}
\caption{Spectra of two 12-fold fermions under a strain tensor field $(\epsilon_{yz},\epsilon_{xz},\epsilon_{xy})$.
(a)
12-fold fermion
with symmetry group $O_h\times{Z}_2^T$ only realizable in SSGs, coming from
projective class $\rho_1$ in Eq.\eqref{12dfermion1}. Under coupling (\ref{qp1h}), the 12-fold fermion splits into three 4-fold bands colored in red.
(b)
symmetry group is $O_h\times{Z}_2^T\times\mathrm{SU}(2)$,
a 6-dimensional irRep \eqref{12dfermion3} of $O_{h}\times Z_{2}^{T}$ splits into 4-fold and 2-fold bands under coupling (\ref{qp2h}).
Due to extra $\mathrm{SU}(2)$ spin symmetry, the 12-fold fermion splits into
$4\times 2$-fold and $2\times 2$-fold bands colored in blue. }
\label{fig:6}
\end{figure}

We compare the two 12-fold fermions coming from two different projective classes \eqref{12dfermion1} and \eqref{12dfermion3}.
Following the general argument, $\rho_{1}\otimes \rho^{\ast}_{1}$ and $\rho_{3}\otimes \rho^{\ast}_{3}$ are decomposed into linear irReps of $O_{h}\times Z_{2}^{T}$:
\beq
\rho_{1}\otimes \rho^{\ast}_{1}
&=&
\cdots \oplus  2T_{2g}^{+}\oplus \cdots,
\label{12dcg}
\\
\rho_{3}\otimes \rho^{\ast}_{3}
&=&
\cdots \oplus  T_{2g}^{+}\oplus \cdots .
\label{6dcg}
\eeq
In Eqs.\eqref{12dcg},\eqref{6dcg},
$N_{\rho_{1}}(T_{2g}^{+})\neq N_{\rho_{3}}(T_{2g}^{+})$,
so the two quasiparticles have different spectra under an external $\mu=T_{2g}^{+}$-field ($+$ means even under time reversal), which can be a strain tensor field $(\epsilon_{yz},\epsilon_{xz},\epsilon_{xy})$ with the components $\phi^{\mu}_{1}=\epsilon_{yz}$, $\phi^{\mu}_{2}=\epsilon_{xz}$, $\phi^{\mu}_{3}=\epsilon_{xy}$.
To be more specific, according to Eq.\eqref{eq:20}, for projective class $\rho_{1}$ \eqref{12dfermion1} the
coupling takes the form
\beq\label{qp1h}
\hat{H}_{1}
&=&
v_{1}\Big( \epsilon_{yz}\tau_{0}\sigma_{0}\{L_{y},L_{z}\}+\epsilon_{zx}\tau_{0}\sigma_{0}\{L_{z},L_{x}\}
\nonumber\\
&&
+\epsilon_{xy}\tau_{0}\sigma_{0}\{L_{x},L_{y}\} \Big)+
v_{2}\Big( \epsilon_{yz}\tau_{x}\sigma_{0}L_{x}
\nonumber\\
&&
+\epsilon_{zx}\tau_{x}\sigma_{0}L_{y}
+\epsilon_{xy}\tau_{x}\sigma_{0}L_{z} \Big),
\eeq
for projective class $\rho_{3}$ \eqref{12dfermion3} we have
\beq\label{qp2h}
\hat{H}_{3}
&=&
v_{3}\Big( \epsilon_{yz}\tau_{0}\{L_{y},L_{z}\}+\epsilon_{zx}\tau_{0}\{L_{z},L_{x}\}
\nonumber\\
&&
+\epsilon_{xy}\tau_{0}\{L_{x},L_{y}\} \Big).
\eeq
We plot the spectrum of band splitting of $\hat{H}_{1}$ (colored in red) and $\hat{H}_{3}$ (colored in blue) under ($\epsilon_{yz}$,$\epsilon_{xz}$,$\epsilon_{xy}$)$=$ ($\epsilon$,$\epsilon$,$\epsilon$) for $v_{1}=v_{3}=1$, $v_{2}=\frac{1}{3}$ and $\epsilon\in[0,\frac{\sqrt{3}}{3}]$.
From Fig.\ref{fig:6}, we see that the two 12-fold fermions are qualitatively different under an external strain tensor field.

\section{Material candidates}

It would be helpful for experimentalists if we can list a few candidate materials where any one of the 218 SSG-only fermions may be discovered.
In fact, a comprehensive search is possible using (i) Supplementary Tables~\ref{C4hdrztspin}-\ref{Ohdrztinv} of this work, (ii) the database MAGNDATA where over 2000 experimentally discovered magnetic structures are registered, and (iii) the enumeration of all SSGs in Ref.\cite{jiang2023enumeration}.
First, we sort out all realistic magnetic structures having non-coplanar magnetic structures, which amount to 206 in total.
Next, for any one of them, we ask if the hosting SSG is isomorphic to a type-IV MSG.
Then, we ask if the point-group part of the SSG is any one of the following: $P\times Z_{2}^{T}$ with $P=C_{3h,4h,6h},D_{3,4,6,3d,3h,4h,6h},O,O_h$.
Finally, we compute the symmetry invariants contributed by the spin operations.
If any SSG contains the invariants marked as RED in Supplementary Tables~\ref{C4hdrztspin}-\ref{Ohdrztspin}, the corresponding material hosts an SSG-only fermion according to our theory.
A comprehensive search following the above protocol has been performed, yielding one material: Ce$_3$NIn.
The SSG-only co-representation in this material is discussed in Supplementary Note \ref{sec:corepce3nin}.
Interestingly, during the revision process of this work, this same material has been proposed as an unconventional p-wave magnet\cite{ce3inn2023}.
The scarcity of candidate materials is related to the fact that only a small fraction of magnetic materials have their magnetic structures detected.
There are only 2000 materials with known magnetic structures, in contrast to the number of 200000 materials with known crystal structures.

\section{Discussion}

We use symmetry invariants to label projective classes of $P\times Z_{2}^{T}$ and find 218 new types of quasiparticles only realized in the electronic bands of non-coplanar ($S_{0}=\{E\}$) SSGs whose magnetic unit cells are only 2 times the size of the crystal unit cells.
The non-coplanar SSGs are isomorphic to type IV MSGs and the little co-groups at high-symmetry points are of the form $G_{\mathbf{K}}=P\times Z_{2}^{T}$.
For each projective class of $P\times Z_{2}^{T}$, the irRep $\rho$ and the Clebsch-Gordon coefficients of the reducible Rep $\rho\otimes\rho^{\ast}$ are obtained by the eigenfunction method \cite{YangLiu, YangYFL_CG2021}.

Our procedure can be extended to study Rep theory of other types of SSGs\cite{jiang2023enumeration, ssg_website}, including the non-coplanar SSGs isomorphic to type I or type III MSGs, and the SSGs with nontrivial spin-only groups $S_0$, namely the collinear and coplanar SSGs.
In the following, we comment on these two cases separately.

The type I MSGs are 230 SGs, and the little co-groups at high-symmetry points are 32 crystallographic point groups.
Since all the projective classes of the 32 point groups have already been realized in the 230 SGs, one cannot obtain any new symmetry invariants (new quasiparticle types).

For the non-coplanar SSGs isomorphic to type III MSGs, the little co-groups $G_{\mathbf{K}}$ at high-symmetry point $\mathbf{K}$ can be 58 type III magnetic point groups of the form $H+T(P-H)$, where $H$ is a subgroup of the point group $P$ containing half of the elements in $P$, the other half $P-H$ followed by time reversal $T$.
There are in total 380 projective classes for the 58 type III magnetic point groups. From Eqs.\eqref{eq:chiLp1p2}, \eqref{eq:chiLpt}, \eqref{eq:chiSp1p2} and \eqref{eq:chiSpt}, we calculate lattice-part invariants $\eta^L$ and spin-part invariants $\eta^S$, then obtain the full invariants $\eta=\eta^L\eta^S$.
Using the symmetry invariants, we have confirmed that 260 classes are realizable in type III MSGs or at non-time-reversal-invariant high-symmetry points of type II and IV MSGs.
There are 120 sets of invariants that cannot be realized in MSGs.
In Supplementary Table~\ref{tab:invrtstypeIII}, we give definitions of symmetry invariants of 58 type III magnetic point groups.
If the symmetry invariants of a group cannot be completely realized in MSGs, the values of symmetry invariants are marked in BLUE if they are realizable via MSGs, and are marked in RED if they are realizable by SSGs only.
The BLACK sets of invariants cannot be realized by either MSGs or SSGs.
We find additional 108 classes of new quasiparticles with type III magnetic point group symmetries only realizable in the electronic bands of SSGs.
If the minimal dimension of irReps of the same class is FOUR or higher, we provide the degree of degeneracy, the lowest-order dispersion, and the direction of nodal lines.
An example is a four-fold degenerate 7-nodal-line-nexus fermion protected by $T_{d}\times Z_2^{IT}$ (No.57 type III magnetic point group in Supplementary Table~\ref{tab:invrtstypeIII}), only realizable in SSGs, with three nodal lines along $k_x, k_y, k_z$ directions and four along $k_x+k_y+k_z$, $k_x+k_y-k_z$, $k_x-k_y+k_z$, and $k_x-k_y-k_z$ directions.

For coplanar SSGs, the spin only group is $S_0=Z_2^{C_{2z}T}$ $=\{E,C_{2z}T\}$, and for collinear SSGs, the spin only group is $S_0=\mathrm{SO}(2)\rtimes Z_2^{C_{2x}T}$ with $Z_2^{C_{2x}T}=\{E,C_{2x}T\}$.
The nontrivial $S_0$ will generally give rise to extra symmetry invariants and may also affect the dimensions of irReps (see Supplementary Note \ref{sec:invcpclssg} for more discussions).
Due to $S_0$ and the fact that the magnetic unit cell of a general SSG is enlarged comparing to the original crystal unit cell, the little co-group of SSG at a high-symmetry point $\mathbf{K}$ is generally not isomorphic to $P\times Z_{2}^{T}$ or type III magnetic point groups.
Our approach also applies to these SSGs.

\normalem

%

\onecolumngrid

\begin{center}
\textbf{\center{\large{Supplementary Information}}}
\end{center}

\beginsupplement

\section{Projective Reps and 2$^{\rm nd}$ group cohomology}

For a projective Rep of a finite group $G$, an element  $g\in G$ is represented by $\rho(g)$ if $g$ is a unitary element and is represented by $\rho(g)K$ if $g$ is anti-unitary, where $K$ is the complex-conjugate operator satisfying $KU=U^*K$ with $U$ an arbitrary matrix and $U^*$ its complex conjugation. 

The multiplication of (projective) Reps of $g_1, g_2$ depends on if they are unitary or anti-unitary.
If we define a unitarity indicator $\zeta_g$
\beq
\zeta_g=\left\{
\begin{aligned}
&0,& &{\ \rm if} \ g \ {\rm is\ unitary,\ \ \ }   \\
&1,& &{\ \rm if}\ g \ {\rm is\
\text{anti-unitary},
\ \ \ }
\end{aligned}
\right.
\eeq
and the corresponding operator $K^{\zeta_g}$
\beq
K^{\zeta_g}=\left\{
\begin{aligned}
&{\rm identity \ operator},& &{\ \rm if\ } \zeta_g=0,\  \  \\
&K,& &{\ \rm if\ }\zeta_g=1,
\end{aligned}
\right.
\eeq
then we have the multiplication rule of a projective Rep,
\beq
\rho(g_1)K^{\zeta_{g_1}}\rho(g_2)K^{\zeta_{g_2}} = \rho(g_1g_2)e^{i\theta_2(g_1,g_2)}K^{\zeta_{g_1g_2}},
\eeq
where the $\mathrm{U(1)}$ phase factor $\omega_2(g_1,g_2) \equiv e^{i\theta_2(g_1,g_2)}$ is a function of two group variables and is called the factor system. If $\omega_2(g_1,g_2) = 1$ for any $g_1,g_2\in G$, then above projective Rep becomes a linear Rep.

Substituting above results into the associativity relation of the sequence of operations $g_1\times g_2\times g_3$, we can obtain
\beq
&&\rho(g_1)K^{\zeta_{g_1}}\rho(g_2)K^{\zeta_{g_2}}\rho(g_3)K^{\zeta_{g_3}}
\nonumber\\
&=& \rho(g_1g_2g_3)\omega_2(g_1,g_2)\omega_2(g_1g_2,g_3) K^{\zeta_{g_1g_2g_3}}
\nonumber
\\
&=& \rho(g_1g_2g_3)\omega_2(g_1,g_2g_3)\omega_2^{(-1)^{\zeta_{g_1}}}(g_2,g_3)K^{\zeta_{g_1g_2g_3}},
\nonumber
\\
\eeq
namely,
\begin{equation}\label{2cocyl}
\omega_2(g_1,g_2)\omega_2(g_1g_2,g_3) = \omega_2^{(-1)^{\zeta_{g_1}}}(g_2,g_3)\omega_2(g_1,g_2g_3).
\end{equation}
Supplementary Equation
(SEq.)(\ref{2cocyl}) is the general relation that the factor systems of any finite group (no matter unitary or anti-unitary) should satisfy. The solution of the above equations are called 2-cocyles. If we introduce a gauge transformation $\rho'(g)K^{\zeta_g}=\rho(g)\Omega_1(g)K^{\zeta_g}$, where the phase factor $\Omega_1(g)=e^{i\theta_1(g)}$ depends on a single group variable, then the factor system changes into
\beq\label{gaugeomega2}
\omega'_2(g_1,g_2)=\omega_2(g_1,g_2)\Omega_2(g_1,g_2),
\eeq
with
\beq\label{2cob2}
\Omega_2(g_1,g_2) = {\Omega_1(g_1)\Omega_1^{(-1)^{\zeta_{g_1}}}(g_2)\over \Omega_1(g_1g_2)}.
\eeq
The quantity $\Omega_2(g_1,g_2)$ defined above are called 2-coboundaries. The equivalent relations (\ref{gaugeomega2}) and (\ref{2cob2}) define the equivalent classes of the solutions of (\ref{2cocyl}). The number of equivalent classes for a finite group is usually finite.

Usually, the so-called  standard gauge choice is adopted, where $\omega_2(E,g)=\omega_2(g,E)=1$ for any $g\in G$. This is equivalent to require that $E$ is always represented as the identity matrix. 

Two 2-cocycles $\omega_2'(g_1,g_2)$ and $\omega_2(g_1,g_2)$ are equivalent if they differ by a 2-coboundary, see Eq.(\ref{gaugeomega2}). The equivalent classes of the 2-cocycles $\omega_2(g_1,g_2)$ form the second group cohomology $\mathcal H^2(G, \mathrm{U(1)})$.

Writing  $\omega_{2}(g_{1},g_{2})=e^{i\theta_{2}(g_{1},g_{2})}$, where $\theta_{2}(g_{1},g_{2})\in[0,2\pi)$, then the cocycle Eqs.(\ref{2cocyl})
can be written in terms of linear equations,
\begin{eqnarray}
&&(-1)^{\zeta_{g_1}}\theta_{2}(g_{2},g_{3})-\theta_{2}(g_{1}g_{2},g_{3})+\theta_{2}(g_{1},g_{2}g_{3})\nonumber\\
&&  -\theta_{2}(g_{1},g_{2})=0.
\label{2cocycle}
\end{eqnarray}
Similarly, if we write $\Omega_1(g_1)=e^{i\theta_1(g_1)}$ and $\Omega_2(g_1,g_2)=e^{i\Theta_2(g_1,g_2)}$, then the 2-coboundary (\ref{2cob2})
can be written as
\beq
\Theta_{2}(g_{1},g_{2})= (-1)^{\zeta_{g_1}}\theta_{1}(g_{2})-\theta_{1}(g_{1}g_{2})+\theta_{1}(g_{1}) .
  \label{2coboundary}
\end{eqnarray}
The equal sign in Eqs.(\ref{2cocycle}) and (\ref{2coboundary}) means equal mod $2\pi$. From these linear equations, we can obtain the solution space of the cocycle equations, as well as the classes that the solutions belong to. The set of classes form a finite Abelian group, which labels the classification of the projective Reps (the ``projective classes'' of $G$).

\section{Symmetry Invariants}
The second group-cohomology group is generated by a set of invariants, dubbed projective symmetry invariants, or symmetry invariants for short. The invariants are special combinations of the cocycles variables $\omega_{2}(g_{1},g_{2})$. For any cocycle solutions, the invariants are, by definition, invariant under the gauge transformation (\ref{gaugeomega2}) and (\ref{2cob2}), and their values must be equal to roots of 1. Therefore, the values of the invariants are fixed for a given class of factor systems $\omega_{2}(g_{1},g_{2})$. On the other hand, the values of the complete set of symmetry invariants uniquely determine the projective class to  which a factor system (and the corresponding projective Reps) belongs.

For { crystallographic point groups}, all the invariants take $\mathbb Z_2$ values, namely $\pm1$. Once the values for the complete set of symmetry invariants are given, the factor systems in the corresponding projective class can be easily obtained (up to gauge transformations). In the following we give three examples.

(I) The unitary group $D_2=\{E,C_{2x}\}\times \{E,C_{2y}\}$.  For the unitary Abelian group $D_2=\{E,C_{2x}\}\times \{E,C_{2y}\}=\{E, C_{2x}, C_{2y}, C_{2z}\}$, the classification of 2-cocycles is
\beq
\mathcal H^2(D_2, \mathrm{U(1)})=\mathbb Z_2,
\eeq
where the 2-cocycle equations can be simplified into $\big({\omega_2(C_{2x},C_{2y})\over \omega_2(C_{2y},C_{2x})}\big)^2=1$ plus some coboundary relations. Therefore, there  is only one independent  symmetry invariant, namely,
\beq
\eta_{C_{2x},C_{2y}} = {\omega_2(C_{2x},C_{2y})\over \omega_2(C_{2y},C_{2x})}=\pm1.
\eeq

The projective class with invariant
$\eta_{C_{2x},C_{2y}}=-1$ is nontrivial, which indicates that the projective Rep matrices of $C_{2x}$ and $C_{2y}$ are anti-commuting, $\rho(C_{2x})\rho(C_{2y})=-\rho(C_{2y})\rho(C_{2x})$ although $C_{2x}$ and $C_{2y}$ are commuting as group elements of $D_2$. As an example, the factor system in this class can be chosen as: $\omega_2(C_{2y}, C_{2x}) = \omega_2(C_{2y}, C_{2z}) = \omega_2(C_{2z}, C_{2x}) = \omega_2(C_{2z}, C_{2z})=-1$, and all the other components are equal to 1. The projective irReps belonging to this class are 2-dimensional.

On the other hand,  the projective class with invariant $\eta_{C_{2x},C_{2y}}=1$ is trivial. A typical factor system in this class is the one with all components equal to 1. In the trivial class, all the irReps, including the linear Reps as a special case, are 1-dimensional.

It should be mentioned that the choices of symmetry invariants may not be unique. For the $D_2$ group, since the relation $\eta_{C_{2x},C_{2y}}=\eta_{C_{2x},C_{2z}}=\eta_{C_{2y},C_{2z}}$ is always valid, any one of $\eta_{C_{2x},C_{2y}}, \eta_{C_{2x},C_{2z}}, \eta_{C_{2y},C_{2z}}$
can be chosen as the symmetry invariant.

(II) The anti-unitary group $Z_2^T= \{E,{T}\}$.  The simplest anti-unitary group is {the time-reversal group} $Z_2^T= \{E,{T}\}$, the classification of its 2-cocycles is
\beq
\mathcal H^2(Z_2^T, \mathrm{U(1)})=\mathbb Z_2,
\eeq
where the 2-cocycle equations can be simplified into $\big( \omega_2({{T}},{T}) \big)^2=1$ under the standard gauge. Therefore, the symmetry invariant is given by
\beq
\eta_T=\omega_2({{T}},{T})=\pm1.
\eeq
Here projective class with $\eta_T=-1$ stands for the Kramers class which guarantees the double-degeneracy, while projective class with $\eta_T=1$ stands for the trivial class where no degeneracy is guaranteed.

(III) The anti-unitary group $D_2\times Z_2^T$.  Now we consider the group $D_2\times Z_2^T$, the classification of 2-cocycles is
\beq
\mathcal H^2(D_2\times  Z_2^T, \mathrm{U(1)})=\mathbb Z_2^4.
\eeq
Under the standard gauge, the 2-cocycle equations can be simplified into $\big({\omega_2(C_{2x},C_{2y})\over \omega_2(C_{2y},C_{2x})}\Big)^2=1, \big( \omega_2({{T}},{T}) \big)^2=1, \big( \omega_2(TC_{2x},TC_{2x}) \big)^2=1, \big( \omega_2(TC_{2y},TC_{2y}) \big)^2=1$ plus some coboundary relations. Therefore, the resultant symmetry invariants are
\beq
&& \eta_{C_{2x},C_{2y}} = {\omega_2(C_{2x},C_{2y})\over \omega_2(C_{2y},C_{2x})}=\pm1,
\nonumber\\
&& \eta_T=\omega_2({{T}},{T})=\pm1,
\nonumber\\
&& \eta_{TC_{2x}}=\omega_2( TC_{2x},TC_{2x})=\pm1,
\nonumber\\
&& \eta_{TC_{2y}}=\omega_2( TC_{2y},TC_{2y})=\pm1.
\eeq
There are totally 16 projective classes, 15 of them are nontrivial and the rest one, the trivial class with all the invariants equal to 1, is gauge equivalent to linear Reps.

\section{Supplementary Table~\ref{tab:invrts}: Symmetry invariants of anti-unitary groups $P\times Z_{2}^{T}$}\label{regularProj}

{\bf Symmetry invariants for groups with structure $G=P\times Z_{2}^{T}$.}
In Supplementary Table~\ref{tab:invrts}, we list a complete set of symmetry invariants for groups with structure $G= P\times Z^T_{2}$, where $P$ is any one of the 32 crystallographic point groups and  $Z_{2}^{T}$ is the time-reversal group $Z_{2}^{T}=\{E,T\},\ T^2=E$.
The symmetry invariants generate a group ${Z}_2^m$, giving rise to $2^m$ distinct classes of quasiparticles. Each class corresponds to a unique $m$-component $\pm1$-valued vector $\vec\eta=(\eta_1,\dots,\eta_m)$  or a Boolean vector $(1-\vec\eta)/2$.

{\it Notation:} $C_{2m}, C_{4m}^{\pm}$ with $m=x,y,z$ label the 2-fold or 4-fold axis along $\hat x, \hat y, \hat z$ respectively; $C_{3j}^{\pm}$ with $j=1,2,3,4$ label four different 3-fold axes along
$\hat x + \hat y + \hat z,\hat x + \hat y - \hat z,-\hat x + \hat y + \hat z,\hat x - \hat y + \hat z$, respectively; $C_{2}$
labels 2-fold axis along $\hat z$,
$C'_{2i}$ with $i=1,2,3$ label three different 2-fold axes $\hat x,\hat x -\sqrt{3}\hat y, \hat x +\sqrt{3}\hat y$ respectively,
$C''_{2i}$ with $i=1,2,3$ label three different 2-fold axes $\hat y,\sqrt{3}\hat x+\hat y,\sqrt{3}\hat x-\hat y$ respectively;
$C_{2p}$ with $p=a,b,c,d,e,f$ label six different 2-fold axes along $\hat x +\hat y,\hat x-\hat y, \hat x +\hat z,
\hat y +\hat z, \hat x -\hat z, \hat y -\hat z$, respectively. These notations can be found in Fig.1.1, Fig.1.2 and Fig.1.3 of
Supplementary Reference (SRef.)\cite{Bradley2010}. Furthermore, $I$ stands for spacial  inversion, $M$ denotes  planar mirror reflection. Some combined operations containing the spacial inversion $I$ are replaced by brief notations: $IC_{2}=M_{h}$, $IC_{2m}=M_{m}$, $IC'_{2i}=M_{di}$, $IC''_{2i}=M_{vi}$, $IC_{2p}=M_{dp}$, $IC_{3}^{\pm}=S_{6}^{\mp}$, $IC_{4m}^{\pm}=S_{4m}^{\mp}$, $IC_{6}^{\pm}=S_{3}^{\mp}$, $IC_{3j}^{\pm}=S_{6j}^{\mp}$.

{\bf Obtaining projective irreducible co-representations (irReps).} For a given factor system $\omega_2(g_1,g_2)$, all the irReps of $G$ can be obtained by reducing the regular projective Rep using the eigenfunction method. The details of the method are provided in
SRefs.\cite{YangLiu,YangYFL_CG2021} and will not be repeated here. The complete results of irReps of some anti-unitary groups for given factor systems are provided in SRef.\cite{Yang2021}.

{\bf Realization of symmetry invariants via double valued Reps for Magnetic point groups and Spin point groups.}
For a spin-space group $\mathcal{G}$,
the little co-group $G_\mathbf{K}$ at momentum $\mathbf{K}$ of Brillouin zone (BZ)
is a spin point group \cite{Litvin1977,schiff2023spin}.
Since we study spin-space groups isomorphic to magnetic space groups, we have $G_{\mathbf{K}}\cong M_{\mathbf{K}}$,
where the magnetic point group $M_{\mathbf{K}}$ is
the little co-group of isomorphic magnetic space group $\mathcal{M}$.

Recall that a projective irRep of a group $G$ is also a linear irRep of the cover group of $G$.
A group $G$ may have more than one classes of cover groups,
therefore the group $G$ has more than one classes of projective irReps.
As an example, every magnetic point group $M_{\mathbf{K}}$ has a familiar cover group-the double point group
$\tilde{M}_{\mathbf{K}}$. Some linear irReps of $\tilde{M}_{\mathbf{K}}$,
called the double valued irReps of $M_{\mathbf{K}}$, are projective irReps of $M_{\mathbf{K}}$.
In the double point group $\tilde{M}_{\mathbf{K}}$, each lattice operation
is associated with the corresponding spin operation as a result of spin-orbit coupling.
However, for a spin point group $G_{\mathbf{K}}\cong M_{\mathbf{K}}$, the double group of
$G_{\mathbf{K}}$ (noted as $\tilde{G}_{\mathbf{K}}$) is not necessarily the same with the double group
$\tilde{M}_{\mathbf{K}}$, because in $\tilde{G}_{\mathbf{K}}$ a lattice operation
is not necessarily associated with the corresponding spin operation.
As a result, the double valued irReps of a spin point group $G_{\mathbf{K}}$
are not necessarily the same with the double valued irReps of a
magnetic point group $M_{\mathbf{K}}$ even if $G_{\mathbf{K}}\cong M_{\mathbf{K}}$.

In the following we will illustrate the above conclusion via a concrete example-the
magnetic point group $M_{\mathbf{K}}=D_{2}\times Z_{2}^{T}$. This group
has 4 $1$-dimensional linear irReps,
and 15 classes of nontrivial projective irReps
(the linear Reps belong to the trivial class)
characterized by the four $\pm 1$-valued symmetry invariants
($\eta_{C_{2x},C_{2y}}$,$\eta_{T}$,$\eta_{TC_{2x}}$,$\eta_{TC_{2y}}$).
Since each lattice operation
is associated with the corresponding spin operation: $\varphi_E=E$, $\varphi_{C_{2x}}=C_{2x}$, $\varphi_{C_{2y}}=C_{2y}$, $\varphi_{C_{2z}}=C_{2z}$, $\varphi_T=T$,
the double valued irRep of the
magnetic point group $M_{\mathbf{K}}$
\beq
d(\varphi_{T})K=-i\sigma_{y}K,
d(\varphi_{C_{2x}})=-i\sigma_{x},
d(\varphi_{C_{2y}})=-i\sigma_{y},
d(\varphi_{C_{2z}})=-i\sigma_{z}
\eeq
belongs to a nontrivial class of the projective irReps whose symmetry invariants take the following values
$\eta_{C_{2x},C_{2y}}=[d(\varphi_{C_{2x}})d(\varphi_{C_{2y}})][d(\varphi_{C_{2y}})d(\varphi_{C_{2x}})]^{-1}=-1$, $\eta_{T}=[d(\varphi_{T})K]^{2}=-1$,
$\eta_{TC_{2x}}=[d(\varphi_{TC_{2x}})K]^{2}=1$,
$\eta_{TC_{2y}}=[d(\varphi_{TC_{2y}})K]^{2}=1$
from Eqs.\eqref{eq:chiSp1p2},\eqref{eq:chiSpt} of the main text.

Now we consider a spin point group
$G_{\mathbf{K}}\cong M_{\mathbf{K}}=D_{2}\times Z_{2}^{T}$ where the elements $E$ and $C_{2x}$
are associated with trivial spin operation $\varphi_{E,C_{2x}}=E$, while
$C_{2y}$ and $C_{2z}$ are associated with the spin rotation $\exp(-i\frac{\sigma_{z}}{2}\pi)$,
i.e. $\varphi_{C_{2y},C_{2z}}=C_{2z}$. Then
the double valued irRep of $G_{\mathbf{K}}$
\beq
d(\varphi_{T})K=-i\sigma_{y}K,
d(\varphi_{C_{2x}})=-1_{2\times 2},
d(\varphi_{C_{2y}})=-i\sigma_{z},
d(\varphi_{C_{2z}})=-i\sigma_{z}
\eeq
belongs to another nontrivial class of the projective irReps of $G_{\mathbf{K}}\cong M_{\mathbf{K}}$
with invariants $\eta_{C_{2x},C_{2y}}=1$, $\eta_{T}=-1$, $\eta_{TC_{2x}}=-1$, $\eta_{TC_{2y}}=1$
from Eqs.\eqref{eq:chiSp1p2},\eqref{eq:chiSpt} of the main text.
Hence, the difference between spin point groups and magnetic point groups
(spin-space groups and magnetic space groups) relies on the spin degrees of freedom.

Besides the spin degrees of freedom, the non-symmorphic fractional translations associated with the point group operations
can also contribute nontrivial factor system to the projective Rep of the little co-groups for $\mathbf{K}$ points at the BZ boundary. The final factor system is a product of the factor system contributed from the lattice part and the one from the spin part
(i.e. the double valued Rep). Hence, the concept of projective Rep is more powerful than the double valued Rep
since it can treat the spin degrees of freedom and the sub-lattice degrees of freedom in an unified framework.

\section{Supplementary Tables \ref{Pdrztlat} $\sim$ \ref{Ohdrztinv}: Symmetry invariants and quasiparticle dispersions in
Spin-Space Groups}\label{sec:invssg}

For point groups with $P = C_{1,s,i,2,3,4,6,2v,3v,4v,6v,2h,3i}, S_{4}, D_{2,2d,2h}, T, T_{d,h}$,  all the projective classes of $P\times{Z}_2^T$ can be realized within Shubnikov's magnetic space group (MSG) $\mathcal{M}$ with $P\times{Z}_2^T$ being the little co-group at certain high-symmetry points
at the boundary of BZ.
For $P=C_{3h,4h,6h},D_{3,4,6,3d,3h,4h,6h},O,O_{h}$, partial projective classes can only be realized in spin-space groups (SSGs)
\cite{Brinkman1966, Litvin1974, jiang2023enumeration, ssg_website, xiao2023spin, ren2023enumeration,
watanabe2023symmetry,shinohara2024algorithm}
where spin rotations are decoupled with lattice rotations. We will focus on SSGs that can realize more projective classes of $P\times{Z}^T_{2}$ which cannot be realized in MSG.
Especially, we study SSGs which are isomorphic to type IV MSGs.

All of the results are presented in forms of tables as attached in this supplementary information. In the following we summarize the contents and notations in these tables.



\begin{enumerate}
\item[$\bullet$] {\bf Supplementary Table~\ref{Pdrztlat}: The Lattice Invariants $\vec{\eta}^L$ of $G_\mathbf{K}\cong M_\mathbf{K}= P\times Z_{2}^{T}$.}\\
We first enumerate all the combinations $(\mathcal{M},\mathbf{K})$ having the little co-groups $M_\mathbf{K}={P}\times{Z}_2^T$ with $P$ being one of the groups $C_{3h,4h,6h},D_{3,4,6,3d,3h,4h,6h},O,O_{h}$. Then we explicitly give the lattice invariants $\vec{\eta}^L$ contributed solely from the lattice operations. These invariants are labeled by a Boolean vector $(1-\vec{\eta}^L)/2$ in the last column of Supplementary Table~\ref{Pdrztlat}.

{\it Notation:} We use $G_{\mathbf K}$, $M_{\mathbf K}$ and $L_{\mathbf K}$ to denote the little co-groups at ${\mathbf K}$ for the SSG $\mathcal G$, the MSG $\mathcal{M}$ and the lattice part $\mathcal{L}$ of MSG $\mathcal{M}$, respectively.
Here $\mathcal{M}$ is a type IV magnetic group, which has exactly the same lattice opereation as the SSG $\mathcal{G}$. It follows that $\mathcal{G}\cong\mathcal{M}\cong\mathcal{L}$ and $G_\mathbf{K}\cong{M}_\mathbf{K}\cong{L}_\mathbf{K}$,
and that the lattice invariants $\vec{\eta}^L$ of $G_{\mathbf K}$ are the same with those of $M_{\mathbf K}$.

\item[$\bullet$] {\bf Supplementary Tables~\ref{C4hdrztspin} $\sim$ \ref{Ohdrztspin}: The Spin Invariants $\vec{\eta}^S$ of $G_\mathbf{K}\cong M_\mathbf{K} = P\times Z_{2}^{T}$.}\\
The spin invariants $\vec{\eta}^{S}$ of $G_{\mathbf{K}}\cong M_{\mathbf K} = P\times Z_2^T$ are obtained in three steps. (1) Specify $L_{0\mathbf K} \lhd P$ with $L_{0\mathbf K}$ the group formed by pure lattice point-operations. To this end, we enumerate all the normal subgroups of $P$.  (2) Determine the quotient group $\varphi(M_{\mathbf K}) \cong M_{\mathbf K}/L_{0\mathbf K}$ constituted by spin operations. As an abstract group, $M_{\mathbf K}/L_{0\mathbf K}$ is  uniquely determined by $M_{\mathbf K}$ and $L_{0\mathbf K}$. But the concrete group $\varphi(M_{\mathbf K})$  and the method of the mapping $\varphi$ may have more than one choice. Each choice of $\varphi$ may give rise to distinct values of the spin invariants $\vec{\eta}^S$. (3) Finally, once the map $\varphi$ is determined, then the spin operation $\varphi_g$ associated with each lattice operation $g$ is determined, the values of the spin invariants can be easily calculated from the double valued Rep of $\varphi_g$.

The spin invariants for $G_\mathbf{K}\cong  P\times Z_{2}^{T}$ with $P=C_{3h,4h,6h},D_{3,4,6,3d,3h,4h,6h},O,O_h$ are listed in Supplementary Tables~\ref{C4hdrztspin} $\sim$ \ref{Ohdrztspin}. For easy reference, the $\vec{\eta}^{S}$ leading to invariants only realized in SSG are labeled by Boolean vectors $(1{-}\vec{\eta}^{S})/2$ colored in RED.

{\it Notation and remark:}  $L_{0\mathbf{K}}$ is actually the little co-group of the group $\mathcal L_0$. Here $\mathcal L_0$ is formed by the {\it pure} lattice operations (namely the spin operations associated with these lattice operations are trivial). Since $T$ has nontrivial operations on spins, all anti-unitary  operations have nontrivial spin part. Therefore, $\mathcal L_0$ must be a unitary group. An important property of $\mathcal L_0$ is that it is a normal subgroup of $\mathcal G$, namely $\mathcal L_0 \lhd \mathcal G\cong\mathcal{L}$. Since $\mathcal M$ is isomorphic to the lattice part of $\mathcal G$, it follows that $\mathcal L_0 \lhd\mathcal M\cong\mathcal{L}$  and that $L_{0\mathbf K} \lhd M_{\mathbf K}= P\times Z_{2}^{T}$. Remembering that $L_{0\mathbf K}$ is unitary, it follows that $L_{0\mathbf K} \lhd P$. In the present work, we do not explicitly enumerate $\mathcal L_0$. Instead, we just list all possible $L_{0\mathbf K}$ for given $P$ (see the first column of Supplementary Tables~\ref{C4hdrztspin} $\sim$ \ref{Ohdrztspin}).

\item[$\bullet$] {\bf Supplementary Tables~\ref{C4hdrztinv} $\sim$ \ref{Ohdrztinv}: The total Invariants $\vec{\eta}$  of $G_\mathbf{K}\cong M_\mathbf{K} = P\times Z_{2}^{T}$.}\\
The total invariants $\vec{\eta}$ are combinations of the lattice invariants and the spin invariants with components  $\eta_i = \eta^L_i\eta^S_i$. For $G_\mathbf{K}\cong  P\times Z_{2}^{T}$ with $P=C_{3h,4h,6h},D_{3,4,6,3d,3h,4h,6h},O,O_h$,  the total invariants  are listed in Supplementary Tables~\ref{C4hdrztinv} $\sim$ \ref{Ohdrztinv}. Those colored in GREEN are type II MSG-realizable, those in BLUE are type IV MSG-realizable, and those in RED are only SSG-realizable. The BLACK sets of invariants cannot be realized in MSG, or even SSG.

For every set of symmetry invariants that is only realizable in SSG colored in RED, the second column lists all possible pairs ($\vec{\eta}^{L}$,$\vec{\eta}^{S}$), the following three columns summarize all band degeneracies
(if minimal degree$\ge4$) , including the degree, the lowest-order dispersion, and the direction of the nodal lines (if any) meeting at $\mathbf{K}$.

{\it Notation:} In the third column, if dim=$n$ with $n\geq4$, then it means that all the irReps are of $n$-dimensional.  For the cases with `dim=$m|n$' ($m,n$ are two integers with $m\leq n$), $m$ stands for the lowest dimension of the irReps of the same projective class and $n$ is the dimension of the present irRep. Take $G_{\mathbf K} \cong M_{\mathbf K}=O_{h}\times Z_{2}^{T}$ (see Supplementary Table \ref{Ohdrztinv}) as an example, the projective class ($+1,+1,-1,-1,-1$) has one $12$-dim irRep and three inequivalent $4$-dim irReps. For ``dim$=4|12$'', the first number 4 stands for the lowest dimension of the irReps of $O_{h}\times Z_{2}^{T}$ and the second number 12 is the dimension of the irRep discussed at present (we are discussing the dispersion and nodal line the 12-dim irRep may lead to). Similarly, ``dim$=4|4_{1}$'' shows that we are discussing the dispersion and nodal line the one 4-dim irRep may lead to, and ``dim$=4|4_{2,3}$'' shows that we are discussing the dispersion and nodal line the two 4-dim irReps may lead to.

In the fourth column, we list some dispersions, which are polynomials of {$\mathbf{k}$}.
Different dispersion terms of {$\mathbf{k}$} belong to different linear irReps of
$G_{\mathbf K}$\cite{Yang2021,YangYFL_CG2021}.
When one linear irRep of $G_{\mathbf K}$ corresponds to linear dispersion term of {$\mathbf{k}$}, if it is $3$-dimensional, we denote it as $[k_{x},k_{y},k_{z}]$; if it is $2$-dimensional in $(k_{x},k_{y})$ plane, we denote it as $[k_{x},k_{y}]$; if it is $1$-dimensional only in the direction of $k_{z}$, we denote it as $k_{z}$. Here, $k_{x},k_{y},k_{z}$ are bases of linear irRep of $G_{\mathbf K}$.
When the linear irRep of $G_{\mathbf K}$ corresponds to quadratic or cubic dispersion term of {$\mathbf{k}$}, the bases of linear irRep can be $k_{x}k_{y}$, $k_{x}k_{z}$, $k_{y}k_{z}$, $k_{x}k_{y}k_{z}$,$[k_{z}k_{x},k_{z}k_{y}]$,
$[k_{x}k_{y},k_{x}k_{z},k_{y}k_{z}]$,$\cdots$ . The bracketed terms correspond to bases of $2$-dim or $3$-dim linear irRep of $G_{\mathbf K}$.
The bases of linear irRep of $G_{\mathbf K}$ are denoted as $\phi^{\mu}_{i}$ in Eq.\eqref{eq:20} of the main text.
\end{enumerate}

In the following we provide an instruction to the tables for spin invariants and total invariants via concrete examples.

\subsection{Example: The Spin Invariants for $G_{\mathbf K}\cong M_{\mathbf K}=D_{4h} \times Z_2^T$} \label{sec:tabinstructspininv}

In the following we will illustrate the above procedure via a concrete example of  $G_\mathbf{K}\cong M_\mathbf{K} = P\times Z_{2}^{T}$ with $P=D_{4h}$, whose spin invariants are collected in Supplementary Table~\ref{D4hdrztspin}.  We denote the spin operation associated with the lattice operation $g\in M_{\mathbf K}=P\times Z_2^T$ as $\varphi_g\equiv \varphi(g)$. The set of spin operations $\varphi(M_{\mathbf K})$,  which form a subgroup of $\mathrm{SO(3)}\times Z_2^T$,  is isomorphic to the quotient group $M_{\mathbf K}/L_{0\mathbf K}$, namely $\varphi(M_{\mathbf K})\cong M_{\mathbf K}/L_{0\mathbf K}$ with $\varphi(L_{0\mathbf K})=E$.

For the group $M_{\mathbf K}=D_{4h} \times Z_2^T$ with $P=D_{4h}$, there are 17 different choices of $L_{0\mathbf K}\lhd P$, as listed in the first column of Supplementary Table \ref{D4hdrztspin}. We will discuss in details for the cases $L_{0\mathbf K} = D_2 = \{E,C_{2x,2y,2z}\}$ and $L_{0\mathbf K} = D_4=\{E, C_{4z}^{\pm}, C_{2x,2y,2z}, C_{2a,2b}\}$.

\begin{enumerate}
\item[$\bullet$] $L_{0\mathbf K}=D_2 =\{E,C_{2x,2y,2z}\}$\\
For $L_{0\mathbf K}=D_2\lhd D_{4h}, \ M_{\mathbf K}=D_{4h}\times Z_2^T$, the group $\varphi(M_{\mathbf K})\cong D_{4h}/D_2$ is uniquely determined as $\varphi(M_{\mathbf K}) = D_2\times Z_2^T =\{E, C_{2x}, C_{2y},  C_{2z}, T, TC_{2x}, TC_{2y},  TC_{2z}\}$  (up to transformation of the spin frames).

However, there are four different maps from $M_{\mathbf K}$ to the quotient group $D_2\times Z_2^T$. Noticing that the group $D_{4h}$ can be decomposed as cosets $D_{4h}=D_2 + ID_2 + C_{4z}^+D_2 + S_{4z}^{-} D_2$, the four maps are shown in the following:

\begin{enumerate}
\item[(1)] $\varphi(L_{0\mathbf K})=E, \ \varphi(IL_{0\mathbf K})=C_{2z},\ \varphi(C_{4z}^+ L_{0\mathbf K})=C_{2x},\  \varphi(TL_{0\mathbf K})=T$.

These relations are reflected in the SSG elements $(\varphi_g || g)$ for $g\in M_{\mathbf K}$. It will be sufficient to list the generators of $\varphi(M_{\mathbf K})$, namely, $(C_{2z}||I,M_{x,y,z}),\ (C_{2x}||C_{4z}^{\pm}, C_{2a,2b}),\ (T||T,TC_{2x,2y,2z})$, where the elements separated by the commas share the same spin operation.  By group multiplication, one can easily obtain all the other SSG elements, $(C_{2y}||S_{4z}^{\mp},M_{da,db})$, $(TC_{2z}||IT,TM_{x,y,z})$,  $(TC_{2x}||TC_{4z}^{\pm},TC_{2a,2b})$ and $(TC_{2y}||TS_{4z}^{\mp},TM_{da,db})$.

From the above spin operations $\varphi_{g}$, we can write down the double valued Reps $d(\varphi_{g})K^{\zeta_{g}}$:
\begin{eqnarray}
d(\varphi_{C_{2x}})&=&d(\varphi_{C_{2y}})=1_{2\times 2}~,~
d(\varphi_{C_{2a}})=-i\sigma_{x}~,~ d(\varphi_{I})=-i\sigma_{z},   \nonumber\\
d(\varphi_{T})K&=&d(\varphi_{TC_{2x}})K=i\sigma_{y}K~,~d(\varphi_{IT})K=-i\sigma_x K~,~
d(\varphi_{TC_{2a}})K=i\sigma_{z}K.
\end{eqnarray}
Since $\eta^S_{p_1,p_2}{\cdot1_{2\times 2}} = d({\varphi_{p_1}})d({\varphi_{p_2}})d^{-1}({\varphi_{p_1}})d^{-1}({\varphi_{p_2}})$
for unitary elements $p_{1},p_{2}$ with $p_{1}p_{2}=p_{2}p_{1}$ and
$\eta^S_{p}{\cdot1_{2\times 2}}=[d(\varphi_p) {K}]^2$ for anti-unitary element $p$ with $p^{2}=E$, the values of the spin invariants $\vec{\eta}^{S}=(\eta_{C_{2x},C_{2y}}^{S}, \eta_T^{S}, \eta_{IT}^{S}, \eta_{TC_{2x}}^{S}, \eta_{TC_{2a}}^{S}, \eta_{I,C_{2x}}^{S}, \eta_{I,C_{2a}}^{S})$ can be obtained:
\begin{equation}\label{}
\eta_{C_{2x},C_{2y}}^{S}=+1~,~\eta_T^{S}=-1~,~\eta_{IT}^{S}=+1~,~\eta_{TC_{2x}}^{S}=-1~,~
\eta_{TC_{2a}}^{S}=+1~,~\eta_{I,C_{2x}}^{S}=+1~,~\eta_{I,C_{2a}}^{S}=-1.
\end{equation}

\item[ (2)] $\varphi(L_{0\mathbf K})=E,\ \varphi(IL_{0\mathbf K})=C_{2z},\ \varphi(C_{4z}^+ L_{0\mathbf K})=C_{2x},\ \varphi(TL_{0\mathbf K})=TC_{2z}$. Similar to previous procedure, the spin invariants can be obtained as
$
\eta_{C_{2x},C_{2y}}^{S}=+1~,~\eta_T^{S}=+1~,~\eta_{IT}^{S}=-1~,~\eta_{TC_{2x}}^{S}=+1~,~
\eta_{TC_{2a}}^{S}=+1~,~\eta_{I,C_{2x}}^{S}=+1~,~\eta_{I,C_{2a}}^{S}=-1.
$
This set of invariants can only be realized in SSG, so is labeled by the Boolean vector  {\R $0010001$} colored in RED.\\

\item[ (3)] $\varphi(L_{0\mathbf K})=E,\ \varphi(IL_{0\mathbf K})=C_{2z},\ \varphi(C_{4z}^+ L_{0\mathbf K})=C_{2x},\ \varphi(TL_{0\mathbf K})=TC_{2x}$. Similar to previous procedure, the spin invariants can be obtained as
$
\eta_{C_{2x},C_{2y}}^{S}=+1~,~\eta_T^{S}=+1~,~\eta_{IT}^{S}=+1~,~\eta_{TC_{2x}}^{S}=+1~,~
\eta_{TC_{2a}}^{S}=-1~,~\eta_{I,C_{2x}}^{S}=+1~,~\eta_{I,C_{2a}}^{S}=-1.
$
This set of invariants can only be realized in SSG, so is labeled by the Boolean vector  {\R $0000101$} colored in RED.\\

\item[ (4)] $\varphi(L_{0\mathbf K})=E,\ \varphi(IL_{0\mathbf K})=C_{2z},\ \varphi(C_{4z}^+ L_{0\mathbf K})=C_{2x},\ \varphi(TL_{0\mathbf K})=TC_{2y}$.
Similar to previous procedure, the spin invariants can be obtained as
$
\eta_{C_{2x},C_{2y}}^{S}=+1~,~\eta_T^{S}=+1~,~\eta_{IT}^{S}=+1~,~\eta_{TC_{2x}}^{S}=+1~,~
\eta_{TC_{2a}}^{S}=+1~,~\eta_{I,C_{2x}}^{S}=+1~,~\eta_{I,C_{2a}}^{S}=-1.
$
\end{enumerate}

\item[$\bullet$] $L_{0\mathbf K}=D_4=\{E, C_{4z}^{\pm}, C_{2x,2y,2z}, C_{2a,2b}\}$

For $L_{0\mathbf K}=D_4\lhd D_{4h}, \ M_{\mathbf K}=D_{4h}\times Z_2^T$, up to transformation of the spin frames there are two possible choices of quotient groups $\varphi(M_{\mathbf K})\cong D_{4h}/D_4$, namely, $\varphi(M_{\mathbf K})=Z_2\times Z_2^T=\{E, C_{2z}, T, TC_{2z}\}$ or $\varphi(M_{\mathbf K}) =\{E, C_{2z}, TC_{2x}, TC_{2y}\}$. We will discuss them separately.

\begin{enumerate}
\item[(A)] $\varphi(M_{\mathbf K})=Z_2\times Z_2^T=\{E, C_{2z}, T, TC_{2z}\}$.\\
 For this fixed quotient group $\varphi(M_{\mathbf K})$, there are still two possible maps from $M_{\mathbf K}$ to $\varphi(M_{\mathbf K})$.

\begin{enumerate}
\item[ (1)] $\varphi(L_{0\mathbf K})=E, \varphi(IL_{0\mathbf K})=C_{2z}, \varphi(TL_{0\mathbf K})=T$. Similar to previous procedure, the spin invariants can be obtained as
$
\eta_{C_{2x},C_{2y}}^{S}=+1~,~\eta_T^{S}=-1~,~\eta_{IT}^{S}=+1~,~\eta_{TC_{2x}}^{S}=-1~,~
\eta_{TC_{2a}}^{S}=-1~,~\eta_{I,C_{2x}}^{S}=+1~,~\eta_{I,C_{2a}}^{S}=+1.
$
This set of invariants can only be realized in SSG, so is labeled by the Boolean vector  {\R $0101100$} colored in RED.

\item[(2)] $\varphi(L_{0\mathbf K})=E, \varphi(IL_{0\mathbf K})=C_{2z}, \varphi(TL_{0\mathbf K})=TC_{2z}$.
Similar to previous procedure, the spin invariants can be obtained as
$
\eta_{C_{2x},C_{2y}}^{S}=+1~,~\eta_T^{S}=+1~,~\eta_{IT}^{S}=-1~,~\eta_{TC_{2x}}^{S}=+1~,~
\eta_{TC_{2a}}^{S}=+1~,~\eta_{I,C_{2x}}^{S}=+1~,~\eta_{I,C_{2a}}^{S}=+1.
$
This set of invariants can only be realized in SSG, so is labeled by the Boolean vector  {\R $0010000$} colored in RED.
\end{enumerate}

\item[ (B)] $\varphi(M_{\mathbf K}) =\{E, C_{2z}, TC_{2x}, TC_{2y}\}$.\\
 For this fixed quotient group $\varphi(M_{\mathbf K})$, there is only one map from $M_{\mathbf K}$ to $\varphi(M_{\mathbf K})$: $\varphi(L_{0\mathbf K})=E, \varphi(IL_{0\mathbf K})=C_{2z}, \varphi(TL_{0\mathbf K})=TC_{2y}$.
Similar to previous procedure, the spin invariants can be obtained as
$
\eta_{C_{2x},C_{2y}}^{S}=+1~,~\eta_T^{S}=+1~,~\eta_{IT}^{S}=+1~,~\eta_{TC_{2x}}^{S}=+1~,~
\eta_{TC_{2a}}^{S}=+1~,~\eta_{I,C_{2x}}^{S}=+1~,~\eta_{I,C_{2a}}^{S}=+1.
$
\end{enumerate}
\end{enumerate}

Finally, it should be noted that there may exist more than one choices of $L_{0\mathbf K}$ which are isomorphic to each other but their spin invariants are not the same. For instance, $M_{\mathbf K}=D_{4h}\times Z_2^T$ has two $D_{2d}$ normal subgroups, namely $L_{0\mathbf K}=D_{2d}^1 = \{E, S_{4z}^\pm, C_{2x,2y, 2z}, M_{da, db}\}$
and $L_{0\mathbf K}=D_{2d}^2 = \{E, S_{4z}^\pm, C_{2a,2b, 2z}, M_{x, y}\}$. The former can realize the invariants  ($\eta_{C_{2x},C_{2y}}^{S}$,$\eta_T^{S}$,$\eta_{IT}^{S}$,$\eta_{TC_{2x}}^{S}$,
$\eta_{TC_{2a}}^{S}$,$\eta_{I,C_{2x}}^{S}$,$\eta_{I,C_{2a}}^{S}$)$=$
($+1$,$-1$,$+1$,$-1$,$+1$,$+1$,$+1$),
($+1$,$+1$,$-1$,$+1$,$-1$,$+1$,$+1$),
($+1$,$+1$,$+1$,$+1$,$+1$,$+1$,$+1$),
while the latter  can  realize invariants
($\eta_{C_{2x},C_{2y}}^{S}$,$\eta_T^{S}$,$\eta_{IT}^{S}$,$\eta_{TC_{2x}}^{S}$,
$\eta_{TC_{2a}}^{S}$,$\eta_{I,C_{2x}}^{S}$,$\eta_{I,C_{2a}}^{S}$)$=$
($+1$,$-1$,$+1$,$+1$,$-1$,$+1$,$+1$),
($+1$,$+1$,$-1$,$-1$,$+1$,$+1$,$+1$),
($+1$,$+1$,$+1$,$+1$,$+1$,$+1$,$+1$).

\subsection{Example: The total invariants for $G_{\mathbf K}\cong M_{\mathbf K}=D_{4h}\times Z_2^T$}\label{sec:tabinstructinv}

We illustrate that each set of  only-SSG-realizable invariants  may be realized by different combinations of lattice invariants and spin invariants. The lattice part comes from the high-symmetry momentum of certain type IV MSG ($\mathcal{M}$,$\mathbf{K}$), and the spin invariants are realized by properly choosing the normal subgroup $L_{0\mathbf K}$ and the homomorphic map $\varphi$ from $M_{\mathbf K}$ to $\varphi(M_{\mathbf K})$.

As an example, the only SSG-realizable invariants (${\color{red}+1,-1,-1,-1,+1,+1,-1}$) of $G_{\mathbf K}\cong M_{\mathbf K}=D_{4h}\times Z_2^T$ can be realized in eight different ways, as listed in Supplementary Table~\ref{D4hdrztinv}.
The lattice invariants are realized by the combinations ($\mathcal{M}$,$\mathbf{K}$) in Supplementary Table~\ref{Pdrztlat} and
the spin invariants are shown in Supplementary Table~\ref{D4hdrztspin}:

\begin{enumerate}
\item[ (1)]($0001001$,{\color{red}$0110000$})

Lattice part:
(126.385,A),(131.445,A)

Spin part:

 \ \ \ (a)$L_{0\mathbf K}=\{E,C_{2z},I,M_{z}\}$, $M_{\mathbf K}/L_{0\mathbf K}=\{E,C_{2x,2y,2z}\}\times Z_{2}^{T}$, $\varphi_{C_{2x,2y},M_{x,y}}=C_{2x}$, $\varphi_{C_{2a,2b},M_{da,db}}=C_{2y}$, $\varphi_{T,TC_{2z},IT,TM_{z}}=T$

\ \ \ \ \ \!(b)$L_{0\mathbf K}=C_{4h}$, $M_{\mathbf K}/L_{0\mathbf K}=\{E,C_{2z}\}\times Z_{2}^{T}$, $\varphi_{C_{2x,2y},C_{2a,2b},M_{x,y},M_{da,db}}=C_{2z}$, $\varphi_{T,TC_{4z}^{\pm},TC_{2z},IT,TS_{4z}^{\mp},TM_{z}}=T$


\item[ (2)]($0000001$,{\color{red}$0111000$})

Lattice part:
(130.433,A),(131.445,Z),(133.469,Z),(135.493,Z),
(135.493,A),(137.517,Z),(141.560,Z),(142.570,Z)

Spin part:
$L_{0\mathbf K}=D_{2h}$, $M_{\mathbf K}/L_{0\mathbf K}=\{E,C_{2z}\}\times Z_{2}^{T}$,
$\varphi_{C_{4z}^{\pm},C_{2a,2b},S_{4z}^{\mp},M_{da,db}}=C_{2z}$,
$\varphi_{T,TC_{2x,2y,2z},IT,TM_{x,y,z}}=T$


\item[ (3)] ($0101001$,{\color{red}$0010000$})

Lattice part:
(126.386,A),(130.432,A),(131.446,A),(135.492,A)

Spin part:
$L_{0\mathbf K}=D_{4}$, $M_{\mathbf K}/L_{0\mathbf K}=\{E,C_{2z}\}\times Z_{2}^{T}$,
$\varphi_{I,S_{4z}^{\mp},M_{z},M_{x,y},M_{da,db}}=C_{2z}$,
$\varphi_{T,TC_{4z}^{\pm},TC_{2z},TC_{2x,2y},TC_{2a,2b}}=TC_{2z}$


\item[ (4)]($0100001$,{\color{red}$0011000$})

Lattice part:
(126.384,A),(130.434,A),(131.444,Z),(131.444,A),
(131.446,Z),(133.468,Z),

~~~~~~~~~~~~~~~~~
(133.470,Z),(135.492,Z),
(135.494,Z),(135.494,A),(137.516,Z),(137.518,Z)

Spin part:
$L_{0\mathbf K}=D_{2d}^{2}$, $M_{\mathbf K}/L_{0\mathbf K}=\{E,C_{2z}\}\times Z_{2}^{T}$,
$\varphi_{I,C_{4z}^{\pm},M_{z},C_{2x,2y},M_{da,db}}=C_{2z}$,
$\varphi_{T,TS_{4z}^{\mp},TC_{2z},TM_{x,y},TC_{2a,2b}}=TC_{2z}$


\item[ (5)]($0101010$,{\color{red}$0010011$})

Lattice part:
(125.374,A),(129.420,A),(132.458,A),(136.504,A)

Spin part:

\ \ \ (a)$L_{0\mathbf K}=\{E,C_{4z}^{\pm},C_{2z}\}$,$M_{\mathbf K}/L_{0\mathbf K}=\{E,C_{2x,2y,2z}\}\times Z_{2}^{T}$, $\varphi_{I,S_{4z}^{\mp},M_{z}}=C_{2z}$,$\varphi_{C_{2x,2y},C_{2a,2b}}=C_{2x}$,
$\varphi_{T,TC_{4z}^{\pm},TC_{2z}}=TC_{2z}$

\ \ \ (b)$L_{0\mathbf K}=\{E,S_{4z}^{\mp},C_{2z}\}$,$M_{\mathbf K}/L_{0\mathbf K}=\{E,C_{2x,2y,2z}\}\times Z_{2}^{T}$, $\varphi_{I,C_{4z}^{\pm},M_{z}}=C_{2z}$, $\varphi_{C_{2x,2y},M_{da,db}}=C_{2x}$, $\varphi_{T,TS_{4z}^{\mp},TC_{2z}}=TC_{2z}$

\item[ (6)]($0101000$,{\color{red}$0010001$})

Lattice part:
(123.350,A),(127.396,A),(134.482,A),(138.528,A)

Spin part:
$L_{0\mathbf K}=\{E,C_{2x,2y,2z}\}$,$M_{\mathbf K}/L_{0\mathbf K}=\{E,C_{2x,2y,2z}\}\times Z_{2}^{T}$,
$\varphi_{I,M_{x,y,z}}=C_{2z}$,
$\varphi_{C_{4z}^{\pm},C_{2a,2b}}=C_{2x}$,
$\varphi_{T,TC_{2x,2y,2z}}=TC_{2z}$


\item[ (7)]($0100000$,{\color{red}$0011001$})

Lattice part:
(123.348,Z),(123.348,A),(123.350,Z),(125.372,Z),
(125.374,Z),(127.396,Z),

~~~~~~~~~~~~~~~~~
(127.398,Z),(127.398,A),
(129.420,Z),(129.422,Z),(134.480,A),(138.530,A)

Spin part:
$L_{0\mathbf K}=\{E,M_{x,y},C_{2z}\}$,$M_{\mathbf K}/L_{0\mathbf K}=\{E,C_{2x,2y,2z}\}\times Z_{2}^{T}$,
$\varphi_{I,C_{2x,2y},M_{z}}=C_{2z}$,
$\varphi_{S_{4z}^{\mp},C_{2a,2b}}=C_{2x}$,
$\varphi_{T,TM_{x,y},TC_{2z}}=TC_{2z}$


\item[ (8)]($0101011$,{\color{red}$0010010$})

Lattice part:
(124.362,A),(128.408,A),(133.470,A),(137.516,A)

Spin part:
$L_{0\mathbf K}=\{E,C_{2b,2a},C_{2z}\}$,$M_{\mathbf K}/L_{0\mathbf K}=\{E,C_{2x,2y,2z}\}\times Z_{2}^{T}$,
$\varphi_{I,M_{db,da},M_{z}}=C_{2z}$,
$\varphi_{C_{4z}^{\pm},C_{2x,2y}}=C_{2x}$,
$\varphi_{T,TC_{2b,2a},TC_{2z}}=TC_{2z}$

\end{enumerate}

\section{Theorem: different projective classes lead to inequivalent direct-product representations}\label{sec:prffusiontheory}

In this section we prove in detail the key theorem in the main text, which states that if two projective irReps $\rho_{1,2}$ of $P\times{Z}_2^T$ belong to different projective classes $\vec\eta(\rho_1)\neq\vec\eta(\rho_2)$, then $\rho_1\otimes\rho^\ast_1$ and $\rho_2\otimes\rho^\ast_2$ must be inequivalent.

As $\rho_1$ and $\rho_2$ belong to two different projective classes, at least one symmetry invariant must take different values for $\rho_1$ and $\rho_2$.
Then there are two possibilities:

(i) there is one anti-unitary invariant $\eta_a$ such that $\eta_a(\rho_1)\neq\eta_a(\rho_2)$, and

(ii) all anti-unitary invariants are same respectively, but one or more unitary invariants take different values, or symbolically $\eta_a(\rho_1)=\eta_a(\rho_2)$ for all anti-unitary elements $a$, but $\eta_{u,v}(\rho_1)\neq\eta_{u,v}(\rho_2)$.

In fact, one can prove that for $P$ being any crystallographic point group other than $C_{4h}$, each unitary invariant can be expressed in terms of several anti-unitary invariants.
This is because, according to Supplementary Table \ref{tab:invrts}, if $P\neq{C}_{4h}$, for unitary invariant $\eta_{u,v}$,
we have $u^2=v^2=(uv)^2=E$
and can easily prove that
\begin{equation}
\rho(uT)\rho^{\ast}(uT)\rho(vT)\rho^{\ast}(vT)\rho(T)\rho^{\ast}(T)=
\rho(u)\rho(u)\rho(v)\rho(v)\rho(T)\rho^{\ast}(T)=
\eta_{u,v}(\rho)\rho(uvT)\rho^{\ast}(uvT),
\end{equation}
which entails
\begin{equation}\label{unitaryinv}
\eta_{u,v}=\eta_{uT}\eta_{vT}\eta_{T}\eta_{uvT}.
\end{equation}
For example, if $P=D_{4h}$, one set of symmetry invariants can be defined by
$\vec{\eta}^{}=$
($\eta_{C_{2x},C_{2y}}^{}$,
$\eta_T^{}$,
$\eta_{IT}^{}$,
$\eta_{TC_{2x}}^{}$,
$\eta_{TC_{2a}}^{}$,
$\eta_{I,C_{2x}}^{}$,
$\eta_{I,C_{2a}}^{}$)
in Supplementary Table \ref{tab:invrts}.
From SEq.(\ref{unitaryinv}),
\begin{eqnarray}
\eta_{C_{2x},C_{2y}}&=&\eta_{TC_{2x}}\eta_{TC_{2y}}\eta_{T}\eta_{TC_{2z}}
=\eta_{TC_{2x}}^{2}\eta_{T}\eta_{TC_{2z}}, \\
\eta_{I,C_{2x}}&=&\eta_{IT}\eta_{TC_{2x}}\eta_{T}\eta_{TM_{x}},  \\
\eta_{I,C_{2a}}&=&\eta_{IT}\eta_{TC_{2a}}\eta_{T}\eta_{TM_{da}},
\end{eqnarray}
so the unitary invariants $\eta_{C_{2x},C_{2y}}$, $\eta_{I,C_{2x}}$ and
$\eta_{I,C_{2a}}$ can be replaced by the anti-unitary invariants $\eta_{TC_{2z}}$, $\eta_{TM_{x}}$
and $\eta_{TM_{da}}$, respectively.
We obtain a new definition of symmetry invariants $\vec{\eta}^{}=$($\eta_{TC_{2z}}$,$\eta_T$,$\eta_{IT}$,$\eta_{TC_{2x}}$,
$\eta_{TC_{2a}}$,$\eta_{TM_{x}}$,$\eta_{TM_{da}}$) of $D_{4h}\times Z_{2}^{T}$.

Therefore, for $P\neq{C}_{4h}$, if $\vec\eta(\rho_1)\neq\vec\eta(\rho_2)$, then there must be
an anti-unitary element $a$ ($a^2=E$) such that $\eta_a(\rho_1)\neq\eta_a(\rho_2)$
since all the symmetry invariants can be anti-unitary invariants.
This belongs to case-(i).

Now we want to prove that $\rho_1\otimes\rho_1^\ast$ and $\rho_2\otimes\rho_2^\ast$ are inequivalent Reps of $P\times{Z}_2^T$.
To do this, we first observe that we show that they are inequivalent Reps of $P\times{Z}_2^T$ if they are inequivalent Reps of any subgroup of $P\times{Z}_2^T$.
Consider the subgroup $H=\{E,a\}$, which is the smallest group containing $a$.
The hermitian fermion bilinear operators considered in the main text can be combined into two linear Reps of $H$: (i) those that are even under $a$, namely $D^{(+)}(a)=1$; (ii) those that are odd under $a$, namely $D^{(-)}(a)=-1$.
Without loss of generality, suppose $\eta_a(\rho_1)=-\eta_a(\rho_2)=1$, then one can prove that\cite{YangYFL_CG2021}
\begin{eqnarray}\label{eq:Z2a}
\rho_1\otimes\rho_1^\ast&=&\frac{d_1(d_1+1)}{2}D^{(+)}(a)\oplus\frac{d_1(d_1-1)}{2}D^{(-)}(a),\\
\nonumber
\rho_2\otimes\rho_2^\ast&=&\frac{d_2(d_2-1)}{2}D^{(+)}(a)\oplus\frac{d_2(d_2+1)}{2}D^{(-)}(a),
\end{eqnarray}
where $d_{1,2}$ is the dimension of $\rho_{1,2}$.
 From SEq.(\ref{eq:Z2a}), we see that the two direct-product Reps are inequivalent, independent of the values of $d_{1,2}$.
Following the previous argument, they must also be inequivalent Reps of $P\times{Z}_2^T$.

Now, if $P=C_{4h}$, there is a unitary invariant $\eta_{I,C^{+}_{4z}}$ that cannot be expressed in terms of $\eta_a$.
So if $\rho_1$ and $\rho_2$ have $\eta_a(\rho_1)=\eta_a(\rho_2)$ for all anti-unitary $a$ ($a^2=E$), it is still possible that $\eta_{I,C^{+}_{4z}}(\rho_1)\neq\eta_{I,C^{+}_{4z}}(\rho_2)$.
For this case, we have computed all such pairs of $\rho_{1,2}$, and checked explicitly that, for all the pairs, $\rho_1\otimes\rho_1^\ast$ and $\rho_2\otimes\rho_2^\ast$ are inequivalent Reps of $C_{4h}\times{Z}_2^T$.

We conclude that the theorem holds for all $P\times{Z}_2^T$, where $P$ is any crystallographic point group.

However, although this conclusion can be rigorously proved for groups $P\times Z_2^T$, it is not always valid if the symmetry group is not of the form $P\times Z_2^T$.
In this case, one may distinguish different projective classes
by lowering the symmetry with a background field.

\section{Comparison with 8-fold fermions in Magnetic space groups}

For little co-group $G_\mathbf{K}=D_{4h}\times Z_{2}^{T}$, ${\rm Invariants}\equiv$ ($\eta_{C_{2x},C_{2y}}$,$\eta_T$,$\eta_{IT}$,$\eta_{TC_{2x}}$,$\eta_{TC_{2a}}$,$\eta_{I,C_{2x}}$,$\eta_{I,C_{2a}}$), we compare three different 8-dimensional irReps coming from three projective classes of
$D_{4h}\times Z_{2}^{T}$.

From Supplementary Table \ref{D4hdrztinv},  for the set of invariants $\vec{\eta}=(-1,-1,-1,-1,+1,+1,-1)$ colored in GREEN, the corresponding projective class can be realized at A in BZ of type II MSGs 130.424, 135.484 with spin-orbit coupling (SOC) and type IV MSGs 126.385,131.445 with SOC;
for the set of invariants $\vec{\eta}=(-1,+1,-1,-1,+1,-1,+1)$ colored in BLUE, the corresponding projective class can be realized at A in BZ of type IV MSGs 125.374, 129.420,132.458,136.504 with SOC. The two sets of invariants lead to two different 8-fold fermions discussed in SRefs.\cite{Bradlyn2016} and \cite{Cano2019}.

While the set of invariants $\vec{\eta}=$($-1$,$-1$,$+1$,$-1$,$+1$,$-1$, $+1$) colored in RED in Supplementary Table \ref{D4hdrztinv}
also allow a 8-dimensional irRep which can be realized at A in BZ of SSGs
\href{https://cmpdc.iphy.ac.cn/ssg/ssgs/124.2.8.20}{124.2.8.20} (isomorphic to type IV MSG $\mathcal{M}=126.385$),
\href{https://cmpdc.iphy.ac.cn/ssg/ssgs/132.2.8.28}{132.2.8.28} (isomorphic to type IV MSG $\mathcal{M}=131.445$),
\href{https://cmpdc.iphy.ac.cn/ssg/ssgs/140.2.8.35}{140.2.8.35} (isomorphic to type IV MSG $\mathcal{M}=125.374$),
\href{https://cmpdc.iphy.ac.cn/ssg/ssgs/129.2.8.15}{129.2.8.15} (isomorphic to type IV MSG $\mathcal{M}=129.420$),
\href{https://cmpdc.iphy.ac.cn/ssg/ssgs/140.2.8.28}{140.2.8.28} (isomorphic to type IV MSG $\mathcal{M}=132.458$),
\href{https://cmpdc.iphy.ac.cn/ssg/ssgs/127.2.8.14}{127.2.8.14} (isomorphic to type IV MSG $\mathcal{M}=136.504$).
The elements of space group (SG) $\mathcal{L}_0\equiv Pb$ (7) are pure lattice operations
of SSG 124.2.8.20,
and the elements of SG $\mathcal{L}_0\equiv Pm$ (6) are pure lattice operations
of SSG 132.2.8.28.
The little co-group of SSGs 124.2.8.20 and 132.2.8.28 at high-symmetry point A
is $G_{\mathbf{K}}=D_{4h}\times Z_{2}^{T}$, which is generated by
$(C_{2z}||C_{2z},I)$, $(C_{2x}||C_{2x},M_{y})$, $(C_{2a}||C_{2a},M_{db})$, $(C_{2y}||C_{2y},M_{x})$ and
$(T||T,TM_{z})$ as listed in Supplementary Table \ref{D4hdrztspin}.
The elements of SG $\mathcal{L}_{0}\equiv P\bar{1}$ (2) are pure lattice operations of SSGs
140.2.8.35, 129.2.8.15, 140.2.8.28 and 127.2.8.14,
whose little co-group at high-symmetry point A
is $G_{\mathbf{K}}=D_{4h}\times Z_{2}^{T}$, which is generated by
$(C_{2x}||C_{2x},M_{x})$, $(C_{2a}||C_{2a},M_{da})$, $(C_{2y}||C_{2y},M_{y})$ and
$(TC_{2z}||T,IT)$ as listed in Supplementary Table \ref{D4hdrztspin}.

We compare (i)
the 8-dimensional irRep allowed by
the set of invariants $\vec{\eta}=(-1,-1,-1,-1,+1,+1,-1)$ which can be realized at A in BZ of MSGs 130.424, 135.484
\cite{Bradlyn2016} and  MSGs 126.385,131.445
\cite{Cano2019}
\beq\label{8dfermion}
\rho_{1}(C_{2x})
&=&
\tau_{0}\sigma_{y}\sigma_{x},
\rho_{1}(C_{2y})=\tau_{0}\sigma_{y}\sigma_{y},
\nonumber\\
\rho_{1}
(C_{2a})&=&\tau_{0}\sigma_{0}
\exp(-i\sigma_{xy}\pi/2),
\rho_{1}
(I)=\tau_{0}\sigma_{x}\sigma_{z},
\nonumber\\
\rho_{1}
(T)K&=&\tau_{y}\sigma_{y}\sigma_{y}K,
\eeq
and (ii) the 8-dimensional irRep
allowed by
the set of invariants $\vec{\eta}=(-1,+1,-1,-1,+1,-1,+1)$ which can be realized at A in BZ of  MSGs 125.374, 129.420,132.458,136.504
\cite{Cano2019}
\beq\label{8dfermion2}
\rho_{2}
(C_{2x})&=&
i\tau_{z}\sigma_{0}\sigma_{x},
\rho_{2}
(C_{2y})=i\tau_{z}\sigma_{0}\sigma_{y},
\nonumber\\
\rho_{2}
(C_{2a})&=&i\tau_{0}\sigma_{y}
\exp(-i\sigma_{xy}\pi/2),
\rho_{2}
(I)=i\tau_{z}\sigma_{z}\sigma_{z},
\nonumber\\
\rho_{2}
(T)K&=&\tau_{x}\sigma_{x}\sigma_{x}K,
\eeq
with (iii) the 8-dimensional irRep allowed by
the set of invariants $\vec{\eta}=(-1,-1,+1,-1,+1,-1,+1)$ which can be realized at A in BZ
of SSGs 124.2.8.20, 132.2.8.28, 140.2.8.35, 129.2.8.15, 140.2.8.28, 127.2.8.14
\beq\label{8dfermion3}
\rho_{3}
(C_{2x})
&=&
\tau_{0}\sigma_{0}\sigma_{x},
\rho_{3}
(C_{2y})=\tau_{0}\sigma_{0}\sigma_{y},
\nonumber\\
\rho_{3}
(C_{2a})
&=&\tau_{0}\sigma_{z}
\exp(-i\sigma_{xy}\pi/2),
\rho_{3}
(I)=i\tau_{0}\sigma_{x}\sigma_{z},
\nonumber\\
\rho_{3}
(T)K
&=&\tau_{y}\sigma_{x}\sigma_{x}K,
\eeq
where $\sigma_{xy}\equiv (\sigma_{x}+\sigma_{y})/\sqrt{2}$.

\begin{figure}
\centering
\includegraphics[width=0.48\textwidth]{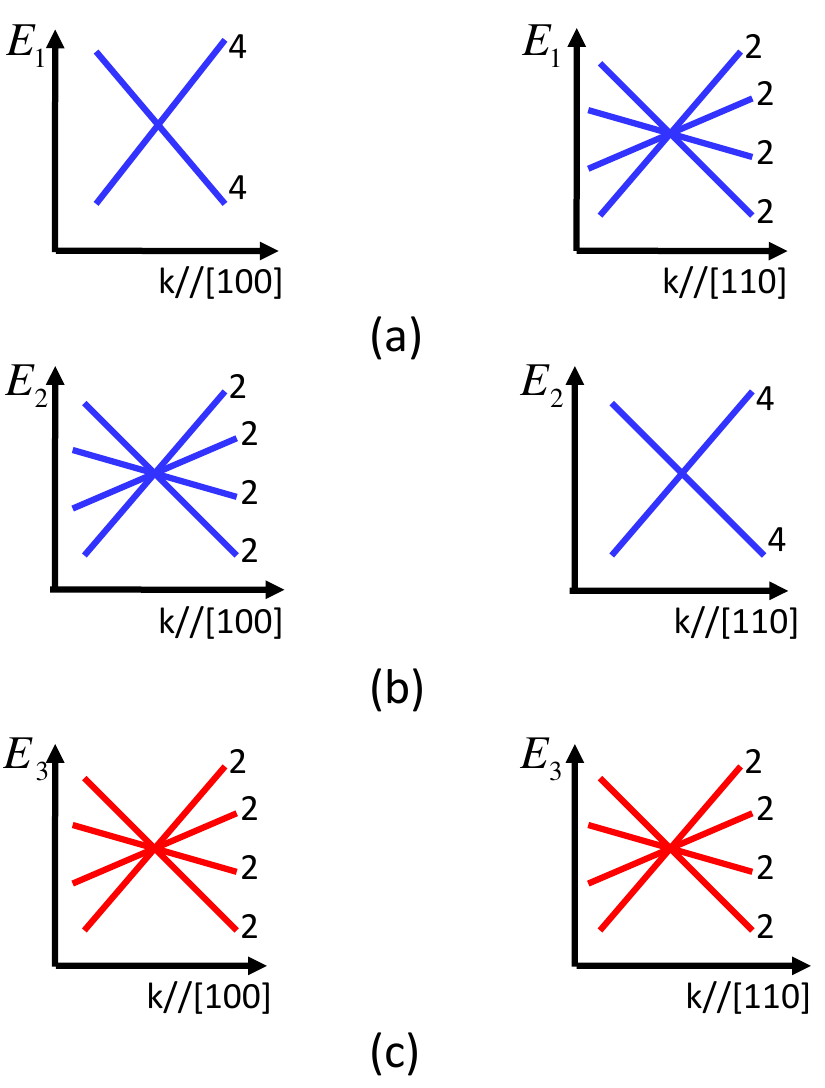}
\caption{
Band splitting of 8-fold fermions protected by
$G_\mathbf{K}=D_{4h}\times Z_{2}^{T}$
along high-symmetry lines.
(a)
Under linear dispersion in SEq.\eqref{d4hqp1h}, a
8-fold fermion, realizable at A in BZ of MSGs 130.424, 135.484, 126.385, 131.445 with SOC, splits into
two 4-fold bands colored in blue along $[100]$, and
four 2-fold bands colored in blue along $[110]$.
(b)
Under linear dispersion in SEq.\eqref{d4hqp2h}, a
8-fold fermion, realizable at A in BZ of MSGs 125.374, 129.420, 132.458, 136.504 with SOC, splits into
four 2-fold bands colored in blue along $[100]$, and
two 4-fold bands colored in blue along $[110]$.
(c)
Under linear dispersion in SEq.\eqref{s34}, a
8-fold fermion, realizable at A in BZ of SSGs 124.2.8.20, 132.2.8.28, 140.2.8.35, 129.2.8.15, 140.2.8.28, 127.2.8.14, splits into
four 2-fold bands colored in red along $[100]$ and $[110]$.}\label{sfig1}
\end{figure}

We decompose $\rho_{1}\otimes \rho^{\ast}_{1}$,
$\rho_{2}\otimes \rho^{\ast}_{2}$ and
$\rho_{3}\otimes \rho^{\ast}_{3}$
into irReps of $D_{4h}\times Z_{2}^{T}$
\beq
\rho_{1}\otimes \rho^{\ast}_{1}&=&\cdots \oplus  4E_{u}^{-}\oplus \cdots ,  \label{8dcg1}\\
\rho_{2}\otimes \rho^{\ast}_{2}&=&\cdots \oplus  2E_{u}^{-}\oplus \cdots ,
\label{8dcg2}\\
\rho_{3}\otimes \rho^{\ast}_{3}&=&\cdots \oplus  6E_{u}^{-}\oplus \cdots ,  \label{8dcg3}
\eeq
where $N_{\rho_{1}}(E_{u}^{-})=4$,
$N_{\rho_{2}}(E_{u}^{-})=2$,
$N_{\rho_{3}}(E_{u}^{-})=6$,
so the three 8-fold fermions have different response
to dispersion $E_{u}^{-}$, which can be linear dispersion
$[k_x,k_y]$.
From Eq.\eqref{eq:20} of the main text,
linear irRep $\mu=E_{u}^-$,
$\phi^{\mu}_{1}=k_{x}$,
$\phi^{\mu}_{2}=k_{y}$.
The linear dispersion of 8-fold fermion (i)
takes the form
\beq\label{d4hqp1h}
\hat{H}_{1}
&=&
v_{11}
\tau_{0}
(k_{x}\sigma_{0}\sigma_{x}+
k_{y}\sigma_{0}\sigma_{y})
+
(
v_{12}
\tau_{x}
+
v_{13}
\tau_{y}
+
v_{14}
\tau_{z}
)
(k_{x}\sigma_{x}
\sigma_{y}+
k_{y}\sigma_{x}
\sigma_{x}),
\eeq
where the coupling parameters in Eq.\eqref{eq:20} of the main text:
$\lambda^{(\mu)}_{1}=v_{11}$,
$\lambda^{(\mu)}_{2}=v_{12}$,
$\lambda^{(\mu)}_{3}=v_{13}$,
$\lambda^{(\mu)}_{4}=v_{14}$.
The linear dispersion of 8-fold fermion (ii)
takes the form
\beq\label{d4hqp2h}
\hat{H}_{2}
&=&
v_{21}
\tau_{z}
(k_{x}\sigma_{0}\sigma_{x}+
k_{y}\sigma_{0}\sigma_{y})
+
v_{22}
\tau_{0}
(k_{x}\sigma_{z}
\sigma_{x}-
k_{y}\sigma_{z}
\sigma_{y}),
\eeq
where the coupling parameters in Eq.\eqref{eq:20} of the main text:
$\lambda^{(\mu)}_{1}=v_{21}$,
$\lambda^{(\mu)}_{2}=v_{22}$.
And
the linear dispersion of 8-fold fermion (iii)
takes the form
\beq
\hat{H}_{3}=(v_{31}\tau_{x}+v_{32}\tau_{y}+v_{33}\tau_{z})
(k_{x}\sigma_{0}\sigma_{x}+k_{y}\sigma_{0}\sigma_{y})+
(v_{34}\tau_{x}+v_{35}\tau_{y}+v_{36}\tau_{z})
(k_{x}\sigma_{x}\sigma_{x}-k_{y}\sigma_{x}\sigma_{y}),
\label{s34}
\eeq
where the coupling parameters in Eq.\eqref{eq:20} of the main text:
$\lambda^{(\mu)}_{1}=v_{31}$,
$\lambda^{(\mu)}_{2}=v_{32}$,
$\lambda^{(\mu)}_{3}=v_{33}$,
$\lambda^{(\mu)}_{4}=v_{34}$,
$\lambda^{(\mu)}_{5}=v_{35}$,
$\lambda^{(\mu)}_{6}=v_{36}$.
Along high-symmetry lines $[100]$ and $[110]$,
the schematic linear dispersions  of the 8-fold fermion $\hat{H}_{1}$ (colored in blue) in SRefs.\cite{Bradlyn2016} and \cite{Cano2019}, the 8-fold fermion $\hat{H}_{2}$ (colored in blue) in SRef.\cite{Cano2019}, and the 8-fold fermion $\hat{H}_{3}$ (colored in red) only realizable in SSGs are given in Supplementary Fig.S1.
Since the three 8-fold fermions come from three different projective classes in
SEqs.(S26),\eqref{8dfermion2},\eqref{8dfermion3}
of $D_{4h}\times Z_{2}^{T}$,
they have different physical properties, such
as different behaviours under linear dispersion in Supplementary Fig.S1.

\section{The SSG-only representation in candidate material $\textbf{Ce}_{3}\textbf{NIn}$}\label{sec:corepce3nin}

Here we propose an SSG candidate material to realize the invariants of $G_{\mathbf{K}}=D_{4h}\times Z_2^T$,
namely ($\eta_{C_{2x},C_{2y}}$,$\eta_T$,$\eta_{IT}$,
$\eta_{TC_{2x}}$,$\eta_{TC_{2a}}$,$\eta_{I,C_{2x}}$,$\eta_{I,C_{2a}}$).
From Supplementary Table \ref{D4hdrztinv}, for the set of invariants $\vec{\eta}=(-1,-1,+1,+1,+1,-1,-1)$
colored in RED,
the corresponding projective class is only realizable
in spin-space groups.
One possible SSG is
\href{https://cmpdc.iphy.ac.cn/ssg/ssgs/123.2.8.35}{123.2.8.35} in the database \cite{ssg_website},
which is isomorphic to type IV MSG $\mathcal{M}=127.397$.
The elements of SG $\mathcal{L}_0=Pm$(6) are pure lattice operations.
The little co-group at high-symmetry points $\Gamma$,M,Z,A
is $G_{\mathbf{K}}=D_{4h}\times Z_{2}^{T}$, which is generated by
$(E||E,M_{z})$,
$(C_{2x}||C_{2x},M_{y})$,$(C_{2a}||C_{2a},M_{db})$,$(C_{2z}||C_{2z},I)$ and
$(T||T,TM_{z})$, where $I$ is spacial inversion, $M$ is mirror reflection plane.
The full symmetry of magnetic structure of $\mathrm{Ce}_{3}\mathrm{NIn}$ \cite{ce3inn2023} in
Supplementary Fig.(S2) is exactly the non-coplanar SSG 123.2.8.35.
The symmetry invariants $\vec{\eta}=(-1,-1,+1,+1,+1,-1,-1)$ can be realized at
$\Gamma$,M,Z,A points in BZ and support 2-dimensional irRep generated by
\beq
\label{2dfermion}
\rho(T)K&=&i\sigma_{y}K,\rho(I)=i\sigma_{z},
\rho(C_{2x})=i\sigma_{x},\rho(C_{2y})=i\sigma_{y},
\nonumber\\
\rho(C_{2a})&=&\exp[-\frac{i(\sigma_{x}+\sigma_{y})}
{\sqrt{2}}\frac{\pi}{2}]
=-\frac{i(\sigma_{x}+\sigma_{y})}{\sqrt{2}}.
\eeq
From SEq.\eqref{2dfermion}, $\rho(IT)K=i\sigma_{x}K$,
$\eta_{IT}=[\rho(IT)K]^2=1$.
The band splitting around these points are linear dispersion in $k_{x},k_{y}$ plane
\beq
\hat{H}(\mathbf{k})=k_x\sigma_{x}+k_{y}\sigma_{y}.
\eeq
The $\eta_{IT}=1$ two-fold spinful fermion with linear dispersion in $k_{x},k_{y}$ plane is unique in SSGs
since $\eta_{IT}=1$ fermions in MSGs are spinless and carry single-valued irReps which cannot have linear band splitting in $k_{x},k_{y}$ directions.

\begin{figure}
\centering
\includegraphics[width=0.5\textwidth]{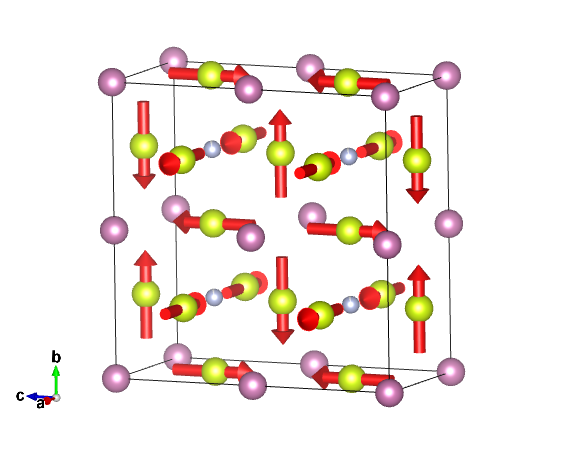}
\caption{\label{magce3nin}Magnetic structure of $\mathrm{Ce}_{3}\mathrm{NIn}$ with MAGNDATA
label \href{http://webbdcrista1.ehu.es/magndata/index.php?this_label=1.152}{1.152}
\cite{gallego2016magndataI,gallego2016magndataII}.
The red arrows on yellow balls indicate the directions of magnetic moments of $\mathrm{Ce}$ atoms.
The SSG symmetry
is 123.2.8.35 in the database \cite{ssg_website}.}
\end{figure}

\section{Quasiparticles in other spin-space groups}

\subsection{Supplementary Table \ref{tab:invrtstypeIII}: Symmetry invariants for SSGs isomorphic to type III MSGs}
For the non-coplanar SSGs isomorphic to type III MSGs, the little co-groups $G_{\mathbf{K}}$ at high-symmetry point $\bf K$
 can be 58 type III magnetic point groups of the form
$H+T(P-H)$, where $H$ is a subgroup of the point group $P$ containing half of the elements in $P$, the other half $P-H$ followed by time reversal $T$.
Following Eqs.
\eqref{eq:chiLp1p2}-\eqref{eq:chiSpt}
of the main text, we calculate lattice-part invariants $\eta^L$ and spin-part invariants $\eta^S$, then obtain the full invariants $\eta=\eta^L\eta^S$.
In Supplementary Table \ref{tab:invrtstypeIII}, we give definitions of symmetry invariants of
58 type III magnetic point groups.
For a type III magnetic point group, if not all the sets of symmetry invariants can be realized
in type III MSGs or at non-time-reversal-invariant high-symmetry points of type II and IV MSGs,
we list the values of all symmetry invariants,
where MSG-realizable in BLUE, and only SSG-realizable in RED.
The BLACK sets of invariants cannot be realized by either MSGs or SSGs.
We find additional 108 classes of new quasiparticles with type III magnetic point group symmetries only realizable in the  electronic bands of SSGs.
If the minimal dimension of irReps of the same class is FOUR or higher,
we provide the degree of degeneracy, the lowest-order dispersion, and the direction of nodal lines. An example is a four-fold degenerate 7-nodal-line-nexus fermion protected by $T_{d}\times Z_2^{IT}$ (No.57 type III magnetic point group in Supplementary Table \ref{tab:invrtstypeIII}), only realizable in SSGs, with three nodal lines along $k_x, k_y, k_z$ directions and four along $k_x+k_y+k_z$, $k_x+k_y-k_z$, $k_x-k_y+k_z$, and $k_x-k_y-k_z$ directions.

\subsection{Coplanar and collinear SSGs with nontrivial spin only groups $S_0$}\label{sec:invcpclssg}

The spin only group $S_0$ may have two consequences:\\
(I) $S_0$ may give rise to extra nontrivial symmetry invariants;\\
(II) $S_0$ may change (enlarge) the dimensions of some projective irReps, even if no nontrivial values of invariants are introduced by $S_0$.

{\bf More symmetry invariants.} Now we address the issue about the symmetry invariants for little co-groups with nontrivial spin only group $S_{0}$.
The little co-group $G_\mathbf{K}$ at momentum $\bf K$ has the following structure,
\beq
G_\mathbf{K} /S_{0\mathbf K}\cong  M'_{\mathbf K}
\eeq
where $S_{0\mathbf K}$ is the spin-only group at $\mathbf K$ (which is not necesarily the same with $S_0$ since anti-unitary elements in $S_0$ reverse the lattice momentum) and $M'_{\mathbf K}$ stands for the group formed by pure lattice operations or `spin associated' lattice operations. In most cases, the little co-group $G_\mathbf{K}$ is a direct product group with
$G_{\bf K}=S_{0\bf K}\times M'_{\bf K}$.
In other cases, $G_\mathbf{K}$ is a semidirect product group $G_{\bf K}=S_{0\bf K}\rtimes M'_{\bf K}$.

For unitary product groups of the form $G_1\times G_2$, we have the Kunneth formula (see SRef.\cite{PhysRevB.87.165107})
$\mathcal H^d(G_1\times G_2, \mathrm{U}(1)) = \bigoplus\limits_{p=0}^d \mathcal H^{d-p}[G_1,\mathcal H^p(G_2,\mathrm{U}(1))]$,
where $\mathcal{H}^{d}(G,\mathcal{A})$ is $d$th group cohomology with $\mathcal{A}$-coefficient
(usually the Abelian group $\mathcal{A}$ is a subgroup of $\mathrm{U}(1)$).
Similar formula exists for semidirect product groups $G_1\rtimes G_2$. For $d=2$, one has
\beq
\mathcal H^2(G_1\times G_2,\mathrm{U}(1)) = \mathcal H^2(G_1, \mathrm{U}(1)) \oplus \mathcal H^2(G_2,\mathrm{U}(1))\oplus
\mathcal H^1(G_1,\mathcal H^1(G_2, \mathrm{U}(1))).
\eeq
With slight modifications, this formular can be applied to anti-unitary groups, for instance,
\beq
\mathcal H^2(G\times Z_2^T, \mathrm{U}(1))= \mathcal H^2(Z_2^T, \mathrm{U}(1)) \oplus \mathcal H^1[\mathcal H^2(G, \mathrm{U}(1)),  Z_2] \oplus \mathcal H^1(G,Z_2).
\eeq

Generally, for the group $G_{\bf K}=S_{0\bf K}\times M'_{\bf K} $ or $G_{\bf K}=S_{0\bf K}\rtimes M'_{\bf K} $,
with $G_{\bf K}$ either unitary or anti-unitary, the symmetry invariants contain three parts: \\
\indent (1) invariants of $M'_{\mathbf K}$ from $\mathcal H^2(M'_{\mathbf K}, \mathrm{U}(1))$, \\
\indent (2) invariants of $S_{0\mathbf K}$ from $\mathcal H^2(S_{0\mathbf K},\mathrm{U}(1))$, \\
\indent (3) mutual invariants involving both $M'_{\mathbf K}$ and $S_{0\mathbf K}$.

In the following we discuss possible new invariants related to $S_0$, and will discuss the collinear and coplanar SSGs separately.

{\bf \noindent(I), Collinear SSG $\mathcal{G}$ with $S_0=\mathrm{SO}(2)\rtimes Z_{2}^{C_{2x}T}$}
\begin{itemize}
  \item[(a)] $(T||T|\pmb{\tau}) \in \mathcal{G}$, with the fractional translation $\pmb{\tau}\neq\mathbf{0}$.

Noticing that $(T||T|\pmb{\tau})$ acts on both lattice momentum and spin, and that
$(T||T|\pmb{\tau})\cdot(C_{2x}T||T|\mathbf{0})=(C_{2x}||E|\pmb{\tau})$, we can choose the spin-only group at $\bf K$ as the unitary group
\beq
  S_{0\bf K}=\mathrm{SO}(2)\rtimes \mathscr C_{2x}.
\eeq
Then we have
$
\mathcal H^2(S_{0\bf K}, \mathrm{U}(1))=\mathbb Z_2,
$
the corresponding invariant is
\beq
\eta_{x,z}=\eta^{S}_{x,z}=(\hat{C}_{2x}||E)(\hat{C}_{2z}||E)(\hat{C}_{2x}||E)^{-1}(\hat{C}_{2z}||E)^{-1}=(-1)^{2s},
\eeq
where $(C_{2z}||E)\in S_{0\bf K}$,
the hatted quantity are the representation of operators acting on single-particle Hilbert space,
$s$ is spin quantum number.
For spin-1/2, one has $\eta^{S}_{x,z} =-1$, indicating the 2-fold degeneracy in the whole BZ in the spin sector. This case belongs to the conventional antiferromagnetism(AFM).

For simplicity, we denote $M'_{\bf K}$ as the group formed by pure lattice operations (except for anti-unitary elements which act on spin in the same way as $T$), then $G_{\bf K}=S_{0\bf K}\times M'_{\bf K} $. Since $T$ acts on both lattice and spin degrees of freedom, there may exist some nontrivial mutual invariants
contributed from $S_{0\bf K}$ and $M'_{\bf K}$. For instance, for
$(C_{2m}||E)\in S_{0\bf K}$ with $m=x,y,z$, an SSG operation $(C_{2m}T||T|\pmb{\tau})$,
the mutual invariant
\beq
\eta_{(C_{2m}T||T)}=(\hat{C}_{2m}\hat{T}||\hat{T})^2=
\eta_{(C_{2m}T||T)}^{S}\eta_{(C_{2m}T||T)}^{L}=\eta^{S}_{C_{2m}T}\eta_{T}^{L}=\eta_{T}^{L}
\eeq
may take nontrivial value, here we have used the result
$\eta^{S}_{C_{2m}T}=(\hat{C}_{2m}\hat{T})^{2}=1$
for both integer and half-odd-integer spin,
and $\eta^L_{T}=\exp(-i\mathbf{K}_{T}\cdot \pmb{\tau})$ from
Eq.\eqref{eq:chiLpt} of the main text.
Hence all nontrivial mutual invariants are originating from the lattice invariant
$\eta_{T}^{L}$.
Notice that the invariant
$\eta_{(C_{2m}T||T)}=\eta_{T}^L$
is not the same as the invariant $\eta_{(T||T)}$
since
$\eta_{(T||T)}=(\hat{T}||\hat{T})^2=\eta_{T}^{S}\eta_{T}^{L}=(\hat{T})^{2}\eta_{T}^{L}=(-1)^{2s}\eta_{T}^{L}$.





  \item[(b)] $(T||IT|\pmb{\tau})\in \mathcal{G}$ with $|\pmb{\tau}|\geq0$.

   In the operation $(T||IT|\pmb{\tau})$, spacial inversion
   $I$ commutes with all the other spatial operations and it acts trivially on the spin sector, the discussion in case (a) can be generalized to the case $(T||IT|\pmb{\tau})\in\mathcal{G}$. Since
   $(T||IT|\pmb{\tau})\cdot(C_{2x}T||T|\mathbf{0})=(C_{2x}||I|\pmb{\tau})$, we just define the `spin-only group' at $\bf K$ as
  \beq
  S_{0\bf K}=\mathrm{SO}(2)\rtimes \mathscr C'_{2x}
  \eeq
  with $\mathscr C'_{2x}=\{E, ( C_{2x}||I)\}$. Then we still have
$
  \mathcal H^2(S_{0\bf K}, \mathrm{U}(1))=\mathbb Z_2,
$
with the invariant
  \beq 
  \eta_{x,z} = (\hat{C}_{2x}||\hat{I})(\hat{C}_{2z}||E)(\hat{C}_{2x}||\hat{I})^{-1}(\hat{C}_{2z}||E)^{-1}
  =(-1)^{2s}.
  \eeq
  For spin-1/2, $\eta_{x,z} =-1$, one also has 2-fold degeneracy in the whole BZ in the spin sector, which also
  belongs to the conventional AFM.
  Notice that the spin degeneracy is also characterized by
  $\eta_{IT} =(\hat{T}||\hat{I}\hat{T})^2=\eta_{T}^{S}\eta_{IT}^{L}=\eta_{T}^{S}=\hat{T}^{2}=(-1)^{2s}=\eta_{x,z}=-1$
  since $\eta^L_{IT}=\exp(-i\mathbf{K}_{IT}\cdot \pmb{\tau})=1$ from
  Eq.\eqref{eq:chiLpt} of the main text.

We also denote $M'_{\bf K}$ as the group formed by pure lattice operations (except for anti-unitary elements which act on spin in the same way as $T$), then we still have the structure $G_{\bf K}=S_{0\bf K}\times M'_{\bf K} $. There are no nontrivial mutual invariants in this case.

%

    \item[(c)] $(T||T|\pmb{\tau})\notin \mathcal{G}$ and $(T | |I T|\pmb{\tau}) \notin \mathcal{G}$.

    If $\bf K$ is a time-reversal-invariant momentum(TRIM),  then we choose
    \beq
    S_{0\bf K}=\mathrm{SO}(2)\rtimes Z_2^{C_{2x}T}=S_0.
    \eeq
    In this case we have $\mathcal H^2(S_{0\bf K}, \mathrm{U}(1))=\mathbb Z_2$
    which is generated by the symmetry invariant
    \beq
    \eta_{(C_{2x}T||T)}=(\hat{C}_{2x}\hat{T}||\hat T)^2
    =\eta_{C_{2x}T}^{S}\eta_{T}^{L}=\eta_{C_{2x}T}^{S}
    =1,
    \eeq
    which always takes trivial values for both integer and half-odd-integer spins.
    Here, $\eta_{T}^{L}=1$ from Eq.\eqref{eq:chiLpt} of the main text because the fractional translation of $T$ is $\mathbf{0}$.

    If $\bf K$ is not a TRIM,  we choose
    \beq
    S_{0\bf K}=\mathrm{SO}(2).
    \eeq
    In this case we have  $\mathcal H^2(S_{0\bf K}, \mathrm{U}(1))=\mathbb Z_1$, so there is no invariants for $S_{0\bf K}$ alone, indicating that the Rep in the spin sector
    belongs to the trivial class and generally no degeneracy is protected.

    Hence, spin splitting exists for general points in the BZ except for some special points (see below).
    Especially, if $\tilde g=(C_{2x}||g|\pmb{\tau}_{g})\in \mathcal{G}$ where $g\neq I$ is unitary, then the magnetic conuration is antiferromagnetic and corresponds to the {\it altermagnetism} (ALM) in literature.
    Otherwise, if such $\tilde g$ does not exist, then the magnetic conuration is either ferromagnetic or ferrimagnetic.

    For momentum $\bf K$ no matter TRIM or not,
    we denote $M'_{\bf K}$ as the group formed by `spin associated' lattice operations. In the following we discuss two different situations of $M'_{\bf K}$.

   (1) $M'_{\bf K}$ does not contain $C_{2x}$ spin operation. In this case, we still have $G_{\bf K}=S_{0\bf K}\times M'_{\bf K} $. There are no nontrivial mutual invariants.

   (2) $M'_{\bf K}$ contains $C_{2x}$ spin operation. In this case $G_{\bf K}=S_{0\bf K}\rtimes M'_{\bf K} $. If there exists an element $g' = (C_{2x}||g)\in M'_{\bf K}$ with $g$ unitary ($\bf K$ is invariant under the action of $g$),
   then for $(C_{2z}||E)\in S_{0\bf K}$ there exists nontrivial mutual invariant
   \beq
   \eta_{(C_{2x}||g), (C_{2z}||E)}
   =(\hat{C}_{2x}||\hat{g})(\hat{C}_{2z}||E)
    (\hat{C}_{2x}||\hat{g}|)^{-1}(\hat{C}_{2z}||E)^{-1}
   =(-1)^{2s}.
   \eeq
   This nontrivial invariant protects the spin degeneracy at $\bf K$ for $s=1/2$ quasiparticles.
   Otherwise, there are not nontrivial mutual invariants.


\end{itemize}

{\bf \noindent(II), Coplanar SSG $\mathcal{G}$ with $S_0=Z_{2}^{C_{2z}T}$}

In this case $M'_{\bf K}$ denotes the group formed by `spin associated' lattice operations.
\begin{itemize}
  \item[(a)] $(T || T | \pmb{\tau}) \in \mathcal{G}$ with the fractional translation $\pmb{\tau}\neq \mathbf{0}$.

  In this case, from $(T||T|\pmb{\tau})\cdot(C_{2z}T||T|\mathbf{0})=(C_{2z}||E|\pmb{\tau})$ we can define the spin-only group at $\bf K$ as a unitary group
  \beq
  S_{0\bf K}=\mathscr C_{2z}=\{E,(C_{2z}||E)\}.
  \eeq
  Since $\mathcal H^2(S_{0\bf K}, \mathrm{U}(1))=\mathbb Z_1$, there are no symmetry invariant for $S_{0\bf K}$ only.

  If the SSG contains other spin operations such as $C_{2x}$ or $C_{2y}$ (namely if there exist some lattice operation $g$ such that $(C_{2x}||g|)\in M'_{\bf K}$ or $(C_{2y}||g)\in M'_{\bf K}$), then there will exist nontrivial mutual invariant:
   \beq
   \eta_{(C_{2x}||g), (C_{2z}||E)}
   =(\hat{C}_{2x}||\hat{g})(\hat{C}_{2z}||E)
    (\hat{C}_{2x}||\hat{g})^{-1}(\hat{C}_{2z}||E)^{-1}
   =\eta_{x,z}^{S}
   =(-1)^{2s},
   \eeq
  which originates from the spin invariant $\eta^{S}_{x,z}$ and takes nontrivial value $(-1)^{2s}=-1$ for spin-1/2.


 \item[(b)] $(T || IT |\pmb{\tau}) \in \mathcal{G}$ with $|\pmb{\tau}|\geq0$.

 In this case, from $(T||IT|\pmb{\tau})\cdot(C_{2z}T||T|\mathbf{0})=(C_{2z}||I|\pmb{\tau})$ we can define the spin-only group at $\bf K$ as a unitary group
 \beq
 S_{0\bf K}=\mathscr C'_{2z}=\{E,(C_{2z}||I)\}.
 \eeq
 Since $\mathcal H^2(S_{0\bf K}, \mathrm{U}(1))=\mathbb Z_1$, there are no symmetry invariants for $S_{0\bf K}$ only.

 If the SSG contains other spin operations such as $C_{2x}$ or $C_{2y}$ (namely if there exist some lattice operation $g$ such that
 $(C_{2x}||g)\in M'_{\bf K}$ or $(C_{2y}||g)\in M'_{\bf K}$), then there will exist nontrivial mutual invariant
  \beq
   \eta_{(C_{2x}||g), (C_{2z}||I)}
   =(\hat{C}_{2x}||\hat{g})(\hat{C}_{2z}||\hat{I})
    (\hat{C}_{2x}||\hat{g})^{-1}(\hat{C}_{2z}||\hat{I})^{-1}
   =\eta_{x,z}^{S}\eta_{I,g}^{L}
   =(-1)^{2s}\eta_{I,g}^{L},
  \eeq
   which may contain contribution from both spin and lattice invariants.
   Here $\eta_{I,g}^{L}$ can be calculated from Eq.\eqref{eq:chiLp1p2} of the main text.

%


  \item[(c)]$(T||T | \pmb{\tau})\notin \mathcal{G}$ and $(T || IT |\pmb{\tau}) \notin \mathcal{G}$.

  If $\bf K$ is not a time-reversal-invariant point, then
  \beq
  S_{0\bf K}=\{E\}
  \eeq
  is trivial. So the symmetry invariants are completely determined by $M'_{\bf K}$.

  If $\bf K$ is a time-reversal-invariant point, we can choose
  \beq
  S_{0\bf K}=S_0=Z_{2}^{C_{2z}T}.
  \eeq
  In this case $\mathcal H^2(S_{0\bf K}, \mathrm{U}(1))=\mathbb Z_2$,
  but the corresponding symmetry invariant
    \beq
    \eta_{(C_{2z}T||T)}=(\hat{C}_{2z}\hat{T}||\hat T)^2
    =\eta_{C_{2z}T}^{S}\eta_{T}^{L}=\eta_{C_{2z}T}^{S}
    =1,
    \eeq
    which always takes trivial value for both integer and half-odd-integer spins.
    Here, $\eta_{T}^{L}=1$ from Eq.\eqref{eq:chiLpt} of the main text because the fractional translation of $T$ is $\mathbf{0}$.
If there is a pure lattice operation $(E||g)\in M'_{\bf K}$ with $g^2=E$, from $(E||g|\pmb{\tau}_g)(C_{2z}T||T|\mathbf{0})=(C_{2z}T||gT|\pmb{\tau}_g)$,
then there exists nontrivial mutual invariant
\beq
\eta_{(C_{2z}T||gT)}=(\hat C_{2z}\hat{T}||\hat g\hat{T})^2
=\eta_{C_{2z}T}^{S}\eta_{gT}^{L}
=\eta_{gT}^{L},
\eeq
which originates from the lattice invariant $\eta_{gT}^{L}=\exp(-i\mathbf{K}_{gT}\cdot \pmb{\tau}_g)$ from
  Eq.\eqref{eq:chiLpt} of the main text.


 \end{itemize}


{\bf \noindent(III), Non-magnetic SSG $\mathcal{G}$ with no SOC and with $S_0=\mathrm{SO}(3)$}

In this case, the spin sector and the lattice (orbital) sector are completely separated. The spin sector has $(2s+1)$-fold degeneracy, and all the other symmetry invariants of unitary symmetry elements are contributed from the non-symmorphic lattice operations. The spin $\mathrm{SO}(3)$ group (for $s=1/2$ it is actually a $\mathrm{SU}(2)$ group) may indeed have nontrivial physical consequence. An example is given in the main text when comparing the 12-fold fermions with or without $\mathrm{SU}(2)$ spin symmetry.

 In summary, introducing of $S_0$ can give rise to new symmetry invariants with nontrivial values: (I) in collinear SSGs for conventional antiferromagnets, invariant from $S_{0\bf K}$ alone can take nontrivial value, giving rise to spin degeneracy in the whole Brillouin zone; (II) generally there may exist mutual invariants from $S_{0\bf K}$ and $M'_{\bf K}$ taking nontrivial values.

{\bf Higher dimensional irReps.} Since $S_0$ enlarges the group, the dimension of irReps may be enlarged even if
the symmetry invariants remain unchanged. For instance, the three SSGs
\href{https://cmpdc.iphy.ac.cn/ssg/ssgs/229.2.1.9} {229.2.1.9} ({non-coplanar}), \href{https://cmpdc.iphy.ac.cn/ssg/ssgs/229.2.1.6.P}{229.2.1.6.P} ({coplanar}) and \href{https://cmpdc.iphy.ac.cn/ssg/ssgs/229.2.1.3.L}{229.2.1.3.L} (collinear)
differ by their spin-only groups, namely ${ S_0=\{E\}}$, ${ S_0'=Z_2^{C_{2z}T}=\{E, {C_{2z}T}\}}$ and ${S_0''=\mathrm{SO}(2)\rtimes Z_2^{C_{2x} T} }$ respectively.
At the $R=(\pi, \pi, \pi)$ point, the little co-groups of the three SSGs are $G_\mathbf{K} = { O_h\times Z_2^T},\ { O_h\times Z_2^T\times S_0'}$ and ${ O_h\times Z_2^T\times S_0''}$ respectively. As far as the common subgroup $O_h\times Z_2^T$ is concerned, the five symmetry invariants $(\eta_{C_{2x},C_{2y}}, \eta_T, \eta_{IT}, \eta_{TC_{2a}}, \eta_{I,C_{2a}})$ take the same values $(+1,+1,-1,-1,-1)$ in the three SSGs. However, at the $R$ point the irReps for $O_h\times Z_2^T$ are not the same for the three SSGs:  the group { 229.2.1.9} has 12-dimensional (12-D) and  4-D irReps, but both of { 229.2.1.6.P} and { 229.2.1.3.L} have  12-D, 8-D and 4-D irReps. The three groups all have a 12-D irRep and have the same $k\cdot p$ theory around the 12-fold degenerate point ($R$ point).

\begin{table*}[htbp]
\caption{The definitions of symmetry invariants of { $G=P\times Z^T_{2}$} 
are given in terms of factor system $\omega$,
where $P$ is any one of the 32 crystallographic point groups,
$I={\rm spacial\ inversion}$, $M={\rm mirror\ reflection\ plane}$, $T={\rm time\ reversal}$,
$Z_{2}^{T}=\{E, T\}$.
The invariants are interpreted as the following: $\eta_T\equiv\omega_{2}(T,T)$,
$\eta_{IT}\equiv\omega_{2}(IT,IT)$,
$\eta_{TC_{2}}\equiv\omega_{2}(TC_{2},TC_{2})$,
$\eta_{TM}\equiv\omega_{2}(TM,TM)$,
$\eta_{I,C_{2z}}\equiv\frac{\omega_{2}(I,C_{2z})}{\omega_{2}(C_{2z},I)}$,
$\eta_{I,C_{2x}}\equiv\frac{\omega_{2}(I,C_{2x})}{\omega_{2}(C_{2x},I)}$,
$\eta_{I,C_{2y}}\equiv\frac{\omega_{2}(I,C_{2y})}{\omega_{2}(C_{2y},I)}$,
$\eta_{I,C_{2a}}\equiv\frac{\omega_{2}(I,C_{2a})}{\omega_{2}(C_{2a},I)}$,
$\eta_{I,C'_{21}}\equiv\frac{\omega_{2}(I,C'_{21})}{\omega_{2}(C'_{21},I)}$,
$\eta_{I,C''_{21}}\equiv\frac{\omega_{2}(I,C''_{21})}{\omega_{2}(C''_{21},I)}$,
$\eta_{I,C^{+}_{4z}}\equiv\frac{\omega_{2}(I,C^{+}_{4z})}{\omega_{2}(C^{+}_{4z},I)}$,
$\eta_{I,C^{}_{2}}\equiv\frac{\omega_{2}(I,C^{}_{2})}{\omega_{2}(C^{}_{2},I)}$,
$\eta_{C_{2x},C_{2y}}\equiv\frac{\omega_{2}(C_{2x},C_{2y})}{\omega_{2}(C_{2y},C_{2x})}$,
$\eta_{M_{x},M_{y}}\equiv\frac{\omega_{2}(M_{x},M_{y})}{\omega_{2}(M_{y},M_{x})}$,
$\eta_{C_{2},C'_{21}}\equiv\frac{\omega_{2}(C_{2},C'_{21})}{\omega_{2}(C'_{21},C_{2})}$,
$\eta_{C_{2},M_{d1}}\equiv\frac{\omega_{2}(C_{2},M_{d1})}{\omega_{2}(M_{d1},C_{2})}$,
$\eta_{M_{h},C'_{21}}\equiv\frac{\omega_{2}(M_{h},C'_{21})}{\omega_{2}(C'_{21},M_{h})}$,
$\eta_{M_{h},M_{d1}}\equiv\frac{\omega_{2}(M_{h},M_{d1})}{\omega_{2}(M_{d1},M_{h})}$,
$\eta_{C'_{21},C''_{21}}\equiv\frac{\omega_{2}(C'_{21},C''_{21})}{\omega_{2}(C''_{21},C'_{21})}$.} \label{tab:invrts}
\centering
\begin{tabular}{ |c|c|c| }
\hline
 $G$ & Generators & Invariants\\ 
\hline
$C_{1}^{}\times Z_{2}^{T}$& $T$ &
$\eta_{T}$\\



\hline

$C_{i}^{}\times Z_{2}^{T}$& $I,T$ &
($\eta_{T}$,$\eta_{IT}$) \\

\hline

$\mathscr{C}_{2}^{}\times Z_{2}^{T}$& $C_{2z},T$ &
($\eta_{T}$,$\eta_{TC_{2z}}$) \\





\hline

$C_{1h}^{}\times Z_{2}^{T}$& $M_{z},T$ &
($\eta_{T}$,$\eta_{TM_{z}}$) \\



\hline

$C_{2h}^{}\times Z_{2}^{T}$& $C_{2z},I,T$&
($\eta_{I,C_{2z}}$,$\eta_T$ ,$\eta_{TC_{2z}}$, $\eta_{IT}$)  \\








\hline

$D_{2}^{}\times Z_{2}^{T}$& $C_{2x},C_{2y},T$ &
($\eta_{C_{2x},C_{2y}}$,$\eta_{T}$,
$\eta_{TC_{2x}}$,$\eta_{TC_{2y}}$)  \\


\hline

$C_{2v}^{}\times Z_{2}^{T}$& $M_{x},M_{y},T$ &
($\eta_{M_{x},M_{y}}$,$\eta_{T}$,
$\eta_{TM_{x}}$,$\eta_{TM_{y}}$) \\


\hline

$D_{2h}^{}\times Z_{2}^{T}$& $C_{2x}, C_{2y},I,T$ &
($\eta_{C_{2x},C_{2y}}$, $\eta_T$,$\eta_{IT}$,
$\eta_{TC_{2x}}$,$\eta_{TC_{2y}}$, $\eta_{I,C_{2x}}$,
$\eta_{I,C_{2y}}$) \\



\hline

$C_{4}\times Z_{2}^{T}$& $C_{4z}^{+},T$ &
($\eta_{T}$,$\eta_{TC_{2z}}$) \\

\hline

$S_{4}\times Z_{2}^{T}$& $S_{4z}^{-},T$ &
($\eta_{T}$,$\eta_{TC_{2z}}$) \\

\hline

$C_{4h}\times Z_{2}^{T}$& $C_{4z}^{+},I,T$ &
($\eta_{I,C_{4z}^{+}}$,$\eta_T$,$\eta_{IT}$,
$\eta_{TM_{z}}$) \\
\hline

$D_{4}^{}\times Z_{2}^{T}$& $C_{4z}^{+},C_{2x},T$ &
($\eta_{C_{2x},C_{2y}}$,$\eta_{T}$,$\eta_{TC_{2x}}$,$\eta_{TC_{2a}}$)  \\



\hline

$C_{4v}\times Z_{2}^{T}$& $C_{4z}^{+},M_{x},T$ &
($\eta_{M_{x},M_{y}}$,$\eta_{T}$,
$\eta_{TM_{x}}$,$\eta_{TM_{da}}$) \\

\hline

$D_{2d}^{1}\times Z_{2}^{T}$& $S_{4z}^{-},C_{2x},T$ &
($\eta_{C_{2x},C_{2y}}$,$\eta_{T}$,$\eta_{TC_{2x}}$,$\eta_{TM_{da}}$)  \\

$D_{2d}^{2}\times Z_{2}^{T}$& $S_{4z}^{-},M_{x},T$ &
($\eta_{M_{x},M_{y}}$,
$\eta_T$, $\eta_{TM_{x}}$,$\eta_{TC_{2a}}$)\\




\hline

$D_{4h}^{}\times Z_{2}^{T}$&
$C_{4z}^{+},C_{2x},I,T$ &
($\eta_{C_{2x},C_{2y}}$, $\eta_T$,$\eta_{IT}$,
$\eta_{TC_{2x}}$,$\eta_{TC_{2a}}$, $\eta_{I,C_{2x}}$,
$\eta_{I,C_{2a}}$)  \\


\hline

$C_{3}\times Z_{2}^{T}$& $TC_{3}^{+}$ &$\eta_{T}$ \\



\hline

$C_{3i}\times Z_{2}^{T}$& $S_{6}^{+},T$ &($\eta_{T}$,$\eta_{IT}$) \\

\hline

$D_{3}^{1}\times Z_{2}^{T}$ & $TC_{3}^{+},C'_{21}$ &
($\eta_{T}$, $\eta_{TC'_{21}}$) \\

$D_{3}^{2}\times Z_{2}^{T}$ & $TC_{3}^{+},C''_{21}$ &
($\eta_{T}$, $\eta_{TC''_{21}}$) \\


\hline

$C_{3v}^{1}\times Z_{2}^{T}$ & $TC_{3}^{+},M_{d1}$ &
($\eta_{T}$, $\eta_{TM_{d1}}$) \\

$C_{3v}^{2}\times Z_{2}^{T}$ & $TC_{3}^{+},M_{v1}$ &
($\eta_{T}$, $\eta_{TM_{v1}}$) \\

\hline

$D_{3d}^{1}\times Z_{2}^{T}$& $S_{6}^{+},C'_{21},T$ &
($\eta_{I,C'_{21}}$,$\eta_T$,$\eta_{IT}$,$\eta_{TC'_{21}}$) \\

$D_{3d}^{2}\times Z_{2}^{T}$& $S_{6}^{+},C''_{21},T$ &
($\eta_{I,C''_{21}}$,$\eta_T$,$\eta_{IT}$,$\eta_{TC''_{21}}$)  \\




\hline

%

$C_{6}\times Z_{2}^{T}$& $C_{6}^{+},T$ &($\eta_{T}$,$\eta_{TC_{2}}$) \\

\hline

$C_{3h}\times Z_{2}^{T}$& $S_{3}^{-},T$ &($\eta_{T}$,$\eta_{TM_{h}}$) \\

\hline

$C_{6h}\times Z_{2}^{T}$ & $C_{6}^{+},I, T$ &
($\eta_{I,C_{2}^{}}$,$\eta_T$,$\eta_{IT}$,
$\eta_{TM_{h}}$)  \\
\hline

$D_{6}\times Z_{2}^{T}$ & $C_{6}^{+},C'_{21},T$ & ($\eta_{C_{2},C'_{21}}$,
$\eta_{T}$, $\eta_{TC_{2}}$ ,$\eta_{TC'_{21}}$)  \\

\hline

$C_{6v}\times Z_{2}^{T}$& $C_{6}^{+},M_{d1},T$ &
($\eta_{C_{2},M_{d1}}$,
$\eta_{T}$,$\eta_{TC_{2}}$ ,$\eta_{TM_{d1}}$) \\

\hline

$D_{3h}^{1}\times Z_{2}^{T}$ & $S_{3}^{-},C'_{21},T$ &
($\eta_{M_{h},C'_{21}}$,
$\eta_{T}$, $\eta_{TM_{h}}$,$\eta_{TC'_{21}}$)  \\

$D_{3h}^{2}\times Z_{2}^{T}$ & $S_{3}^{-},M_{d1},T$ &
($\eta_{M_{h},M_{d1}}$,
$\eta_{T}$, $\eta_{TM_{h}}$ ,$\eta_{TM_{d1}}$)  \\


\hline

$D_{6h}^{}\times Z_{2}^T$
& $C_{6}^{+},C'_{21},I,T$
&($\eta_{C'_{21},C''_{21}}$, $\eta_T$,$\eta_{IT}$,
$\eta_{TC'_{21}}$,$\eta_{TC''_{21}}$, $\eta_{I,C'_{21}}$,$\eta_{I,C''_{21}}$)  \\
\hline

$T$\footnote{Here, $T$ is point group $T\equiv\{E,C_{31,32,33,34}^{\pm},C_{2x,2y,2z}\}$.} $\times Z_{2}^{T}$
&$C_{31}^{-},C_{33}^{+},T$
&($\eta_{T},\eta_{TC_{2z}}$)\\

\hline

$T_{h}\times Z_{2}^T$
&$C_{31}^{-},C_{33}^{+},I,T$
&($\eta_{C_{2x},C_{2y}}$,
$\eta_T$,$\eta_{IT}$)\\
\hline

$O\times Z_{2}^{T}$
&$C_{2x}$,$C_{2y}$,$C_{2a}$,$C_{2f}$,$T$
&($\eta_{T},\eta_{TC_{2z}},\eta_{TC_{2f}}$)\\

\hline

$T_{d}\times Z_{2}^{T}$
&$C_{2x}$,$C_{2y}$,$M_{da}$,$M_{df}$,$T$
&($\eta_{T},\eta_{TC_{2z}},\eta_{TM_{df}}$)\\


\hline

$O_{h}\times Z_{2}^T$
&$C_{2x}$,$C_{2y}$,$C_{2a}$,$C_{2f}$,$I$,$T$
&($\eta_{C_{2x},C_{2y}}$, $\eta_T$,$\eta_{IT}$,$\eta_{TC_{2a}}$,$\eta_{I,C_{2a}}$)\\
\hline

\end{tabular}
\end{table*}


\begin{table*}[htbp]
\caption{Lattice invariants of { $G_{\mathbf K}\cong M_{\mathbf K}=P\times Z_{2}^{T}$}
with $P=C_{3h,4h,6h},D_{3,4,6,3d,3h,4h,6h},O,O_{h}$. Here $G_{\mathbf K}$ and $M_{\mathbf K}$ stand for the little co-groups of SSG $\mathcal G$ and type IV MSG $\mathcal{M}$ respectively, with $\mathcal{M}$ the lattice part of $\mathcal{G}$. It follows that $\mathcal{G}\cong\mathcal{M}$ and $M_\mathbf{K}\cong{G}_\mathbf{K}$.
The lattice invariants $\vec{\eta}^L$ of $G_{\mathbf K}$ are same as those of $M_{\mathbf K}$, which can be computed as $\eta^L_{p_1,p_2}=e^{-i(\mathbf{K}_{p_{1}}\cdot\mathbf{t}_{p_2} - \mathbf{K}_{p_{2}}\cdot\mathbf{t}_{p_1} )}$ if $p_1, p_2$ are unitary with $p_{1}p_{2}=p_{2}p_{1}$ and $\eta^L_{p}=e^{-i\mathbf{K}_{p}\cdot\mathbf{t}_{p}}$ if $p$ is anti-unitary with $p^2=E$,
and $\omega_2(g_1,g_2)\equiv e^{-i\mathbf{K}_{g_1}\cdot\mathbf{t}_{g_2}}(g_{1},g_{2}\in M_{\mathbf K})$
is the factor system of projective Rep of $M_{\mathbf{K}}$ at high-symmetry point in BZ \cite{Yang2021}.}\label{Pdrztlat}
\centering
\begin{tabular}{ |c|cc|c|c| }
\hline
Shubnikov space group &high-symmetry point $\mathbf{K}$& $M_{\mathbf K}$ & lattice invariants $\vec{\eta}^{L}$ &\\
\hline

83.48&$\Gamma$&$C_{4h}\times Z_{2}^T$&($+1,+1,+1,+1$)&$0000$ \\

&M&$C_{4h}\times Z_{2}^T$&($+1,+1,+1,+1$)&$0000$ \\

 &Z&$C_{4h}\times Z_{2}^T$&($+1,-1,+1,+1$)&$0100$ \\

 &A&$C_{4h}\times Z_{2}^T$&($+1,-1,+1,+1$)&$0100$ \\

%
\hline
83.49&$\Gamma$&$C_{4h}\times Z_{2}^T$&($+1,+1,+1,+1$)&$0000$ \\

&M&$C_{4h}\times Z_{2}^T$& ($+1,+1,+1,+1$)&$0000$ \\

&Z&$C_{4h}\times Z_{2}^T$& ($+1,+1,+1,+1$)&$0000$ \\
\hline
\end{tabular}
\end{table*}

\begin{table*}[htbp]
(Extension of Supplementary Table \ref{Pdrztlat})\\
\centering
\begin{tabular}{ |c|cc|c|c| }
\hline
Shubnikov space group &high-symmetry point $\mathbf{K}$& $M_{\mathbf K}$ & lattice invariants $\vec{\eta}^{L}$&\\
\hline

89.92&$\Gamma$&$D_{4}^{}\times Z_{2}^T$&($+1,+1,+1,+1$) & $0000$ \\

&M&$D_{4}^{}\times Z_{2}^T$&($+1,+1,+1,+1$) & $0000$ \\

&Z&$D_{4}^{}\times Z_{2}^T$&($+1,-1,+1,+1$) & $0100$ \\

&A&$D_{4}^{}\times Z_{2}^T$&($+1,-1,+1,+1$) & $0100$ \\
\hline

89.93&$\Gamma$&$D_{4}^{}\times Z_{2}^T$&($+1,+1,+1,+1$) & $0000$ \\

&M&$D_{4}^{}\times Z_{2}^T$&($+1,+1,-1,+1$) & $0010$ \\

&Z&$D_{4}^{}\times Z_{2}^T$&($+1,+1,+1,+1$) & $0000$ \\
\hline
\end{tabular}
\end{table*}

\begin{table*}[htbp]
(Extension of Supplementary Table \ref{Pdrztlat})\\
\centering
\begin{tabular}{ |c|cc|c|c| }
\hline
Shubnikov space group &high-symmetry point $\mathbf{K}$& $M_{\mathbf K}$ & lattice invariants $\vec{\eta}^{L}$&\\
\hline

93.126&$\Gamma$&$D_{4}^{}\times Z_{2}^T$&($+1,+1,+1,+1$) & $0000$ \\

&M&$D_{4}^{}\times Z_{2}^T$&($+1,+1,-1,+1$) & $0010$ \\

&Z&$D_{4}^{}\times Z_{2}^T$&($+1,-1,+1,+1$) & $0100$ \\

&A&$D_{4}^{}\times Z_{2}^T$&($+1,-1,-1,+1$) & $0110$ \\


\hline

94.132&$\Gamma$&$D_{4}^{}\times Z_{2}^T$&($+1,+1,+1,+1$) & $0000$ \\

&M&$D_{4}^{}\times Z_{2}^T$&($+1,+1,-1,+1$) & $0010$ \\

&Z&$D_{4}^{}\times Z_{2}^T$&($+1,-1,+1,+1$) & $0100$ \\
\hline
\end{tabular}
\end{table*}

\begin{table*}[htbp]
(Extension of Supplementary Table \ref{Pdrztlat})\\
\centering
\begin{tabular}{ |c|cc|c|c| }
\hline
Shubnikov space group &high-symmetry point $\mathbf{K}$& $M_{\mathbf K}$ & lattice invariants $\vec{\eta}^{L}$&\\
\hline

124.360&$\Gamma$&$D_{4h}^{}\times Z_{2}^T$&($+1,+1,+1,+1,+1,+1,+1$)&$0000000$ \\

&M&$D_{4h}^{}\times Z_{2}^T$&($+1,+1,+1,+1,+1,+1,+1$)&$0000000$ \\

&Z&$D_{4h}^{}\times Z_{2}^T$&($+1,-1,+1,+1,+1,-1,-1$)&$0100011$ \\

&A&$D_{4h}^{}\times Z_{2}^T$&($+1,-1,+1,+1,+1,-1,-1$)&$0100011$ \\


\hline

124.361&$\Gamma$&$D_{4h}^{}\times Z_{2}^T$&($+1,+1,+1,+1,+1,+1,+1$)&$0000000$ \\

&M&$D_{4h}^{}\times Z_{2}^T$&($+1,+1,+1,-1,+1,+1,+1$)&$0001000$ \\

&Z&$D_{4h}^{}\times Z_{2}^T$&($+1,+1,+1,+1,+1,-1,-1$)&$0000011$ \\
\hline

\end{tabular}
\end{table*}

\begin{table*}[htbp]
(Extension of Supplementary Table \ref{Pdrztlat})\\
\centering
\begin{tabular}{ |c|cc|c|c| }
\hline
Shubnikov space group &high-symmetry point $\mathbf{K}$& $M_{\mathbf K}$ & lattice invariants $\vec{\eta}^{L}$& \\
\hline

128.410&$\Gamma$&$D_{4h}^{}\times Z_{2}^T$&($+1,+1,+1,+1,+1,+1,+1$)&$0000000$ \\

&M&$D_{4h}^{}\times Z_{2}^T$&($+1,+1,+1,+1,+1,+1,+1$)&$0000000$ \\

&Z&$D_{4h}^{}\times Z_{2}^T$&($+1,-1,+1,+1,+1,-1,-1$)&$0100011$ \\

&A&$D_{4h}^{}\times Z_{2}^T$&($+1,-1,+1,+1,+1,-1,-1$)&$0100011$ \\


\hline

129.420&$\Gamma$&$D_{4h}^{}\times Z_{2}^T$&($+1,+1,+1,+1,+1,+1,+1$)&$0000000$ \\

&M&$D_{4h}^{}\times Z_{2}^T$&($+1,+1,+1,-1,+1,-1,+1$)&$0001010$ \\

&Z&$D_{4h}^{}\times Z_{2}^T$&($+1,-1,+1,+1,+1,+1,+1$)&$0100000$ \\
\hline

\end{tabular}
\end{table*}

\begin{table*}[htbp]
(Extension of Supplementary Table \ref{Pdrztlat})\\
\centering
\begin{tabular}{ |c|cc|c|c| }
\hline
Shubnikov space group &high-symmetry point $\mathbf{K}$& $M_{\mathbf K}$ & lattice invariants $\vec{\eta}^{L}$& \\
\hline

133.469&$\Gamma$&$D_{4h}^{}\times Z_{2}^T$&($+1,+1,+1,+1,+1,+1,+1$)&$0000000$ \\

&M&$D_{4h}^{}\times Z_{2}^T$&($+1,+1,+1,-1,+1,-1,+1$)&$0001010$ \\

&Z&$D_{4h}^{}\times Z_{2}^T$&($+1,+1,+1,+1,+1,+1,-1$)&$0000001$ \\

&A&$D_{4h}^{}\times Z_{2}^T$&($+1,+1,+1,-1,+1,-1,-1$)&$0001011$ \\


\hline
133.470&$\Gamma$&$D_{4h}^{}\times Z_{2}^T$&($+1,+1,+1,+1,+1,+1,+1$)&$0000000$ \\

&M&$D_{4h}^{}\times Z_{2}^T$&($+1,+1,+1,-1,+1,-1,+1$)&$0001010$ \\

&Z&$D_{4h}^{}\times Z_{2}^T$&($+1,-1,+1,+1,+1,+1,-1$)&$0100001$ \\
\hline
\end{tabular}
\end{table*}

\begin{table*}[htbp]
(Extension of Supplementary Table \ref{Pdrztlat})\\
\centering
\begin{tabular}{ |c|cc|c|c| }
\hline
Shubnikov space group &high-symmetry point $\mathbf{K}$& $M_{\mathbf K}$ & lattice invariants $\vec{\eta}^{L}$& \\
\hline

138.528&$\Gamma$&$D_{4h}^{}\times Z_{2}^T$&($+1,+1,+1,+1,+1,+1,+1$)&$0000000$ \\

&M&$D_{4h}^{}\times Z_{2}^T$&($+1,+1,+1,-1,+1,-1,+1$)&$0001010$ \\

&Z&$D_{4h}^{}\times Z_{2}^T$&($+1,-1,+1,+1,+1,-1,+1$)&$0100010$ \\

&A&$D_{4h}^{}\times Z_{2}^T$&($+1,-1,+1,-1,+1,+1,+1$)&$0101000$ \\


\hline

138.529&$\Gamma$&$D_{4h}^{}\times Z_{2}^T$&($+1,+1,+1,+1,+1,+1,+1$)&$0000000$ \\

&M&$D_{4h}^{}\times Z_{2}^T$&($+1,+1,+1,+1,+1,-1,+1$)&$0000010$ \\

&Z&$D_{4h}^{}\times Z_{2}^T$&($+1,+1,+1,+1,+1,-1,+1$)&$0000010$ \\
\hline
\end{tabular}
\end{table*}

\begin{table*}[htbp]
(Extension of Supplementary Table \ref{Pdrztlat})\\
\centering
\begin{tabular}{ |c|cc|c|c| }
\hline
Shubnikov space group &high-symmetry point $\mathbf{K}$& $M_{\mathbf K}$ & lattice invariants $\vec{\eta}^{L}$& \\
\hline

179.166&$\Gamma$&$D_{6}\times Z_{2}^T$& ($+1,+1,+1,+1$) & $0000$\\

&A&$D_{6}\times Z_{2}^T$&($-1,-1,+1,+1$) & $1100$ \\ 
\hline

180.172&$\Gamma$&$D_{6}\times Z_{2}^T$& ($+1,+1,+1,+1$) & $0000$\\

&A&$D_{6}\times Z_{2}^T$& ($+1,-1,-1,+1$) & $0110$\\ 
\hline

181.178&$\Gamma$&$D_{6}\times Z_{2}^T$&($+1,+1,+1,+1$) & $0000$ \\

&A&$D_{6}\times Z_{2}^T$& ($+1,-1,-1,+1$) & $0110$\\ 
\hline

182.184&$\Gamma$&$D_{6}\times Z_{2}^T$& ($+1,+1,+1,+1$) & $0000$\\
\hline

\end{tabular}
\end{table*}

\begin{table*}[htbp]
\caption{
For ($G_{\mathbf K}\cong M_{\mathbf K}=C_{4h}\times{Z}_2^T,L_{0\mathbf{K}},M_{\mathbf K}/L_{0\mathbf{K}}$),
we list generators of spin point group $(\varphi_{g}||g_{1},\ldots,g_{n})$, $(g_{1},\ldots,g_{n})$ is a coset of $L_{0\mathbf{K}}$,
$g\in (g_{1},\ldots,g_{n}),\varphi_{g}\in M_{\mathbf K}/L_{0\mathbf{K}}$,
$n$ is order of $L_{0\mathbf{K}}$, and
spin invariants $\vec{\eta}^{S}$ of $G_{\mathbf K}$ is
$\eta^S_{p_1,p_2}{\cdot1_{2\times 2}} = d({\varphi_{p_1}})d({\varphi_{p_2}})d^{-1}({\varphi_{p_1}})d^{-1}({\varphi_{p_2}})$
for unitary elements $p_{1},p_{2}$ with $p_{1}p_{2}=p_{2}p_{1}$ and
$\eta^S_{p}{\cdot1_{2\times 2}}=[d(\varphi_p) {K}]^2$
for anti-unitary element $p$ with $p^{2}=E$,
where irRep of $g\in G_{\mathbf K}$ is given by
$d(\varphi_g)K^{\zeta_g}=u\Big( s( T^{\zeta_g}\varphi_g)\Big)(i\sigma_{y}K)^{\zeta_g}$,
$s( T^{\zeta_g}\varphi_g)$ is vector Rep of $T^{\zeta_g}\varphi_{g}\in\mathrm{SO(3)}$,
$u\Big( s( T^{\zeta_g}\varphi_g)\Big)$ is $\mathrm{SU(2)}$ Rep of $T^{\zeta_g}\varphi_{g}$,
$\zeta_{g}=0(1)$ for unitary (anti-unitary) $g$.
The $\vec{\eta}^{S}$ leading to invariants only realized in SSG are labeled by Boolean vectors
colored in {\color{red}RED}.} \label{C4hdrztspin}
\centering
\begin{tabular}{ |c|cccc|}
\hline
$P=C_{4h}$& & spin invariants $\vec{\eta}^{S}$ &  &\\

$G_{\mathbf K}\cong M_{\mathbf K}=C_{4h}\times Z_2^T$
&$\eta_{I,C_{4z}^{+}}^{S}$
&$\eta_{T}^{S}$&$\eta_{IT}^{S}$ &$\eta_{TM_{z}}^{S}$\\
\hline

$L_{0\mathbf K}=\{E,I\},M_{\mathbf K}/L_{0\mathbf K}=\{E,C_{4z}^{+},C_{2z},C_{4z}^{-}\}\times Z_{2}^{T}$,
& $+1$ &$-1$ &$-1$ &$+1$ \\ 

$(C_{4z}^{+}||C_{4z}^{+},S_{4z}^{-}),(C_{2z}||C_{2z},M_{z}),(T||T,IT)$
&&&& \\  
\hline

$L_{0\mathbf K}=\{E,I\},M_{\mathbf K}/L_{0\mathbf K}=\{E,C_{4z}^{+},C_{2z},C_{4z}^{-}\}\times Z_{2}^{T}$,
& $+1$ &$+1$ &$+1$ &$-1$ \\ 

$(C_{4z}^{+}||C_{4z}^{+},S_{4z}^{-}),(C_{2z}||C_{2z},M_{z}),(TC_{2z}||T,IT)$
&& {\color{red}$0001$}  && \\  
\hline

$L_{0\mathbf K}=\{E,M_{z}\},M_{\mathbf K}/L_{0\mathbf K}=\{E,C_{4z}^{+},C_{2z},C_{4z}^{-}\}\times Z_{2}^{T}$,
& $+1$ &$-1$ &$+1$ &$-1$ \\ 

$(C_{4z}^{+}||C_{4z}^{+},S_{4z}^{+}),(C_{2z}||C_{2z},I),(T||T,TM_{z})$
&& {\color{red}$0101$}  && \\  
\hline

$L_{0\mathbf K}=\{E,M_{z}\},M_{\mathbf K}/L_{0\mathbf K}=\{E,C_{4z}^{+},C_{2z},C_{4z}^{-}\}\times Z_{2}^{T}$,
& $+1$ &$+1$ &$-1$ &$+1$ \\ 

$(C_{4z}^{+}||C_{4z}^{+},S_{4z}^{+}),(C_{2z}||C_{2z},I),(TC_{2z}||T,TM_{z})$
&&&& \\  
\hline

$L_{0\mathbf K}=\{E,C_{2z}\},
M_{\mathbf K}/L_{0\mathbf K}=\{E,C_{2z},C_{2x},C_{2y}\}\times Z_{2}^{T}$,
& $-1$ &$-1$ &$+1$ &$+1$ \\ 

$(C_{2z}||M_{z},I),(C_{2x}||C_{4z}^{+},C_{4z}^{-}),(T||T,TC_{2z})$
&&&& \\   
\hline

$L_{0\mathbf K}=\{E,C_{2z}\},
M_{\mathbf K}/L_{0\mathbf K}=\{E,C_{2z},C_{2x},C_{2y}\}\times Z_{2}^{T}$,
& $-1$ &$+1$ &$-1$ &$-1$ \\ 

$(C_{2z}||M_{z},I),(C_{2x}||C_{4z}^{+},C_{4z}^{-}),(TC_{2z}||T,TC_{2z})$
&& {\color{red}$1011$}  && \\  
\hline

$L_{0\mathbf K}=\{E,C_{2z}\},
M_{\mathbf K}/L_{0\mathbf K}=\{E,C_{2z},C_{2x},C_{2y}\}\times Z_{2}^{T}$,
& $-1$ &$+1$ &$+1$ &$+1$ \\ 

$(C_{2z}||M_{z},I),(C_{2x}||C_{4z}^{+},C_{4z}^{-}),(TC_{2x}||T,TC_{2z})$
&&&& \\  
\hline

$L_{0\mathbf K}=\{E,C_{2z}\},
M_{\mathbf K}/L_{0\mathbf K}=\{E,C_{2z},C_{2x},C_{2y}\}\times Z_{2}^{T}$,
& $-1$ &$+1$ &$+1$ &$+1$ \\ 

$(C_{2z}||M_{z},I),(C_{2x}||C_{4z}^{+},C_{4z}^{-}),(TC_{2y}||T,TC_{2z})$
&&&& \\   
\hline

$L_{0\mathbf K}=\{E,C_{2z},I,M_{z}\},M_{\mathbf K}/L_{0\mathbf K}=\{E,C_{2z}\}\times Z_{2}^{T}$,
& $+1$ &$-1$ &$-1$ &$-1$ \\ 

$(C_{2z}||C_{4z}^{+},C_{4z}^{-},S_{4z}^{-},S_{4z}^{+}),(T||T,TC_{2z},IT,TM_{z})$
&& {\color{red}$0111$} && \\   
\hline

$L_{0\mathbf K}=\{E,C_{2z},I,M_{z}\},M_{\mathbf K}/L_{0\mathbf K}=\{E,C_{2z}\}\times Z_{2}^{T}$,
& $+1$ &$+1$ &$+1$ &$+1$ \\ 

$(C_{2z}||C_{4z}^{+},C_{4z}^{-},S_{4z}^{-},S_{4z}^{+}),(TC_{2z}||T,TC_{2z},IT,TM_{z})$
&&&& \\  
\hline

$L_{0\mathbf K}=\{E,C_{2z},I,M_{z}\},M_{\mathbf K}/L_{0\mathbf K}=\{E,C_{2z},TC_{2x},TC_{2y}\}$,
& $+1$ &$+1$ &$+1$ &$+1$ \\ 

$(C_{2z}||C_{4z}^{+},C_{4z}^{-},S_{4z}^{-},S_{4z}^{+}),(TC_{2y}||T,TC_{2z},IT,TM_{z})$
&&&& \\  
\hline

$L_{0\mathbf K}=\{E,C_{4z}^{+},C_{2z},C_{4z}^{-}\},M_{\mathbf K}/L_{0\mathbf K}=\{E,C_{2z}\}\times Z_{2}^{T}$,
& $+1$ &$-1$ &$+1$ &$+1$ \\ 

$(C_{2z}||I,S_{4z}^{-},M_{z},S_{4z}^{+}),(T||T,TC_{4z}^{+},TC_{2z},TC_{4z}^{-})$
&&&& \\   
\hline

$L_{0\mathbf K}=\{E,C_{4z}^{+},C_{2z},C_{4z}^{-}\},M_{\mathbf K}/L_{0\mathbf K}=\{E,C_{2z}\}\times Z_{2}^{T}$,
& $+1$ &$+1$ &$-1$ &$-1$ \\ 

$(C_{2z}||I,S_{4z}^{-},M_{z},S_{4z}^{+}),(TC_{2z}||T,TC_{4z}^{+},TC_{2z},TC_{4z}^{-})$
&& {\color{red}$0011$}  && \\  
\hline

$L_{0\mathbf K}=\{E,C_{4z}^{+},C_{2z},C_{4z}^{-}\},M_{\mathbf K}/L_{0\mathbf K}=\{E,C_{2z},TC_{2x},TC_{2y}\}$,
& $+1$ &$+1$ &$+1$ &$+1$ \\ 

$(C_{2z}||I,S_{4z}^{-},M_{z},S_{4z}^{+}),(TC_{2y}||T,TC_{4z}^{+},TC_{2z},TC_{4z}^{-})$
&&&& \\   
\hline

$L_{0\mathbf K}=\{E,S_{4z}^{-},C_{2z},S_{4z}^{+}\},M_{\mathbf K}/L_{0\mathbf K}=\{E,C_{2z}\}\times Z_{2}^{T}$,
& $+1$ &$-1$ &$+1$ &$+1$ \\ 

$(C_{2z}||I,C_{4z}^{+},M_{z},C_{4z}^{-}),(T||T,TS_{4z}^{-},TC_{2z},TS_{4z}^{+})$
&&&& \\   
\hline

$L_{0\mathbf K}=\{E,S_{4z}^{-},C_{2z},S_{4z}^{+}\},M_{\mathbf K}/L_{0\mathbf K}=\{E,C_{2z}\}\times Z_{2}^{T}$,
& $+1$ &$+1$ &$-1$ &$-1$ \\ 

$(C_{2z}||I,C_{4z}^{+},M_{z},C_{4z}^{-}),(TC_{2z}||T,TS_{4z}^{-},TC_{2z},TS_{4z}^{+})$
&& {\color{red}$0011$}  && \\  
\hline

$L_{0\mathbf K}=\{E,S_{4z}^{-},C_{2z},S_{4z}^{+}\},
M_{\mathbf K}/L_{0\mathbf K}=\{E,C_{2z},TC_{2x},TC_{2y}\}$,
& $+1$ &$+1$ &$+1$ &$+1$ \\ 

$(C_{2z}||I,C_{4z}^{+},M_{z},C_{4z}^{-}),(TC_{2y}||T,TS_{4z}^{-},TC_{2z},TS_{4z}^{+})$
&&&& \\   
\hline

$L_{0\mathbf K}=C_{4h},
M_{\mathbf K}/L_{0\mathbf K}=Z_{2}^{T},$
& $+1$ &$-1$ &$-1$ &$-1$ \\ 

$(T||T,TC_{4z}^{+},TC_{2z},TC_{4z}^{-},IT,TS_{4z}^{-},TM_{z},TS_{4z}^{+})$
&& {\color{red}$0111$} && \\  
\hline

$L_{0\mathbf K}=C_{4h},
M_{\mathbf K}/L_{0\mathbf K}=\{E,TC_{2y}\},$
& $+1$ &$+1$ &$+1$ &$+1$ \\ 

$(TC_{2y}||T,TC_{4z}^{+},TC_{2z},TC_{4z}^{-},IT,TS_{4z}^{-},TM_{z},TS_{4z}^{+})$
&&&& \\ 
\hline

\end{tabular}
\end{table*}

\begin{table*}[htbp]
\caption{
For ($G_{\mathbf K}\cong M_{\mathbf K}=D_{4}\times{Z}_2^T,L_{0\mathbf{K}},M_{\mathbf K}/L_{0\mathbf{K}}$),
we list generators of spin point group $(\varphi_{g}||g_{1},\ldots,g_{n})$, $(g_{1},\ldots,g_{n})$ is a coset of $L_{0\mathbf{K}}$,
$g\in (g_{1},\ldots,g_{n}),\varphi_{g}\in M_{\mathbf K}/L_{0\mathbf{K}}$,
$n$ is order of $L_{0\mathbf{K}}$, and
spin invariants $\vec{\eta}^{S}$ of $G_{\mathbf K}$ is
$\eta^S_{p_1,p_2}{\cdot1_{2\times 2}} = d({\varphi_{p_1}})d({\varphi_{p_2}})d^{-1}({\varphi_{p_1}})d^{-1}({\varphi_{p_2}})$
for unitary elements $p_{1},p_{2}$ with $p_{1}p_{2}=p_{2}p_{1}$ and
$\eta^S_{p}{\cdot1_{2\times 2}}=[d(\varphi_p) {K}]^2$
for anti-unitary element $p$ with $p^{2}=E$,
where irRep of $g\in G_{\mathbf K}$ is given by
$d(\varphi_g)K^{\zeta_g}=u\Big( s( T^{\zeta_g}\varphi_g)\Big)(i\sigma_{y}K)^{\zeta_g}$,
$s( T^{\zeta_g}\varphi_g)$ is vector Rep of $T^{\zeta_g}\varphi_{g}\in\mathrm{SO(3)}$,
$u\Big( s( T^{\zeta_g}\varphi_g)\Big)$ is $\mathrm{SU(2)}$ Rep of $T^{\zeta_g}\varphi_{g}$,
$\zeta_{g}=0(1)$ for unitary (anti-unitary) $g$.
The $\vec{\eta}^{S}$ leading to invariants only realized in SSG are labeled by Boolean vectors
colored in {\color{red}RED}.} \label{D4drztspin}
\centering
\begin{tabular}{ |c|cccc|}
\hline
$P=D_{4}^{}$& & spin invariants $\vec{\eta}^{S}$ &  &\\

$G_{\mathbf K}\cong M_{\mathbf K}=D_{4}^{}\times Z_2^T$
&$\eta_{C_{2x},C_{2y}}^{S}$
&$\eta_{T}^{S}$&$\eta_{TC_{2x}}^{S}$ &$\eta_{TC_{2a}}^{S}$\\
\hline

$L_{0\mathbf K}=E,M_{\mathbf K}/L_{0\mathbf K}=D_{4}^{}\times Z_{2}^{T},
(C_{2x}||C_{2x}),(C_{2y}||C_{2y})$,
&$-1$ &$-1$ &$+1$ &$+1$ \\ 

$(C_{2a}||C_{2a}),(T||T)$
&&&& \\  
\hline

$L_{0\mathbf K}=E,M_{\mathbf K}/L_{0\mathbf K}=D_{4}^{}\times Z_{2}^{T},
(C_{2x}||C_{2x}),(C_{2y}||C_{2y})$,
&$-1$ &$+1$ &$+1$ &$+1$ \\ 

$(C_{2a}||C_{2a}),(TC_{2z}||T)$
&&&& \\  
\hline

$L_{0\mathbf K}=\{E,C_{2z}\},M_{\mathbf K}/L_{0\mathbf K}=\{E,C_{2z},C_{2x},C_{2y}\}\times Z_{2}^{T}$,
& $+1$ &$-1$ &$+1$ &$+1$ \\ 

$(C_{2x}||C_{2x},C_{2y}),(C_{2y}||C_{2a},C_{2b}),(T||T,TC_{2z})$
&&&& \\  
\hline

$L_{0\mathbf K}=\{E,C_{2z}\},M_{\mathbf K}/L_{0\mathbf K}=\{E,C_{2z},C_{2x},C_{2y}\}\times Z_{2}^{T}$,
& $+1$ &$+1$ &$+1$ &$+1$ \\ 

$(C_{2x}||C_{2x},C_{2y}),(C_{2y}||C_{2a},C_{2b}),(TC_{2z}||T,TC_{2z})$
&&&& \\   
\hline

$L_{0\mathbf K}=\{E,C_{2z}\},M_{\mathbf K}/L_{0\mathbf K}=\{E,C_{2z},C_{2x},C_{2y}\}\times Z_{2}^{T}$,
& $+1$ &$+1$ &$-1$ &$+1$ \\ 

$(C_{2x}||C_{2x},C_{2y}),(C_{2y}||C_{2a},C_{2b}),(TC_{2x}||T,TC_{2z}),$
&&&& \\  
\hline

$L_{0\mathbf K}=\{E,C_{2z}\},M_{\mathbf K}/L_{0\mathbf K}=\{E,C_{2z},C_{2x},C_{2y}\}\times Z_{2}^{T}$,
& $+1$ &$+1$ &$+1$ &$-1$ \\ 

$(C_{2x}||C_{2x},C_{2y}),(C_{2y}||C_{2a},C_{2b}),(TC_{2y}||T,TC_{2z})$
&& {\color{red}$0001$}  && \\  
\hline

$L_{0\mathbf K}=\{E,C_{4z}^{+},C_{2z},C_{4z}^{-}\},
M_{\mathbf K}/L_{0\mathbf K}=\{E,C_{2z}\}\times Z_{2}^{T}$,
& $+1$ &$-1$ &$+1$ &$+1$ \\ 

$(C_{2z}||C_{2x},C_{2a},C_{2y},C_{2b}),(T||T,TC_{4z}^{+},TC_{2z},TC_{4z}^{-})$
&&&& \\  
\hline

$L_{0\mathbf K}=\{E,C_{4z}^{+},C_{2z},C_{4z}^{-}\},
M_{\mathbf K}/L_{0\mathbf K}=\{E,C_{2z}\}\times Z_{2}^{T}$,
& $+1$ &$+1$ &$-1$ &$-1$ \\ 

$(C_{2z}||C_{2x},C_{2a},C_{2y},C_{2b}),(TC_{2z}||T,TC_{4z}^{+},TC_{2z},TC_{4z}^{-})$
&& {\color{red}$0011$}   && \\   
\hline

$L_{0\mathbf K}=\{E,C_{4z}^{+},C_{2z},C_{4z}^{-}\},
M_{\mathbf K}/L_{0\mathbf K}=\{E,C_{2z},TC_{2x},TC_{2y}\}$,
& $+1$ &$+1$ &$+1$ &$+1$ \\ 

$(C_{2z}||C_{2x},C_{2a},C_{2y},C_{2b}),(TC_{2y}||T,TC_{4z}^{+},TC_{2z},TC_{4z}^{-})$
&&&& \\  
\hline

$L_{0\mathbf K}=\{E,C_{2x},C_{2y},C_{2z}\},
M_{\mathbf K}/L_{0\mathbf K}=\{E,C_{2z}\}\times Z_{2}^{T}$,
& $+1$ &$-1$ &$-1$ &$+1$ \\ 

$(C_{2z}||C_{4z}^{+},C_{4z}^{-},C_{2a},C_{2b}),(T||T,TC_{2x},TC_{2y},TC_{2z})$
&&&& \\   
\hline

$L_{0\mathbf K}=\{E,C_{2x},C_{2y},C_{2z}\},
M_{\mathbf K}/L_{0\mathbf K}=\{E,C_{2z}\}\times Z_{2}^{T}$,
& $+1$ &$+1$ &$+1$ &$-1$ \\ 

$(C_{2z}||C_{4z}^{+},C_{4z}^{-},C_{2a},C_{2b}),(TC_{2z}||T,TC_{2x},TC_{2y},TC_{2z})$
&& {\color{red}$0001$}  && \\  
\hline

$L_{0\mathbf K}=\{E,C_{2x},C_{2y},C_{2z}\},
M_{\mathbf K}/L_{0\mathbf K}=\{E,C_{2z},TC_{2x},TC_{2y}\}$,
& $+1$ &$+1$ &$+1$ &$+1$ \\ 

$(C_{2z}||C_{4z}^{+},C_{4z}^{-},C_{2a},C_{2b}),(TC_{2y}||T,TC_{2x},TC_{2y},TC_{2z})$
&&&& \\  
\hline

$L_{0\mathbf K}=\{E,C_{2b},C_{2a},C_{2z}\},
M_{\mathbf K}/L_{0\mathbf K}=\{E,C_{2z}\}\times Z_{2}^{T}$,
& $+1$ &$-1$ &$+1$ &$-1$ \\ 

$(C_{2z}||C_{4z}^{+},C_{4z}^{-},C_{2x},C_{2y}),(T||T,TC_{2b},TC_{2a},TC_{2z})$
&& {\color{red}$0101$}  && \\  
\hline

$L_{0\mathbf K}=\{E,C_{2b},C_{2a},C_{2z}\},
M_{\mathbf K}/L_{0\mathbf K}=\{E,C_{2z}\}\times Z_{2}^{T}$,
& $+1$ &$+1$ &$-1$ &$+1$ \\ 

$(C_{2z}||C_{4z}^{+},C_{4z}^{-},C_{2x},C_{2y}),(TC_{2z}||T,TC_{2b},TC_{2a},TC_{2z})$
&&&& \\  
\hline

$L_{0\mathbf K}=\{E,C_{2b},C_{2a},C_{2z}\},
M_{\mathbf K}/L_{0\mathbf K}=\{E,C_{2z},TC_{2x},TC_{2y}\}$,
& $+1$ &$+1$ &$+1$ &$+1$ \\ 

$(C_{2z}||C_{4z}^{+},C_{4z}^{-},C_{2x},C_{2y}),(TC_{2y}||T,TC_{2b},TC_{2a},TC_{2z})$
&&&& \\   
\hline

$L_{0\mathbf K}=D_{4}^{},
M_{\mathbf K}/L_{0\mathbf K}=Z_{2}^{T}$,
& $+1$ &$-1$ &$-1$ &$-1$ \\ 

$(T||T,TC_{4z}^{+},TC_{2z},TC_{4z}^{-},TC_{2x},TC_{2a},TC_{2y},TC_{2b})$
&& {\color{red}$0111$} && \\  
\hline

$L_{0\mathbf K}=D_{4}^{},
M_{\mathbf K}/L_{0\mathbf K}=\{E,TC_{2y}\}$,
& $+1$ &$+1$ &$+1$ &$+1$ \\ 

$(TC_{2y}||T,TC_{4z}^{+},TC_{2z},TC_{4z}^{-},TC_{2x},TC_{2a},TC_{2y},TC_{2b})$
&&&& \\  
\hline

\end{tabular}
\end{table*}

\begin{table*}[htbp]
\caption{
For ($G_{\mathbf K}\cong M_{\mathbf K}=D_{4h}\times{Z}_2^T,L_{0\mathbf{K}},M_{\mathbf K}/L_{0\mathbf{K}}$),
we list generators of spin point group $(\varphi_{g}||g_{1},\ldots,g_{n})$, $(g_{1},\ldots,g_{n})$ is a coset of $L_{0\mathbf{K}}$,
$g\in (g_{1},\ldots,g_{n}),\varphi_{g}\in M_{\mathbf K}/L_{0\mathbf{K}}$,
$n$ is order of $L_{0\mathbf{K}}$, and
spin invariants $\vec{\eta}^{S}$ of $G_{\mathbf K}$ is
$\eta^S_{p_1,p_2}{\cdot1_{2\times 2}} = d({\varphi_{p_1}})d({\varphi_{p_2}})d^{-1}({\varphi_{p_1}})d^{-1}({\varphi_{p_2}})$
for unitary elements $p_{1},p_{2}$ with $p_{1}p_{2}=p_{2}p_{1}$ and
$\eta^S_{p}{\cdot1_{2\times 2}}=[d(\varphi_p) {K}]^2$
for anti-unitary element $p$ with $p^{2}=E$,
where irRep of $g\in G_{\mathbf K}$ is given by
$d(\varphi_g)K^{\zeta_g}=u\Big( s( T^{\zeta_g}\varphi_g)\Big)(i\sigma_{y}K)^{\zeta_g}$,
$s( T^{\zeta_g}\varphi_g)$ is vector Rep of $T^{\zeta_g}\varphi_{g}\in\mathrm{SO(3)}$,
$u\Big( s( T^{\zeta_g}\varphi_g)\Big)$ is $\mathrm{SU(2)}$ Rep of $T^{\zeta_g}\varphi_{g}$,
$\zeta_{g}=0(1)$ for unitary (anti-unitary) $g$.
The $\vec{\eta}^{S}$ leading to invariants only realized in SSG are labeled by Boolean vectors
colored in {\color{red}RED}.} \label{D4hdrztspin}
\centering
\begin{tabular}{ |c|ccccccc|}
\hline
 $P=D_{4h}^{}$& && & spin invariants $\vec{\eta}^{S}$&&&\\

$G_{\mathbf K}\cong M_{\mathbf K}=D_{4h}^{}\times Z_2^T$
&$\eta_{C_{2x},C_{2y}}^{S}$
&$\eta_{T}^{S}$&$\eta_{IT}^{S}$ &$\eta_{TC_{2x}}^{S}$&$\eta_{TC_{2a}}^{S}$
&$\eta_{I,C_{2x}}^{S}$&
$\eta_{I,C_{2a}}^{S}$\\
\hline

$L_{0\mathbf K}=\{E,I\},M_{\mathbf K}/L_{0\mathbf K}=D_{4}^{}\times Z_{2}^{T},(C_{2x}||C_{2x},M_{x})$,
& $-1$ &$-1$ &$-1$ &$+1$ &$+1$ &$+1$ &$+1$ \\ 

$(C_{2a}||C_{2a},M_{da}),(C_{2y}||C_{2y},M_{y}),(T||T,IT)$
&&&&&&& \\

%

\hline

$L_{0\mathbf K}=\{E,I\},M_{\mathbf K}/L_{0\mathbf K}=D_{4}^{}\times Z_{2}^{T},(C_{2x}||C_{2x},M_{x})$,
& $-1$ &$+1$ &$+1$ &$+1$ &$+1$ &$+1$ &$+1$ \\ 

$(C_{2a}||C_{2a},M_{da}),(C_{2y}||C_{2y},M_{y}),(TC_{2z}||T,IT)$
&&&& {\color{red}$1000000$} &&& \\


%

\hline

$L_{0\mathbf K}=\{E,M_{z}\},M_{\mathbf K}/L_{0\mathbf K}=D_{4}^{}\times Z_{2}^{T},
(C_{2z}||C_{2z},I)$,
& $-1$ &$-1$ &$+1$ &$+1$ &$+1$ &$-1$ &$-1$ \\ 

$(C_{2x}||C_{2x},M_{y}),(C_{2a}||C_{2a},M_{db}),(C_{2y}||C_{2y},M_{x})$,
&&&& {\color{red}$1100011$} &&& \\

$(T||T,TM_{z})$
&&&&&&& \\  

%

\hline

$L_{0\mathbf K}=\{E,M_{z}\},M_{\mathbf K}/L_{0\mathbf K}=D_{4}^{}\times Z_{2}^{T},
(C_{2z}||C_{2z},I)$,
& $-1$ &$+1$ &$-1$ &$+1$ &$+1$ &$-1$ &$-1$ \\ 

$(C_{2x}||C_{2x},M_{y}),(C_{2a}||C_{2a},M_{db}),(C_{2y}||C_{2y},M_{x})$,
&&&&&&& \\

$(TC_{2z}||T,TM_{z})$
&&&&&&& \\ 

%

\hline

$L_{0\mathbf K}=\{E,C_{2z},I,M_{z}\},M_{\mathbf K}/L_{0\mathbf K}=\{E,C_{2x,2y,2z}\}\times Z_{2}^{T}$,
& $+1$ &$-1$ &$-1$ &$+1$ &$+1$ &$+1$ &$+1$ \\ 

$(C_{2x}||C_{2x,2y},M_{x,y}),(C_{2y}||C_{2a,2b},M_{da,db})$,
&&&& {\color{red}$0110000$} &&& \\

$(T||T,TC_{2z},IT,TM_{z})$
&&&&&&& \\ 


\hline

$L_{0\mathbf K}=\{E,C_{2z},I,M_{z}\},M_{\mathbf K}/L_{0\mathbf K}=\{E,C_{2x,2y,2z}\}\times Z_{2}^{T}$,
& $+1$ &$+1$ &$+1$ &$+1$ &$+1$ &$+1$ &$+1$ \\ 

$(C_{2x}||C_{2x,2y},M_{x,y}),(C_{2y}||C_{2a,2b},M_{da,db})$,
&&&&&&& \\

$(TC_{2z}||T,TC_{2z},IT,TM_{z})$
&&&&&&& \\ 


\hline

$L_{0\mathbf K}=\{E,C_{2z},I,M_{z}\},M_{\mathbf K}/L_{0\mathbf K}=\{E,C_{2x,2y,2z}\}\times Z_{2}^{T}$,
& $+1$ &$+1$ &$+1$ &$-1$ &$+1$ &$+1$ &$+1$ \\ 

$(C_{2x}||C_{2x,2y},M_{x,y}),(C_{2y}||C_{2a,2b},M_{da,db})$,
&&&&&&& \\

$(TC_{2x}||T,TC_{2z},IT,TM_{z})$
&&&&&&& \\ 


\hline

$L_{0\mathbf K}=\{E,C_{2z},I,M_{z}\},M_{\mathbf K}/L_{0\mathbf K}=\{E,C_{2x,2y,2z}\}\times Z_{2}^{T}$,
& $+1$ &$+1$ &$+1$ &$+1$ &$-1$ &$+1$ &$+1$ \\ 

$(C_{2x}||C_{2x,2y},M_{x,y}),(C_{2y}||C_{2a,2b},M_{da,db})$,
&&&& {\color{red}$0000100$} &&& \\

$(TC_{2y}||T,TC_{2z},IT,TM_{z})$
&&&&&&& \\ 


\hline

$L_{0\mathbf K}=\{E,C_{2x,2y,2z}\},M_{\mathbf K}/L_{0\mathbf K}=\{E,C_{2x,2y,2z}\}\times Z_{2}^{T}$,
& $+1$ &$-1$ &$+1$ &$-1$ &$+1$ &$+1$ &$-1$ \\ 

$(C_{2z}||I,M_{x,y,z}),(C_{2x}||C_{4z}^{\pm},C_{2a,2b}),(T||T,TC_{2x,2y,2z})$
&&&&&&& \\



\hline

$L_{0\mathbf K}=\{E,C_{2x,2y,2z}\},M_{\mathbf K}/L_{0\mathbf K}=\{E,C_{2x,2y,2z}\}\times Z_{2}^{T}$,
& $+1$ &$+1$ &$-1$ &$+1$ &$+1$ &$+1$ &$-1$ \\ 

$(C_{2z}||I,M_{x,y,z}),(C_{2x}||C_{4z}^{\pm},C_{2a,2b}),(TC_{2z}||T,TC_{2x,2y,2z})$
&&&& {\color{red}$0010001$} &&& \\



\hline

$L_{0\mathbf K}=\{E,C_{2x,2y,2z}\},M_{\mathbf K}/L_{0\mathbf K}=\{E,C_{2x,2y,2z}\}\times Z_{2}^{T}$,
& $+1$ &$+1$ &$+1$ &$+1$ &$-1$ &$+1$ &$-1$ \\ 

$(C_{2z}||I,M_{x,y,z}),(C_{2x}||C_{4z}^{\pm},C_{2a,2b}),(TC_{2x}||T,TC_{2x,2y,2z})$
&&&& {\color{red}$0000101$} &&& \\



\hline

$L_{0\mathbf K}=\{E,C_{2x,2y,2z}\},M_{\mathbf K}/L_{0\mathbf K}=\{E,C_{2x,2y,2z}\}\times Z_{2}^{T}$,
& $+1$ &$+1$ &$+1$ &$+1$ &$+1$ &$+1$ &$-1$ \\ 

$(C_{2z}||I,M_{x,y,z}),(C_{2x}||C_{4z}^{\pm},C_{2a,2b}),(TC_{2y}||T,TC_{2x,2y,2z})$
&&&&&&& \\



\hline

$L_{0\mathbf K}=\{E,C_{2b,2a},C_{2z}\},M_{\mathbf K}/L_{0\mathbf K}=\{E,C_{2x,2y,2z}\}\times Z_{2}^{T}$,
& $+1$ &$-1$ &$+1$ &$+1$ &$-1$ &$-1$ &$+1$ \\ 

$(C_{2z}||I,M_{db,da},M_{z}),(C_{2x}||C_{4z}^{\pm},C_{2x,2y})$,
&&&& {\color{red}$0100110$} &&& \\

$(T||T,TC_{2b,2a},TC_{2z})$ 
&&&&&&& \\
\hline

$L_{0\mathbf K}=\{E,C_{2b,2a},C_{2z}\},M_{\mathbf K}/L_{0\mathbf K}=\{E,C_{2x,2y,2z}\}\times Z_{2}^{T}$,
& $+1$ &$+1$ &$-1$ &$+1$ &$+1$ &$-1$ &$+1$ \\ 

$(C_{2z}||I,M_{db,da},M_{z}),(C_{2x}||C_{4z}^{\pm},C_{2x,2y})$,
&&&& {\color{red}$0010010$} &&& \\

$(TC_{2z}||T,TC_{2b,2a},TC_{2z})$ 
&&&&&&& \\
\hline

$L_{0\mathbf K}=\{E,C_{2b,2a},C_{2z}\},M_{\mathbf K}/L_{0\mathbf K}=\{E,C_{2x,2y,2z}\}\times Z_{2}^{T}$,
& $+1$ &$+1$ &$+1$ &$-1$ &$+1$ &$-1$ &$+1$ \\ 

$(C_{2z}||I,M_{db,da},M_{z}),(C_{2x}||C_{4z}^{\pm},C_{2x,2y})$,
&&&&&&& \\

$(TC_{2x}||T,TC_{2b,2a},TC_{2z})$  
&&&&&&& \\
\hline

$L_{0\mathbf K}=\{E,C_{2b,2a},C_{2z}\},M_{\mathbf K}/L_{0\mathbf K}=\{E,C_{2x,2y,2z}\}\times Z_{2}^{T}$,
& $+1$ &$+1$ &$+1$ &$+1$ &$+1$ &$-1$ &$+1$ \\ 

$(C_{2z}||I,M_{db,da},M_{z}),(C_{2x}||C_{4z}^{\pm},C_{2x,2y})$,
&&&&&&& \\

$(TC_{2y}||T,TC_{2b,2a},TC_{2z})$ 
&&&&&&& \\
\hline

$L_{0\mathbf K}=\{E,M_{x,y},C_{2z}\},M_{\mathbf K}/L_{0\mathbf K}=\{E,C_{2x,2y,2z}\}\times Z_{2}^{T}$,
& $+1$ &$-1$ &$+1$ &$+1$ &$+1$ &$+1$ &$-1$ \\ 

$(C_{2z}||I,C_{2x,2y},M_{z}),(C_{2x}||S_{4z}^{\mp},C_{2a,2b})$,
&&&&&&& \\

$(T||T,TM_{x,y},TC_{2z})$  
&&&&&&& \\
\hline

$L_{0\mathbf K}=\{E,M_{x,y},C_{2z}\},M_{\mathbf K}/L_{0\mathbf K}=\{E,C_{2x,2y,2z}\}\times Z_{2}^{T}$,
& $+1$ &$+1$ &$-1$ &$-1$ &$+1$ &$+1$ &$-1$ \\ 

$(C_{2z}||I,C_{2x,2y},M_{z}),(C_{2x}||S_{4z}^{\mp},C_{2a,2b})$,
&&&& {\color{red}$0011001$} &&& \\

$(TC_{2z}||T,TM_{x,y},TC_{2z})$ 
&&&&&&& \\
\hline

$L_{0\mathbf K}=\{E,M_{x,y},C_{2z}\},M_{\mathbf K}/L_{0\mathbf K}=\{E,C_{2x,2y,2z}\}\times Z_{2}^{T}$,
& $+1$ &$+1$ &$+1$ &$+1$ &$-1$ &$+1$ &$-1$ \\ 

$(C_{2z}||I,C_{2x,2y},M_{z}),(C_{2x}||S_{4z}^{\mp},C_{2a,2b})$,
&&&& {\color{red}$0000101$} &&& \\

$(TC_{2x}||T,TM_{x,y},TC_{2z})$  
&&&&&&& \\
\hline

\end{tabular}
\end{table*}

\begin{table*}
(Extension of Supplementary Table \ref{D4hdrztspin})\\
\centering
\begin{tabular}{ |c|ccccccc|}
\hline
$P=D_{4h}^{}$& && & spin invariants $\vec{\eta}^{S}$&&&\\

$G_{\mathbf K}\cong M_{\mathbf K}=D_{4h}^{}\times Z_2^T$
&$\eta_{C_{2x},C_{2y}}^{S}$
&$\eta_{T}^{S}$&$\eta_{IT}^{S}$ &$\eta_{TC_{2x}}^{S}$&$\eta_{TC_{2a}}^{S}$
&$\eta_{I,C_{2x}}^{S}$&
$\eta_{I,C_{2a}}^{S}$\\
\hline

$L_{0\mathbf K}=\{E,M_{x,y},C_{2z}\},M_{\mathbf K}/L_{0\mathbf K}=\{E,C_{2x,2y,2z}\}\times Z_{2}^{T}$,
& $+1$ &$+1$ &$+1$ &$+1$ &$+1$ &$+1$ &$-1$ \\ 

$(C_{2z}||I,C_{2x,2y},M_{z}),(C_{2x}||S_{4z}^{\mp},C_{2a,2b})$,
&&&&&&& \\

$(TC_{2y}||T,TM_{x,y},TC_{2z})$  
&&&&&&& \\
\hline

$L_{0\mathbf K}=\{E,M_{db,da},C_{2z}\},M_{\mathbf K}/L_{0\mathbf K}=\{E,C_{2x,2y,2z}\}\times Z_{2}^{T}$,
& $+1$ &$-1$ &$+1$ &$+1$ &$+1$ &$-1$ &$+1$ \\ 

$(C_{2z}||I,C_{2b,2a},M_{z}),(C_{2x}||S_{4z}^{\mp},C_{2x,2y})$,
&&&&&&& \\

$(T||T,TM_{db,da},TC_{2z})$    
&&&&&&& \\
\hline

$L_{0\mathbf K}=\{E,M_{db,da},C_{2z}\},M_{\mathbf K}/L_{0\mathbf K}=\{E,C_{2x,2y,2z}\}\times Z_{2}^{T}$,
& $+1$ &$+1$ &$-1$ &$+1$ &$-1$ &$-1$ &$+1$ \\ 

$(C_{2z}||I,C_{2b,2a},M_{z}),(C_{2x}||S_{4z}^{\mp},C_{2x,2y})$,
&&&& {\color{red}$0010110$} &&& \\

$(TC_{2z}||T,TM_{db,da},TC_{2z})$  
&&&&&&& \\
\hline

$L_{0\mathbf K}=\{E,M_{db,da},C_{2z}\},M_{\mathbf K}/L_{0\mathbf K}=\{E,C_{2x,2y,2z}\}\times Z_{2}^{T}$,
& $+1$ &$+1$ &$+1$ &$-1$ &$+1$ &$-1$ &$+1$ \\ 

$(C_{2z}||I,C_{2b,2a},M_{z}),(C_{2x}||S_{4z}^{\mp},C_{2x,2y})$,
&&&&&&& \\

$(TC_{2x}||T,TM_{db,da},TC_{2z})$ 
&&&&&&& \\
\hline

$L_{0\mathbf K}=\{E,M_{db,da},C_{2z}\},M_{\mathbf K}/L_{0\mathbf K}=\{E,C_{2x,2y,2z}\}\times Z_{2}^{T}$,
& $+1$ &$+1$ &$+1$ &$+1$ &$+1$ &$-1$ &$+1$ \\ 

$(C_{2z}||I,C_{2b,2a},M_{z}),(C_{2x}||S_{4z}^{\mp},C_{2x,2y})$,
&&&&&&& \\

$(TC_{2y}||T,TM_{db,da},TC_{2z})$ 
&&&&&&& \\
\hline

$L_{0\mathbf K}=\{E,C_{4z}^{\pm},C_{2z}\},M_{\mathbf K}/L_{0\mathbf K}=\{E,C_{2x,2y,2z}\}\times Z_{2}^{T}$,
& $+1$ &$-1$ &$+1$ &$+1$ &$+1$ &$-1$ &$-1$ \\ 

$(C_{2z}||I,S_{4z}^{\mp},M_{z}),(C_{2x}||C_{2x,2y},C_{2a,2b})$,
&&&&&&& \\

$(T||T,TC_{4z}^{\pm},TC_{2z})$  
&&&&&&& \\
\hline

$L_{0\mathbf K}=\{E,C_{4z}^{\pm},C_{2z}\},M_{\mathbf K}/L_{0\mathbf K}=\{E,C_{2x,2y,2z}\}\times Z_{2}^{T}$,
& $+1$ &$+1$ &$-1$ &$+1$ &$+1$ &$-1$ &$-1$ \\ 

$(C_{2z}||I,S_{4z}^{\mp},M_{z}),(C_{2x}||C_{2x,2y},C_{2a,2b})$,
&&&& {\color{red}$0010011$} &&& \\

$(TC_{2z}||T,TC_{4z}^{\pm},TC_{2z})$  
&&&&&&& \\
\hline

$L_{0\mathbf K}=\{E,C_{4z}^{\pm},C_{2z}\},M_{\mathbf K}/L_{0\mathbf K}=\{E,C_{2x,2y,2z}\}\times Z_{2}^{T}$,
& $+1$ &$+1$ &$+1$ &$-1$ &$-1$ &$-1$ &$-1$ \\ 

$(C_{2z}||I,S_{4z}^{\mp},M_{z}),(C_{2x}||C_{2x,2y},C_{2a,2b})$,
&&&& {\color{red}$0001111$} &&& \\

$(TC_{2x}||T,TC_{4z}^{\pm},TC_{2z})$ 
&&&&&&& \\
\hline

$L_{0\mathbf K}=\{E,C_{4z}^{\pm},C_{2z}\},M_{\mathbf K}/L_{0\mathbf K}=\{E,C_{2x,2y,2z}\}\times Z_{2}^{T}$,
& $+1$ &$+1$ &$+1$ &$+1$ &$+1$ &$-1$ &$-1$ \\ 

$(C_{2z}||I,S_{4z}^{\mp},M_{z}),(C_{2x}||C_{2x,2y},C_{2a,2b})$,
&&&&&&& \\

$(TC_{2y}||T,TC_{4z}^{\pm},TC_{2z})$  
&&&&&&& \\
\hline

$L_{0\mathbf K}=\{E,S_{4z}^{\mp},C_{2z}\},M_{\mathbf K}/L_{0\mathbf K}=\{E,C_{2x,2y,2z}\}\times Z_{2}^{T}$,
& $+1$ &$-1$ &$+1$ &$+1$ &$+1$ &$-1$ &$-1$ \\ 

$(C_{2z}||I,C_{4z}^{\pm},M_{z}),(C_{2x}||C_{2x,2y},M_{da,db})$,
&&&&&&& \\

$(C_{2y}||C_{2a,2b},M_{x,y}),(T||T,TS_{4z}^{\mp},TC_{2z})$  
&&&&&&& \\
\hline

$L_{0\mathbf K}=\{E,S_{4z}^{\mp},C_{2z}\},M_{\mathbf K}/L_{0\mathbf K}=\{E,C_{2x,2y,2z}\}\times Z_{2}^{T}$,
& $+1$ &$+1$ &$-1$ &$+1$ &$+1$ &$-1$ &$-1$ \\ 

$(C_{2z}||I,C_{4z}^{\pm},M_{z}),(C_{2x}||C_{2x,2y},M_{da,db})$,
&&&& {\color{red}$0010011$} &&& \\

$(C_{2y}||C_{2a,2b},M_{x,y}),(TC_{2z}||T,TS_{4z}^{\mp},TC_{2z})$  
&&&&&&& \\
\hline

$L_{0\mathbf K}=\{E,S_{4z}^{\mp},C_{2z}\},M_{\mathbf K}/L_{0\mathbf K}=\{E,C_{2x,2y,2z}\}\times Z_{2}^{T}$,
& $+1$ &$+1$ &$+1$ &$-1$ &$+1$ &$-1$ &$-1$ \\ 

$(C_{2z}||I,C_{4z}^{\pm},M_{z}),(C_{2x}||C_{2x,2y},M_{da,db})$,
&&&&&&& \\

$(C_{2y}||C_{2a,2b},M_{x,y}),(TC_{2x}||T,TS_{4z}^{\mp},TC_{2z})$  
&&&&&&& \\
\hline

$L_{0\mathbf K}=\{E,S_{4z}^{\mp},C_{2z}\},M_{\mathbf K}/L_{0\mathbf K}=\{E,C_{2x,2y,2z}\}\times Z_{2}^{T}$,
& $+1$ &$+1$ &$+1$ &$+1$ &$-1$ &$-1$ &$-1$ \\ 

$(C_{2z}||I,C_{4z}^{\pm},M_{z}),(C_{2x}||C_{2x,2y},M_{da,db})$,
&&&& {\color{red}$0000111$} &&& \\

$(C_{2y}||C_{2a,2b},M_{x,y}),(TC_{2y}||T,TS_{4z}^{\mp},TC_{2z})$  
&&&&&&& \\
\hline

$L_{0\mathbf K}=D_{4}^{}, M_{\mathbf K}/L_{0\mathbf K}=\{E,C_{2z}\}\times Z_{2}^{T}$,
& $+1$ &$-1$ &$+1$ &$-1$ &$-1$ &$+1$ &$+1$ \\ 

$(C_{2z}||I,S_{4z}^{\mp},M_{z},M_{x,y},M_{da,db})$,
&&&& {\color{red}$0101100$} &&& \\

$(T||T,TC_{4z}^{\pm},TC_{2z},TC_{2x,2y},TC_{2a,2b})$
&&&&&&& \\  
\hline

$L_{0\mathbf K}=D_{4}^{}, M_{\mathbf K}/L_{0\mathbf K}=\{E,C_{2z}\}\times Z_{2}^{T}$,
& $+1$ &$+1$ &$-1$ &$+1$ &$+1$ &$+1$ &$+1$ \\ 

$(C_{2z}||I,S_{4z}^{\mp},M_{z},M_{x,y},M_{da,db})$,
&&&& {\color{red}$0010000$} &&& \\

$(TC_{2z}||T,TC_{4z}^{\pm},TC_{2z},TC_{2x,2y},TC_{2a,2b})$
&&&&&&& \\   
\hline

$L_{0\mathbf K}=D_{4}^{}, M_{\mathbf K}/L_{0\mathbf K}=\{E,C_{2z},TC_{2x},TC_{2y}\}$,
& $+1$ &$+1$ &$+1$ &$+1$ &$+1$ &$+1$ &$+1$ \\ 

$(C_{2z}||I,S_{4z}^{\mp},M_{z},M_{x,y},M_{da,db})$,
&&&&&&& \\

$(TC_{2y}||T,TC_{4z}^{\pm},TC_{2z},TC_{2x,2y},TC_{2a,2b})$
&&&&&&& \\   
\hline

$L_{0\mathbf K}=C_{4h}^{}, M_{\mathbf K}/L_{0\mathbf K}=\{E,C_{2z}\}\times Z_{2}^{T}$,
& $+1$ &$-1$ &$-1$ &$+1$ &$+1$ &$+1$ &$+1$ \\ 

$(C_{2z}||C_{2x,2y},C_{2a,2b},M_{x,y},M_{da,db})$,
&&&& {\color{red}$0110000$} &&& \\

$(T||T,TC_{4z}^{\pm},TC_{2z},IT,TS_{4z}^{\mp},TM_{z})$
&&&&&&& \\ 
\hline

$L_{0\mathbf K}=C_{4h}^{}, M_{\mathbf K}/L_{0\mathbf K}=\{E,C_{2z}\}\times Z_{2}^{T}$,
& $+1$ &$+1$ &$+1$ &$-1$ &$-1$ &$+1$ &$+1$ \\ 

$(C_{2z}||C_{2x,2y},C_{2a,2b},M_{x,y},M_{da,db})$,
&&&& {\color{red}$0001100$} &&& \\

$(TC_{2z}||T,TC_{4z}^{\pm},TC_{2z},IT,TS_{4z}^{\mp},TM_{z})$
&&&&&&& \\  
\hline

$L_{0\mathbf K}=C_{4h}^{}, M_{\mathbf K}/L_{0\mathbf K}=\{E,C_{2z},TC_{2x},TC_{2y}\}$,
& $+1$ &$+1$ &$+1$ &$+1$ &$+1$ &$+1$ &$+1$ \\ 

$(C_{2z}||C_{2x,2y},C_{2a,2b},M_{x,y},M_{da,db})$,
&&&&&&& \\

$(TC_{2y}||T,TC_{4z}^{\pm},TC_{2z},IT,TS_{4z}^{\mp},TM_{z})$
&&&&&&& \\   
\hline

\end{tabular}
\end{table*}

\begin{table*}
(Extension of Supplementary Table \ref{D4hdrztspin})\\
\centering
\begin{tabular}{ |c|ccccccc|}
\hline
$P=D_{4h}^{}$& && & spin invariants $\vec{\eta}^{S}$&&&\\

$G_{\mathbf K}\cong M_{\mathbf K}=D_{4h}^{}\times Z_2^T$
&$\eta_{C_{2x},C_{2y}}^{S}$
&$\eta_{T}^{S}$&$\eta_{IT}^{S}$ &$\eta_{TC_{2x}}^{S}$&$\eta_{TC_{2a}}^{S}$
&$\eta_{I,C_{2x}}^{S}$&
$\eta_{I,C_{2a}}^{S}$\\
\hline

$L_{0\mathbf K}=C_{4v}^{}, M_{\mathbf K}/L_{0\mathbf K}=\{E,C_{2z}\}\times Z_{2}^{T}$,
& $+1$ &$-1$ &$+1$ &$+1$ &$+1$ &$+1$ &$+1$ \\ 

$(C_{2z}||I,S_{4z}^{\mp},M_{z},C_{2x,2y},C_{2a,2b})$,
&&&&&&& \\

$(T||T,TC_{4z}^{\pm},TC_{2z},TM_{x,y},TM_{da,db})$
&&&&&&& \\ 
\hline

$L_{0\mathbf K}=C_{4v}^{}, M_{\mathbf K}/L_{0\mathbf K}=\{E,C_{2z}\}\times Z_{2}^{T}$,
& $+1$ &$+1$ &$-1$ &$-1$ &$-1$ &$+1$ &$+1$ \\ 

$(C_{2z}||I,S_{4z}^{\mp},M_{z},C_{2x,2y},C_{2a,2b})$,
&&&& {\color{red}$0011100$} &&& \\

$(TC_{2z}||T,TC_{4z}^{\pm},TC_{2z},TM_{x,y},TM_{da,db})$
&&&&&&& \\ 
\hline

$L_{0\mathbf K}=C_{4v}^{}, M_{\mathbf K}/L_{0\mathbf K}=\{E,C_{2z},TC_{2x},TC_{2y}\}$,
& $+1$ &$+1$ &$+1$ &$+1$ &$+1$ &$+1$ &$+1$ \\ 

$(C_{2z}||I,S_{4z}^{\mp},M_{z},C_{2x,2y},C_{2a,2b})$,
&&&&&&& \\

$(TC_{2y}||T,TC_{4z}^{\pm},TC_{2z},TM_{x,y},TM_{da,db})$
&&&&&&& \\  
\hline

$L_{0\mathbf K}=D_{2d}^{1}, M_{\mathbf K}/L_{0\mathbf K}=\{E,C_{2z}\}\times Z_{2}^{T}$,
& $+1$ &$-1$ &$+1$ &$-1$ &$+1$ &$+1$ &$+1$ \\ 

$(C_{2z}||I,C_{4z}^{\pm},M_{z},M_{x,y},C_{2a,2b})$,
&&&&&&& \\

$(T||T,TS_{4z}^{\mp},TC_{2z},TC_{2x,2y},TM_{da,db})$
&&&&&&&  \\  
\hline

$L_{0\mathbf K}=D_{2d}^{1}, M_{\mathbf K}/L_{0\mathbf K}=\{E,C_{2z}\}\times Z_{2}^{T}$,
& $+1$ &$+1$ &$-1$ &$+1$ &$-1$ &$+1$ &$+1$ \\ 

$(C_{2z}||I,C_{4z}^{\pm},M_{z},M_{x,y},C_{2a,2b})$,
&&&& {\color{red}$0010100$} &&& \\

$(TC_{2z}||T,TS_{4z}^{\mp},TC_{2z},TC_{2x,2y},TM_{da,db})$
&&&&&&&  \\  
\hline

$L_{0\mathbf K}=D_{2d}^{1}, M_{\mathbf K}/L_{0\mathbf K}=\{E,C_{2z},TC_{2x},TC_{2y}\}$,
& $+1$ &$+1$ &$+1$ &$+1$ &$+1$ &$+1$ &$+1$ \\ 

$(C_{2z}||I,C_{4z}^{\pm},M_{z},M_{x,y},C_{2a,2b})$,
&&&&&&& \\

$(TC_{2y}||T,TS_{4z}^{\mp},TC_{2z},TC_{2x,2y},TM_{da,db})$
&&&&&&&  \\  
\hline

$L_{0\mathbf K}=D_{2d}^{2}, M_{\mathbf K}/L_{0\mathbf K}=\{E,C_{2z}\}\times Z_{2}^{T}$,
& $+1$ &$-1$ &$+1$ &$+1$ &$-1$ &$+1$ &$+1$ \\ 

$(C_{2z}||I,C_{4z}^{\pm},M_{z},C_{2x,2y},M_{da,db})$,
&&&& {\color{red}$0100100$} &&& \\

$(T||T,TS_{4z}^{\mp},TC_{2z},TM_{x,y},TC_{2a,2b})$
&&&&&&& \\  
\hline

$L_{0\mathbf K}=D_{2d}^{2}, M_{\mathbf K}/L_{0\mathbf K}=\{E,C_{2z}\}\times Z_{2}^{T}$,
& $+1$ &$+1$ &$-1$ &$-1$ &$+1$ &$+1$ &$+1$ \\ 

$(C_{2z}||I,C_{4z}^{\pm},M_{z},C_{2x,2y},M_{da,db})$,
&&&& {\color{red}$0011000$} &&& \\

$(TC_{2z}||T,TS_{4z}^{\mp},TC_{2z},TM_{x,y},TC_{2a,2b})$
&&&&&&& \\  
\hline

$L_{0\mathbf K}=D_{2d}^{2}, M_{\mathbf K}/L_{0\mathbf K}=\{E,C_{2z},TC_{2x},TC_{2y}\}$,
& $+1$ &$+1$ &$+1$ &$+1$ &$+1$ &$+1$ &$+1$ \\ 

$(C_{2z}||I,C_{4z}^{\pm},M_{z},C_{2x,2y},M_{da,db})$,
&&&&&&& \\

$(TC_{2y}||T,TS_{4z}^{\mp},TC_{2z},TM_{x,y},TC_{2a,2b})$
&&&&&&& \\  
\hline

$L_{0\mathbf K}=D_{2h}^{}, M_{\mathbf K}/L_{0\mathbf K}=\{E,C_{2z}\}\times Z_{2}^{T}$,
& $+1$ &$-1$ &$-1$ &$-1$ &$+1$ &$+1$ &$+1$ \\ 

$(C_{2z}||C_{4z}^{\pm},C_{2a,2b},S_{4z}^{\mp},M_{da,db})$,
&&&& {\color{red}$0111000$} &&& \\

$(T||T,TC_{2x,2y,2z},IT,TM_{x,y,z})$
&&&&&&& \\  
\hline

$L_{0\mathbf K}=D_{2h}^{}, M_{\mathbf K}/L_{0\mathbf K}=\{E,C_{2z}\}\times Z_{2}^{T}$,
& $+1$ &$+1$ &$+1$ &$+1$ &$-1$ &$+1$ &$+1$ \\ 

$(C_{2z}||C_{4z}^{\pm},C_{2a,2b},S_{4z}^{\mp},M_{da,db})$,
&&&& {\color{red}$0000100$} &&& \\

$(TC_{2z}||T,TC_{2x,2y,2z},IT,TM_{x,y,z})$
&&&&&&& \\  
\hline

$L_{0\mathbf K}=D_{2h}^{}, M_{\mathbf K}/L_{0\mathbf K}=\{E,C_{2z},TC_{2x},TC_{2y}\}$,
& $+1$ &$+1$ &$+1$ &$+1$ &$+1$ &$+1$ &$+1$ \\ 

$(C_{2z}||C_{4z}^{\pm},C_{2a,2b},S_{4z}^{\mp},M_{da,db})$,
&&&&&&& \\

$(TC_{2y}||T,TC_{2x,2y,2z},IT,TM_{x,y,z})$
&&&&&&& \\  
\hline

$L_{0\mathbf K}=\{E,C_{2b,2a},C_{2z},I,M_{db,da},M_{z}\}$,
& $+1$ &$-1$ &$-1$ &$+1$ &$-1$ &$+1$ &$+1$ \\ 

$M_{\mathbf K}/L_{0\mathbf K}=\{E,C_{2z}\}\times Z_{2}^{T}$,
&&&& {\color{red}$0110100$} &&& \\

$(C_{2z}||C_{4z}^{\pm},C_{2x,2y},S_{4z}^{\mp},M_{x,y})$,
&&&&&&& \\

$(T||T,TC_{2b,2a},TC_{2z},IT,TM_{db,da},TM_{z})$
&&&&&&&  \\  
\hline

$L_{0\mathbf K}=\{E,C_{2b,2a},C_{2z},I,M_{db,da},M_{z}\}$,
& $+1$ &$+1$ &$+1$ &$-1$ &$+1$ &$+1$ &$+1$ \\ 

$M_{\mathbf K}/L_{0\mathbf K}=\{E,C_{2z}\}\times Z_{2}^{T}$,
&&&&&&& \\

$(C_{2z}||C_{4z}^{\pm},C_{2x,2y},S_{4z}^{\mp},M_{x,y})$,
&&&&&&& \\

$(TC_{2z}||T,TC_{2b,2a},TC_{2z},IT,TM_{db,da},TM_{z})$
&&&&&&&  \\  
\hline

$L_{0\mathbf K}=\{E,C_{2b,2a},C_{2z},I,M_{db,da},M_{z}\}$,
& $+1$ &$+1$ &$+1$ &$+1$ &$+1$ &$+1$ &$+1$ \\ 

$M_{\mathbf K}/L_{0\mathbf K}=\{E,C_{2z},TC_{2x},TC_{2y}\}$,
&&&&&&& \\

$(C_{2z}||C_{4z}^{\pm},C_{2x,2y},S_{4z}^{\mp},M_{x,y})$,
&&&&&&& \\

$(TC_{2y}||T,TC_{2b,2a},TC_{2z},IT,TM_{db,da},TM_{z})$
&&&&&&&  \\  
\hline

$L_{0\mathbf K}=D_{4h}^{},
M_{\mathbf K}/L_{0\mathbf K}=Z_{2}^{T}$
& $+1$ &$-1$ &$-1$ &$-1$ &$-1$ &$+1$ &$+1$ \\ 

$(T||T,TC_{4z}^{\pm},TC_{2z},TC_{2x,2y},TC_{2a,2b},$
&&&& {\color{red}$0111100$} &&& \\

$IT,TS_{4z}^{\mp},TM_{z},TM_{x,y},TM_{da,db})$
&&&&&&& \\ 
\hline

$L_{0\mathbf K}=D_{4h}^{},
M_{\mathbf K}/L_{0\mathbf K}=\{E,TC_{2y}\}$
& $+1$ &$+1$ &$+1$ &$+1$ &$+1$ &$+1$ &$+1$ \\ 

$(TC_{2y}||T,TC_{4z}^{\pm},TC_{2z},TC_{2x,2y},TC_{2a,2b},$
&&&&&&& \\

$IT,TS_{4z}^{\mp},TM_{z},TM_{x,y},TM_{da,db})$
&&&&&&& \\ 
\hline

\end{tabular}
\end{table*}

\begin{table*}[htbp]
\caption{
For ($G_{\mathbf K}\cong M_{\mathbf K}=D_{3}\times{Z}_2^T,L_{0\mathbf{K}},M_{\mathbf K}/L_{0\mathbf{K}}$),
we list generators of spin point group $(\varphi_{g}||g_{1},\ldots,g_{n})$, $(g_{1},\ldots,g_{n})$ is a coset of $L_{0\mathbf{K}}$,
$g\in (g_{1},\ldots,g_{n}),\varphi_{g}\in M_{\mathbf K}/L_{0\mathbf{K}}$,
$n$ is order of $L_{0\mathbf{K}}$, and
spin invariants $\vec{\eta}^{S}$ of $G_{\mathbf K}$ is
$\eta^S_{p_1,p_2}{\cdot1_{2\times 2}} = d({\varphi_{p_1}})d({\varphi_{p_2}})d^{-1}({\varphi_{p_1}})d^{-1}({\varphi_{p_2}})$
for unitary elements $p_{1},p_{2}$ with $p_{1}p_{2}=p_{2}p_{1}$ and
$\eta^S_{p}{\cdot1_{2\times 2}}=[d(\varphi_p) {K}]^2$
for anti-unitary element $p$ with $p^{2}=E$,
where irRep of $g\in G_{\mathbf K}$ is given by
$d(\varphi_g)K^{\zeta_g}=u\Big( s( T^{\zeta_g}\varphi_g)\Big)(i\sigma_{y}K)^{\zeta_g}$,
$s( T^{\zeta_g}\varphi_g)$ is vector Rep of $T^{\zeta_g}\varphi_{g}\in\mathrm{SO(3)}$,
$u\Big( s( T^{\zeta_g}\varphi_g)\Big)$ is $\mathrm{SU(2)}$ Rep of $T^{\zeta_g}\varphi_{g}$,
$\zeta_{g}=0(1)$ for unitary (anti-unitary) $g$.
The $\vec{\eta}^{S}$ leading to invariants only realized in SSG are labeled by Boolean vectors
colored in {\color{red}RED}.} \label{D3drztspin}
\centering
\begin{tabular}{ |c|ccc|}
\hline
& & spin invariants $\vec{\eta}^{S}$& \\

$P=D_{3}^{1},G_{\mathbf K}\cong M_{\mathbf K}=D_{3}^{1}\times Z_2^T$
& $\eta_{T}^{S}$ & $\eta_{TC'_{21}}^{S}$& \\

$P=D_{3}^{2},G_{\mathbf K}\cong M_{\mathbf K}=D_{3}^{2}\times Z_2^T$& $\eta_{T}^{S}$ & $\eta_{TC''_{21}}^{S}$& \\
\hline

$M_{\mathbf K}=D_{3}^{1}\times Z_2^T,L_{0\mathbf K}=E,
M_{\mathbf K}/L_{0\mathbf K}=D_{3}^{1}\times Z_{2}^{T}$,
&$-1$ & $+1$& \\

$(C'_{21}||C'_{21}),(C'_{22}||C'_{22}),(T||T)$
&&& \\  
\hline

$M_{\mathbf K}=D_{3}^{1}\times Z_2^T,L_{0\mathbf K}=E$,
&$+1$ & $+1$& \\

$M_{\mathbf K}/L_{0\mathbf K}=\{E,C_{3}^{+},C_{3}^{-},C'_{21,22,23},TC_{2},TC_{6}^{-},TC_{6}^{+},TC''_{21,22,23}\}$,
&&& \\

$(C'_{21}||C'_{21}),(C'_{22}||C'_{22}),(TC_{2}||T)$
&&&\\ 
\hline

$M_{\mathbf K}=D_{3}^{2}\times Z_2^T,L_{0\mathbf K}=E,
M_{\mathbf K}/L_{0\mathbf K}=D_{3}^{2}\times Z_{2}^{T}$,
&$-1$ & $+1$& \\

$(C''_{21}||C''_{21}),(C''_{22}||C''_{22}),(T||T)$
&&& \\  
\hline

$M_{\mathbf K}=D_{3}^{2}\times Z_2^T,L_{0\mathbf K}=E$,
&$+1$ & $+1$& \\

$M_{\mathbf K}/L_{0\mathbf K}=\{E,C_{3}^{+},C_{3}^{-},C''_{21,22,23},TC_{2},TC_{6}^{-},TC_{6}^{+},TC'_{21,22,23}\}$,
&&& \\

$(C''_{21}||C''_{21}),(C''_{22}||C''_{22}),(TC_{2}||T)$
&&&\\ 
\hline

$M_{\mathbf K}=D_{3}^{1}\times Z_2^T,L_{0\mathbf K}=C_{3},
M_{\mathbf K}/L_{0\mathbf K}=\{E,C_{2z}\}\times Z_{2}^{T}$,
&$-1$ & $+1$& \\

$(C_{2z}||C'_{21,22,23}),(T||T,TC_{3}^{+},TC_{3}^{-})$
&&&   \\  
\hline

$M_{\mathbf K}=D_{3}^{1}\times Z_2^T,L_{0\mathbf K}=C_{3},
M_{\mathbf K}/L_{0\mathbf K}=\{E,C_{2z}\}\times Z_{2}^{T}$,
&$+1$ & $-1$& {\color{red}$01$} \\

$(C_{2z}||C'_{21,22,23}),(TC_{2z}||T,TC_{3}^{+},TC_{3}^{-})$
&&&  \\  
\hline

$M_{\mathbf K}=D_{3}^{1}\times Z_2^T,L_{0\mathbf K}=C_{3},
M_{\mathbf K}/L_{0\mathbf K}=\{E,C_{2z},TC_{2x},TC_{2y}\}$,
&$+1$ & $+1$& \\

$(C_{2z}||C'_{21,22,23}),(TC_{2y}||T,TC_{3}^{+},TC_{3}^{-})$
&&& \\ 
\hline

$M_{\mathbf K}=D_{3}^{2}\times Z_2^T,L_{0\mathbf K}=C_{3},
M_{\mathbf K}/L_{0\mathbf K}=\{E,C_{2z}\}\times Z_{2}^{T}$,
&$-1$ & $+1$& \\

$(C_{2z}||C''_{21,22,23}),(T||T,TC_{3}^{+},TC_{3}^{-})$
&&&   \\  
\hline

$M_{\mathbf K}=D_{3}^{2}\times Z_2^T,L_{0\mathbf K}=C_{3},
M_{\mathbf K}/L_{0\mathbf K}=\{E,C_{2z}\}\times Z_{2}^{T}$,
&$+1$ & $-1$& {\color{red}$01$} \\

$(C_{2z}||C''_{21,22,23}),(TC_{2z}||T,TC_{3}^{+},TC_{3}^{-})$
&&&  \\  
\hline

$M_{\mathbf K}=D_{3}^{2}\times Z_2^T,L_{0\mathbf K}=C_{3},
M_{\mathbf K}/L_{0\mathbf K}=\{E,C_{2z},TC_{2x},TC_{2y}\}$,
&$+1$ & $+1$& \\

$(C_{2z}||C''_{21,22,23}),(TC_{2y}||T,TC_{3}^{+},TC_{3}^{-})$
&&& \\ 
\hline

$L_{0\mathbf K}=D_{3}^{1},
M_{\mathbf K}/L_{0\mathbf K}=Z_{2}^{T},
(T||T,TC_{3}^{+},TC_{3}^{-},TC'_{21,22,23})$
& $-1$ &$-1$&{\color{red}$11$} \\ 
\hline

$L_{0\mathbf K}=D_{3}^{1},
M_{\mathbf K}/L_{0\mathbf K}=\{E,TC_{2y}\},
(TC_{2y}||T,TC_{3}^{+},TC_{3}^{-},TC'_{21,22,23})$
& $+1$ &$+1$&\\ 
\hline

$L_{0\mathbf K}=D_{3}^{2},
M_{\mathbf K}/L_{0\mathbf K}=Z_{2}^{T},
(T||T,TC_{3}^{+},TC_{3}^{-},TC''_{21,22,23})$
& $-1$ &$-1$&{\color{red}$11$} \\ 
\hline

$L_{0\mathbf K}=D_{3}^{2},
M_{\mathbf K}/L_{0\mathbf K}=\{E,TC_{2y}\},
(TC_{2y}||T,TC_{3}^{+},TC_{3}^{-},TC''_{21,22,23})$
& $+1$ &$+1$&\\ 
\hline

\end{tabular}
\end{table*}

\clearpage

\begin{table*}[htbp]
\caption{
For ($G_{\mathbf K}\cong M_{\mathbf K}=D_{3d}\times{Z}_2^T,L_{0\mathbf{K}},M_{\mathbf K}/L_{0\mathbf{K}}$),
we list generators of spin point group $(\varphi_{g}||g_{1},\ldots,g_{n})$, $(g_{1},\ldots,g_{n})$ is a coset of $L_{0\mathbf{K}}$,
$g\in (g_{1},\ldots,g_{n}),\varphi_{g}\in M_{\mathbf K}/L_{0\mathbf{K}}$,
$n$ is order of $L_{0\mathbf{K}}$, and
spin invariants $\vec{\eta}^{S}$ of $G_{\mathbf K}$ is
$\eta^S_{p_1,p_2}{\cdot1_{2\times 2}} = d({\varphi_{p_1}})d({\varphi_{p_2}})d^{-1}({\varphi_{p_1}})d^{-1}({\varphi_{p_2}})$
for unitary elements $p_{1},p_{2}$ with $p_{1}p_{2}=p_{2}p_{1}$ and
$\eta^S_{p}{\cdot1_{2\times 2}}=[d(\varphi_p) {K}]^2$
for anti-unitary element $p$ with $p^{2}=E$,
where irRep of $g\in G_{\mathbf K}$ is given by
$d(\varphi_g)K^{\zeta_g}=u\Big( s( T^{\zeta_g}\varphi_g)\Big)(i\sigma_{y}K)^{\zeta_g}$,
$s( T^{\zeta_g}\varphi_g)$ is vector Rep of $T^{\zeta_g}\varphi_{g}\in\mathrm{SO(3)}$,
$u\Big( s( T^{\zeta_g}\varphi_g)\Big)$ is $\mathrm{SU(2)}$ Rep of $T^{\zeta_g}\varphi_{g}$,
$\zeta_{g}=0(1)$ for unitary (anti-unitary) $g$.
The $\vec{\eta}^{S}$ leading to invariants only realized in SSG are labeled by Boolean vectors
colored in {\color{red}RED}.} \label{D3ddrztspin}
\centering
\begin{tabular}{ |c|cccc|}
\hline
& & spin invariants $\vec{\eta}^{S}$  & &\\

$P=D_{3d}^{1},G_{\mathbf K}\cong M_{\mathbf K}=D_{3d}^{1}\times Z_2^T$
&$\eta_{I,C'_{21}}^{S}$
&$\eta_{T}^{S}$&$\eta_{IT}^{S}$ &$\eta_{TC'_{21}}^{S}$\\

$P=D_{3d}^{2},G_{\mathbf K}\cong M_{\mathbf K}=D_{3d}^{2}\times Z_2^T$
&$\eta_{I,C''_{21}}^{S}$
&$\eta_{T}^{S}$&$\eta_{IT}^{S}$ &$\eta_{TC''_{21}}^{S}$\\
\hline

$M_{\mathbf K}=D_{3d}^{1}\times Z_2^T,L_{0\mathbf K}=E,
M_{\mathbf K}/L_{0\mathbf K}=D_{6}\times Z_{2}^{T}$,
& $-1$ &$-1$ &$+1$ &$+1$ \\  

$(C_{2}||I),(C'_{21}||C'_{21}),(C'_{22}||C'_{22}),(T||T)$
&&&& \\  

\hline

$M_{\mathbf K}=D_{3d}^{1}\times Z_2^T,L_{0\mathbf K}=E,
M_{\mathbf K}/L_{0\mathbf K}=D_{6}\times Z_{2}^{T}$,
& $-1$ &$+1$ &$-1$ &$+1$ \\  

$(C_{2}||I),(C'_{21}||C'_{21}),(C'_{22}||C'_{22}),(TC_{2}||T)$
&&&& \\  

\hline

$M_{\mathbf K}=D_{3d}^{2}\times Z_2^T,L_{0\mathbf K}=E,
M_{\mathbf K}/L_{0\mathbf K}=D_{6}\times Z_{2}^{T}$,
& $-1$ &$-1$ &$+1$ &$+1$ \\  

$(C_{2}||I),(C''_{21}||C''_{21}),(C''_{22}||C''_{22}),(T||T)$
&&&& \\  

\hline

$M_{\mathbf K}=D_{3d}^{2}\times Z_2^T,L_{0\mathbf K}=E,
M_{\mathbf K}/L_{0\mathbf K}=D_{6}\times Z_{2}^{T}$,
& $-1$ &$+1$ &$-1$ &$+1$ \\  

$(C_{2}||I),(C''_{21}||C''_{21}),(C''_{22}||C''_{22}),(TC_{2}||T)$
&&&& \\  

\hline

$M_{\mathbf K}=D_{3d}^{1}\times Z_2^T,L_{0\mathbf K}=\{E,I\},
M_{\mathbf K}/L_{0\mathbf K}=D_{3}^{1}\times Z_{2}^{T}$,
& $+1$ &$-1$ &$-1$ &$+1$ \\  

$(C'_{21}||C'_{21},M_{d1}),(C'_{22}||C'_{22},M_{d2}),(T||T,IT)$
&&&& \\  

\hline

$M_{\mathbf K}=D_{3d}^{1}\times Z_2^T,L_{0\mathbf K}=\{E,I\}$,
& $+1$ &$+1$ &$+1$ &$+1$ \\  

$M_{\mathbf K}/L_{0\mathbf K}=\{E,C_{3}^{+},C_{3}^{-},C'_{21,22,23},TC_{2},TC_{6}^{-},TC_{6}^{+},TC''_{21,22,23}\}$,
&&&& \\

$(C'_{21}||C'_{21},M_{d1}),(C'_{22}||C'_{22},M_{d2}),(TC_{2}||T,IT)$
&&&& \\ 

\hline

$M_{\mathbf K}=D_{3d}^{2}\times Z_2^T,L_{0\mathbf K}=\{E,I\},
M_{\mathbf K}/L_{0\mathbf K}=D_{3}^{2}\times Z_{2}^{T}$,
& $+1$ &$-1$ &$-1$ &$+1$ \\  

$(C''_{21}||C''_{21},M_{v1}),(C''_{22}||C''_{22},M_{v2}),(T||T,IT)$
&&&& \\  

\hline

$M_{\mathbf K}=D_{3d}^{2}\times Z_2^T,L_{0\mathbf K}=\{E,I\}$,
& $+1$ &$+1$ &$+1$ &$+1$ \\  

$M_{\mathbf K}/L_{0\mathbf K}=\{E,C_{3}^{+},C_{3}^{-},C''_{21,22,23},TC_{2},TC_{6}^{-},TC_{6}^{+},TC'_{21,22,23}\}$,
&&&& \\

$(C''_{21}||C''_{21},M_{v1}),(C''_{22}||C''_{22},M_{v2}),(TC_{2}||T,IT)$
&&&& \\ 

\hline

$M_{\mathbf K}=D_{3d}^{1}\times Z_2^T,L_{0\mathbf K}=C_{3},
M_{\mathbf K}/L_{0\mathbf K}=\{E,C'_{21},C''_{21},C_{2}\}\times Z_{2}^{T}$,
& $-1$ &$-1$ &$+1$ &$+1$ \\ 

$(C_{2}||I,S_{6}^{-},S_{6}^{+}),(C'_{21}||C'_{21,22,23}),(T||T,TC_{3}^{+},TC_{3}^{-})$
&&&& \\  
\hline

$M_{\mathbf K}=D_{3d}^{1}\times Z_2^T,L_{0\mathbf K}=C_{3},
M_{\mathbf K}/L_{0\mathbf K}=\{E,C'_{21},C''_{21},C_{2}\}\times Z_{2}^{T}$,
& $-1$ &$+1$ &$-1$ &$+1$ \\ 

$(C_{2}||I,S_{6}^{-},S_{6}^{+}),(C'_{21}||C'_{21,22,23}),(TC_{2}||T,TC_{3}^{+},TC_{3}^{-})$
&&&& \\  
\hline

$M_{\mathbf K}=D_{3d}^{1}\times Z_2^T,L_{0\mathbf K}=C_{3},
M_{\mathbf K}/L_{0\mathbf K}=\{E,C'_{21},C''_{21},C_{2}\}\times Z_{2}^{T}$,
& $-1$ &$+1$ &$+1$ &$-1$ \\ 

$(C_{2}||I,S_{6}^{-},S_{6}^{+}),(C'_{21}||C'_{21,22,23}),(TC'_{21}||T,TC_{3}^{+},TC_{3}^{-})$
&& {\color{red}$1001$}  && \\  
\hline

$M_{\mathbf K}=D_{3d}^{1}\times Z_2^T,L_{0\mathbf K}=C_{3},
M_{\mathbf K}/L_{0\mathbf K}=\{E,C'_{21},C''_{21},C_{2}\}\times Z_{2}^{T}$,
& $-1$ &$+1$ &$+1$ &$+1$ \\ 

$(C_{2}||I,S_{6}^{-},S_{6}^{+}),(C'_{21}||C'_{21,22,23}),(TC''_{21}||T,TC_{3}^{+},TC_{3}^{-})$
&&&& \\   
\hline

$M_{\mathbf K}=D_{3d}^{2}\times Z_2^T,L_{0\mathbf K}=C_{3},
M_{\mathbf K}/L_{0\mathbf K}=\{E,C'_{21},C''_{21},C_{2}\}\times Z_{2}^{T}$,
& $-1$ &$-1$ &$+1$ &$+1$ \\ 

$(C_{2}||I,S_{6}^{-},S_{6}^{+}),(C''_{21}||C''_{21,22,23}),(T||T,TC_{3}^{+},TC_{3}^{-})$
&&&& \\   
\hline

$M_{\mathbf K}=D_{3d}^{2}\times Z_2^T,L_{0\mathbf K}=C_{3},
M_{\mathbf K}/L_{0\mathbf K}=\{E,C'_{21},C''_{21},C_{2}\}\times Z_{2}^{T}$,
& $-1$ &$+1$ &$-1$ &$+1$ \\ 

$(C_{2}||I,S_{6}^{-},S_{6}^{+}),(C''_{21}||C''_{21,22,23}),(TC_{2}||T,TC_{3}^{+},TC_{3}^{-})$
&&&& \\  
\hline

$M_{\mathbf K}=D_{3d}^{2}\times Z_2^T,L_{0\mathbf K}=C_{3},
M_{\mathbf K}/L_{0\mathbf K}=\{E,C'_{21},C''_{21},C_{2}\}\times Z_{2}^{T}$,
& $-1$ &$+1$ &$+1$ &$+1$ \\ 

$(C_{2}||I,S_{6}^{-},S_{6}^{+}),(C''_{21}||C''_{21,22,23}),(TC'_{21}||T,TC_{3}^{+},TC_{3}^{-})$
&&&& \\  
\hline

$M_{\mathbf K}=D_{3d}^{2}\times Z_2^T,L_{0\mathbf K}=C_{3},
M_{\mathbf K}/L_{0\mathbf K}=\{E,C'_{21},C''_{21},C_{2}\}\times Z_{2}^{T}$,
& $-1$ &$+1$ &$+1$ &$-1$ \\ 

$(C_{2}||I,S_{6}^{-},S_{6}^{+}),(C''_{21}||C''_{21,22,23}),(TC''_{21}||T,TC_{3}^{+},TC_{3}^{-})$
&& {\color{red}$1001$} && \\  
\hline

$M_{\mathbf K}=D_{3d}^{1}\times Z_2^T,L_{0\mathbf K}=C_{3i},
M_{\mathbf K}/L_{0\mathbf K}=\{E,C_{2}\}\times Z_{2}^{T}$,
& $+1$ &$-1$ &$-1$ &$+1$ \\ 

$(C_{2}||C'_{21,22,23},M_{d1,d2,d3}),(T||T,TC_{3}^{+},TC_{3}^{-},IT,TS_{6}^{-},TS_{6}^{+})$
&&&& \\ 
\hline

$M_{\mathbf K}=D_{3d}^{1}\times Z_2^T,L_{0\mathbf K}=C_{3i},
M_{\mathbf K}/L_{0\mathbf K}=\{E,C_{2}\}\times Z_{2}^{T}$,
& $+1$ &$+1$ &$+1$ &$-1$ \\ 

$(C_{2}||C'_{21,22,23},M_{d1,d2,d3}),(TC_{2}||T,TC_{3}^{+},TC_{3}^{-},IT,TS_{6}^{-},TS_{6}^{+})$
&& {\color{red}$0001$}  && \\ 
\hline

$M_{\mathbf K}=D_{3d}^{1}\times Z_2^T,L_{0\mathbf K}=C_{3i},
M_{\mathbf K}/L_{0\mathbf K}=\{E,C_{2},TC'_{21},TC''_{21}\}$,
& $+1$ &$+1$ &$+1$ &$+1$ \\ 

$(C_{2}||C'_{21,22,23},M_{d1,d2,d3}),(TC''_{21}||T,TC_{3}^{+},TC_{3}^{-},IT,TS_{6}^{-},TS_{6}^{+})$
&&&& \\  
\hline

$M_{\mathbf K}=D_{3d}^{2}\times Z_2^T,L_{0\mathbf K}=C_{3i},
M_{\mathbf K}/L_{0\mathbf K}=\{E,C_{2}\}\times Z_{2}^{T}$,
& $+1$ &$-1$ &$-1$ &$+1$ \\ 

$(C_{2}||C''_{21,22,23},M_{v1,v2,v3}),(T||T,TC_{3}^{+},TC_{3}^{-},IT,TS_{6}^{-},TS_{6}^{+})$
&&&& \\  
\hline

$M_{\mathbf K}=D_{3d}^{2}\times Z_2^T,L_{0\mathbf K}=C_{3i},
M_{\mathbf K}/L_{0\mathbf K}=\{E,C_{2}\}\times Z_{2}^{T}$,
& $+1$ &$+1$ &$+1$ &$-1$ \\ 

$(C_{2}||C''_{21,22,23},M_{v1,v2,v3}),(TC_{2}||T,TC_{3}^{+},TC_{3}^{-},IT,TS_{6}^{-},TS_{6}^{+})$
&& {\color{red}$0001$} && \\  
\hline

$M_{\mathbf K}=D_{3d}^{2}\times Z_2^T,L_{0\mathbf K}=C_{3i},
M_{\mathbf K}/L_{0\mathbf K}=\{E,C_{2},TC'_{21},TC''_{21}\}$,
& $+1$ &$+1$ &$+1$ &$+1$ \\ 

$(C_{2}||C''_{21,22,23},M_{v1,v2,v3}),(TC''_{21}||T,TC_{3}^{+},TC_{3}^{-},IT,TS_{6}^{-},TS_{6}^{+})$
&&&& \\   
\hline

$M_{\mathbf K}=D_{3d}^{1}\times Z_2^T,L_{0\mathbf K}=D_{3}^{1},
M_{\mathbf K}/L_{0\mathbf K}=\{E,C_{2}\}\times Z_{2}^{T}$,
& $+1$ &$-1$ &$+1$ &$-1$ \\ 

$(C_{2}||I,S_{6}^{-},S_{6}^{+},M_{d1,d2,d3}),(T||T,TC_{3}^{+},TC_{3}^{-},TC'_{21,22,23})$
&& {\color{red}$0101$}  && \\  
\hline

$M_{\mathbf K}=D_{3d}^{1}\times Z_2^T,L_{0\mathbf K}=D_{3}^{1},
M_{\mathbf K}/L_{0\mathbf K}=\{E,C_{2}\}\times Z_{2}^{T}$,
& $+1$ &$+1$ &$-1$ &$+1$ \\ 

$(C_{2}||I,S_{6}^{-},S_{6}^{+},M_{d1,d2,d3}),(TC_{2}||T,TC_{3}^{+},TC_{3}^{-},TC'_{21,22,23})$
&&&& \\  
\hline

\end{tabular}
\end{table*}

\begin{table*}
(Extension of Supplementary Table \ref{D3ddrztspin})\\
\centering
\begin{tabular}{ |c|cccc|}
\hline
& & spin invariants $\vec{\eta}^{S}$  & &\\

$P=D_{3d}^{1},G_{\mathbf K}\cong M_{\mathbf K}=D_{3d}^{1}\times Z_2^T$
&$\eta_{I,C'_{21}}^{S}$
&$\eta_{T}^{S}$&$\eta_{IT}^{S}$ &$\eta_{TC'_{21}}^{S}$\\

$P=D_{3d}^{2},G_{\mathbf K}\cong M_{\mathbf K}=D_{3d}^{2}\times Z_2^T$
&$\eta_{I,C''_{21}}^{S}$
&$\eta_{T}^{S}$&$\eta_{IT}^{S}$ &$\eta_{TC''_{21}}^{S}$\\
\hline

$M_{\mathbf K}=D_{3d}^{1}\times Z_2^T,L_{0\mathbf K}=D_{3}^{1},
M_{\mathbf K}/L_{0\mathbf K}=\{E,C_{2},TC'_{21},TC''_{21}\}$,
& $+1$ &$+1$ &$+1$ &$+1$ \\ 

$(C_{2}||I,S_{6}^{-},S_{6}^{+},M_{d1,d2,d3}),(TC''_{21}||T,TC_{3}^{+},TC_{3}^{-},TC'_{21,22,23})$
&&&& \\ 
\hline

$M_{\mathbf K}=D_{3d}^{2}\times Z_2^T,L_{0\mathbf K}=D_{3}^{2},
M_{\mathbf K}/L_{0\mathbf K}=\{E,C_{2}\}\times Z_{2}^{T}$,
& $+1$ &$-1$ &$+1$ &$-1$ \\ 

$(C_{2}||I,S_{6}^{-},S_{6}^{+},M_{v1,v2,v3}),(T||T,TC_{3}^{+},TC_{3}^{-},TC''_{21,22,23})$
&& {\color{red}$0101$}  && \\  
\hline

$M_{\mathbf K}=D_{3d}^{2}\times Z_2^T,L_{0\mathbf K}=D_{3}^{2},
M_{\mathbf K}/L_{0\mathbf K}=\{E,C_{2}\}\times Z_{2}^{T}$,
& $+1$ &$+1$ &$-1$ &$+1$ \\ 

$(C_{2}||I,S_{6}^{-},S_{6}^{+},M_{v1,v2,v3}),(TC_{2}||T,TC_{3}^{+},TC_{3}^{-},TC''_{21,22,23})$
&&&& \\  
\hline

$M_{\mathbf K}=D_{3d}^{2}\times Z_2^T,L_{0\mathbf K}=D_{3}^{2},
M_{\mathbf K}/L_{0\mathbf K}=\{E,C_{2},TC'_{21},TC''_{21}\}$,
& $+1$ &$+1$ &$+1$ &$+1$ \\ 

$(C_{2}||I,S_{6}^{-},S_{6}^{+},M_{v1,v2,v3}),(TC''_{21}||T,TC_{3}^{+},TC_{3}^{-},TC''_{21,22,23})$
&&&& \\  
\hline

$M_{\mathbf K}=D_{3d}^{1}\times Z_2^T,L_{0\mathbf K}=C_{3v}^{1},
M_{\mathbf K}/L_{0\mathbf K}=\{E,C_{2}\}\times Z_{2}^{T}$,
& $+1$ &$-1$ &$+1$ &$+1$ \\ 

$(C_{2}||I,S_{6}^{-},S_{6}^{+},C'_{21,22,23}),(T||T,TC_{3}^{+},TC_{3}^{-},TM_{d1,d2,d3})$
&&&& \\  
\hline

$M_{\mathbf K}=D_{3d}^{1}\times Z_2^T,L_{0\mathbf K}=C_{3v}^{1},
M_{\mathbf K}/L_{0\mathbf K}=\{E,C_{2}\}\times Z_{2}^{T}$,
& $+1$ &$+1$ &$-1$ &$-1$ \\ 

$(C_{2}||I,S_{6}^{-},S_{6}^{+},C'_{21,22,23}),(TC_{2}||T,TC_{3}^{+},TC_{3}^{-},TM_{d1,d2,d3})$
&& {\color{red}$0011$}  && \\ 
\hline

$M_{\mathbf K}=D_{3d}^{1}\times Z_2^T,L_{0\mathbf K}=C_{3v}^{1},
M_{\mathbf K}/L_{0\mathbf K}=\{E,C_{2},TC'_{21},TC''_{21}\}$,
& $+1$ &$+1$ &$+1$ &$+1$ \\ 

$(C_{2}||I,S_{6}^{-},S_{6}^{+},C'_{21,22,23}),(TC''_{21}||T,TC_{3}^{+},TC_{3}^{-},TM_{d1,d2,d3})$
&&&& \\ 
\hline

$M_{\mathbf K}=D_{3d}^{2}\times Z_2^T,L_{0\mathbf K}=C_{3v}^{2},
M_{\mathbf K}/L_{0\mathbf K}=\{E,C_{2}\}\times Z_{2}^{T}$,
& $+1$ &$-1$ &$+1$ &$+1$ \\ 

$(C_{2}||I,S_{6}^{-},S_{6}^{+},C''_{21,22,23}),(T||T,TC_{3}^{+},TC_{3}^{-},TM_{v1,v2,v3})$
&&&& \\  
\hline

$M_{\mathbf K}=D_{3d}^{2}\times Z_2^T,L_{0\mathbf K}=C_{3v}^{2},
M_{\mathbf K}/L_{0\mathbf K}=\{E,C_{2}\}\times Z_{2}^{T}$,
& $+1$ &$+1$ &$-1$ &$-1$ \\ 

$(C_{2}||I,S_{6}^{-},S_{6}^{+},C''_{21,22,23}),(TC_{2}||T,TC_{3}^{+},TC_{3}^{-},TM_{v1,v2,v3})$
&& {\color{red}$0011$} && \\  
\hline

$M_{\mathbf K}=D_{3d}^{2}\times Z_2^T,L_{0\mathbf K}=C_{3v}^{2},
M_{\mathbf K}/L_{0\mathbf K}=\{E,C_{2},TC'_{21},TC''_{21}\}$,
& $+1$ &$+1$ &$+1$ &$+1$ \\ 

$(C_{2}||I,S_{6}^{-},S_{6}^{+},C''_{21,22,23}),(TC''_{21}||T,TC_{3}^{+},TC_{3}^{-},TM_{v1,v2,v3})$
&&&& \\  
\hline

$L_{0\mathbf K}=D_{3d}^{1},
M_{\mathbf K}/L_{0\mathbf K}=Z_{2}^{T}$,
& $+1$ &$-1$ &$-1$ &$-1$ \\ 

$(T||T,TC_{3}^{+},TC_{3}^{-},TC'_{21,22,23},IT,TS_{6}^{-},TS_{6}^{+},TM_{d1,d2,d3})$
&& {\color{red}$0111$} && \\ 
\hline

$L_{0\mathbf K}=D_{3d}^{1},
M_{\mathbf K}/L_{0\mathbf K}=\{E,TC_{2y}\}$,
& $+1$ &$+1$ &$+1$ &$+1$ \\ 

$(TC_{2y}||T,TC_{3}^{+},TC_{3}^{-},TC'_{21,22,23},IT,TS_{6}^{-},TS_{6}^{+},TM_{d1,d2,d3})$
&&&& \\ 
\hline

$L_{0\mathbf K}=D_{3d}^{2},
M_{\mathbf K}/L_{0\mathbf K}=Z_{2}^{T}$,
& $+1$ &$-1$ &$-1$ &$-1$ \\ 

$(T||T,TC_{3}^{+},TC_{3}^{-},TC''_{21,22,23},IT,TS_{6}^{-},TS_{6}^{+},TM_{v1,v2,v3})$
&& {\color{red}$0111$} && \\ 
\hline

$L_{0\mathbf K}=D_{3d}^{2},
M_{\mathbf K}/L_{0\mathbf K}=\{E,TC_{2y}\}$,
& $+1$ &$+1$ &$+1$ &$+1$ \\ 

$(TC_{2y}||T,TC_{3}^{+},TC_{3}^{-},TC''_{21,22,23},IT,TS_{6}^{-},TS_{6}^{+},TM_{v1,v2,v3})$
&&&& \\ 
\hline
\end{tabular}
\end{table*}

\begin{table*}[htbp]
\caption{
For ($G_{\mathbf K}\cong M_{\mathbf K}=C_{3h}\times{Z}_2^T,L_{0\mathbf{K}},M_{\mathbf K}/L_{0\mathbf{K}}$),
we list generators of spin point group $(\varphi_{g}||g_{1},\ldots,g_{n})$, $(g_{1},\ldots,g_{n})$ is a coset of $L_{0\mathbf{K}}$,
$g\in (g_{1},\ldots,g_{n}),\varphi_{g}\in M_{\mathbf K}/L_{0\mathbf{K}}$,
$n$ is order of $L_{0\mathbf{K}}$, and
spin invariants $\vec{\eta}^{S}$ of $G_{\mathbf K}$ is
$\eta^S_{p_1,p_2}{\cdot1_{2\times 2}} = d({\varphi_{p_1}})d({\varphi_{p_2}})d^{-1}({\varphi_{p_1}})d^{-1}({\varphi_{p_2}})$
for unitary elements $p_{1},p_{2}$ with $p_{1}p_{2}=p_{2}p_{1}$ and
$\eta^S_{p}{\cdot1_{2\times 2}}=[d(\varphi_p) {K}]^2$
for anti-unitary element $p$ with $p^{2}=E$,
where irRep of $g\in G_{\mathbf K}$ is given by
$d(\varphi_g)K^{\zeta_g}=u\Big( s( T^{\zeta_g}\varphi_g)\Big)(i\sigma_{y}K)^{\zeta_g}$,
$s( T^{\zeta_g}\varphi_g)$ is vector Rep of $T^{\zeta_g}\varphi_{g}\in\mathrm{SO(3)}$,
$u\Big( s( T^{\zeta_g}\varphi_g)\Big)$ is $\mathrm{SU(2)}$ Rep of $T^{\zeta_g}\varphi_{g}$,
$\zeta_{g}=0(1)$ for unitary (anti-unitary) $g$.
The $\vec{\eta}^{S}$ leading to invariants only realized in SSG are labeled by Boolean vectors
colored in {\color{red}RED}.} \label{C3hdrztspin}
\centering
\begin{tabular}{ |c|ccc|}
\hline
$P=C_{3h}$& & spin invariants $\vec{\eta}^{S}$ &  \\

$G_{\mathbf K}\cong M_{\mathbf K}=C_{3h}\times Z_2^T$
& $\eta_{T}^{S}$ & $\eta_{TM_{h}}^{S}$& \\
\hline

$L_{0\mathbf K}=E,M_{\mathbf K}/L_{0\mathbf K}=C_{6}\times Z_{2}^{T},
(C_{6}^{+}||S_{3}^{-}),(C_{2}||M_{h}),(T||T)$
&$-1$ & $+1$& \\  
\hline

$L_{0\mathbf K}=E,M_{\mathbf K}/L_{0\mathbf K}=C_{6}\times Z_{2}^{T},
(C_{6}^{+}||S_{3}^{-}),(C_{2}||M_{h}),(TC_{2}||T)$
&$+1$ & $-1$ & {\color{red}$01$}\\ 
\hline

$L_{0\mathbf K}=\{E,M_{h}\},M_{\mathbf K}/L_{0\mathbf K}=C_{3}\times Z_{2}^{T},
(C_{3}^{+}||C_{3}^{+},S_{3}^{+}),(T||T,TM_{h})$
& $-1$ &$-1$ & {\color{red}$11$}\\  
\hline

$L_{0\mathbf K}=\{E,M_{h}\},
M_{\mathbf K}/L_{0\mathbf K}=\{E,C_{3}^{+},C_{3}^{-},TC_{2},TC_{6}^{-},TC_{6}^{+}\}$,
& $+1$ &$+1$ &\\   

$(C_{3}^{+}||C_{3}^{+},S_{3}^{+}),(TC_{2}||T,TM_{h})$
&&& \\
\hline

$L_{0\mathbf K}=C_{3},
M_{\mathbf K}/L_{0\mathbf K}=\{E,C_{2}\}\times Z_{2}^{T}$,
& $-1$ &$+1$& \\

$(C_{2}||M_{h},S_{3}^{-},S_{3}^{+}),(T||T,TC_{3}^{+},TC_{3}^{-})$
&&& \\ 
\hline

$L_{0\mathbf K}=C_{3},
M_{\mathbf K}/L_{0\mathbf K}=\{E,C_{2}\}\times Z_{2}^{T}$,
& $+1$ &$-1$& {\color{red}$01$} \\

$(C_{2}||M_{h},S_{3}^{-},S_{3}^{+}),(TC_{2}||T,TC_{3}^{+},TC_{3}^{-})$
&&& \\ 
\hline

$L_{0\mathbf K}=C_{3},
M_{\mathbf K}/L_{0\mathbf K}=\{E,C_{2},TC'_{21},TC''_{21}\}$
& $+1$ &$+1$ &\\

$(C_{2}||M_{h},S_{3}^{-},S_{3}^{+}),(TC''_{21}||T,TC_{3}^{+},TC_{3}^{-})$
&&& \\ 
\hline

$L_{0\mathbf K}=C_{3h},M_{\mathbf K}/L_{0\mathbf K}=Z_{2}^{T}$,
& $-1$ &$-1$ & {\color{red}$11$} \\

$(T||T,TC_{3}^{+},TC_{3}^{-},TM_{h},TS_{3}^{-},TS_{3}^{+})$
&&& \\ 
\hline

$L_{0\mathbf K}=C_{3h},
M_{\mathbf K}/L_{0\mathbf K}=\{E,TC_{2y}\}$,
& $+1$ &$+1$ & \\

$(TC_{2y}||T,TC_{3}^{+},TC_{3}^{-},TM_{h},TS_{3}^{-},TS_{3}^{+})$
&&&  \\ 
\hline
\end{tabular}
\end{table*}

\begin{table*}[htbp]
\caption{
For ($G_{\mathbf K}\cong M_{\mathbf K}=C_{6h}\times{Z}_2^T,L_{0\mathbf{K}},M_{\mathbf K}/L_{0\mathbf{K}}$),
we list generators of spin point group $(\varphi_{g}||g_{1},\ldots,g_{n})$, $(g_{1},\ldots,g_{n})$ is a coset of $L_{0\mathbf{K}}$,
$g\in (g_{1},\ldots,g_{n}),\varphi_{g}\in M_{\mathbf K}/L_{0\mathbf{K}}$,
$n$ is order of $L_{0\mathbf{K}}$, and
spin invariants $\vec{\eta}^{S}$ of $G_{\mathbf K}$ is
$\eta^S_{p_1,p_2}{\cdot1_{2\times 2}} = d({\varphi_{p_1}})d({\varphi_{p_2}})d^{-1}({\varphi_{p_1}})d^{-1}({\varphi_{p_2}})$
for unitary elements $p_{1},p_{2}$ with $p_{1}p_{2}=p_{2}p_{1}$ and
$\eta^S_{p}{\cdot1_{2\times 2}}=[d(\varphi_p) {K}]^2$
for anti-unitary element $p$ with $p^{2}=E$,
where irRep of $g\in G_{\mathbf K}$ is given by
$d(\varphi_g)K^{\zeta_g}=u\Big( s( T^{\zeta_g}\varphi_g)\Big)(i\sigma_{y}K)^{\zeta_g}$,
$s( T^{\zeta_g}\varphi_g)$ is vector Rep of $T^{\zeta_g}\varphi_{g}\in\mathrm{SO(3)}$,
$u\Big( s( T^{\zeta_g}\varphi_g)\Big)$ is $\mathrm{SU(2)}$ Rep of $T^{\zeta_g}\varphi_{g}$,
$\zeta_{g}=0(1)$ for unitary (anti-unitary) $g$.
The $\vec{\eta}^{S}$ leading to invariants only realized in SSG are labeled by Boolean vectors
colored in {\color{red}RED}.}\label{C6hdrztspin}
\centering
\begin{tabular}{ |c|cccc|}
\hline
$P=C_{6h}$& & spin invariants $\vec{\eta}^{S}$  & &\\

$G_{\mathbf K}\cong M_{\mathbf K}=C_{6h}\times Z_2^T$
&$\eta_{I,C_{2}}^{S}$
&$\eta_{T}^{S}$&$\eta_{IT}^{S}$ &$\eta_{TM_{h}}^{S}$\\
\hline

$L_{0\mathbf K}=\{E,I\},
M_{\mathbf K}/L_{0\mathbf K}=C_{6}\times Z_{2}^{T}$,
& $+1$ &$-1$ &$-1$ &$+1$ \\ 

$(C_{6}^{+}||C_{6}^{+},S_{3}^{-}),(C_{2}||C_{2},M_{h}),(T||T,IT)$
&&&& \\  
\hline

$L_{0\mathbf K}=\{E,I\},
M_{\mathbf K}/L_{0\mathbf K}=C_{6}\times Z_{2}^{T}$,
& $+1$ &$+1$ &$+1$ &$-1$ \\ 

$(C_{6}^{+}||C_{6}^{+},S_{3}^{-}),(C_{2}||C_{2},M_{h}),(TC_{2}||T,IT)$
&& {\color{red}$0001$} && \\  
\hline

$L_{0\mathbf K}=\{E,C_{2}\},
M_{\mathbf K}/L_{0\mathbf K}=C_{6}\times Z_{2}^{T}$,
& $+1$ &$-1$ &$+1$ &$+1$ \\ 

$(C_{3}^{-}||C_{6}^{+},C_{3}^{-}),(C_{2}||I,M_{h}),(T||T,TC_{2})$
&&&& \\  
\hline

$L_{0\mathbf K}=\{E,C_{2}\},
M_{\mathbf K}/L_{0\mathbf K}=C_{6}\times Z_{2}^{T}$,
& $+1$ &$+1$ &$-1$ &$-1$ \\ 

$(C_{3}^{-}||C_{6}^{+},C_{3}^{-}),(C_{2}||I,M_{h}),(TC_{2}||T,TC_{2})$
&& {\color{red}$0011$} && \\  
\hline

$L_{0\mathbf K}=\{E,M_{h}\},
M_{\mathbf K}/L_{0\mathbf K}=C_{6}\times Z_{2}^{T}$,
& $+1$ &$-1$ &$+1$ &$-1$ \\ 

$(C_{6}^{+}||C_{6}^{+},S_{6}^{+}),(C_{2}||I,C_{2}),(T||T,TM_{h})$
&& {\color{red}$0101$} && \\  
\hline

$L_{0\mathbf K}=\{E,M_{h}\},
M_{\mathbf K}/L_{0\mathbf K}=C_{6}\times Z_{2}^{T}$,
& $+1$ &$+1$ &$-1$ &$+1$ \\ 

$(C_{6}^{+}||C_{6}^{+},S_{6}^{+}),(C_{2}||I,C_{2}),(TC_{2}||T,TM_{h})$
&&&& \\  
\hline

$L_{0\mathbf K}=\{E,C_{2},I,M_{h}\},
M_{\mathbf K}/L_{0\mathbf K}=C_{3}\times Z_{2}^{T}$,
& $+1$ &$-1$ &$-1$ &$-1$ \\ 

$(C_{3}^{-}||C_{6}^{+},C_{3}^{-},S_{3}^{-},S_{6}^{+}),
(T||T,TC_{2},IT,TM_{h})$
&& {\color{red}$0111$}  && \\  
\hline

$L_{0\mathbf K}=\{E,C_{2},I,M_{h}\}$,
& $+1$ &$+1$ &$+1$ &$+1$ \\ 

$M_{\mathbf K}/L_{0\mathbf K}=\{E,C_{3}^{+},C_{3}^{-},TC_{2},TC_{6}^{-},TC_{6}^{+}\}$,
&&&& \\

$(C_{3}^{-}||C_{6}^{+},C_{3}^{-},S_{3}^{-},S_{6}^{+}),
(TC_{2}||T,TC_{2},IT,TM_{h})$
&&&& \\  
\hline

$L_{0\mathbf K}=C_{3},
M_{\mathbf K}/L_{0\mathbf K}=\{E,C'_{21},C''_{21},C_{2}\}\times Z_{2}^{T}$,
& $-1$ &$-1$ &$+1$ &$+1$ \\ 

$(C_{2}||C_{2},C_{6}^{+},C_{6}^{-}),(C'_{21}||I,S_{6}^{-},S_{6}^{+})$,
&&&& \\

$(C''_{21}||M_{h},S_{3}^{-},S_{3}^{+}),(T||T,TC_{3}^{+},TC_{3}^{-})$
&&&& \\ 
\hline

$L_{0\mathbf K}=C_{3},
M_{\mathbf K}/L_{0\mathbf K}=\{E,C'_{21},C''_{21},C_{2}\}\times Z_{2}^{T}$,
& $-1$ &$+1$ &$+1$ &$+1$ \\  

$(C_{2}||C_{2},C_{6}^{+},C_{6}^{-}),(C'_{21}||I,S_{6}^{-},S_{6}^{+})$,
&&&& \\

$(C''_{21}||M_{h},S_{3}^{-},S_{3}^{+}),(TC_{2}||T,TC_{3}^{+},TC_{3}^{-})$
&&&& \\  
\hline

$L_{0\mathbf K}=C_{3},
M_{\mathbf K}/L_{0\mathbf K}=\{E,C'_{21},C''_{21},C_{2}\}\times Z_{2}^{T}$,
& $-1$ &$+1$ &$-1$ &$+1$ \\ 

$(C_{2}||C_{2},C_{6}^{+},C_{6}^{-}),(C'_{21}||I,S_{6}^{-},S_{6}^{+})$,
&&&& \\

$(C''_{21}||M_{h},S_{3}^{-},S_{3}^{+}),(TC'_{21}||T,TC_{3}^{+},TC_{3}^{-})$
&&&& \\  
\hline

$L_{0\mathbf K}=C_{3},
M_{\mathbf K}/L_{0\mathbf K}=\{E,C'_{21},C''_{21},C_{2}\}\times Z_{2}^{T}$,
& $-1$ &$+1$ &$+1$ &$-1$ \\ 

$(C_{2}||C_{2},C_{6}^{+},C_{6}^{-}),(C'_{21}||I,S_{6}^{-},S_{6}^{+})$,
&&  {\color{red}$1001$} && \\

$(C''_{21}||M_{h},S_{3}^{-},S_{3}^{+}),(TC''_{21}||T,TC_{3}^{+},TC_{3}^{-})$
&&&& \\ 
\hline

$L_{0\mathbf K}=C_{6},
M_{\mathbf K}/L_{0\mathbf K}=\{E,C_{2}\}\times Z_{2}^{T}$,
& $+1$ &$-1$ &$+1$ &$+1$ \\ 

$(C_{2}||I,M_{h},S_{3}^{-},S_{6}^{-},S_{6}^{+},S_{3}^{+})$,
&&&& \\

$(T||T,TC_{6}^{+},TC_{6}^{-},TC_{3}^{+},TC_{3}^{-},TC_{2})$
&&&& \\ 
\hline

$L_{0\mathbf K}=C_{6},
M_{\mathbf K}/L_{0\mathbf K}=\{E,C_{2}\}\times Z_{2}^{T}$,
& $+1$ &$+1$ &$-1$ &$-1$ \\ 

$(C_{2}||I,M_{h},S_{3}^{-},S_{6}^{-},S_{6}^{+},S_{3}^{+})$,
&& {\color{red}$0011$}  && \\

$(TC_{2}||T,TC_{6}^{+},TC_{6}^{-},TC_{3}^{+},TC_{3}^{-},TC_{2})$
&&&& \\ 
\hline

$L_{0\mathbf K}=C_{6},
M_{\mathbf K}/L_{0\mathbf K}=\{E,C_{2},TC'_{21},TC''_{21}\}$,
& $+1$ &$+1$ &$+1$ &$+1$ \\ 

$(C_{2}||I,M_{h},S_{3}^{-},S_{6}^{-},S_{6}^{+},S_{3}^{+})$,
&&&& \\

$(TC''_{21}||T,TC_{6}^{+},TC_{6}^{-},TC_{3}^{+},TC_{3}^{-},TC_{2})$
&&&& \\  
\hline

$L_{0\mathbf K}=C_{3i},
M_{\mathbf K}/L_{0\mathbf K}=\{E,C_{2}\}\times Z_{2}^{T}$,
& $+1$ &$-1$ &$-1$ &$+1$ \\ 

$(C_{2}||C_{2},C_{6}^{-},C_{6}^{+},M_{h},S_{3}^{+},S_{3}^{-})$,
&&&& \\

$(T||T,TC_{3}^{+},TC_{3}^{-},IT,TS_{6}^{-},TS_{6}^{+})$
&&&& \\
\hline

$L_{0\mathbf K}=C_{3i},
M_{\mathbf K}/L_{0\mathbf K}=\{E,C_{2}\}\times Z_{2}^{T}$,
& $+1$ &$+1$ &$+1$ &$-1$ \\ 

$(C_{2}||C_{2},C_{6}^{-},C_{6}^{+},M_{h},S_{3}^{+},S_{3}^{-})$,
&& {\color{red}$0001$}  && \\

$(TC_{2}||T,TC_{3}^{+},TC_{3}^{-},IT,TS_{6}^{-},TS_{6}^{+})$
&&&& \\ 
\hline

$L_{0\mathbf K}=C_{3i},
M_{\mathbf K}/L_{0\mathbf K}=\{E,C_{2},TC'_{21},TC''_{21}\}$,
& $+1$ &$+1$ &$+1$ &$+1$ \\ 

$(C_{2}||C_{2},C_{6}^{-},C_{6}^{+},M_{h},S_{3}^{+},S_{3}^{-})$,
&&&& \\

$(TC''_{21}||T,TC_{3}^{+},TC_{3}^{-},IT,TS_{6}^{-},TS_{6}^{+})$
&&&& \\  
\hline

$L_{0\mathbf K}=C_{3h},
M_{\mathbf K}/L_{0\mathbf K}=\{E,C_{2}\}\times Z_{2}^{T}$,
& $+1$ &$-1$ &$+1$ &$-1$ \\ 

$(C_{2}||I,S_{6}^{-},S_{6}^{+},C_{2},C_{6}^{-},C_{6}^{+})$,
&& {\color{red}$0101$} && \\

$(T||T,TS_{3}^{-},TS_{3}^{+},TC_{3}^{+},TC_{3}^{-},TM_{h})$
&&&&  \\ 
\hline

\end{tabular}
\end{table*}

\begin{table*}
(Extension of Supplementary Table \ref{C6hdrztspin})\\
\centering
\begin{tabular}{ |c|cccc|}
\hline
$P=C_{6h}$& & spin invariants $\vec{\eta}^{S}$  & &\\

$G_{\mathbf K}\cong M_{\mathbf K}=C_{6h}\times Z_2^T$
&$\eta_{I,C_{2}}^{S}$
&$\eta_{T}^{S}$&$\eta_{IT}^{S}$ &$\eta_{TM_{h}}^{S}$\\
\hline

$L_{0\mathbf K}=C_{3h},
M_{\mathbf K}/L_{0\mathbf K}=\{E,C_{2}\}\times Z_{2}^{T}$,
& $+1$ &$+1$ &$-1$ &$+1$ \\ 

$(C_{2}||I,S_{6}^{-},S_{6}^{+},C_{2},C_{6}^{-},C_{6}^{+})$,
&&&& \\

$(TC_{2}||T,TS_{3}^{-},TS_{3}^{+},TC_{3}^{+},TC_{3}^{-},TM_{h})$
&&&& \\ 
\hline

$L_{0\mathbf K}=C_{3h},
M_{\mathbf K}/L_{0\mathbf K}=\{E,C_{2},TC'_{21},TC''_{21}\}$,
& $+1$ &$+1$ &$+1$ &$+1$ \\ 

$(C_{2}||I,S_{6}^{-},S_{6}^{+},C_{2},C_{6}^{-},C_{6}^{+})$,
&&&& \\

$(TC''_{21}||T,TS_{3}^{-},TS_{3}^{+},TC_{3}^{+},TC_{3}^{-},TM_{h})$
&&&& \\ 
\hline

$L_{0\mathbf K}=C_{6h},M_{\mathbf K}/L_{0\mathbf K}=Z_{2}^{T}$,
& $+1$ &$-1$ &$-1$ &$-1$ \\ 

$(T||T,TC_{6}^{+},TC_{6}^{-},TC_{3}^{+},TC_{3}^{-},TC_{2},$
&& {\color{red}$0111$} && \\

$IT,TS_{3}^{-},TS_{3}^{+},TS_{6}^{-},TS_{6}^{+},TM_{h})$
&&&& \\ 
\hline

$L_{0\mathbf K}=C_{6h},M_{\mathbf K}/L_{0\mathbf K}=\{E,TC_{2y}\}$,
& $+1$ &$+1$ &$+1$ &$+1$ \\ 

$(TC_{2y}||T,TC_{6}^{+},TC_{6}^{-},TC_{3}^{+},TC_{3}^{-},TC_{2},$
&&&& \\

$IT,TS_{3}^{-},TS_{3}^{+},TS_{6}^{-},TS_{6}^{+},TM_{h})$
&&&& \\  
\hline

\end{tabular}
\end{table*}

\clearpage

\begin{table*}[htbp]
\caption{
For ($G_{\mathbf K}\cong M_{\mathbf K}=D_{6}\times{Z}_2^T,L_{0\mathbf{K}},M_{\mathbf K}/L_{0\mathbf{K}}$),
we list generators of spin point group $(\varphi_{g}||g_{1},\ldots,g_{n})$, $(g_{1},\ldots,g_{n})$ is a coset of $L_{0\mathbf{K}}$,
$g\in (g_{1},\ldots,g_{n}),\varphi_{g}\in M_{\mathbf K}/L_{0\mathbf{K}}$,
$n$ is order of $L_{0\mathbf{K}}$, and
spin invariants $\vec{\eta}^{S}$ of $G_{\mathbf K}$ is
$\eta^S_{p_1,p_2}{\cdot1_{2\times 2}} = d({\varphi_{p_1}})d({\varphi_{p_2}})d^{-1}({\varphi_{p_1}})d^{-1}({\varphi_{p_2}})$
for unitary elements $p_{1},p_{2}$ with $p_{1}p_{2}=p_{2}p_{1}$ and
$\eta^S_{p}{\cdot1_{2\times 2}}=[d(\varphi_p) {K}]^2$
for anti-unitary element $p$ with $p^{2}=E$,
where irRep of $g\in G_{\mathbf K}$ is given by
$d(\varphi_g)K^{\zeta_g}=u\Big( s( T^{\zeta_g}\varphi_g)\Big)(i\sigma_{y}K)^{\zeta_g}$,
$s( T^{\zeta_g}\varphi_g)$ is vector Rep of $T^{\zeta_g}\varphi_{g}\in\mathrm{SO(3)}$,
$u\Big( s( T^{\zeta_g}\varphi_g)\Big)$ is $\mathrm{SU(2)}$ Rep of $T^{\zeta_g}\varphi_{g}$,
$\zeta_{g}=0(1)$ for unitary (anti-unitary) $g$.
The $\vec{\eta}^{S}$ leading to invariants only realized in SSG are labeled by Boolean vectors
colored in {\color{red}RED}.} \label{D6drztspin}
\centering
\begin{tabular}{ |c|cccc|}
\hline
$P=D_{6}$& & spin invariants $\vec{\eta}^{S}$  & &\\

$G_{\mathbf K}\cong M_{\mathbf K}=D_{6}\times Z_2^T$
&$\eta_{C_{2},C'_{21}}^{S}$
&$\eta_{T}^{S}$&$\eta_{TC_{2}}^{S}$ &$ \eta_{TC'_{21}}^{S}$\\
\hline

$L_{0\mathbf K}=E,M_{\mathbf K}/L_{0\mathbf K}=D_{6}\times Z_{2}^{T},
(C_{2}||C_{2})$,
&$-1$ &$-1$ &$+1$ &$+1$ \\ 

$(C'_{21}||C'_{21}),(C'_{22}||C'_{22}),(T||T)$
&&&& \\ 
\hline

$L_{0\mathbf K}=E,M_{\mathbf K}/L_{0\mathbf K}=D_{6}\times Z_{2}^{T},
(C_{2}||C_{2})$,
&$-1$ &$+1$ &$-1$ &$+1$ \\ 

$(C'_{21}||C'_{21}),(C'_{22}||C'_{22}),(TC_{2}||T)$
&&&& \\ 
\hline

$L_{0\mathbf K}=\{E,C_{2}\},
M_{\mathbf K}/L_{0\mathbf K}=D_{3}^{1}\times Z_{2}^{T}$,
& $+1$ &$-1$ &$-1$ &$+1$ \\ 

$(C'_{21}||C'_{21},C''_{21}),(C'_{22}||C'_{22},C''_{22}),
(T||T,TC_{2})$
&&&& \\  
\hline

$L_{0\mathbf K}=\{E,C_{2}\}$,
& $+1$ &$+1$ &$+1$ &$+1$ \\ 

$M_{\mathbf K}/L_{0\mathbf K}=\{E,C_{3}^{+},C_{3}^{-},C'_{21,22,23},TC_{2},TC_{6}^{-},TC_{6}^{+},TC''_{21,22,23}\}$,
&&&& \\

$(C'_{21}||C'_{21},C''_{21}),(C'_{22}||C'_{22},C''_{22}),
(TC_{2}||T,TC_{2})$
&&&& \\ 
\hline

$L_{0\mathbf K}=C_{3},
M_{\mathbf K}/L_{0\mathbf K}=\{E,C'_{21},C''_{21},C_{2}\}\times Z_{2}^{T}$,
& $-1$ &$-1$ &$+1$ &$+1$ \\ 

$(C_{2}||C_{2},C_{6}^{-},C_{6}^{+}),(C'_{21}||C'_{21,22,23}),(T||T,TC_{3}^{+},TC_{3}^{-})$
&&&& \\   
\hline

$L_{0\mathbf K}=C_{3},
M_{\mathbf K}/L_{0\mathbf K}=\{E,C'_{21},C''_{21},C_{2}\}\times Z_{2}^{T}$,
& $-1$ &$+1$ &$-1$ &$+1$ \\ 

$(C_{2}||C_{2},C_{6}^{-},C_{6}^{+}),(C'_{21}||C'_{21,22,23}),(TC_{2}||T,TC_{3}^{+},TC_{3}^{-})$
&&&& \\  
\hline

$L_{0\mathbf K}=C_{3},
M_{\mathbf K}/L_{0\mathbf K}=\{E,C'_{21},C''_{21},C_{2}\}\times Z_{2}^{T}$,
& $-1$ &$+1$ &$+1$ &$-1$ \\ 

$(C_{2}||C_{2},C_{6}^{-},C_{6}^{+}),(C'_{21}||C'_{21,22,23}),(TC'_{21}||T,TC_{3}^{+},TC_{3}^{-})$
&& {\color{red}$1001$}  && \\  
\hline

$L_{0\mathbf K}=C_{3},
M_{\mathbf K}/L_{0\mathbf K}=\{E,C'_{21},C''_{21},C_{2}\}\times Z_{2}^{T}$,
& $-1$ &$+1$ &$+1$ &$+1$ \\ 

$(C_{2}||C_{2},C_{6}^{-},C_{6}^{+}),(C'_{21}||C'_{21,22,23}),(TC''_{21}||T,TC_{3}^{+},TC_{3}^{-})$
&& {\color{red}$1000$} && \\  
\hline

$L_{0\mathbf K}=C_{6},
M_{\mathbf K}/L_{0\mathbf K}=\{E,C_{2}\}\times Z_{2}^{T}$,
& $+1$ &$-1$ &$-1$ &$+1$ \\ 

$(C_{2}||C'_{21,22,23},C''_{21,22,23}),(T||T,TC_{6}^{+},TC_{6}^{-},TC_{3}^{+},TC_{3}^{-},TC_{2})$
&&&& \\  
\hline

$L_{0\mathbf K}=C_{6},
M_{\mathbf K}/L_{0\mathbf K}=\{E,C_{2}\}\times Z_{2}^{T}$,
& $+1$ &$+1$ &$+1$ &$-1$ \\ 

$(C_{2}||C'_{21,22,23},C''_{21,22,23}),(TC_{2}||T,TC_{6}^{+},TC_{6}^{-},TC_{3}^{+},TC_{3}^{-},TC_{2})$
&& {\color{red}$0001$}  && \\  
\hline

$L_{0\mathbf K}=C_{6},
M_{\mathbf K}/L_{0\mathbf K}=\{E,C_{2},TC'_{21},TC''_{21}\}$,
& $+1$ &$+1$ &$+1$ &$+1$ \\ 

$(C_{2}||C'_{21,22,23},C''_{21,22,23}),(TC''_{21}||T,TC_{6}^{+},TC_{6}^{-},TC_{3}^{+},TC_{3}^{-},TC_{2})$
&&&& \\  
\hline

$L_{0\mathbf K}=D_{3}^{1},
M_{\mathbf K}/L_{0\mathbf K}=\{E,C_{2}\}\times Z_{2}^{T}$,
& $+1$ &$-1$ &$+1$ &$-1$ \\ 

$(C_{2}||C_{2},C_{6}^{+},C_{6}^{-},C''_{21,22,23}),(T||T,TC_{3}^{+},TC_{3}^{-},TC'_{21,22,23})$
&& {\color{red}$0101$}  && \\  
\hline

$L_{0\mathbf K}=D_{3}^{1},
M_{\mathbf K}/L_{0\mathbf K}=\{E,C_{2}\}\times Z_{2}^{T}$,
& $+1$ &$+1$ &$-1$ &$+1$ \\ 

$(C_{2}||C_{2},C_{6}^{+},C_{6}^{-},C''_{21,22,23}),(TC_{2}||T,TC_{3}^{+},TC_{3}^{-},TC'_{21,22,23})$
&& {\color{red}$0010$}  && \\  
\hline

$L_{0\mathbf K}=D_{3}^{1},
M_{\mathbf K}/L_{0\mathbf K}=\{E,C_{2},TC'_{21},TC''_{21}\}$,
& $+1$ &$+1$ &$+1$ &$+1$ \\ 

$(C_{2}||C_{2},C_{6}^{+},C_{6}^{-},C''_{21,22,23}),(TC''_{21}||T,TC_{3}^{+},TC_{3}^{-},TC'_{21,22,23})$
&&&& \\  
\hline

$L_{0\mathbf K}=D_{3}^{2},
M_{\mathbf K}/L_{0\mathbf K}=\{E,C_{2}\}\times Z_{2}^{T}$,
& $+1$ &$-1$ &$+1$ &$+1$ \\ 

$(C_{2}||C_{2},C_{6}^{+},C_{6}^{-},C'_{21,22,23}),
(T||T,TC_{3}^{+},TC_{3}^{-},TC''_{21,22,23})$
&&  {\color{red}$0100$}  && \\  
\hline

$L_{0\mathbf K}=D_{3}^{2},
M_{\mathbf K}/L_{0\mathbf K}=\{E,C_{2}\}\times Z_{2}^{T}$,
& $+1$ &$+1$ &$-1$ &$-1$ \\ 

$(C_{2}||C_{2},C_{6}^{+},C_{6}^{-},C'_{21,22,23}),
(TC_{2}||T,TC_{3}^{+},TC_{3}^{-},TC''_{21,22,23})$
&&  {\color{red}$0011$}  && \\  
\hline

$L_{0\mathbf K}=D_{3}^{2},
M_{\mathbf K}/L_{0\mathbf K}=\{E,C_{2},TC'_{21},TC''_{21}\}$,
& $+1$ &$+1$ &$+1$ &$+1$ \\ 

$(C_{2}||C_{2},C_{6}^{+},C_{6}^{-},C'_{21,22,23}),
(TC''_{21}||T,TC_{3}^{+},TC_{3}^{-},TC''_{21,22,23})$
&&&& \\  
\hline

$L_{0\mathbf K}=D_{6},
M_{\mathbf K}/L_{0\mathbf K}=Z_{2}^{T}$,
& $+1$ &$-1$ &$-1$ &$-1$ \\ 

$(T||T,TC_{6}^{+},TC_{6}^{-},TC_{3}^{+},TC_{3}^{-},TC_{2},TC'_{21,22,23},TC''_{21,22,23})$
&& {\color{red}$0111$} && \\

\hline

$L_{0\mathbf K}=D_{6},
M_{\mathbf K}/L_{0\mathbf K}=\{E,TC_{2y}\}$,
& $+1$ &$+1$ &$+1$ &$+1$ \\ 

$(TC_{2y}||T,TC_{6}^{+},TC_{6}^{-},TC_{3}^{+},TC_{3}^{-},TC_{2},TC'_{21,22,23},TC''_{21,22,23})$
&&&& \\
\hline

\end{tabular}
\end{table*}

\begin{table*}[htbp]
\caption{
For ($G_{\mathbf K}\cong M_{\mathbf K}=D_{3h}\times{Z}_2^T,L_{0\mathbf{K}},M_{\mathbf K}/L_{0\mathbf{K}}$),
we list generators of spin point group $(\varphi_{g}||g_{1},\ldots,g_{n})$, $(g_{1},\ldots,g_{n})$ is a coset of $L_{0\mathbf{K}}$,
$g\in (g_{1},\ldots,g_{n}),\varphi_{g}\in M_{\mathbf K}/L_{0\mathbf{K}}$,
$n$ is order of $L_{0\mathbf{K}}$, and
spin invariants $\vec{\eta}^{S}$ of $G_{\mathbf K}$ is
$\eta^S_{p_1,p_2}{\cdot1_{2\times 2}} = d({\varphi_{p_1}})d({\varphi_{p_2}})d^{-1}({\varphi_{p_1}})d^{-1}({\varphi_{p_2}})$
for unitary elements $p_{1},p_{2}$ with $p_{1}p_{2}=p_{2}p_{1}$ and
$\eta^S_{p}{\cdot1_{2\times 2}}=[d(\varphi_p) {K}]^2$
for anti-unitary element $p$ with $p^{2}=E$,
where irRep of $g\in G_{\mathbf K}$ is given by
$d(\varphi_g)K^{\zeta_g}=u\Big( s( T^{\zeta_g}\varphi_g)\Big)(i\sigma_{y}K)^{\zeta_g}$,
$s( T^{\zeta_g}\varphi_g)$ is vector Rep of $T^{\zeta_g}\varphi_{g}\in\mathrm{SO(3)}$,
$u\Big( s( T^{\zeta_g}\varphi_g)\Big)$ is $\mathrm{SU(2)}$ Rep of $T^{\zeta_g}\varphi_{g}$,
$\zeta_{g}=0(1)$ for unitary (anti-unitary) $g$.
The $\vec{\eta}^{S}$ leading to invariants only realized in SSG are labeled by Boolean vectors
colored in {\color{red}RED}.} \label{D3hdrztspin}
\centering
\begin{tabular}{ |c|cccc|}
\hline
& & spin invariants $\vec{\eta}^{S}$  & &\\

$P=D_{3h}^{1},G_{\mathbf K}\cong M_{\mathbf K}=D_{3h}^{1}\times Z_2^T$
&$\eta_{M_{h},C'_{21}}^{S}$
&$\eta_{T}^{S}$&$\eta_{TM_{h}}^{S}$ &$\eta_{TC'_{21}}^{S} $\\

$P=D_{3h}^{2},G_{\mathbf K}\cong M_{\mathbf K}=D_{3h}^{2}\times Z_2^T$
&$\eta_{M_{h},M_{d1}}^{S}$
&$\eta_{T}^{S}$&$\eta_{TM_{h}}^{S}$ &$\eta_{TM_{d1}}^{S}$\\
\hline

$M_{\mathbf K}=D_{3h}^{1}\times Z_2^T,L_{0\mathbf K}=E,
M_{\mathbf K}/L_{0\mathbf K}=D_{6}\times Z_{2}^{T}$,
& $-1$ &$-1$ &$+1$ &$+1$ \\  

$(C_{6}^{+}||S_{3}^{-}),(C_{2}||M_{h}),(C'_{21}||C'_{21}),(T||T)$
&&&& \\ 

\hline

$M_{\mathbf K}=D_{3h}^{1}\times Z_2^T,L_{0\mathbf K}=E,
M_{\mathbf K}/L_{0\mathbf K}=D_{6}\times Z_{2}^{T}$,
& $-1$ &$+1$ &$-1$ &$+1$ \\  

$(C_{6}^{+}||S_{3}^{-}),(C_{2}||M_{h}),(C'_{21}||C'_{21}),(TC_{2}||T)$
&& {\color{red}$1010$}  && \\  

\hline

$M_{\mathbf K}=D_{3h}^{2}\times Z_2^T,L_{0\mathbf K}=E,
M_{\mathbf K}/L_{0\mathbf K}=D_{6}\times Z_{2}^{T}$,
& $-1$ &$-1$ &$+1$ &$+1$ \\  

$(C_{6}^{+}||S_{3}^{-}),(C_{2}||M_{h}),(C'_{21}||M_{d1}),(T||T)$
&&&& \\   

\hline

$M_{\mathbf K}=D_{3h}^{2}\times Z_2^T,L_{0\mathbf K}=E,
M_{\mathbf K}/L_{0\mathbf K}=D_{6}\times Z_{2}^{T}$,
& $-1$ &$+1$ &$-1$ &$+1$ \\  

$(C_{6}^{+}||S_{3}^{-}),(C_{2}||M_{h}),(C'_{21}||M_{d1}),(TC_{2}||T)$
&& {\color{red}$1010$}  && \\  

\hline

$M_{\mathbf K}=D_{3h}^{1}\times Z_2^T,L_{0\mathbf K}=\{E,M_{h}\},
M_{\mathbf K}/L_{0\mathbf K}=D_{3}^{1}\times Z_{2}^{T}$,
& $+1$ &$-1$ &$-1$ &$+1$ \\  

$(C_{3}^{+}||C_{3}^{+},S_{3}^{+}),(C'_{21}||C'_{21},M_{v1}),(T||T,TM_{h})$
&& {\color{red}$0110$}  && \\  

\hline

$M_{\mathbf K}=D_{3h}^{1}\times Z_2^T,L_{0\mathbf K}=\{E,M_{h}\}$,
& $+1$ &$+1$ &$+1$ &$+1$ \\  

$M_{\mathbf K}/L_{0\mathbf K}=\{E,C_{3}^{+},C_{3}^{-},C'_{21,22,23},TC_{2},TC_{6}^{-},TC_{6}^{+},TC''_{21,22,23}\}$,
&&&& \\

$(C_{3}^{+}||C_{3}^{+},S_{3}^{+}),(C'_{21}||C'_{21},M_{v1}),(TC_{2}||T,TM_{h})$
&&&& \\  

\hline

$M_{\mathbf K}=D_{3h}^{2}\times Z_2^T,L_{0\mathbf K}=\{E,M_{h}\},
M_{\mathbf K}/L_{0\mathbf K}=D_{3}^{2}\times Z_{2}^{T}$,
& $+1$ &$-1$ &$-1$ &$+1$ \\  

$(C_{3}^{+}||C_{3}^{+},S_{3}^{+}),(C''_{21}||C''_{21},M_{d1}),(T||T,TM_{h})$
&& {\color{red}$0110$}  && \\ 

\hline

$M_{\mathbf K}=D_{3h}^{2}\times Z_2^T,L_{0\mathbf K}=\{E,M_{h}\}$,
& $+1$ &$+1$ &$+1$ &$+1$ \\  

$M_{\mathbf K}/L_{0\mathbf K}=\{E,C_{3}^{+},C_{3}^{-},C''_{21,22,23},TC_{2},TC_{6}^{-},TC_{6}^{+},TC'_{21,22,23}\}$,
&&&& \\

$(C_{3}^{+}||C_{3}^{+},S_{3}^{+}),(C''_{21}||C''_{21},M_{d1}),(TC_{2}||T,TM_{h})$
&&&& \\  

\hline

$M_{\mathbf K}=D_{3h}^{1}\times Z_2^T,L_{0\mathbf K}=C_{3},
M_{\mathbf K}/L_{0\mathbf K}=\{E,C'_{21},C''_{21},C_{2}\}\times Z_{2}^{T}$,
& $-1$ &$-1$ &$+1$ &$+1$ \\ 

$(C_{2}||M_{h},S_{3}^{-},S_{3}^{+}),(C'_{21}||C'_{21,22,23}),(T||T,TC_{3}^{+},TC_{3}^{-})$
&&&& \\  
\hline

$M_{\mathbf K}=D_{3h}^{1}\times Z_2^T,L_{0\mathbf K}=C_{3},
M_{\mathbf K}/L_{0\mathbf K}=\{E,C'_{21},C''_{21},C_{2}\}\times Z_{2}^{T}$,
& $-1$ &$+1$ &$-1$ &$+1$ \\ 

$(C_{2}||M_{h},S_{3}^{-},S_{3}^{+}),(C'_{21}||C'_{21,22,23}),(TC_{2}||T,TC_{3}^{+},TC_{3}^{-})$
&&  {\color{red}$1010$}  && \\  
\hline

$M_{\mathbf K}=D_{3h}^{1}\times Z_2^T,L_{0\mathbf K}=C_{3},
M_{\mathbf K}/L_{0\mathbf K}=\{E,C'_{21},C''_{21},C_{2}\}\times Z_{2}^{T}$,
& $-1$ &$+1$ &$+1$ &$-1$ \\ 

$(C_{2}||M_{h},S_{3}^{-},S_{3}^{+}),(C'_{21}||C'_{21,22,23}),(TC'_{21}||T,TC_{3}^{+},TC_{3}^{-})$
&& {\color{red}$1001$}  && \\  
\hline

$M_{\mathbf K}=D_{3h}^{1}\times Z_2^T,L_{0\mathbf K}=C_{3},
M_{\mathbf K}/L_{0\mathbf K}=\{E,C'_{21},C''_{21},C_{2}\}\times Z_{2}^{T}$,
& $-1$ &$+1$ &$+1$ &$+1$ \\ 

$(C_{2}||M_{h},S_{3}^{-},S_{3}^{+}),(C'_{21}||C'_{21,22,23}),(TC''_{21}||T,TC_{3}^{+},TC_{3}^{-})$
&&&& \\  
\hline

$M_{\mathbf K}=D_{3h}^{2}\times Z_2^T,L_{0\mathbf K}=C_{3},
M_{\mathbf K}/L_{0\mathbf K}=\{E,C'_{21},C''_{21},C_{2}\}\times Z_{2}^{T}$,
& $-1$ &$-1$ &$+1$ &$+1$ \\ 

$(C_{2}||M_{h},S_{3}^{-},S_{3}^{+}),(C'_{21}||M_{d1,d2,d3}),(T||T,TC_{3}^{+},TC_{3}^{-})$
&&&& \\  
\hline

$M_{\mathbf K}=D_{3h}^{2}\times Z_2^T,L_{0\mathbf K}=C_{3},
M_{\mathbf K}/L_{0\mathbf K}=\{E,C'_{21},C''_{21},C_{2}\}\times Z_{2}^{T}$,
& $-1$ &$+1$ &$-1$ &$+1$ \\ 

$(C_{2}||M_{h},S_{3}^{-},S_{3}^{+}),(C'_{21}||M_{d1,d2,d3}),(TC_{2}||T,TC_{3}^{+},TC_{3}^{-})$
&& {\color{red}$1010$}  && \\  
\hline

$M_{\mathbf K}=D_{3h}^{2}\times Z_2^T,L_{0\mathbf K}=C_{3},
M_{\mathbf K}/L_{0\mathbf K}=\{E,C'_{21},C''_{21},C_{2}\}\times Z_{2}^{T}$,
& $-1$ &$+1$ &$+1$ &$-1$ \\ 

$(C_{2}||M_{h},S_{3}^{-},S_{3}^{+}),(C'_{21}||M_{d1,d2,d3}),(TC'_{21}||T,TC_{3}^{+},TC_{3}^{-})$
&&&& \\  
\hline

$M_{\mathbf K}=D_{3h}^{2}\times Z_2^T,L_{0\mathbf K}=C_{3},
M_{\mathbf K}/L_{0\mathbf K}=\{E,C'_{21},C''_{21},C_{2}\}\times Z_{2}^{T}$,
& $-1$ &$+1$ &$+1$ &$+1$ \\ 

$(C_{2}||M_{h},S_{3}^{-},S_{3}^{+}),(C'_{21}||M_{d1,d2,d3}),(TC''_{21}||T,TC_{3}^{+},TC_{3}^{-})$
&& {\color{red}$1000$} && \\  
\hline

$M_{\mathbf K}=D_{3h}^{1}\times Z_2^T,L_{0\mathbf K}=C_{3h},
M_{\mathbf K}/L_{0\mathbf K}=\{E,C_{2}\}\times Z_{2}^{T}$,
& $+1$ &$-1$ &$-1$ &$+1$ \\ 

$(C_{2}||C'_{21,22,23},M_{v1,v2,v3}),
(T||T,TC_{3}^{+},TC_{3}^{-},TS_{3}^{+},TS_{3}^{-},TM_{h})$
&& {\color{red}$0110$}   && \\  
\hline

$M_{\mathbf K}=D_{3h}^{1}\times Z_2^T,L_{0\mathbf K}=C_{3h},
M_{\mathbf K}/L_{0\mathbf K}=\{E,C_{2}\}\times Z_{2}^{T}$,
& $+1$ &$+1$ &$+1$ &$-1$ \\ 

$(C_{2}||C'_{21,22,23},M_{v1,v2,v3}),
(TC_{2}||T,TC_{3}^{+},TC_{3}^{-},TS_{3}^{+},TS_{3}^{-},TM_{h})$
&& {\color{red}$0001$}   && \\  
\hline

$M_{\mathbf K}=D_{3h}^{1}\times Z_2^T,L_{0\mathbf K}=C_{3h},
M_{\mathbf K}/L_{0\mathbf K}=\{E,C_{2},TC'_{21},TC''_{21}\}$,
& $+1$ &$+1$ &$+1$ &$+1$ \\ 

$(C_{2}||C'_{21,22,23},M_{v1,v2,v3}),
(TC''_{21}||T,TC_{3}^{+},TC_{3}^{-},TS_{3}^{+},TS_{3}^{-},TM_{h})$
&&&& \\  
\hline

$M_{\mathbf K}=D_{3h}^{2}\times Z_2^T,L_{0\mathbf K}=C_{3h},
M_{\mathbf K}/L_{0\mathbf K}=\{E,C_{2}\}\times Z_{2}^{T}$,
& $+1$ &$-1$ &$-1$ &$+1$ \\ 

$(C_{2}||C''_{21,22,23},M_{d1,d2,d3}),(T||T,TC_{3}^{+},TC_{3}^{-},TS_{3}^{+},TS_{3}^{-},TM_{h})$
&& {\color{red}$0110$}   && \\ 
\hline

$M_{\mathbf K}=D_{3h}^{2}\times Z_2^T,L_{0\mathbf K}=C_{3h},
M_{\mathbf K}/L_{0\mathbf K}=\{E,C_{2}\}\times Z_{2}^{T}$,
& $+1$ &$+1$ &$+1$ &$-1$ \\ 

$(C_{2}||C''_{21,22,23},M_{d1,d2,d3}),
(TC_{2}||T,TC_{3}^{+},TC_{3}^{-},TS_{3}^{+},TS_{3}^{-},TM_{h})$
&& {\color{red}$0001$}  && \\ 
\hline

$M_{\mathbf K}=D_{3h}^{2}\times Z_2^T,L_{0\mathbf K}=C_{3h},
M_{\mathbf K}/L_{0\mathbf K}=\{E,C_{2},TC'_{21},TC''_{21}\}$,
& $+1$ &$+1$ &$+1$ &$+1$ \\ 

$(C_{2}||C''_{21,22,23},M_{d1,d2,d3}),
(TC''_{21}||T,TC_{3}^{+},TC_{3}^{-},TS_{3}^{+},TS_{3}^{-},TM_{h})$
&&&& \\  
\hline

$M_{\mathbf K}=D_{3h}^{1}\times Z_2^T,L_{0\mathbf K}=D_{3}^{1},
M_{\mathbf K}/L_{0\mathbf K}=\{E,C_{2}\}\times Z_{2}^{T}$,
& $+1$ &$-1$ &$+1$ &$-1$ \\ 

$(C_{2}||M_{h},S_{3}^{-},S_{3}^{+},M_{v1,v2,v3}),(T||T,TC_{3}^{+},TC_{3}^{-},TC'_{21,22,23})$
&& {\color{red}$0101$}  && \\  
\hline

\end{tabular}
\end{table*}

\begin{table*}
(Extension of Supplementary Table \ref{D3hdrztspin})\\
\centering
\begin{tabular}{ |c|cccc|}
\hline
& & spin invariants $\vec{\eta}^{S}$  & &\\

$P=D_{3h}^{1},G_{\mathbf K}\cong M_{\mathbf K}=D_{3h}^{1}\times Z_2^T$
&$\eta_{M_{h},C'_{21}}^{S}$
&$\eta_{T}^{S}$&$\eta_{TM_{h}}^{S}$ &$\eta_{TC'_{21}}^{S} $\\

$P=D_{3h}^{2},G_{\mathbf K}\cong M_{\mathbf K}=D_{3h}^{2}\times Z_2^T$
&$\eta_{M_{h},M_{d1}}^{S}$
&$\eta_{T}^{S}$&$\eta_{TM_{h}}^{S}$ &$\eta_{TM_{d1}}^{S}$\\
\hline

$M_{\mathbf K}=D_{3h}^{1}\times Z_2^T,L_{0\mathbf K}=D_{3}^{1},
M_{\mathbf K}/L_{0\mathbf K}=\{E,C_{2}\}\times Z_{2}^{T}$,
& $+1$ &$+1$ &$-1$ &$+1$ \\ 

$(C_{2}||M_{h},S_{3}^{-},S_{3}^{+},M_{v1,v2,v3}),(TC_{2}||T,TC_{3}^{+},TC_{3}^{-},TC'_{21,22,23})$
&&  {\color{red}$0010$}  && \\  
\hline

$M_{\mathbf K}=D_{3h}^{1}\times Z_2^T,L_{0\mathbf K}=D_{3}^{1},
M_{\mathbf K}/L_{0\mathbf K}=\{E,C_{2},TC'_{21},TC''_{21}\}$,
& $+1$ &$+1$ &$+1$ &$+1$ \\ 

$(C_{2}||M_{h},S_{3}^{-},S_{3}^{+},M_{v1,v2,v3}),(TC''_{21}||T,TC_{3}^{+},TC_{3}^{-},TC'_{21,22,23})$
&&&& \\  
\hline

$M_{\mathbf K}=D_{3h}^{2}\times Z_2^T,L_{0\mathbf K}=D_{3}^{2},
M_{\mathbf K}/L_{0\mathbf K}=\{E,C_{2}\}\times Z_{2}^{T}$,
& $+1$ &$-1$ &$+1$ &$+1$ \\ 

$(C_{2}||M_{h},S_{3}^{-},S_{3}^{+},M_{d1,d2,d3}),
(T||T,TC_{3}^{+},TC_{3}^{-},TC''_{21,22,23})$
&& {\color{red}$0100$}  && \\  
\hline

$M_{\mathbf K}=D_{3h}^{2}\times Z_2^T,L_{0\mathbf K}=D_{3}^{2},
M_{\mathbf K}/L_{0\mathbf K}=\{E,C_{2}\}\times Z_{2}^{T}$,
& $+1$ &$+1$ &$-1$ &$-1$ \\ 

$(C_{2}||M_{h},S_{3}^{-},S_{3}^{+},M_{d1,d2,d3}),
(TC_{2}||T,TC_{3}^{+},TC_{3}^{-},TC''_{21,22,23})$
&& {\color{red}$0011$}  && \\  
\hline

$M_{\mathbf K}=D_{3h}^{2}\times Z_2^T,L_{0\mathbf K}=D_{3}^{2},
M_{\mathbf K}/L_{0\mathbf K}=\{E,C_{2},TC'_{21},TC''_{21}\}$,
& $+1$ &$+1$ &$+1$ &$+1$ \\ 

$(C_{2}||M_{h},S_{3}^{-},S_{3}^{+},M_{d1,d2,d3}),
(TC''_{21}||T,TC_{3}^{+},TC_{3}^{-},TC''_{21,22,23})$
&&&& \\  
\hline

$M_{\mathbf K}=D_{3h}^{1}\times Z_2^T,L_{0\mathbf K}=C_{3v}^{2},
M_{\mathbf K}/L_{0\mathbf K}=\{E,C_{2}\}\times Z_{2}^{T}$,
& $+1$ &$-1$ &$+1$ &$+1$ \\ 

$(C_{2}||M_{h},S_{3}^{-},S_{3}^{+},C'_{21,22,23}),
(T||T,TC_{3}^{+},TC_{3}^{-},TM_{v1,v2,v3})$
&&&& \\  
\hline

$M_{\mathbf K}=D_{3h}^{1}\times Z_2^T,L_{0\mathbf K}=C_{3v}^{2},
M_{\mathbf K}/L_{0\mathbf K}=\{E,C_{2}\}\times Z_{2}^{T}$,
& $+1$ &$+1$ &$-1$ &$-1$ \\ 

$(C_{2}||M_{h},S_{3}^{-},S_{3}^{+},C'_{21,22,23}),
(TC_{2}||T,TC_{3}^{+},TC_{3}^{-},TM_{v1,v2,v3})$
&&  {\color{red}$0011$}  && \\  
\hline

$M_{\mathbf K}=D_{3h}^{1}\times Z_2^T,L_{0\mathbf K}=C_{3v}^{2},
M_{\mathbf K}/L_{0\mathbf K}=\{E,C_{2},TC'_{21},TC''_{21}\}$,
& $+1$ &$+1$ &$+1$ &$+1$ \\ 

$(C_{2}||M_{h},S_{3}^{-},S_{3}^{+},C'_{21,22,23}),
(TC''_{21}||T,TC_{3}^{+},TC_{3}^{-},TM_{v1,v2,v3})$
&&&& \\ 
\hline

$M_{\mathbf K}=D_{3h}^{2}\times Z_2^T,L_{0\mathbf K}=C_{3v}^{1},
M_{\mathbf K}/L_{0\mathbf K}=\{E,C_{2}\}\times Z_{2}^{T}$,
& $+1$ &$-1$ &$+1$ &$-1$ \\ 

$(C_{2}||M_{h},S_{3}^{-},S_{3}^{+},C''_{21,22,23}),
(T||T,TC_{3}^{+},TC_{3}^{-},TM_{d1,d2,d3})$
&&&& \\  
\hline

$M_{\mathbf K}=D_{3h}^{2}\times Z_2^T,L_{0\mathbf K}=C_{3v}^{1},
M_{\mathbf K}/L_{0\mathbf K}=\{E,C_{2}\}\times Z_{2}^{T}$,
& $+1$ &$+1$ &$-1$ &$+1$ \\ 

$(C_{2}||M_{h},S_{3}^{-},S_{3}^{+},C''_{21,22,23}),
(TC_{2}||T,TC_{3}^{+},TC_{3}^{-},TM_{d1,d2,d3})$
&& {\color{red}$0010$}   && \\  
\hline

$M_{\mathbf K}=D_{3h}^{2}\times Z_2^T,L_{0\mathbf K}=C_{3v}^{1},
M_{\mathbf K}/L_{0\mathbf K}=\{E,C_{2},TC'_{21},TC''_{21}\}$,
& $+1$ &$+1$ &$+1$ &$+1$ \\ 

$(C_{2}||M_{h},S_{3}^{-},S_{3}^{+},C''_{21,22,23}),
(TC''_{21}||T,TC_{3}^{+},TC_{3}^{-},TM_{d1,d2,d3})$
&&&& \\  
\hline

$L_{0\mathbf K}=D_{3h}^{1},M_{\mathbf K}/L_{0\mathbf K}=Z_{2}^{T}$,
& $+1$ &$-1$ &$-1$ &$-1$ \\ 

$(T||T,TS_{3}^{-},TS_{3}^{+},TC_{3}^{+},TC_{3}^{-},TM_{h},TC'_{21,22,23},TM_{v1,v2,v3})$
&& {\color{red}$0111$} && \\ 
\hline

$L_{0\mathbf K}=D_{3h}^{1},
M_{\mathbf K}/L_{0\mathbf K}=\{E,TC_{2y}\}$,
& $+1$ &$+1$ &$+1$ &$+1$ \\ 

$(TC_{2y}||T,TS_{3}^{-},TS_{3}^{+},TC_{3}^{+},TC_{3}^{-},TM_{h},TC'_{21,22,23},TM_{v1,v2,v3})$
&&&& \\ 
\hline

$L_{0\mathbf K}=D_{3h}^{2},M_{\mathbf K}/L_{0\mathbf K}=Z_{2}^{T}$,
& $+1$ &$-1$ &$-1$ &$-1$ \\ 

$(T||T,TS_{3}^{-},TS_{3}^{+},TC_{3}^{+},TC_{3}^{-},TM_{h},TC''_{21,22,23},TM_{d1,d2,d3})$
&& {\color{red}$0111$} && \\ 
\hline

$L_{0\mathbf K}=D_{3h}^{2},
M_{\mathbf K}/L_{0\mathbf K}=\{E,TC_{2y}\}$,
& $+1$ &$+1$ &$+1$ &$+1$ \\ 

$(TC_{2y}||T,TS_{3}^{-},TS_{3}^{+},TC_{3}^{+},TC_{3}^{-},TM_{h},TC''_{21,22,23},TM_{d1,d2,d3})$
&&&& \\ 
\hline

\end{tabular}
\end{table*}

\begin{table*}[htbp]
\caption{
For ($G_{\mathbf K}\cong M_{\mathbf K}=D_{6h}\times{Z}_2^T,L_{0\mathbf{K}},M_{\mathbf K}/L_{0\mathbf{K}}$),
we list generators of spin point group $(\varphi_{g}||g_{1},\ldots,g_{n})$, $(g_{1},\ldots,g_{n})$ is a coset of $L_{0\mathbf{K}}$,
$g\in (g_{1},\ldots,g_{n}),\varphi_{g}\in M_{\mathbf K}/L_{0\mathbf{K}}$,
$n$ is order of $L_{0\mathbf{K}}$, and
spin invariants $\vec{\eta}^{S}$ of $G_{\mathbf K}$ is
$\eta^S_{p_1,p_2}{\cdot1_{2\times 2}} = d({\varphi_{p_1}})d({\varphi_{p_2}})d^{-1}({\varphi_{p_1}})d^{-1}({\varphi_{p_2}})$
for unitary elements $p_{1},p_{2}$ with $p_{1}p_{2}=p_{2}p_{1}$ and
$\eta^S_{p}{\cdot1_{2\times 2}}=[d(\varphi_p) {K}]^2$
for anti-unitary element $p$ with $p^{2}=E$,
where irRep of $g\in G_{\mathbf K}$ is given by
$d(\varphi_g)K^{\zeta_g}=u\Big( s( T^{\zeta_g}\varphi_g)\Big)(i\sigma_{y}K)^{\zeta_g}$,
$s( T^{\zeta_g}\varphi_g)$ is vector Rep of $T^{\zeta_g}\varphi_{g}\in\mathrm{SO(3)}$,
$u\Big( s( T^{\zeta_g}\varphi_g)\Big)$ is $\mathrm{SU(2)}$ Rep of $T^{\zeta_g}\varphi_{g}$,
$\zeta_{g}=0(1)$ for unitary (anti-unitary) $g$.
The $\vec{\eta}^{S}$ leading to invariants only realized in SSG are labeled by Boolean vectors
colored in {\color{red}RED}.} \label{D6hdrztspin}
\centering
\begin{tabular}{ |c|ccccccc|}
\hline
$P=D_{6h}$& && & spin invariants $\vec{\eta}^{S}$&&&\\

$G_{\mathbf K}\cong M_{\mathbf K}=D_{6h}\times Z_2^T$
&$\eta_{C'_{21},C''_{21}}^{S}$
&$\eta_{T}^{S}$&$\eta_{IT}^{S}$ &$\eta_{TC'_{21}}^{S}$
&$\eta_{TC''_{21}}^{S}$
&$\eta_{I,C'_{21}}^{S}$&
$\eta_{I,C''_{21}}^{S}$\\
\hline

$L_{0\mathbf K}=\{E,I\},
M_{\mathbf K}/L_{0\mathbf K}=D_{6}\times Z_{2}^{T}$,
& $-1$ &$-1$ &$-1$ &$+1$ &$+1$ &$+1$ &$+1$ \\ 

$(C'_{21}||C'_{21},M_{d1}),(C''_{21}||C''_{21},M_{v1})$,
&&&&&&& \\

$(C'_{22}||C'_{22},M_{d2}),(T||T,IT)$
&&&&&&& \\ 
\hline

$L_{0\mathbf K}=\{E,I\},
M_{\mathbf K}/L_{0\mathbf K}=D_{6}\times Z_{2}^{T}$,
& $-1$ &$+1$ &$+1$ &$+1$ &$+1$ &$+1$ &$+1$ \\ 

$(C'_{21}||C'_{21},M_{d1}),(C''_{21}||C''_{21},M_{v1})$,
&&&& {\color{red}$1000000$} &&& \\

$(C'_{22}||C'_{22},M_{d2}),(TC_{2}||T,IT)$
&&&&&&& \\ 
\hline

$L_{0\mathbf K}=\{E,C_{2}\},
M_{\mathbf K}/L_{0\mathbf K}=D_{6}\times Z_{2}^{T}$,
& $+1$ &$-1$ &$+1$ &$+1$ &$+1$ &$-1$ &$-1$ \\ 

$(C_{2}||I,M_{h}),(C'_{21}||C'_{21},C''_{21})$,
&&&&&&& \\

$(C'_{22}||C'_{22},C''_{22}),(T||T,TC_{2})$
&&&&&&& \\  
\hline

$L_{0\mathbf K}=\{E,C_{2}\},
M_{\mathbf K}/L_{0\mathbf K}=D_{6}\times Z_{2}^{T}$,
& $+1$ &$+1$ &$-1$ &$+1$ &$+1$ &$-1$ &$-1$ \\ 

$(C_{2}||I,M_{h}),(C'_{21}||C'_{21},C''_{21})$,
&&&& {\color{red}$0010011$} &&& \\

$(C'_{22}||C'_{22},C''_{22}),(TC_{2}||T,TC_{2})$
&&&&&&& \\ 
\hline

$L_{0\mathbf K}=\{E,M_{h}\},
M_{\mathbf K}/L_{0\mathbf K}=D_{6}\times Z_{2}^{T}$,
& $-1$ &$-1$ &$+1$ &$+1$ &$+1$ &$-1$ &$-1$ \\ 

$(C_{2}||I,C_{2}),(C'_{21}||C'_{21},M_{v1}),(C''_{21}||C''_{21},M_{d1})$,
&&&& {\color{red}$1100011$} &&& \\

$(C'_{22}||C'_{22},M_{v2}),(T||T,TM_{h})$
&&&&&&& \\  
\hline

$L_{0\mathbf K}=\{E,M_{h}\},
M_{\mathbf K}/L_{0\mathbf K}=D_{6}\times Z_{2}^{T}$,
& $-1$ &$+1$ &$-1$ &$+1$ &$+1$ &$-1$ &$-1$ \\ 

$(C_{2}||I,C_{2}),(C'_{21}||C'_{21},M_{v1}),(C''_{21}||C''_{21},M_{d1})$,
&&&&&&& \\

$(C'_{22}||C'_{22},M_{v2}),(TC_{2}||T,TM_{h})$
&&&&&&& \\ 
\hline

$L_{0\mathbf K}=\{E,C_{2},I,M_{h}\},
M_{\mathbf K}/L_{0\mathbf K}=D_{3}^{1}\times Z_{2}^{T}$,
& $+1$ &$-1$ &$-1$ &$+1$ &$+1$ &$+1$ &$+1$ \\ 

$(C_{3}^{+}||C_{3}^{+},C_{6}^{-},S_{6}^{-},S_{3}^{+}),(C'_{21}||C'_{21},C''_{21},M_{d1},M_{v1})$,
&&&& {\color{red}$0110000$} &&& \\

$(T||T,TC_{2},IT,TM_{h})$
&&&&&&& \\ 
\hline

$L_{0\mathbf K}=\{E,C_{2},I,M_{h}\},
M_{\mathbf K}/L_{0\mathbf K}=\{E,C_{3}^{+},C_{3}^{-},C'_{21,22,23},$
& $+1$ &$+1$ &$+1$ &$+1$ &$+1$ &$+1$ &$+1$ \\ 

$TC_{2},TC_{6}^{-},TC_{6}^{+},TC''_{21,22,23}\}$,
&&&&&&& \\

$(C_{3}^{+}||C_{3}^{+},C_{6}^{-},S_{6}^{-},S_{3}^{+}),(C'_{21}||C'_{21},C''_{21},M_{d1},M_{v1})$,
&&&&&&& \\

$(TC_{2}||T,TC_{2},IT,TM_{h})$
&&&&&&& \\
\hline

$L_{0\mathbf K}=C_{3i},
M_{\mathbf K}/L_{0\mathbf K}=\{E,C'_{21},C''_{21},C_{2}\}\times Z_{2}^{T}$,
& $-1$ &$-1$ &$-1$ &$+1$ &$+1$ &$+1$ &$+1$ \\ 

$(C'_{21}||C'_{21,22,23},M_{d1,d2,d3}),(C''_{21}||C''_{21,22,23},M_{v1,v2,v3})$,
&&&&&&&\\

$(T||T,TC_{3}^{+},TC_{3}^{-},IT,TS_{6}^{-},TS_{6}^{+})$
&&&&&&& \\  
\hline

$L_{0\mathbf K}=C_{3i},
M_{\mathbf K}/L_{0\mathbf K}=\{E,C'_{21},C''_{21},C_{2}\}\times Z_{2}^{T}$,
& $-1$ &$+1$ &$+1$ &$+1$ &$+1$ &$+1$ &$+1$ \\ 

$(C'_{21}||C'_{21,22,23},M_{d1,d2,d3}),(C''_{21}||C''_{21,22,23},M_{v1,v2,v3})$,
&&&& {\color{red}$1000000$} &&&\\

$(TC_{2}||T,TC_{3}^{+},TC_{3}^{-},IT,TS_{6}^{-},TS_{6}^{+})$
&&&&&&& \\ 
\hline

$L_{0\mathbf K}=C_{3i},
M_{\mathbf K}/L_{0\mathbf K}=\{E,C'_{21},C''_{21},C_{2}\}\times Z_{2}^{T}$,
& $-1$ &$+1$ &$+1$ &$-1$ &$+1$ &$+1$ &$+1$ \\ 

$(C'_{21}||C'_{21,22,23},M_{d1,d2,d3}),(C''_{21}||C''_{21,22,23},M_{v1,v2,v3})$,
&&&& {\color{red}$1001000$} &&&\\

$(TC'_{21}||T,TC_{3}^{+},TC_{3}^{-},IT,TS_{6}^{-},TS_{6}^{+})$
&&&&&&& \\  
\hline

$L_{0\mathbf K}=C_{3i},
M_{\mathbf K}/L_{0\mathbf K}=\{E,C'_{21},C''_{21},C_{2}\}\times Z_{2}^{T}$,
& $-1$ &$+1$ &$+1$ &$+1$ &$-1$ &$+1$ &$+1$ \\ 

$(C'_{21}||C'_{21,22,23},M_{d1,d2,d3}),(C''_{21}||C''_{21,22,23},M_{v1,v2,v3})$,
&&&& {\color{red}$1000100$} &&&\\

$(TC''_{21}||T,TC_{3}^{+},TC_{3}^{-},IT,TS_{6}^{-},TS_{6}^{+})$
&&&&&&& \\  
\hline

$L_{0\mathbf K}=C_{3h},
M_{\mathbf K}/L_{0\mathbf K}=\{E,C'_{21},C''_{21},C_{2}\}\times Z_{2}^{T}$,
& $-1$ &$-1$ &$+1$ &$+1$ &$+1$ &$-1$ &$-1$ \\ 

$(C'_{21}||C'_{21,22,23},M_{v1,v2,v3}),(C''_{21}||C''_{21,22,23},M_{d1,d2,d3})$,
&&&& {\color{red}$1100011$} &&&\\

$(C_{2}||I,S_{6}^{-},S_{6}^{+},C_{2},C_{6}^{+},C_{6}^{-})$,
&&&&&&&  \\

$(T||T,TC_{3}^{+},TC_{3}^{-},TM_{h},TS_{3}^{-},TS_{3}^{+})$
&&&&&&& \\   
\hline

$L_{0\mathbf K}=C_{3h},
M_{\mathbf K}/L_{0\mathbf K}=\{E,C'_{21},C''_{21},C_{2}\}\times Z_{2}^{T}$,
& $-1$ &$+1$ &$-1$ &$+1$ &$+1$ &$-1$ &$-1$ \\ 

$(C'_{21}||C'_{21,22,23},M_{v1,v2,v3}),(C''_{21}||C''_{21,22,23},M_{d1,d2,d3})$,
&&&&&&&\\

$(C_{2}||I,S_{6}^{-},S_{6}^{+},C_{2},C_{6}^{+},C_{6}^{-})$,
&&&&&&&  \\

$(TC_{2}||T,TC_{3}^{+},TC_{3}^{-},TM_{h},TS_{3}^{-},TS_{3}^{+})$
&&&&&&& \\  
\hline

$L_{0\mathbf K}=C_{3h},
M_{\mathbf K}/L_{0\mathbf K}=\{E,C'_{21},C''_{21},C_{2}\}\times Z_{2}^{T}$,
& $-1$ &$+1$ &$+1$ &$-1$ &$+1$ &$-1$ &$-1$ \\ 

$(C'_{21}||C'_{21,22,23},M_{v1,v2,v3}),(C''_{21}||C''_{21,22,23},M_{d1,d2,d3})$,
&&&& {\color{red}$1001011$} &&& \\

$(C_{2}||I,S_{6}^{-},S_{6}^{+},C_{2},C_{6}^{+},C_{6}^{-})$,
&&&&&&&  \\

$(TC'_{21}||T,TC_{3}^{+},TC_{3}^{-},TM_{h},TS_{3}^{-},TS_{3}^{+})$
&&&&&&& \\   
\hline

\end{tabular}
\end{table*}

\begin{table*}
(Extension of Supplementary Table \ref{D6hdrztspin})\\
\centering
\begin{tabular}{ |c|ccccccc|}
\hline
$P=D_{6h}$& && & spin invariants $\vec{\eta}^{S}$&&&\\

$G_{\mathbf K}\cong M_{\mathbf K}=D_{6h}\times Z_2^T$
&$\eta_{C'_{21},C''_{21}}^{S}$
&$\eta_{T}^{S}$&$\eta_{IT}^{S}$ &$\eta_{TC'_{21}}^{S}$
&$\eta_{TC''_{21}}^{S}$
&$\eta_{I,C'_{21}}^{S}$&
$\eta_{I,C''_{21}}^{S}$\\
\hline

$L_{0\mathbf K}=C_{3h},
M_{\mathbf K}/L_{0\mathbf K}=\{E,C'_{21},C''_{21},C_{2}\}\times Z_{2}^{T}$,
& $-1$ &$+1$ &$+1$ &$+1$ &$-1$ &$-1$ &$-1$ \\ 

$(C'_{21}||C'_{21,22,23},M_{v1,v2,v3}),(C''_{21}||C''_{21,22,23},M_{d1,d2,d3})$,
&&&& {\color{red}$1000111$} &&&\\

$(C_{2}||I,S_{6}^{-},S_{6}^{+},C_{2},C_{6}^{+},C_{6}^{-})$,
&&&&&&&  \\

$(TC''_{21}||T,TC_{3}^{+},TC_{3}^{-},TM_{h},TS_{3}^{-},TS_{3}^{+})$
&&&&&&& \\  
\hline

$L_{0\mathbf K}=C_{6},
M_{\mathbf K}/L_{0\mathbf K}=\{E,C'_{21},C''_{21},C_{2}\}\times Z_{2}^{T}$,
& $+1$ &$-1$ &$+1$ &$+1$ &$+1$ &$-1$ &$-1$ \\ 

$(C'_{21}||C'_{21,22,23},C''_{21,22,23}),(C_{2}||I,M_{h},S_{3}^{-},S_{6}^{-},S_{6}^{+},S_{3}^{+})$,
&&&&&&&\\

$(T||T,TC_{6}^{+},TC_{6}^{-},TC_{3}^{+},TC_{3}^{-},TC_{2})$
&&&&&&& \\  
\hline

$L_{0\mathbf K}=C_{6},
M_{\mathbf K}/L_{0\mathbf K}=\{E,C'_{21},C''_{21},C_{2}\}\times Z_{2}^{T}$,
& $+1$ &$+1$ &$-1$ &$+1$ &$+1$ &$-1$ &$-1$ \\ 

$(C'_{21}||C'_{21,22,23},C''_{21,22,23}),(C_{2}||I,M_{h},S_{3}^{-},S_{6}^{-},S_{6}^{+},S_{3}^{+})$,
&&&& {\color{red}$0010011$} &&&\\

$(TC_{2}||T,TC_{6}^{+},TC_{6}^{-},TC_{3}^{+},TC_{3}^{-},TC_{2})$
&&&&&&& \\  
\hline

$L_{0\mathbf K}=C_{6},
M_{\mathbf K}/L_{0\mathbf K}=\{E,C'_{21},C''_{21},C_{2}\}\times Z_{2}^{T}$,
& $+1$ &$+1$ &$+1$ &$-1$ &$-1$ &$-1$ &$-1$ \\ 

$(C'_{21}||C'_{21,22,23},C''_{21,22,23}),(C_{2}||I,M_{h},S_{3}^{-},S_{6}^{-},S_{6}^{+},S_{3}^{+})$,
&&&& {\color{red}$0001111$} &&&\\

$(TC'_{21}||T,TC_{6}^{+},TC_{6}^{-},TC_{3}^{+},TC_{3}^{-},TC_{2})$
&&&&&&& \\ 
\hline

$L_{0\mathbf K}=C_{6},
M_{\mathbf K}/L_{0\mathbf K}=\{E,C'_{21},C''_{21},C_{2}\}\times Z_{2}^{T}$,
& $+1$ &$+1$ &$+1$ &$+1$ &$+1$ &$-1$ &$-1$ \\ 

$(C'_{21}||C'_{21,22,23},C''_{21,22,23}),(C_{2}||I,M_{h},S_{3}^{-},S_{6}^{-},S_{6}^{+},S_{3}^{+})$,
&&&&&&&\\

$(TC''_{21}||T,TC_{6}^{+},TC_{6}^{-},TC_{3}^{+},TC_{3}^{-},TC_{2})$
&&&&&&& \\  
\hline

$L_{0\mathbf K}=D_{3}^{1},
M_{\mathbf K}/L_{0\mathbf K}=\{E,C'_{21},C''_{21},C_{2}\}\times Z_{2}^{T}$,
& $+1$ &$-1$ &$+1$ &$-1$ &$+1$ &$+1$ &$-1$ \\ 

$(C'_{21}||I,S_{6}^{-},S_{6}^{+},M_{d1,d2,d3}),(C_{2}||C_{2},C_{6}^{+},C_{6}^{-},C''_{21,22,23})$,
&&&& {\color{red}$0101001$} &&&\\

$(T||T,TC_{3}^{+},TC_{3}^{-},TC'_{21,22,23})$
&&&&&&& \\  
\hline

$L_{0\mathbf K}=D_{3}^{1},
M_{\mathbf K}/L_{0\mathbf K}=\{E,C'_{21},C''_{21},C_{2}\}\times Z_{2}^{T}$,
& $+1$ &$+1$ &$+1$ &$+1$ &$-1$ &$+1$ &$-1$ \\ 

$(C'_{21}||I,S_{6}^{-},S_{6}^{+},M_{d1,d2,d3}),(C_{2}||C_{2},C_{6}^{+},C_{6}^{-},C''_{21,22,23})$,
&&&& {\color{red}$0000101$} &&&\\

$(TC_{2}||T,TC_{3}^{+},TC_{3}^{-},TC'_{21,22,23})$
&&&&&&& \\ 
\hline

$L_{0\mathbf K}=D_{3}^{1},
M_{\mathbf K}/L_{0\mathbf K}=\{E,C'_{21},C''_{21},C_{2}\}\times Z_{2}^{T}$,
& $+1$ &$+1$ &$-1$ &$+1$ &$+1$ &$+1$ &$-1$ \\ 

$(C'_{21}||I,S_{6}^{-},S_{6}^{+},M_{d1,d2,d3}),(C_{2}||C_{2},C_{6}^{+},C_{6}^{-},C''_{21,22,23})$,
&&&&&&&\\

$(TC'_{21}||T,TC_{3}^{+},TC_{3}^{-},TC'_{21,22,23})$
&&&&&&& \\  
\hline

$L_{0\mathbf K}=D_{3}^{1},
M_{\mathbf K}/L_{0\mathbf K}=\{E,C'_{21},C''_{21},C_{2}\}\times Z_{2}^{T}$,
& $+1$ &$+1$ &$+1$ &$+1$ &$+1$ &$+1$ &$-1$ \\ 

$(C'_{21}||I,S_{6}^{-},S_{6}^{+},M_{d1,d2,d3}),(C_{2}||C_{2},C_{6}^{+},C_{6}^{-},C''_{21,22,23})$,
&&&& {\color{red}$0000001$} &&&\\

$(TC''_{21}||T,TC_{3}^{+},TC_{3}^{-},TC'_{21,22,23})$
&&&&&&& \\  
\hline

$L_{0\mathbf K}=D_{3}^{2},
M_{\mathbf K}/L_{0\mathbf K}=\{E,C'_{21},C''_{21},C_{2}\}\times Z_{2}^{T}$,
& $+1$ &$-1$ &$+1$ &$+1$ &$-1$ &$-1$ &$+1$ \\ 

$(C''_{21}||I,S_{6}^{-},S_{6}^{+},M_{v1,v2,v3}),(C_{2}||C_{2},C_{6}^{+},C_{6}^{-},C'_{21,22,23})$,
&&&& {\color{red}$0100110$} &&&\\

$(T||T,TC_{3}^{+},TC_{3}^{-},TC''_{21,22,23})$
&&&&&&& \\ 
\hline

$L_{0\mathbf K}=D_{3}^{2},
M_{\mathbf K}/L_{0\mathbf K}=\{E,C'_{21},C''_{21},C_{2}\}\times Z_{2}^{T}$,
& $+1$ &$+1$ &$+1$ &$-1$ &$+1$ &$-1$ &$+1$ \\ 

$(C''_{21}||I,S_{6}^{-},S_{6}^{+},M_{v1,v2,v3}),(C_{2}||C_{2},C_{6}^{+},C_{6}^{-},C'_{21,22,23})$,
&&&& {\color{red}$0001010$} &&&\\

$(TC_{2}||T,TC_{3}^{+},TC_{3}^{-},TC''_{21,22,23})$
&&&&&&& \\  
\hline

$L_{0\mathbf K}=D_{3}^{2},
M_{\mathbf K}/L_{0\mathbf K}=\{E,C'_{21},C''_{21},C_{2}\}\times Z_{2}^{T}$,
& $+1$ &$+1$ &$+1$ &$+1$ &$+1$ &$-1$ &$+1$ \\ 

$(C''_{21}||I,S_{6}^{-},S_{6}^{+},M_{v1,v2,v3}),(C_{2}||C_{2},C_{6}^{+},C_{6}^{-},C'_{21,22,23})$,
&&&& {\color{red}$0000010$} &&&\\

$(TC'_{21}||T,TC_{3}^{+},TC_{3}^{-},TC''_{21,22,23})$
&&&&&&& \\  
\hline

$L_{0\mathbf K}=D_{3}^{2},
M_{\mathbf K}/L_{0\mathbf K}=\{E,C'_{21},C''_{21},C_{2}\}\times Z_{2}^{T}$,
& $+1$ &$+1$ &$-1$ &$+1$ &$+1$ &$-1$ &$+1$ \\ 

$(C''_{21}||I,S_{6}^{-},S_{6}^{+},M_{v1,v2,v3}),(C_{2}||C_{2},C_{6}^{+},C_{6}^{-},C'_{21,22,23})$,
&&&&&&&\\

$(TC''_{21}||T,TC_{3}^{+},TC_{3}^{-},TC''_{21,22,23})$
&&&&&&& \\  
\hline

$L_{0\mathbf K}=C_{3v}^{1},
M_{\mathbf K}/L_{0\mathbf K}=\{E,C'_{21},C''_{21},C_{2}\}\times Z_{2}^{T}$,
& $-1$ &$-1$ &$+1$ &$+1$ &$+1$ &$+1$ &$-1$ \\ 

$(C'_{21}||I,S_{6}^{-},S_{6}^{+},C'_{21,22,23}),(C_{2}||C_{2},C_{6}^{+},C_{6}^{-},M_{v1,v2,v3})$,
&&&&&&&\\

$(C''_{21}||M_{h},S_{3}^{-},S_{3}^{+},C''_{21,22,23})$,
&&&&&&& \\

$(T||T,TC_{3}^{+},TC_{3}^{-},TM_{d1,d2,d3})$
&&&&&&& \\  
\hline

$L_{0\mathbf K}=C_{3v}^{1},
M_{\mathbf K}/L_{0\mathbf K}=\{E,C'_{21},C''_{21},C_{2}\}\times Z_{2}^{T}$,
& $-1$ &$+1$ &$+1$ &$+1$ &$+1$ &$+1$ &$-1$ \\ 

$(C'_{21}||I,S_{6}^{-},S_{6}^{+},C'_{21,22,23}),(C_{2}||C_{2},C_{6}^{+},C_{6}^{-},M_{v1,v2,v3})$,
&&&&&&&\\

$(C''_{21}||M_{h},S_{3}^{-},S_{3}^{+},C''_{21,22,23})$,
&&&&&&& \\

$(TC_{2}||T,TC_{3}^{+},TC_{3}^{-},TM_{d1,d2,d3})$
&&&&&&& \\  
\hline

$L_{0\mathbf K}=C_{3v}^{1},
M_{\mathbf K}/L_{0\mathbf K}=\{E,C'_{21},C''_{21},C_{2}\}\times Z_{2}^{T}$,
& $-1$ &$+1$ &$-1$ &$-1$ &$+1$ &$+1$ &$-1$ \\ 

$(C'_{21}||I,S_{6}^{-},S_{6}^{+},C'_{21,22,23}),(C_{2}||C_{2},C_{6}^{+},C_{6}^{-},M_{v1,v2,v3})$,
&&&& {\color{red}$1011001$} &&&\\

$(C''_{21}||M_{h},S_{3}^{-},S_{3}^{+},C''_{21,22,23})$,
&&&&&&& \\

$(TC'_{21}||T,TC_{3}^{+},TC_{3}^{-},TM_{d1,d2,d3})$
&&&&&&& \\ 
\hline

$L_{0\mathbf K}=C_{3v}^{1},
M_{\mathbf K}/L_{0\mathbf K}=\{E,C'_{21},C''_{21},C_{2}\}\times Z_{2}^{T}$,
& $-1$ &$+1$ &$+1$ &$+1$ &$-1$ &$+1$ &$-1$ \\ 

$(C'_{21}||I,S_{6}^{-},S_{6}^{+},C'_{21,22,23}),(C_{2}||C_{2},C_{6}^{+},C_{6}^{-},M_{v1,v2,v3})$,
&&&& {\color{red}$1000101$} &&&\\

$(C''_{21}||M_{h},S_{3}^{-},S_{3}^{+},C''_{21,22,23})$,
&&&&&&& \\

$(TC''_{21}||T,TC_{3}^{+},TC_{3}^{-},TM_{d1,d2,d3})$
&&&&&&& \\  
\hline

\end{tabular}
\end{table*}

\begin{table*}
(Extension of Supplementary Table \ref{D6hdrztspin})\\
\centering
\begin{tabular}{ |c|ccccccc|}
\hline
$P=D_{6h}$& && & spin invariants $\vec{\eta}^{S}$&&&\\

$G_{\mathbf K}\cong M_{\mathbf K}=D_{6h}\times Z_2^T$
&$\eta_{C'_{21},C''_{21}}^{S}$
&$\eta_{T}^{S}$&$\eta_{IT}^{S}$ &$\eta_{TC'_{21}}^{S}$
&$\eta_{TC''_{21}}^{S}$
&$\eta_{I,C'_{21}}^{S}$&
$\eta_{I,C''_{21}}^{S}$\\
\hline

$L_{0\mathbf K}=C_{3v}^{2},
M_{\mathbf K}/L_{0\mathbf K}=\{E,C'_{21},C''_{21},C_{2}\}\times Z_{2}^{T}$,
& $-1$ &$-1$ &$+1$ &$+1$ &$+1$ &$-1$ &$+1$ \\ 

$(C''_{21}||I,S_{6}^{-},S_{6}^{+},C''_{21,22,23}),
(C_{2}||C_{2},C_{6}^{+},C_{6}^{-},M_{d1,d2,d3})$,
&&&&&&&\\

$(C'_{21}||M_{h},S_{3}^{-},S_{3}^{+},C'_{21,22,23})$,
&&&&&&& \\

$(T||T,TC_{3}^{+},TC_{3}^{-},TM_{v1,v2,v3})$
&&&&&&& \\  
\hline

$L_{0\mathbf K}=C_{3v}^{2},
M_{\mathbf K}/L_{0\mathbf K}=\{E,C'_{21},C''_{21},C_{2}\}\times Z_{2}^{T}$,
& $-1$ &$+1$ &$+1$ &$+1$ &$+1$ &$-1$ &$+1$ \\ 

$(C''_{21}||I,S_{6}^{-},S_{6}^{+},C''_{21,22,23}),
(C_{2}||C_{2},C_{6}^{+},C_{6}^{-},M_{d1,d2,d3})$,
&&&&&&&\\

$(C'_{21}||M_{h},S_{3}^{-},S_{3}^{+},C'_{21,22,23})$,
&&&&&&& \\

$(TC_{2}||T,TC_{3}^{+},TC_{3}^{-},TM_{v1,v2,v3})$
&&&&&&& \\  
\hline

$L_{0\mathbf K}=C_{3v}^{2},
M_{\mathbf K}/L_{0\mathbf K}=\{E,C'_{21},C''_{21},C_{2}\}\times Z_{2}^{T}$,
& $-1$ &$+1$ &$+1$ &$-1$ &$+1$ &$-1$ &$+1$ \\ 

$(C''_{21}||I,S_{6}^{-},S_{6}^{+},C''_{21,22,23}),
(C_{2}||C_{2},C_{6}^{+},C_{6}^{-},M_{d1,d2,d3})$,
&&&& {\color{red}$1001010$} &&&\\

$(C'_{21}||M_{h},S_{3}^{-},S_{3}^{+},C'_{21,22,23})$,
&&&&&&& \\

$(TC'_{21}||T,TC_{3}^{+},TC_{3}^{-},TM_{v1,v2,v3})$
&&&&&&& \\  
\hline

$L_{0\mathbf K}=C_{3v}^{2},
M_{\mathbf K}/L_{0\mathbf K}=\{E,C'_{21},C''_{21},C_{2}\}\times Z_{2}^{T}$,
& $-1$ &$+1$ &$-1$ &$+1$ &$-1$ &$-1$ &$+1$ \\ 

$(C''_{21}||I,S_{6}^{-},S_{6}^{+},C''_{21,22,23}),
(C_{2}||C_{2},C_{6}^{+},C_{6}^{-},M_{d1,d2,d3})$,
&&&& {\color{red}$1010110$} &&&\\

$(C'_{21}||M_{h},S_{3}^{-},S_{3}^{+},C'_{21,22,23})$,
&&&&&&& \\

$(TC''_{21}||T,TC_{3}^{+},TC_{3}^{-},TM_{v1,v2,v3})$
&&&&&&& \\ 
\hline

$L_{0\mathbf K}=D_{6}, M_{\mathbf K}/L_{0\mathbf K}=\{E,C_{2}\}\times Z_{2}^{T}$,
& $+1$ &$-1$ &$+1$ &$-1$ &$-1$ &$+1$ &$+1$ \\ 

$(C_{2}||I,S_{3}^{-},S_{3}^{+},S_{6}^{-},S_{6}^{+},M_{h},
M_{d1,d2,d3},M_{v1,v2,v3})$,
&&&& {\color{red}$0101100$} &&& \\

$(T||T,TC_{6}^{+},TC_{6}^{-},TC_{3}^{+},TC_{3}^{-},TC_{2},$
&&&&&&&  \\

$TC'_{21,22,23},TC''_{21,22,23})$
&&&&&&& \\ 
\hline

$L_{0\mathbf K}=D_{6}, M_{\mathbf K}/L_{0\mathbf K}=\{E,C_{2}\}\times Z_{2}^{T}$,
& $+1$ &$+1$ &$-1$ &$+1$ &$+1$ &$+1$ &$+1$ \\ 

$(C_{2}||I,S_{3}^{-},S_{3}^{+},S_{6}^{-},S_{6}^{+},M_{h},
M_{d1,d2,d3},M_{v1,v2,v3})$,
&&&& {\color{red}$0010000$} &&& \\

$(TC_{2}||T,TC_{6}^{+},TC_{6}^{-},TC_{3}^{+},TC_{3}^{-},TC_{2},$
&&&&&&&  \\

$TC'_{21,22,23},TC''_{21,22,23})$
&&&&&&& \\ 
\hline

$L_{0\mathbf K}=D_{6}, M_{\mathbf K}/L_{0\mathbf K}=\{E,C_{2},TC'_{21},TC''_{21}\}$,
& $+1$ &$+1$ &$+1$ &$+1$ &$+1$ &$+1$ &$+1$ \\ 

$(C_{2}||I,S_{3}^{-},S_{3}^{+},S_{6}^{-},S_{6}^{+},M_{h},
M_{d1,d2,d3},M_{v1,v2,v3})$,
&&&&&&& \\

$(TC''_{21}||T,TC_{6}^{+},TC_{6}^{-},TC_{3}^{+},TC_{3}^{-},TC_{2},$
&&&&&&&  \\

$TC'_{21,22,23},TC''_{21,22,23})$
&&&&&&& \\
\hline

$L_{0\mathbf K}=C_{6h}, M_{\mathbf K}/L_{0\mathbf K}=\{E,C_{2}\}\times Z_{2}^{T}$,
& $+1$ &$-1$ &$-1$ &$+1$ &$+1$ &$+1$ &$+1$ \\ 

$(C_{2}||C'_{21,22,23},C''_{21,22,23},
M_{d1,d2,d3},M_{v1,v2,v3})$,
&&&& {\color{red}$0110000$} &&& \\

$(T||T,TC_{6}^{+},TC_{6}^{-},TC_{3}^{+},TC_{3}^{-},TC_{2},$
&&&&&&& \\

$IT,TS_{3}^{-},TS_{3}^{+},TS_{6}^{-},TS_{6}^{+},TM_{h})$
&&&&&&& \\ 
\hline

$L_{0\mathbf K}=C_{6h}, M_{\mathbf K}/L_{0\mathbf K}=\{E,C_{2}\}\times Z_{2}^{T}$,
& $+1$ &$+1$ &$+1$ &$-1$ &$-1$ &$+1$ &$+1$ \\ 

$(C_{2}||C'_{21,22,23},C''_{21,22,23},
M_{d1,d2,d3},M_{v1,v2,v3})$,
&&&& {\color{red}$0001100$} &&& \\

$(TC_{2}||T,TC_{6}^{+},TC_{6}^{-},TC_{3}^{+},TC_{3}^{-},TC_{2},$
&&&&&&& \\

$IT,TS_{3}^{-},TS_{3}^{+},TS_{6}^{-},TS_{6}^{+},TM_{h})$
&&&&&&& \\ 
\hline

$L_{0\mathbf K}=C_{6h}, M_{\mathbf K}/L_{0\mathbf K}=\{E,C_{2},TC'_{21},TC''_{21}\}$,
& $+1$ &$+1$ &$+1$ &$+1$ &$+1$ &$+1$ &$+1$ \\ 

$(C_{2}||C'_{21,22,23},C''_{21,22,23},
M_{d1,d2,d3},M_{v1,v2,v3})$,
&&&&&&& \\

$(TC''_{21}||T,TC_{6}^{+},TC_{6}^{-},TC_{3}^{+},TC_{3}^{-},TC_{2},$
&&&&&&& \\

$IT,TS_{3}^{-},TS_{3}^{+},TS_{6}^{-},TS_{6}^{+},TM_{h})$
&&&&&&& \\ 
\hline

$L_{0\mathbf K}=C_{6v}, M_{\mathbf K}/L_{0\mathbf K}=\{E,C_{2}\}\times Z_{2}^{T}$,
& $+1$ &$-1$ &$+1$ &$+1$ &$+1$ &$+1$ &$+1$ \\ 

$(C_{2}||I,S_{3}^{-},S_{3}^{+},S_{6}^{-},S_{6}^{+},M_{h},
C'_{21,22,23},C''_{21,22,23})$,
&&&&&&& \\

$(T||T,TC_{6}^{+},TC_{6}^{-},TC_{3}^{+},TC_{3}^{-},TC_{2},$
&&&&&&& \\

$TM_{d1,d2,d3},TM_{v1,v2,v3})$
&&&&&&& \\ 
\hline

$L_{0\mathbf K}=C_{6v}, M_{\mathbf K}/L_{0\mathbf K}=\{E,C_{2}\}\times Z_{2}^{T}$,
& $+1$ &$+1$ &$-1$ &$-1$ &$-1$ &$+1$ &$+1$ \\ 

$(C_{2}||I,S_{3}^{-},S_{3}^{+},S_{6}^{-},S_{6}^{+},M_{h},
C'_{21,22,23},C''_{21,22,23})$,
&&&& {\color{red}$0011100$} &&& \\

$(TC_{2}||T,TC_{6}^{+},TC_{6}^{-},TC_{3}^{+},TC_{3}^{-},TC_{2},$
&&&&&&& \\

$TM_{d1,d2,d3},TM_{v1,v2,v3})$
&&&&&&& \\ 
\hline

$L_{0\mathbf K}=C_{6v}, M_{\mathbf K}/L_{0\mathbf K}=\{E,C_{2},TC'_{21},TC''_{21}\}$,
& $+1$ &$+1$ &$+1$ &$+1$ &$+1$ &$+1$ &$+1$ \\ 

$(C_{2}||I,S_{3}^{-},S_{3}^{+},S_{6}^{-},S_{6}^{+},M_{h},
C'_{21,22,23},C''_{21,22,23})$,
&&&&&&& \\

$(TC''_{21}||T,TC_{6}^{+},TC_{6}^{-},TC_{3}^{+},TC_{3}^{-},TC_{2},$
&&&&&&& \\

$TM_{d1,d2,d3},TM_{v1,v2,v3})$
&&&&&&& \\ 
\hline

$L_{0\mathbf K}=D_{3d}^{1}, M_{\mathbf K}/L_{0\mathbf K}=\{E,C_{2}\}\times Z_{2}^{T}$,
& $+1$ &$-1$ &$-1$ &$-1$ &$+1$ &$+1$ &$+1$ \\ 

$(C_{2}||C_{2},C_{6}^{+},C_{6}^{-},M_{h},S_{3}^{-},S_{3}^{+},
C''_{21,22,23},M_{v1,v2,v3})$,
&&&& {\color{red}$0111000$} &&& \\

$(T||T,TC_{3}^{+},TC_{3}^{-},IT,TS_{6}^{-},TS_{6}^{+},$
&&&&&&& \\

$TC'_{21,22,23},TM_{d1,d2,d3})$
&&&&&&& \\ 
\hline

\end{tabular}
\end{table*}

\begin{table*}
(Extension of Supplementary Table \ref{D6hdrztspin})\\
\centering
\begin{tabular}{ |c|ccccccc|}
\hline
$P=D_{6h}$& && & spin invariants $\vec{\eta}^{S}$&&&\\

$G_{\mathbf K}\cong M_{\mathbf K}=D_{6h}\times Z_2^T$
&$\eta_{C'_{21},C''_{21}}^{S}$
&$\eta_{T}^{S}$&$\eta_{IT}^{S}$ &$\eta_{TC'_{21}}^{S}$
&$\eta_{TC''_{21}}^{S}$
&$\eta_{I,C'_{21}}^{S}$&
$\eta_{I,C''_{21}}^{S}$\\
\hline

$L_{0\mathbf K}=D_{3d}^{1}, M_{\mathbf K}/L_{0\mathbf K}=\{E,C_{2}\}\times Z_{2}^{T}$,
& $+1$ &$+1$ &$+1$ &$+1$ &$-1$ &$+1$ &$+1$ \\ 

$(C_{2}||C_{2},C_{6}^{+},C_{6}^{-},M_{h},S_{3}^{-},S_{3}^{+},
C''_{21,22,23},M_{v1,v2,v3})$,
&&&& {\color{red}$0000100$} &&& \\

$(TC_{2}||T,TC_{3}^{+},TC_{3}^{-},IT,TS_{6}^{-},TS_{6}^{+},$
&&&&&&& \\

$TC'_{21,22,23},TM_{d1,d2,d3})$
&&&&&&& \\ 
\hline

$L_{0\mathbf K}=D_{3d}^{1}, M_{\mathbf K}/L_{0\mathbf K}=\{E,C_{2},TC'_{21},TC''_{21}\}$,
& $+1$ &$+1$ &$+1$ &$+1$ &$+1$ &$+1$ &$+1$ \\ 

$(C_{2}||C_{2},C_{6}^{+},C_{6}^{-},M_{h},S_{3}^{-},S_{3}^{+},
C''_{21,22,23},M_{v1,v2,v3})$,
&&&&&&& \\

$(TC''_{21}||T,TC_{3}^{+},TC_{3}^{-},IT,TS_{6}^{-},TS_{6}^{+},$
&&&&&&& \\

$TC'_{21,22,23},TM_{d1,d2,d3})$
&&&&&&& \\ 
\hline

$L_{0\mathbf K}=D_{3d}^{2}, M_{\mathbf K}/L_{0\mathbf K}=\{E,C_{2}\}\times Z_{2}^{T}$,
& $+1$ &$-1$ &$-1$ &$+1$ &$-1$ &$+1$ &$+1$ \\ 

$(C_{2}||C_{2},C_{6}^{+},C_{6}^{-},M_{h},S_{3}^{-},S_{3}^{+},
C'_{21,22,23},M_{d1,d2,d3})$,
&&&& {\color{red}$0110100$} &&& \\

$(T||T,TC_{3}^{+},TC_{3}^{-},IT,TS_{6}^{-},TS_{6}^{+},$
&&&&&&& \\

$TC''_{21,22,23},TM_{v1,v2,v3})$
&&&&&&& \\ 
\hline

$L_{0\mathbf K}=D_{3d}^{2}, M_{\mathbf K}/L_{0\mathbf K}=\{E,C_{2}\}\times Z_{2}^{T}$,
& $+1$ &$+1$ &$+1$ &$-1$ &$+1$ &$+1$ &$+1$ \\ 

$(C_{2}||C_{2},C_{6}^{+},C_{6}^{-},M_{h},S_{3}^{-},S_{3}^{+},
C'_{21,22,23},M_{d1,d2,d3})$,
&&&& {\color{red}$0001000$} &&& \\

$(TC_{2}||T,TC_{3}^{+},TC_{3}^{-},IT,TS_{6}^{-},TS_{6}^{+},$
&&&&&&& \\

$TC''_{21,22,23},TM_{v1,v2,v3})$
&&&&&&& \\
\hline

$L_{0\mathbf K}=D_{3d}^{2}, M_{\mathbf K}/L_{0\mathbf K}=\{E,C_{2},TC'_{21},TC''_{21}\}$,
& $+1$ &$+1$ &$+1$ &$+1$ &$+1$ &$+1$ &$+1$ \\ 

$(C_{2}||C_{2},C_{6}^{+},C_{6}^{-},M_{h},S_{3}^{-},S_{3}^{+},
C'_{21,22,23},M_{d1,d2,d3})$,
&&&&&&& \\

$(TC''_{21}||T,TC_{3}^{+},TC_{3}^{-},IT,TS_{6}^{-},TS_{6}^{+},$
&&&&&&& \\

$TC''_{21,22,23},TM_{v1,v2,v3})$
&&&&&&& \\ 
\hline

$L_{0\mathbf K}=D_{3h}^{1}, M_{\mathbf K}/L_{0\mathbf K}=\{E,C_{2}\}\times Z_{2}^{T}$,
& $+1$ &$-1$ &$+1$ &$-1$ &$+1$ &$+1$ &$+1$ \\ 

$(C_{2}||C_{2},C_{6}^{+},C_{6}^{-},I,S_{6}^{-},S_{6}^{+},
M_{d1,d2,d3},C''_{21,22,23})$,
&&&& {\color{red}$0101000$} &&& \\

$(T||T,TS_{3}^{-},TS_{3}^{+},TC_{3}^{+},TC_{3}^{-},TM_{h},$
&&&&&&& \\

$TC'_{21,22,23},TM_{v1,v2,v3})$
&&&&&&& \\ 
\hline

$L_{0\mathbf K}=D_{3h}^{1}, M_{\mathbf K}/L_{0\mathbf K}=\{E,C_{2}\}\times Z_{2}^{T}$,
& $+1$ &$+1$ &$-1$ &$+1$ &$-1$ &$+1$ &$+1$ \\ 

$(C_{2}||C_{2},C_{6}^{+},C_{6}^{-},I,S_{6}^{-},S_{6}^{+},
M_{d1,d2,d3},C''_{21,22,23})$,
&&&& {\color{red}$0010100$} &&& \\

$(TC_{2}||T,TS_{3}^{-},TS_{3}^{+},TC_{3}^{+},TC_{3}^{-},TM_{h},$
&&&&&&& \\

$TC'_{21,22,23},TM_{v1,v2,v3})$
&&&&&&& \\ 
\hline

$L_{0\mathbf K}=D_{3h}^{1}, M_{\mathbf K}/L_{0\mathbf K}=\{E,C_{2},TC'_{21},TC''_{21}\}$,
& $+1$ &$+1$ &$+1$ &$+1$ &$+1$ &$+1$ &$+1$ \\ 

$(C_{2}||C_{2},C_{6}^{+},C_{6}^{-},I,S_{6}^{-},S_{6}^{+},
M_{d1,d2,d3},C''_{21,22,23})$,
&&&&&&& \\

$(TC''_{21}||T,TS_{3}^{-},TS_{3}^{+},TC_{3}^{+},TC_{3}^{-},TM_{h},$
&&&&&&& \\

$TC'_{21,22,23},TM_{v1,v2,v3})$
&&&&&&& \\ 
\hline

$L_{0\mathbf K}=D_{3h}^{2}, M_{\mathbf K}/L_{0\mathbf K}=\{E,C_{2}\}\times Z_{2}^{T}$,
& $+1$ &$-1$ &$+1$ &$+1$ &$-1$ &$+1$ &$+1$ \\ 

$(C_{2}||C_{2},C_{6}^{+},C_{6}^{-},I,S_{6}^{-},S_{6}^{+},
C'_{21,22,23},M_{v1,v2,v3})$,
&&&& {\color{red}$0100100$} &&& \\

$(T||T,TS_{3}^{-},TS_{3}^{+},TC_{3}^{+},TC_{3}^{-},TM_{h},$
&&&&&&& \\

$TC''_{21,22,23},TM_{d1,d2,d3})$
&&&&&&& \\ 
\hline

$L_{0\mathbf K}=D_{3h}^{2}, M_{\mathbf K}/L_{0\mathbf K}=\{E,C_{2}\}\times Z_{2}^{T}$,
& $+1$ &$+1$ &$-1$ &$-1$ &$+1$ &$+1$ &$+1$ \\ 

$(C_{2}||C_{2},C_{6}^{+},C_{6}^{-},I,S_{6}^{-},S_{6}^{+},
C'_{21,22,23},M_{v1,v2,v3})$,
&&&& {\color{red}$0011000$} &&& \\

$(TC_{2}||T,TS_{3}^{-},TS_{3}^{+},TC_{3}^{+},TC_{3}^{-},TM_{h},$
&&&&&&& \\

$TC''_{21,22,23},TM_{d1,d2,d3})$
&&&&&&& \\ 
\hline

$L_{0\mathbf K}=D_{3h}^{2}, M_{\mathbf K}/L_{0\mathbf K}=\{E,C_{2},TC'_{21},TC''_{21}\}$,
& $+1$ &$+1$ &$+1$ &$+1$ &$+1$ &$+1$ &$+1$ \\ 

$(C_{2}||C_{2},C_{6}^{+},C_{6}^{-},I,S_{6}^{-},S_{6}^{+},
C'_{21,22,23},M_{v1,v2,v3})$,
&&&&&&& \\

$(TC''_{21}||T,TS_{3}^{-},TS_{3}^{+},TC_{3}^{+},TC_{3}^{-},TM_{h},$
&&&&&&& \\

$TC''_{21,22,23},TM_{d1,d2,d3})$
&&&&&&& \\ 
\hline

$L_{0\mathbf K}=D_{6h},
M_{\mathbf K}/L_{0\mathbf K}=Z_{2}^{T}$,
& $+1$ &$-1$ &$-1$ &$-1$ &$-1$ &$+1$ &$+1$ \\ 

$(T||T,TC_{6}^{+},TC_{6}^{-},TC_{3}^{+},TC_{3}^{-},TC_{2},$
&&&& {\color{red}$0111100$} &&& \\

$IT,TS_{3}^{-},TS_{3}^{+},TS_{6}^{-},TS_{6}^{+},TM_{h},$
&&&&&&& \\

$TC'_{21,22,23},TC''_{21,22,23},TM_{d1.d2,d3},TM_{v1,v2,v3})$
&&&&&&& \\ 
\hline

$L_{0\mathbf K}=D_{6h},
M_{\mathbf K}/L_{0\mathbf K}=\{E,TC_{2y}\}$,
& $+1$ &$+1$ &$+1$ &$+1$ &$+1$ &$+1$ &$+1$ \\ 

$(TC_{2y}||T,TC_{6}^{+},TC_{6}^{-},TC_{3}^{+},TC_{3}^{-},TC_{2},$
&&&&&&& \\

$IT,TS_{3}^{-},TS_{3}^{+},TS_{6}^{-},TS_{6}^{+},TM_{h},$
&&&&&&& \\

$TC'_{21,22,23},TC''_{21,22,23},TM_{d1.d2,d3},TM_{v1,v2,v3})$
&&&&&&& \\
\hline


\end{tabular}
\end{table*}

\begin{table*}[htbp]
\caption{
For ($G_{\mathbf K}\cong M_{\mathbf K}=O\times{Z}_2^T,L_{0\mathbf{K}},M_{\mathbf K}/L_{0\mathbf{K}}$),
we list generators of spin point group $(\varphi_{g}||g_{1},\ldots,g_{n})$, $(g_{1},\ldots,g_{n})$ is a coset of $L_{0\mathbf{K}}$,
$g\in (g_{1},\ldots,g_{n}),\varphi_{g}\in M_{\mathbf K}/L_{0\mathbf{K}}$,
$n$ is order of $L_{0\mathbf{K}}$, and
spin invariants $\vec{\eta}^{S}$ of $G_{\mathbf K}$ is
$\eta^S_{p_1,p_2}{\cdot1_{2\times 2}} = d({\varphi_{p_1}})d({\varphi_{p_2}})d^{-1}({\varphi_{p_1}})d^{-1}({\varphi_{p_2}})$
for unitary elements $p_{1},p_{2}$ with $p_{1}p_{2}=p_{2}p_{1}$ and
$\eta^S_{p}{\cdot1_{2\times 2}}=[d(\varphi_p) {K}]^2$
for anti-unitary element $p$ with $p^{2}=E$,
where irRep of $g\in G_{\mathbf K}$ is given by
$d(\varphi_g)K^{\zeta_g}=u\Big( s( T^{\zeta_g}\varphi_g)\Big)(i\sigma_{y}K)^{\zeta_g}$,
$s( T^{\zeta_g}\varphi_g)$ is vector Rep of $T^{\zeta_g}\varphi_{g}\in\mathrm{SO(3)}$,
$u\Big( s( T^{\zeta_g}\varphi_g)\Big)$ is $\mathrm{SU(2)}$ Rep of $T^{\zeta_g}\varphi_{g}$,
$\zeta_{g}=0(1)$ for unitary (anti-unitary) $g$.
The $\vec{\eta}^{S}$ leading to invariants only realized in SSG are labeled by Boolean vectors
colored in {\color{red}RED}.} \label{Odrztspin}
\centering
\begin{tabular}{ |c|ccc|}
\hline
$P=O$& & spin invariants $\vec{\eta}^{S}$& \\

$G_{\mathbf K}\cong M_{\mathbf K}=O\times Z_2^T$
&$\eta_{T}^{S}$& $\eta_{TC_{2z}}^{S}$ &$\eta_{TC_{2f}}^{S}$\\
\hline

$L_{0\mathbf K}=E,M_{\mathbf K}/L_{0\mathbf K}=O\times Z_{2}^{T}$,
& $-1$ & $+1$ & $+1$ \\

$(C_{2z}||C_{2z}),(C_{2f}||C_{2f}),(C_{4z}^{+}||C_{4z}^{+}),(T||T)$
&&& \\ 
\hline

$L_{0\mathbf K}=\{E,C_{2x,2y,2z}\},
M_{\mathbf K}/L_{0\mathbf K}=D_{3}^{1}\times Z_{2}^{T}$,
& $-1$&$-1$&$+1$  \\

$(C_{3}^{+}||C_{31,32,33,34}^{+}),(C'_{21}||C_{2f},C_{2d},C_{4x}^{+},C_{4x}^{-})$,
&&&  \\

$(T||T,TC_{2x,2y,2z})$
&&&  \\ 
\hline

$L_{0\mathbf K}=\{E,C_{2x,2y,2z}\}$,
& $+1$&$+1$&$+1$  \\

$M_{\mathbf K}/L_{0\mathbf K}=\{E,C_{3}^{+},C_{3}^{-},C'_{21,22,23},TC_{2},TC_{6}^{-},TC_{6}^{+},TC''_{21,22,23}\}$,
&&& \\

$(C_{3}^{+}||C_{31,32,33,34}^{+}),(C'_{21}||C_{2f},C_{2d},C_{4x}^{+},C_{4x}^{-})$,
&&&  \\

$(TC_{2}||T,TC_{2x,2y,2z})$
&&&  \\ 
\hline

$L_{0\mathbf K}=\{E,C_{31,32,33,34}^{\pm},C_{2x,2y,2z}\},
M_{\mathbf K}/L_{0\mathbf K}=\{E,C_{2z}\}\times Z_{2}^{T}$,
& $-1$&$-1$&$+1$ \\  %

$(C_{2z}||C_{2a,2b,2c,2d,2e,2f},C_{4x,4y,4z}^{\pm})$,
&&& \\

$(T||T,TC_{31,32,33,34}^{\pm},TC_{2x,2y,2z})$
&&& \\ 
\hline

$L_{0\mathbf K}=\{E,C_{31,32,33,34}^{\pm},C_{2x,2y,2z}\},
M_{\mathbf K}/L_{0\mathbf K}=\{E,C_{2z}\}\times Z_{2}^{T}$,
& $+1$&$+1$&$-1$ \\  %

$(C_{2z}||C_{2a,2b,2c,2d,2e,2f},C_{4x,4y,4z}^{\pm})$,
&& {\color{red}$001$} & \\

$(TC_{2z}||T,TC_{31,32,33,34}^{\pm},TC_{2x,2y,2z})$
&&&  \\ 
\hline

$L_{0\mathbf K}=\{E,C_{31,32,33,34}^{\pm},C_{2x,2y,2z}\},
M_{\mathbf K}/L_{0\mathbf K}=\{E,C_{2z},TC_{2x},TC_{2y}\}$,
& $+1$&$+1$&$+1$ \\  %

$(C_{2z}||C_{2a,2b,2c,2d,2e,2f},C_{4x,4y,4z}^{\pm})$,
&&& \\

$(TC_{2y}||T,TC_{31,32,33,34}^{\pm},TC_{2x,2y,2z})$
&&& \\ 

\hline

$L_{0\mathbf K}=O,M_{\mathbf K}/L_{0\mathbf K}=Z_{2}^{T}$,
& $-1$&$-1$&$-1$ \\  %

$(T||T,TC_{31,32,33,34}^{\pm},TC_{2x,2y,2z},TC_{2a,2b,2c,2d,2e,2f},TC_{4x,4y,4z}^{\pm})$
&& {\color{red}$111$}  & \\
\hline

$L_{0\mathbf K}=O,M_{\mathbf K}/L_{0\mathbf K}=\{E,TC_{2y}\}$,
& $+1$&$+1$&$+1$ \\

$(TC_{2y}||T,TC_{31,32,33,34}^{\pm},TC_{2x,2y,2z},TC_{2a,2b,2c,2d,2e,2f},TC_{4x,4y,4z}^{\pm})$
&&& \\ 
\hline

\end{tabular}
\end{table*}

\begin{table*}[htbp]
\caption{
For ($G_{\mathbf K}\cong M_{\mathbf K}=O_{h}\times{Z}_2^T,L_{0\mathbf{K}},M_{\mathbf K}/L_{0\mathbf{K}}$),
we list generators of spin point group $(\varphi_{g}||g_{1},\ldots,g_{n})$, $(g_{1},\ldots,g_{n})$ is a coset of $L_{0\mathbf{K}}$,
$g\in (g_{1},\ldots,g_{n}),\varphi_{g}\in M_{\mathbf K}/L_{0\mathbf{K}}$,
$n$ is order of $L_{0\mathbf{K}}$, and
spin invariants $\vec{\eta}^{S}$ of $G_{\mathbf K}$ is
$\eta^S_{p_1,p_2}{\cdot1_{2\times 2}} = d({\varphi_{p_1}})d({\varphi_{p_2}})d^{-1}({\varphi_{p_1}})d^{-1}({\varphi_{p_2}})$
for unitary elements $p_{1},p_{2}$ with $p_{1}p_{2}=p_{2}p_{1}$ and
$\eta^S_{p}{\cdot1_{2\times 2}}=[d(\varphi_p) {K}]^2$
for anti-unitary element $p$ with $p^{2}=E$,
where irRep of $g\in G_{\mathbf K}$ is given by
$d(\varphi_g)K^{\zeta_g}=u\Big( s( T^{\zeta_g}\varphi_g)\Big)(i\sigma_{y}K)^{\zeta_g}$,
$s( T^{\zeta_g}\varphi_g)$ is vector Rep of $T^{\zeta_g}\varphi_{g}\in\mathrm{SO(3)}$,
$u\Big( s( T^{\zeta_g}\varphi_g)\Big)$ is $\mathrm{SU(2)}$ Rep of $T^{\zeta_g}\varphi_{g}$,
$\zeta_{g}=0(1)$ for unitary (anti-unitary) $g$.
The $\vec{\eta}^{S}$ leading to invariants only realized in SSG are labeled by Boolean vectors
colored in {\color{red}RED}.} \label{Ohdrztspin}
\centering
\begin{tabular}{ |c|ccccc|}
 \hline
$P=O_{h}$& &&  spin invariants $\vec{\eta}^{S}$ &&\\

$G_{\mathbf K}\cong M_{\mathbf K}=O_{h}\times Z_2^T$
&$\eta_{C_{2x},C_{2y}}^{S}$
&$\eta_{T}^{S}$&$\eta_{IT}^{S}$&$\eta_{TC_{2a}}^{S}$
&$\eta_{I,C_{2a}}^{S}$\\
\hline

$L_{0\mathbf K}=\{E,I\},
M_{\mathbf K}/L_{0\mathbf K}=O\times Z_{2}^{T},
(C_{2x}||C_{2x},M_{x})$,
& $-1$&$-1$&$-1$&$+1$&$+1$\\ 

$(C_{2y}||C_{2y},M_{y}),(C_{2a}||C_{2a},M_{da}),(C_{4y}^{+}||C_{4y}^{+},S_{4y}^{-}),(T||T,IT)$
&&&&& \\

\hline

$L_{0\mathbf K}=\{E,C_{2x,2y,2z}\},
M_{\mathbf K}/L_{0\mathbf K}=D_{6}\times Z_{2}^{T}$,
& $+1$&$-1$&$+1$&$+1$&$-1$\\  

$(C_{6}^{+}||S_{61,62,63,64}^{+}),(C_{2}||I,M_{x,y,z}),(C'_{21}||C_{2a,2b},C_{4z}^{\pm})$,
&&&&& \\

$(T||T,TC_{2x,2y,2z})$
&&&&& \\ 
\hline

$L_{0\mathbf K}=\{E,C_{2x,2y,2z}\},
M_{\mathbf K}/L_{0\mathbf K}=D_{6}\times Z_{2}^{T}$,
& $+1$&$+1$&$-1$&$+1$&$-1$\\  

$(C_{6}^{+}||S_{61,62,63,64}^{+}),(C_{2}||I,M_{x,y,z}),(C'_{21}||C_{2a,2b},C_{4z}^{\pm})$,
&&& {\color{red}$00101$} && \\

$(TC_{2}||T,TC_{2x,2y,2z})$
&&&&& \\ 
\hline

$L_{0\mathbf K}=D_{2h}^{},
M_{\mathbf K}/L_{0\mathbf K}=D_{3}^{1}\times Z_{2}^{T}$,

& $+1$&$-1$&$-1$&$+1$&$+1$\\  

$(C_{3}^{+}||C_{31,32,33,34}^{+},S_{61,62,63,64}^{-}),
(C'_{21}||C_{2a,2b},C_{4z}^{\pm},M_{da,db},S_{4z}^{\mp})$,
&&& {\color{red}$01100$} && \\

$(T||T,TC_{2x,2y,2z},IT,TM_{x,y,z})$
&&&&& \\  
\hline

$L_{0\mathbf K}=D_{2h}^{},
M_{\mathbf K}/L_{0\mathbf K}=\{E,C_{3}^{+},C_{3}^{-},C'_{21,22,23},$
& $+1$&$+1$&$+1$&$+1$&$+1$\\  

$TC_{2},TC_{6}^{-},TC_{6}^{+},TC''_{21,22,23}\}$,
&&&&&  \\

$(C_{3}^{+}||C_{31,32,33,34}^{+},S_{61,62,63,64}^{-}),
(C'_{21}||C_{2a,2b},C_{4z}^{\pm},M_{da,db},S_{4z}^{\mp})$,
&&&&& \\

$(TC_{2}||T,TC_{2x,2y,2z},IT,TM_{x,y,z})$
&&&&& \\ 
\hline

$L_{0\mathbf K}=\{E,C_{31,32,33,34}^{\pm},C_{2x,2y,2z}\}$,
& $+1$&$-1$&$+1$&$+1$&$-1$\\  

$M_{\mathbf K}/L_{0\mathbf K}=\{E,C_{2x,2y,2z}\}\times Z_{2}^{T},
(C_{2z}||I,S_{61,62,63,64}^{\mp},M_{x,y,z})$,
&&&&& \\

$(C_{2x}||C_{2a,2b,2c,2d,2e,2f},C_{4x,4y,4z}^{\pm})$,
&&&&& \\

$(T||T,TC_{31,32,33,34}^{\pm},TC_{2x,2y,2z})$
&&&&& \\ 
\hline

$L_{0\mathbf K}=\{E,C_{31,32,33,34}^{\pm},C_{2x,2y,2z}\}$,
& $+1$&$+1$&$-1$&$+1$&$-1$\\  

$M_{\mathbf K}/L_{0\mathbf K}=\{E,C_{2x,2y,2z}\}\times Z_{2}^{T},
(C_{2z}||I,S_{61,62,63,64}^{\mp},M_{x,y,z})$,
&&& {\color{red}$00101$} && \\

$(C_{2x}||C_{2a,2b,2c,2d,2e,2f},C_{4x,4y,4z}^{\pm})$,
&&&&& \\

$(TC_{2z}||T,TC_{31,32,33,34}^{\pm},TC_{2x,2y,2z})$
&&&&& \\  
\hline

$L_{0\mathbf K}=\{E,C_{31,32,33,34}^{\pm},C_{2x,2y,2z}\}$,
& $+1$&$+1$&$+1$&$-1$&$-1$\\  

$M_{\mathbf K}/L_{0\mathbf K}=\{E,C_{2x,2y,2z}\}\times Z_{2}^{T},
(C_{2z}||I,S_{61,62,63,64}^{\mp},M_{x,y,z})$,
&&& {\color{red}$00011$} && \\

$(C_{2x}||C_{2a,2b,2c,2d,2e,2f},C_{4x,4y,4z}^{\pm})$,
&&&&& \\

$(TC_{2x}||T,TC_{31,32,33,34}^{\pm},TC_{2x,2y,2z})$
&&&&& \\  
\hline

$L_{0\mathbf K}=\{E,C_{31,32,33,34}^{\pm},C_{2x,2y,2z}\}$,
& $+1$&$+1$&$+1$&$+1$&$-1$\\  

$M_{\mathbf K}/L_{0\mathbf K}=\{E,C_{2x,2y,2z}\}\times Z_{2}^{T},
(C_{2z}||I,S_{61,62,63,64}^{\mp},M_{x,y,z})$,
&&&&& \\

$(C_{2x}||C_{2a,2b,2c,2d,2e,2f},C_{4x,4y,4z}^{\pm})$,
&&&&& \\

$(TC_{2y}||T,TC_{31,32,33,34}^{\pm},TC_{2x,2y,2z})$
&&&&& \\ 
\hline

$L_{0\mathbf K}=O,M_{\mathbf K}/L_{0\mathbf K}=\{E,C_{2z}\}\times Z_{2}^{T}$,
& $+1$&$-1$&$+1$&$-1$&$+1$\\ 

$(C_{2z}||I,S_{61,62,63,64}^{\mp},M_{x,y,z},
M_{da,db,dc,dd,de,df},S_{4x,4y,4z}^{\mp})$,
&&& {\color{red}$01010$} && \\

$(T||T,TC_{31,32,33,34}^{\pm},TC_{2x,2y,2z},$
&&&&& \\

$TC_{2a,2b,2c,2d,2e,2f},TC_{4x,4y,4z}^{\pm})$
&&&&& \\ 
\hline

$L_{0\mathbf K}=O,M_{\mathbf K}/L_{0\mathbf K}=\{E,C_{2z}\}\times Z_{2}^{T}$,
& $+1$&$+1$&$-1$&$+1$&$+1$\\ 

$(C_{2z}||I,S_{61,62,63,64}^{\mp},M_{x,y,z},
M_{da,db,dc,dd,de,df},S_{4x,4y,4z}^{\mp})$,
&&& {\color{red}$00100$} && \\

$(TC_{2z}||T,TC_{31,32,33,34}^{\pm},TC_{2x,2y,2z},$
&&&&& \\

$TC_{2a,2b,2c,2d,2e,2f},TC_{4x,4y,4z}^{\pm})$
&&&&& \\ 
\hline

$L_{0\mathbf K}=O,M_{\mathbf K}/L_{0\mathbf K}=\{E,C_{2z},TC_{2x},TC_{2y}\}$,
& $+1$&$+1$&$+1$&$+1$&$+1$\\ 

$(C_{2z}||I,S_{61,62,63,64}^{\mp},M_{x,y,z},
M_{da,db,dc,dd,de,df},S_{4x,4y,4z}^{\mp})$,
&&&&& \\

$(TC_{2y}||T,TC_{31,32,33,34}^{\pm},TC_{2x,2y,2z},$
&&&&& \\

$TC_{2a,2b,2c,2d,2e,2f},TC_{4x,4y,4z}^{\pm})$
&&&&& \\ 
\hline

$L_{0\mathbf K}=T_{d},M_{\mathbf K}/L_{0\mathbf K}=\{E,C_{2z}\}\times Z_{2}^{T}$,
& $+1$&$-1$&$+1$&$+1$&$+1$\\ 

$(C_{2z}||I,S_{61,62,63,64}^{\mp},M_{x,y,z},
C_{2a,2b,2c,2d,2e,2f},C_{4x,4y,4z}^{\pm})$,
&&&&& \\

$(T||T,TC_{31,32,33,34}^{\pm},TC_{2x,2y,2z},$
&&&&& \\

$TM_{da,db,dc,dd,de,df},TS_{4x,4y,4z}^{\mp})$
&&&&& \\ 
\hline

$L_{0\mathbf K}=T_{d},M_{\mathbf K}/L_{0\mathbf K}=\{E,C_{2z}\}\times Z_{2}^{T}$,
& $+1$&$+1$&$-1$&$-1$&$+1$\\ 

$(C_{2z}||I,S_{61,62,63,64}^{\mp},M_{x,y,z},
C_{2a,2b,2c,2d,2e,2f},C_{4x,4y,4z}^{\pm})$,
&&& {\color{red}$00110$}  && \\

$(TC_{2z}||T,TC_{31,32,33,34}^{\pm},TC_{2x,2y,2z},$
&&&&& \\

$TM_{da,db,dc,dd,de,df},TS_{4x,4y,4z}^{\mp})$
&&&&& \\ 
\hline

\end{tabular}
\end{table*}

\begin{table*}
(Extension of Supplementary Table \ref{Ohdrztspin})\\
\centering
\begin{tabular}{ |c|ccccc|}
 \hline
$P=O_{h}$& &&  spin invariants $\vec{\eta}^{S}$ &&\\

$G_{\mathbf K}\cong M_{\mathbf K}=O_{h}\times Z_2^T$
&$\eta_{C_{2x},C_{2y}}^{S}$
&$\eta_{T}^{S}$&$\eta_{IT}^{S}$&$\eta_{TC_{2a}}^{S}$
&$\eta_{I,C_{2a}}^{S}$\\
\hline

$L_{0\mathbf K}=T_{d},M_{\mathbf K}/L_{0\mathbf K}=\{E,C_{2z},TC_{2x},TC_{2y}\}$,
& $+1$&$+1$&$+1$&$+1$&$+1$\\ 

$(C_{2z}||I,S_{61,62,63,64}^{\mp},M_{x,y,z},
C_{2a,2b,2c,2d,2e,2f},C_{4x,4y,4z}^{\pm})$
&&&&& \\

$(TC_{2y}||T,TC_{31,32,33,34}^{\pm},TC_{2x,2y,2z},$
&&&&& \\

$TM_{da,db,dc,dd,de,df},TS_{4x,4y,4z}^{\mp})$
&&&&& \\ 
\hline

$L_{0\mathbf K}=T_{h},M_{\mathbf K}/L_{0\mathbf K}=\{E,C_{2z}\}\times Z_{2}^{T}$,
& $+1$&$-1$&$-1$&$+1$&$+1$\\ 

$(C_{2z}||C_{2a,2b,2c,2d,2e,2f},C_{4x,4y,4z}^{\pm},
M_{da,db,dc,dd,de,df},S_{4x,4y,4z}^{\mp})$,
&&& {\color{red}$01100$} && \\

$(T||T,TC_{31,32,33,34}^{\pm},TC_{2x,2y,2z},$
&&&&& \\

$IT,TS_{61,62,63,64}^{\mp},TM_{x,y,z})$
&&&&& \\ 
\hline

$L_{0\mathbf K}=T_{h},M_{\mathbf K}/L_{0\mathbf K}=\{E,C_{2z}\}\times Z_{2}^{T}$,
& $+1$&$+1$&$+1$&$-1$&$+1$\\ 

$(C_{2z}||C_{2a,2b,2c,2d,2e,2f},C_{4x,4y,4z}^{\pm},
M_{da,db,dc,dd,de,df},S_{4x,4y,4z}^{\mp})$,
&&& {\color{red}$00010$} && \\

$(TC_{2z}||T,TC_{31,32,33,34}^{\pm},TC_{2x,2y,2z},$
&&&&& \\

$IT,TS_{61,62,63,64}^{\mp},TM_{x,y,z})$
&&&&& \\ 
\hline

$L_{0\mathbf K}=T_{h},M_{\mathbf K}/L_{0\mathbf K}=\{E,C_{2z},TC_{2x},TC_{2y}\}$,
& $+1$&$+1$&$+1$&$+1$&$+1$\\ 

$(C_{2z}||C_{2a,2b,2c,2d,2e,2f},C_{4x,4y,4z}^{\pm},
M_{da,db,dc,dd,de,df},S_{4x,4y,4z}^{\mp})$,
&&&&& \\

$(TC_{2y}||T,TC_{31,32,33,34}^{\pm},TC_{2x,2y,2z},$
&&&&& \\

$IT,TS_{61,62,63,64}^{\mp},TM_{x,y,z})$
&&&&& \\  
\hline

$L_{0\mathbf K}=O_{h},
M_{\mathbf K}/L_{0\mathbf K}=Z_{2}^{T}$,
& $+1$&$-1$&$-1$&$-1$&$+1$\\   

$(T||T,TC_{31,32,33,34}^{\pm},TC_{2x,2y,2z},
TC_{2a,2b,2c,2d,2e,2f},TC_{4x,4y,4z}^{\pm},$
&&& {\color{red}$01110$} && \\

$IT,TS_{61,62,63,64}^{\mp},TM_{x,y,z},
TM_{da,db,dc,dd,de,df},TS_{4x,4y,4z}^{\mp})$
&&&&& \\ 
\hline

$L_{0\mathbf K}=O_{h},
M_{\mathbf K}/L_{0\mathbf K}=\{E,TC_{2y}\}$,
& $+1$&$+1$&$+1$&$+1$&$+1$\\   

$(TC_{2y}||T,TC_{31,32,33,34}^{\pm},TC_{2x,2y,2z},
TC_{2a,2b,2c,2d,2e,2f},TC_{4x,4y,4z}^{\pm},$
&&&&& \\

$IT,TS_{61,62,63,64}^{\mp},TM_{x,y,z},
TM_{da,db,dc,dd,de,df},TS_{4x,4y,4z}^{\mp})$
&&&&& \\ 
\hline

\end{tabular}
\end{table*}

\clearpage

\begin{table*}[htbp]
\caption{$G_{\mathbf K}\cong M_{\mathbf K}=C_{4h}\times Z_2^T$,
${\rm Invariants}\equiv(\eta_{I,C_{4z}^{+}},\eta_T,\eta_{IT},\eta_{TM_{z}})$,
we list the values of all symmetry invariants,
where type II MSG-realizable are colored in {\color{green}GREEN}, type IV MSG-realizable in {\color{blue}BLUE},
and only SSG-realizable in {\color{red}RED}.
For only SSG-realizable ($\vec{\eta}^{L}$,{\R $\vec{\eta}^{S}$}), dim denotes the lowest dimension of irRep,
disp. are the $k$-terms which can split band degeneracy, n.l. is direction of nodal line in BZ.}
\label{C4hdrztinv}
\centering
\begin{tabular}{ |c|c|c|c|c|}
\hline
Invariants & ($\vec{\eta}^{L}$,{\R $\vec{\eta}^{S}$})&dim &disp.&n.l.\\
\hline

(${\color{green}+1,+1,+1,+1 }$)
&&&& \\

(${\color{green}-1,+1,+1,+1 }$)
&&&& \\

(${\color{green}+1,-1,-1,+1 }$)
&&&& \\

(${\color{green}-1,-1,-1,+1 }$)
&&&& \\   

(${\color{blue}-1,-1,+1,+1 }$)
&&&& \\

(${\color{blue}+1,-1,+1,+1 }$)
&&&& \\

(${\color{blue}-1,+1,-1,+1 }$)
&&&& \\

(${\color{blue}+1,+1,-1,+1 }$)
&&&& \\  
\hline

(${\color{red}+1,-1,+1,-1 }$)
& ($0100$,{\color{red}$0001$}),($0000$,{\color{red}$0101$})

&&& \\ 
\hline

(${\color{red}-1,+1,+1,-1 }$)
&($1000$,{\color{red}$0001$}),($1100$,{\color{red}$0101$})

&$4$ &$[k_{x},k_{y}],k_{z}$& $-$\\ 
\hline

(${\color{red}+1,+1,+1,-1 }$)
&($0000$,{\color{red}$0001$}),($0100$,{\color{red}$0101$})

&&& \\ 
\hline

(${\color{red}-1,-1,+1,-1 }$)
&($1100$,{\color{red}$0001$}),($1000$,{\color{red}$0101$})

&$4$ &$[k_{x},k_{y}],k_{z}$& $-$\\ 
\hline

(${\color{red}+1,-1,-1,-1 }$)
&($0000$,{\color{red}$0111$}),($1100$,{\color{red}$1011$}),
($0100$,{\color{red}$0011$})
&&& \\ 
\hline

(${\color{red}-1,+1,-1,-1 }$)
&($1100$,{\color{red}$0111$}),($0000$,{\color{red}$1011$}),
($1000$,{\color{red}$0011$})
&&& \\ 
\hline

(${\color{red}+1,+1,-1,-1 }$)
&($0100$,{\color{red}$0111$}),($1000$,{\color{red}$1011$}),
($0000$,{\color{red}$0011$})
&&& \\ 
\hline

(${\color{red}-1,-1,-1,-1 }$)
&($1000$,{\color{red}$0111$}),($0100$,{\color{red}$1011$}),
($1100$,{\color{red}$0011$})
&$4$ &$k_{z},k_{x}^{2}-k_{y}^{2},k_{x}k_{y}$&$-$ \\ 
\hline

\end{tabular}
\end{table*}

\begin{table*}[htbp]
\caption{$G_{\mathbf K}\cong M_{\mathbf K}=D_{4}\times Z_2^T$,
${\rm Invariants}\equiv(\eta_{C_{2x},C_{2y}},\eta_{T},\eta_{TC_{2x}},\eta_{TC_{2a}})$,
we list the values of all symmetry invariants,
where type II MSG-realizable are colored in {\color{green}GREEN}, type IV MSG-realizable in {\color{blue}BLUE}, and only SSG-realizable in {\color{red}RED}.
For only SSG-realizable ($\vec{\eta}^{L}$,{\R $\vec{\eta}^{S}$}), dim denotes the lowest dimension of irRep,
disp. are the $k$-terms which can split band degeneracy.} \label{D4drztinv}
\centering
\begin{tabular}{ |c|c|c|c|}
\hline
Invariants & ($\vec{\eta}^{L}$,{\R $\vec{\eta}^{S}$})&dim &disp.\\
\hline

(${\color{green}+1,+1,+1,+1 }$)
&&& \\

(${\color{green}+1,+1,-1,+1 }$)
&&& \\

(${\color{green}-1,+1,+1,+1 }$)
&&& \\

(${\color{green}-1,+1,-1,+1 }$)
&&& \\  

(${\color{green}-1,-1,+1,+1 }$)
&&& \\

(${\color{green}-1,-1,-1,+1 }$)
&&& \\

(${\color{green}+1,-1,+1,+1 }$)
&&& \\

(${\color{green}+1,-1,-1,+1 }$)
&&& \\ 
\hline

(${\color{red}+1,+1,+1,-1 }$)
& ($0000$,{\color{red}$0001$}),($0010$,{\color{red}$0011$}),
($0100$,{\color{red}$0101$}),($0110$,{\color{red}$0111$})

&& \\ 
\hline

(${\color{red}+1,+1,-1,-1 }$)
&($0010$,{\color{red}$0001$}),($0000$,{\color{red}$0011$}),
($0110$,{\color{red}$0101$}),($0100$,{\color{red}$0111$})

&& \\ 
\hline

(${\color{red}+1,-1,+1,-1 }$)
&($0100$,{\color{red}$0001$}),($0110$,{\color{red}$0011$}),
($0000$,{\color{red}$0101$}),($0010$,{\color{red}$0111$})

&& \\ 
\hline

(${\color{red}+1,-1,-1,-1 }$)
&($0110$,{\color{red}$0001$}),($0100$,{\color{red}$0011$}),
($0010$,{\color{red}$0101$}),($0000$,{\color{red}$0111$})

&& \\ 
\hline

(${\color{red}-1,+1,+1,-1 }$)
&($1000$,{\color{red}$0001$}),($1010$,{\color{red}$0011$}),
($1100$,{\color{red}$0101$}),($1110$,{\color{red}$0111$})

&$4$ &$[k_{x},k_{y}],k_{z}$ \\ 
\hline

(${\color{red}-1,+1,-1,-1 }$)
&($1010$,{\color{red}$0001$}),($1000$,{\color{red}$0011$}),
($1110$,{\color{red}$0101$}),($1100$,{\color{red}$0111$})

&$4$ &$[k_{x},k_{y}],k_{z}$ \\ 
\hline

(${\color{red}-1,-1,+1,-1 }$)
&($1100$,{\color{red}$0001$}),($1110$,{\color{red}$0011$}),
($1000$,{\color{red}$0101$}),($1010$,{\color{red}$0111$})

&$4$ &$[k_{x},k_{y}],k_{z}$ \\ 
\hline

(${\color{red}-1,-1,-1,-1 }$)
&($1110$,{\color{red}$0001$}),($1100$,{\color{red}$0011$}),
($1010$,{\color{red}$0101$}),($1000$,{\color{red}$0111$})

&$4$ &$[k_{x},k_{y}],k_{z}$ \\ 
\hline
\end{tabular}
\end{table*}

\begin{table*}[htbp]
\caption{$G_{\mathbf K}\cong M_{\mathbf K}=D_{4h}\times Z_2^T$,
${\rm Invariants}\equiv (\eta_{C_{2x},C_{2y}}, \eta_T,\eta_{IT},
\eta_{TC_{2x}},\eta_{TC_{2a}},\eta_{I,C_{2x}},\eta_{I,C_{2a}})$,
we list the values of all symmetry invariants,
where type II MSG-realizable are colored in {\color{green}GREEN}, type IV MSG-realizable in {\color{blue}BLUE}, and only SSG-realizable in {\color{red}RED}.
For only SSG-realizable ($\vec{\eta}^{L}$,{\R $\vec{\eta}^{S}$}), dim denotes the lowest dimension of irRep,
disp. are the $k$-terms which can split band degeneracy, n.l. is direction of nodal line in BZ.}
\label{D4hdrztinv}
\centering
\begin{tabular}{ |c|c|c|c|c|}
\hline
Invariants & ($\vec{\eta}^{L}$,{\R $\vec{\eta}^{S}$})&dim &disp.&n.l.\\
\hline

({\color{green}$+1,+1,+1,+1,+1,+1,+1$})
&$\cdots$&$\cdots$ & $\cdots$ &$\cdots$ \\
({\color{green}$+1,+1,+1,-1,+1,+1,+1$})
&&&& \\
({\color{green}$+1,+1,+1,+1,+1,-1,-1$})
&&&& \\
({\color{green}$+1,+1,+1,-1,+1,-1,-1$})
&&&& \\  

({\color{green}$+1,+1,+1,+1,+1,+1,-1$})
&&&& \\
({\color{green}$+1,+1,+1,-1,+1,+1,-1$})
&&&& \\
({\color{green}$+1,+1,+1,+1,+1,-1,+1$})
&&&& \\
({\color{green}$+1,+1,+1,-1,+1,-1,+1$})
&&&& \\  

({\color{green}$-1,-1,-1,+1,+1,+1,+1$})
&&&& \\
({\color{green}$-1,-1,-1,-1,+1,+1,+1$})
&&&& \\
({\color{green}$-1,-1,-1,+1,+1,-1,-1$})
&&&& \\
({\color{green}$-1,-1,-1,-1,+1,-1,-1$})
&&&& \\  

({\color{green}$-1,-1,-1,+1,+1,+1,-1$})
&&&& \\
({\color{green}$-1,-1,-1,-1,+1,+1,-1$})
&&&& \\
({\color{green}$-1,-1,-1,+1,+1,-1,+1$})
&&&& \\
({\color{green}$-1,-1,-1,-1,+1,-1,+1$})
&&&& \\  

({\color{blue}$+1,-1,+1,+1,+1,+1,+1$})
&$\cdots$&$\cdots$ & $\cdots$ &$\cdots$ \\

({\color{blue}$+1,-1,+1,-1,+1,+1,+1$})
&&&& \\

({\color{blue}$+1,-1,+1,+1,+1,-1,-1$})
&&&& \\
({\color{blue}$+1,-1,+1,-1,+1,-1,-1$})
&&&& \\ 

({\color{blue}$+1,-1,+1,+1,+1,+1,-1$})
&&&& \\
({\color{blue}$+1,-1,+1,-1,+1,+1,-1$})
&&&& \\
({\color{blue}$+1,-1,+1,+1,+1,-1,+1$})
&&&& \\
({\color{blue}$+1,-1,+1,-1,+1,-1,+1$})
&&&& \\ 

({\color{blue}$-1,+1,-1,+1,+1,+1,+1$})
&&&& \\
({\color{blue}$-1,+1,-1,-1,+1,+1,+1$})
&&&& \\
({\color{blue}$-1,+1,-1,+1,+1,-1,-1$})
&&&& \\
({\color{blue}$-1,+1,-1,-1,+1,-1,-1$})
&&&& \\ 

({\color{blue}$-1,+1,-1,+1,+1,+1,-1$})
&&&& \\
({\color{blue}$-1,+1,-1,-1,+1,+1,-1$})
&&&& \\
({\color{blue}$-1,+1,-1,+1,+1,-1,+1$})
&&&& \\
({\color{blue}$-1,+1,-1,-1,+1,-1,+1$})
&&&& \\ 
\hline

(${\color{red}+1,-1,+1,-1,-1,+1,+1}$)
& ($0000000$,{\color{red}$0101100$}),($0001000$,{\color{red}$0100100$}),
&&& \\ 

& ($0100000$,{\color{red}$0001100$}),($0101000$,{\color{red}$0000100$}),
&&& \\

& ($0100011$,{\color{red}$0001111$}),($0101011$,{\color{red}$0000111$}),
&&& \\

& ($0001010$,{\color{red}$0100110$}),($0101001$,{\color{red}$0000101$})
&&& \\
\hline

(${\color{red}+1,-1,+1,+1,-1,+1,+1}$)
& ($0001000$,{\color{red}$0101100$}),($0000000$,{\color{red}$0100100$}),
&&& \\ 

& ($0101000$,{\color{red}$0001100$}),($0100000$,{\color{red}$0000100$}),
&&& \\

& ($0101011$,{\color{red}$0001111$}),($0100011$,{\color{red}$0000111$}),
&&& \\

& ($0000010$,{\color{red}$0100110$}),($0100001$,{\color{red}$0000101$})
&&& \\
\hline

(${\color{red}+1,-1,+1,-1,-1,-1,-1}$)
& ($0000011$,{\color{red}$0101100$}),($0001011$,{\color{red}$0100100$}),
&$4_{1,2}$ &$k_{z}$&$k_{x},k_{x}\pm k_{y},k_{y}$ \\ 

& ($0100011$,{\color{red}$0001100$}),($0101011$,{\color{red}$0000100$}),
&$4_{3,4}$ &$k_{x}k_{y}k_{z},k_{z}(k_{x}^{2}-k_{y}^{2})$ &$k_{x},k_{x}\pm k_{y},k_{y},k_{z}$ \\

& ($0100000$,{\color{red}$0001111$}),($0101000$,{\color{red}$0000111$}),
&&& \\

& ($0001001$,{\color{red}$0100110$}),($0101010$,{\color{red}$0000101$})
&&& \\
\hline

(${\color{red}+1,-1,+1,+1,-1,-1,-1}$)
& ($0001011$,{\color{red}$0101100$}),($0000011$,{\color{red}$0100100$}),
&$4$ &$k_{z},k_{x}^{2}-k_{y}^{2}$ &$k_{x}\pm k_{y}$ \\ 

& ($0101011$,{\color{red}$0001100$}),($0100011$,{\color{red}$0000100$}),
&&& \\

& ($0101000$,{\color{red}$0001111$}),($0100000$,{\color{red}$0000111$}),
&&& \\

& ($0000001$,{\color{red}$0100110$}),($0100010$,{\color{red}$0000101$})
&&& \\
\hline

(${\color{red}+1,+1,+1,-1,-1,+1,+1}$)
& ($0100000$,{\color{red}$0101100$}),($0101000$,{\color{red}$0100100$}),
&&& \\ 

& ($0000000$,{\color{red}$0001100$}),($0001000$,{\color{red}$0000100$}),
&&& \\

& ($0000011$,{\color{red}$0001111$}),($0001011$,{\color{red}$0000111$}),
&&& \\

& ($0101010$,{\color{red}$0100110$}),($0001001$,{\color{red}$0000101$})
&&& \\
\hline

(${\color{red}+1,+1,+1,+1,-1,+1,+1}$)
& ($0101000$,{\color{red}$0101100$}),($0100000$,{\color{red}$0100100$}),
&&& \\ 

& ($0001000$,{\color{red}$0001100$}),($0000000$,{\color{red}$0000100$}),
&&& \\

& ($0001011$,{\color{red}$0001111$}),($0000011$,{\color{red}$0000111$}),
&&& \\

& ($0100010$,{\color{red}$0100110$}),($0000001$,{\color{red}$0000101$})
&&& \\

\hline
\end{tabular}
\end{table*}

\begin{table*}
(Extension of Supplementary Table \ref{D4hdrztinv},${\rm Invariants}\equiv (\eta_{C_{2x},C_{2y}}, \eta_T,\eta_{IT},
\eta_{TC_{2x}},\eta_{TC_{2a}},\eta_{I,C_{2x}},\eta_{I,C_{2a}})$\\
\centering
\begin{tabular}{ |c|c|c|c|c|}
\hline
Invariants & ($\vec{\eta}^{L}$,{\R $\vec{\eta}^{S}$})&dim &disp.&n.l.\\
\hline

(${\color{red}+1,+1,+1,-1,-1,-1,-1}$)
& ($0100011$,{\color{red}$0101100$}),($0101011$,{\color{red}$0100100$}),
&&& \\ 

& ($0000011$,{\color{red}$0001100$}),($0001011$,{\color{red}$0000100$}),
&&& \\

& ($0000000$,{\color{red}$0001111$}),($0001000$,{\color{red}$0000111$}),
&&& \\

& ($0101001$,{\color{red}$0100110$}),($0001010$,{\color{red}$0000101$})
&&& \\
\hline

(${\color{red}+1,+1,+1,+1,-1,-1,-1}$)
& ($0101011$,{\color{red}$0101100$}),($0100011$,{\color{red}$0100100$}),
&&& \\ 

& ($0001011$,{\color{red}$0001100$}),($0000011$,{\color{red}$0000100$}),
&&& \\

& ($0001000$,{\color{red}$0001111$}),($0000000$,{\color{red}$0000111$}),
&&& \\

& ($0100001$,{\color{red}$0100110$}),($0000010$,{\color{red}$0000101$})
&&& \\
\hline

(${\color{red}+1,-1,+1,-1,-1,+1,-1}$)
& ($0000001$,{\color{red}$0101100$}),($0001001$,{\color{red}$0100100$}),
&$4_{1}$ &$k_{z},k_{x}^{2}-k_{y}^{2}$&$k_{x}\pm k_{y}$ \\ 

& ($0100001$,{\color{red}$0001100$}),($0101001$,{\color{red}$0000100$}),
&$4_{2,3}$ & $k_{x}^{2}-k_{y}^{2},k_{x}k_{y}k_{z}$ &$k_{x}\pm k_{y},k_{z}$ \\

& ($0100010$,{\color{red}$0001111$}),($0101010$,{\color{red}$0000111$}),
&&& \\

& ($0001011$,{\color{red}$0100110$}),($0101000$,{\color{red}$0000101$})
&&& \\
\hline

(${\color{red}+1,-1,+1,+1,-1,+1,-1}$)
& ($0001001$,{\color{red}$0101100$}),($0000001$,{\color{red}$0100100$}),
&$4$ &$k_{z},k_{x}^{2}-k_{y}^{2}$ &$k_{x}\pm k_{y}$ \\ 

& ($0101001$,{\color{red}$0001100$}),($0100001$,{\color{red}$0000100$}),
&&& \\

& ($0101010$,{\color{red}$0001111$}),($0100010$,{\color{red}$0000111$}),
&&& \\

& ($0000011$,{\color{red}$0100110$}),($0100000$,{\color{red}$0000101$})
&&& \\
\hline

(${\color{red}+1,-1,+1,-1,-1,-1,+1}$)
& ($0000010$,{\color{red}$0101100$}),($0001010$,{\color{red}$0100100$}),
&$4_{1}$ &$k_{z},k_{x}k_{y}$&$k_{x},k_{y}$ \\ 

& ($0100010$,{\color{red}$0001100$}),($0101010$,{\color{red}$0000100$}),
&$4_{2,3}$ & $k_{x}k_{y},k_{z}(k_{x}^{2}-k_{y}^{2})$ &$k_{x},k_{y},k_{z}$ \\

& ($0100001$,{\color{red}$0001111$}),($0101001$,{\color{red}$0000111$}),
&&& \\

& ($0001000$,{\color{red}$0100110$}),($0101011$,{\color{red}$0000101$})
&&& \\
\hline

(${\color{red}+1,-1,+1,+1,-1,-1,+1}$)
& ($0001010$,{\color{red}$0101100$}),($0000010$,{\color{red}$0100100$}),
&&& \\ 

& ($0101010$,{\color{red}$0001100$}),($0100010$,{\color{red}$0000100$}),
&&& \\

& ($0101001$,{\color{red}$0001111$}),($0100001$,{\color{red}$0000111$}),
&&& \\

& ($0000000$,{\color{red}$0100110$}),($0100011$,{\color{red}$0000101$})
&&& \\
\hline

(${\color{red}+1,+1,+1,-1,-1,+1,-1}$)
& ($0100001$,{\color{red}$0101100$}),($0101001$,{\color{red}$0100100$}),
&$4_{1}$ &$k_{z},k_{x}^{2}-k_{y}^{2},k_{x}k_{y}$&$-$ \\ 

& ($0000001$,{\color{red}$0001100$}),($0001001$,{\color{red}$0000100$}),
&$4_{2}$ & $k_{x}^{2}-k_{y}^{2},k_{x}k_{y},k_{x}k_{y}k_{z},k_{z}(k_{x}^{2}-k_{y}^{2})$ &$k_{z}$ \\

& ($0000010$,{\color{red}$0001111$}),($0001010$,{\color{red}$0000111$}),
&&& \\

& ($0101011$,{\color{red}$0100110$}),($0001000$,{\color{red}$0000101$})
&&& \\
\hline

(${\color{red}+1,+1,+1,+1,-1,+1,-1}$)
& ($0101001$,{\color{red}$0101100$}),($0100001$,{\color{red}$0100100$}),
&&& \\ 

& ($0001001$,{\color{red}$0001100$}),($0000001$,{\color{red}$0000100$}),
&&& \\

& ($0001010$,{\color{red}$0001111$}),($0000010$,{\color{red}$0000111$}),
&&& \\

& ($0100011$,{\color{red}$0100110$}),($0000000$,{\color{red}$0000101$})
&&& \\
\hline

(${\color{red}+1,+1,+1,-1,-1,-1,+1}$)
& ($0100010$,{\color{red}$0101100$}),($0101010$,{\color{red}$0100100$}),
&$4_{1}$ &$k_{z},k_{x}^{2}-k_{y}^{2},k_{x}k_{y}$&$-$ \\ 

& ($0000010$,{\color{red}$0001100$}),($0001010$,{\color{red}$0000100$}),
&$4_{2}$ & $k_{x}^{2}-k_{y}^{2},k_{x}k_{y},k_{x}k_{y}k_{z},k_{z}(k_{x}^{2}-k_{y}^{2})$ &$k_{z}$ \\

& ($0000001$,{\color{red}$0001111$}),($0001001$,{\color{red}$0000111$}),
&&& \\

& ($0101000$,{\color{red}$0100110$}),($0001011$,{\color{red}$0000101$})
&&& \\
\hline

(${\color{red}+1,+1,+1,+1,-1,-1,+1}$)
& ($0101010$,{\color{red}$0101100$}),($0100010$,{\color{red}$0100100$}),
&$4$ &$k_{z},k_{x}^{2}-k_{y}^{2},k_{x}k_{y}$ &$-$ \\ 

& ($0001010$,{\color{red}$0001100$}),($0000010$,{\color{red}$0000100$}),
&&& \\

& ($0001001$,{\color{red}$0001111$}),($0000001$,{\color{red}$0000111$}),
&&& \\

& ($0100000$,{\color{red}$0100110$}),($0000011$,{\color{red}$0000101$})
&&& \\
\hline

(${\color{red}-1,-1,+1,+1,+1,-1,-1}$)
& ($0000000$,{\color{red}$1100011$}),($0100011$,{\color{red}$1000000$})
&&& \\ 
\hline

(${\color{red}-1,-1,+1,-1,+1,-1,-1}$)
& ($0001000$,{\color{red}$1100011$}),($0101011$,{\color{red}$1000000$})
&$4$ & $[k_{x},k_{y}]$& $k_{z}$ \\  
\hline

(${\color{red}-1,-1,+1,+1,+1,+1,+1}$)
& ($0000011$,{\color{red}$1100011$}),($0100000$,{\color{red}$1000000$})
&$4$ &$[k_{x},k_{y}],k_{z}$& $-$\\  
\hline

(${\color{red}-1,-1,+1,-1,+1,+1,+1}$)
& ($0001011$,{\color{red}$1100011$}),($0101000$,{\color{red}$1000000$})
&$4$ &$[k_{x},k_{y}],[k_{z}k_{x},k_{z}k_{y}]$& $k_{z}$ \\  
\hline

(${\color{red}-1,+1,+1,+1,+1,-1,-1}$)
& ($0100000$,{\color{red}$1100011$}),($0000011$,{\color{red}$1000000$})
&$4$ &$[k_{x},k_{y}],k_{z}$&  $-$\\  
\hline

(${\color{red}-1,+1,+1,-1,+1,-1,-1}$)
& ($0101000$,{\color{red}$1100011$}),($0001011$,{\color{red}$1000000$})
&$4$ &$[k_{x},k_{y}],[k_{z}k_{x},k_{z}k_{y}]$& $k_{z}$ \\  
\hline

(${\color{red}-1,+1,+1,+1,+1,+1,+1}$)
& ($0100011$,{\color{red}$1100011$}),($0000000$,{\color{red}$1000000$})
&&& \\  
\hline

(${\color{red}-1,+1,+1,-1,+1,+1,+1}$)
& ($0101011$,{\color{red}$1100011$}),($0001000$,{\color{red}$1000000$})
&$4$ &$k_{x}k_{y},[k_{z}k_{x},k_{z}k_{y}]$& $k_{x},k_{y},k_{z}$ \\  
\hline

(${\color{red}-1,-1,+1,+1,+1,-1,+1}$)
& ($0000001$,{\color{red}$1100011$}),($0100010$,{\color{red}$1000000$})
&$4$ &$[k_{x},k_{y}],k_{z}$& $-$\\  
\hline

(${\color{red}-1,-1,+1,-1,+1,-1,+1}$)
& ($0001001$,{\color{red}$1100011$}),($0101010$,{\color{red}$1000000$})
&$8$ &$[k_{x},k_{y}],k_{z}$& $-$ \\  
\hline

(${\color{red}-1,-1,+1,+1,+1,+1,-1}$)
& ($0000010$,{\color{red}$1100011$}),($0100001$,{\color{red}$1000000$})
&$4$ &$[k_{x},k_{y}],k_{z}$& $-$\\  
\hline

(${\color{red}-1,-1,+1,-1,+1,+1,-1}$)
& ($0001010$,{\color{red}$1100011$}),($0101001$,{\color{red}$1000000$})
&$4$ &$[k_{x},k_{y}],[k_{z}k_{x},k_{z}k_{y}]$& $k_{z}$ \\  
\hline

(${\color{red}-1,+1,+1,+1,+1,-1,+1}$)
& ($0100001$,{\color{red}$1100011$}),($0000010$,{\color{red}$1000000$})
&$4$ &$[k_{x},k_{y}],k_{z}$& $-$\\  
\hline

(${\color{red}-1,+1,+1,-1,+1,-1,+1}$)
& ($0101001$,{\color{red}$1100011$}),($0001010$,{\color{red}$1000000$})
&$4$ &
$[k_{x},k_{y}],[k_{z}k_{x},k_{z}k_{y}]$& $k_{z}$ \\  
\hline

(${\color{red}-1,+1,+1,+1,+1,+1,-1}$)
& ($0100010$,{\color{red}$1100011$}),($0000001$,{\color{red}$1000000$})
&$4$ &$[k_{x},k_{y}],k_{z}$& $-$\\  
\hline

(${\color{red}-1,+1,+1,-1,+1,+1,-1}$)
& ($0101010$,{\color{red}$1100011$}),($0001001$,{\color{red}$1000000$})
& $8$ &$[k_{x},k_{y}],k_{z}$& $-$ \\  
\hline

(${\color{red}+1,-1,-1,+1,+1,+1,+1}$)
& ($0000000$,{\color{red}$0110000$}),($0001000$,{\color{red}$0111000$}),
&&& \\ 

& ($0100000$,{\color{red}$0010000$}),($0101000$,{\color{red}$0011000$}),
&&& \\

& ($0100011$,{\color{red}$0010011$}),($0100001$,{\color{red}$0010001$}),
&&& \\

& ($0101001$,{\color{red}$0011001$}),($0100010$,{\color{red}$0010010$})
&&& \\

\hline
\end{tabular}
\end{table*}

\begin{table*}
(Extension of Supplementary Table \ref{D4hdrztinv},${\rm Invariants}\equiv (\eta_{C_{2x},C_{2y}},\eta_T,\eta_{IT},
\eta_{TC_{2x}},\eta_{TC_{2a}},\eta_{I,C_{2x}},\eta_{I,C_{2a}})$\\
\centering
\begin{tabular}{ |c|c|c|c|c|}
\hline
Invariants & ($\vec{\eta}^{L}$,{\R $\vec{\eta}^{S}$})&dim &disp.&n.l.\\
\hline

(${\color{red}+1,-1,-1,-1,+1,+1,+1}$)
& ($0001000$,{\color{red}$0110000$}),($0000000$,{\color{red}$0111000$}),
&&& \\ 

& ($0101000$,{\color{red}$0010000$}),($0100000$,{\color{red}$0011000$}),
&&& \\

& ($0101011$,{\color{red}$0010011$}),($0101001$,{\color{red}$0010001$}),
&&& \\

& ($0100001$,{\color{red}$0011001$}),($0101010$,{\color{red}$0010010$})
&&& \\
\hline

(${\color{red}+1,-1,-1,+1,+1,-1,-1}$)
& ($0000011$,{\color{red}$0110000$}),($0001011$,{\color{red}$0111000$}),
&$4_{1}$ &
$k_{z},k_{x}k_{y}k_{z},k_{z}(k_{x}^{2}-k_{y}^{2})$&$k_{x},k_{x}\pm k_{y},k_{y}$ \\ 

& ($0100011$,{\color{red}$0010000$}),($0101011$,{\color{red}$0011000$}),
&$4_{2,3}$ & $k_{z}$&$k_{x},k_{x}\pm k_{y},k_{y}$ \\

& ($0100000$,{\color{red}$0010011$}),($0100010$,{\color{red}$0010001$}),
&&& \\

& ($0101010$,{\color{red}$0011001$}),($0100001$,{\color{red}$0010010$})
&&& \\
\hline

(${\color{red}+1,-1,-1,-1,+1,-1,-1}$)
& ($0001011$,{\color{red}$0110000$}),($0000011$,{\color{red}$0111000$}),
&$4_{1}$ &$k_{z},k_{x}k_{y}k_{z}$&$k_{x},k_{x}\pm k_{y},k_{y}$ \\ 

& ($0101011$,{\color{red}$0010000$}),($0100011$,{\color{red}$0011000$}),
&$4_{2,3}$ & $k_{x}k_{y}k_{z},k_{z}(k_{x}^{2}-k_{y}^{2})$ &$k_{x},k_{x}\pm k_{y},k_{y},k_{z}$ \\

& ($0101000$,{\color{red}$0010011$}),($0101010$,{\color{red}$0010001$}),
&&& \\

& ($0100010$,{\color{red}$0011001$}),($0101001$,{\color{red}$0010010$})
&&& \\
\hline

(${\color{red}+1,+1,-1,+1,+1,+1,+1}$)
& ($0100000$,{\color{red}$0110000$}),($0101000$,{\color{red}$0111000$}),
&&& \\ 

& ($0000000$,{\color{red}$0010000$}),($0001000$,{\color{red}$0011000$}),
&&& \\

& ($0000011$,{\color{red}$0010011$}),($0000001$,{\color{red}$0010001$}),
&&& \\

& ($0001001$,{\color{red}$0011001$}),($0000010$,{\color{red}$0010010$})
&&& \\
\hline

(${\color{red}+1,+1,-1,-1,+1,+1,+1}$)
& ($0101000$,{\color{red}$0110000$}),($0100000$,{\color{red}$0111000$}),
&&& \\ 

& ($0001000$,{\color{red}$0010000$}),($0000000$,{\color{red}$0011000$}),
&&& \\

& ($0001011$,{\color{red}$0010011$}),($0001001$,{\color{red}$0010001$}),
&&& \\

& ($0000001$,{\color{red}$0011001$}),($0001010$,{\color{red}$0010010$})
&&& \\
\hline

(${\color{red}+1,+1,-1,+1,+1,-1,-1}$)
& ($0100011$,{\color{red}$0110000$}),($0101011$,{\color{red}$0111000$}),
&&& \\ 

& ($0000011$,{\color{red}$0010000$}),($0001011$,{\color{red}$0011000$}),
&&& \\

& ($0000000$,{\color{red}$0010011$}),($0000010$,{\color{red}$0010001$}),
&&& \\

& ($0001010$,{\color{red}$0011001$}),($0000001$,{\color{red}$0010010$})
&&& \\
\hline

(${\color{red}+1,+1,-1,-1,+1,-1,-1}$)
& ($0101011$,{\color{red}$0110000$}),($0100011$,{\color{red}$0111000$}),
&$4_{1}$ &
$k_{z},k_{x}k_{y}$&$k_{x},k_{y}$ \\ 

& ($0001011$,{\color{red}$0010000$}),($0000011$,{\color{red}$0011000$}),
&$4_{2}$ & $k_{x}k_{y},k_{x}k_{y}k_{z},k_{z}(k_{x}^{2}-k_{y}^{2})$ &$k_{x},k_{y},k_{z}$ \\

& ($0001000$,{\color{red}$0010011$}),($0001010$,{\color{red}$0010001$}),
&&& \\

& ($0000010$,{\color{red}$0011001$}),($0001001$,{\color{red}$0010010$})
&&& \\
\hline

(${\color{red}+1,-1,-1,+1,+1,+1,-1}$)
& ($0000001$,{\color{red}$0110000$}),($0001001$,{\color{red}$0111000$}),
&$4$ &
$k_{z},k_{x}^{2}-k_{y}^{2}$ &$k_{x}\pm k_{y}$ \\ 

& ($0100001$,{\color{red}$0010000$}),($0101001$,{\color{red}$0011000$}),
&&& \\

& ($0100010$,{\color{red}$0010011$}),($0100000$,{\color{red}$0010001$}),
&&& \\

& ($0101000$,{\color{red}$0011001$}),($0100011$,{\color{red}$0010010$})
&&& \\
\hline

(${\color{red}+1,-1,-1,-1,+1,+1,-1}$)
& ($0001001$,{\color{red}$0110000$}),($0000001$,{\color{red}$0111000$}),
&$4_{1,2}$ &
$k_{z},k_{x}^{2}-k_{y}^{2}$&$k_{x}\pm k_{y}$ \\ 

& ($0101001$,{\color{red}$0010000$}),($0100001$,{\color{red}$0011000$}),
&$4_{3,4}$ & $k_{x}^{2}-k_{y}^{2},k_{x}k_{y}k_{z}$ &$k_{x}\pm k_{y},k_{z}$ \\

& ($0101010$,{\color{red}$0010011$}),($0101000$,{\color{red}$0010001$}),
&&& \\

& ($0100000$,{\color{red}$0011001$}),($0101011$,{\color{red}$0010010$})
&&& \\
\hline

(${\color{red}+1,-1,-1,+1,+1,-1,+1}$)
& ($0000010$,{\color{red}$0110000$}),($0001010$,{\color{red}$0111000$}),
&$4$ &
$k_{z},k_{x}k_{y}$&$k_{x},k_{y}$ \\ 

& ($0100010$,{\color{red}$0010000$}),($0101010$,{\color{red}$0011000$}),
&&& \\

& ($0100001$,{\color{red}$0010011$}),($0100011$,{\color{red}$0010001$}),
&&& \\

& ($0101011$,{\color{red}$0011001$}),($0100000$,{\color{red}$0010010$})
&&& \\
\hline

(${\color{red}+1,-1,-1,-1,+1,-1,+1}$)
& ($0001010$,{\color{red}$0110000$}),($0000010$,{\color{red}$0111000$}),
&$4_{1}$ &
$k_{z},k_{x}k_{y}$&$k_{x},k_{y}$ \\ 

& ($0101010$,{\color{red}$0010000$}),($0100010$,{\color{red}$0011000$}),
&$4_{2}$ & $k_{x}k_{y},k_{x}k_{y}k_{z},k_{z}(k_{x}^{2}-k_{y}^{2})$ &$k_{x},k_{y},k_{z}$ \\

& ($0101001$,{\color{red}$0010011$}),($0101011$,{\color{red}$0010001$}),
&&& \\

& ($0100011$,{\color{red}$0011001$}),($0101000$,{\color{red}$0010010$})
&&& \\
\hline

(${\color{red}+1,+1,-1,+1,+1,+1,-1}$)
& ($0100001$,{\color{red}$0110000$}),($0101001$,{\color{red}$0111000$}),
&&& \\ 

& ($0000001$,{\color{red}$0010000$}),($0001001$,{\color{red}$0011000$}),
&&& \\

& ($0000010$,{\color{red}$0010011$}),($0000000$,{\color{red}$0010001$}),
&&& \\

& ($0001000$,{\color{red}$0011001$}),($0000011$,{\color{red}$0010010$})
&&& \\
\hline

(${\color{red}+1,+1,-1,-1,+1,+1,-1}$)
& ($0101001$,{\color{red}$0110000$}),($0100001$,{\color{red}$0111000$}),
&&& \\ 

& ($0001001$,{\color{red}$0010000$}),($0000001$,{\color{red}$0011000$}),
&&& \\

& ($0001010$,{\color{red}$0010011$}),($0001000$,{\color{red}$0010001$}),
&&& \\

& ($0000000$,{\color{red}$0011001$}),($0001011$,{\color{red}$0010010$})
&&& \\
\hline

(${\color{red}+1,+1,-1,+1,+1,-1,+1}$)
& ($0100010$,{\color{red}$0110000$}),($0101010$,{\color{red}$0111000$}),
&&& \\ 

& ($0000010$,{\color{red}$0010000$}),($0001010$,{\color{red}$0011000$}),
&&& \\

& ($0000001$,{\color{red}$0010011$}),($0000011$,{\color{red}$0010001$}),
&&& \\

& ($0001011$,{\color{red}$0011001$}),($0000000$,{\color{red}$0010010$})
&&& \\
\hline

(${\color{red}+1,+1,-1,-1,+1,-1,+1}$)
& ($0101010$,{\color{red}$0110000$}),($0100010$,{\color{red}$0111000$}),
&$4_{1}$ &
$k_{z},k_{x}k_{y}$&$k_{x},k_{y}$ \\ 

& ($0001010$,{\color{red}$0010000$}),($0000010$,{\color{red}$0011000$}),
&$4_{2,3}$ & $k_{x}k_{y},k_{z}(k_{x}^{2}-k_{y}^{2})$ &$k_{x},k_{y},k_{z}$ \\

& ($0001001$,{\color{red}$0010011$}),($0001011$,{\color{red}$0010001$}),
&&& \\

& ($0000011$,{\color{red}$0011001$}),($0001000$,{\color{red}$0010010$})
&&& \\

\hline
\end{tabular}
\end{table*}

\begin{table*}
(Extension of Supplementary Table \ref{D4hdrztinv},${\rm Invariants}\equiv (\eta_{C_{2x},C_{2y}}, \eta_T,\eta_{IT},
\eta_{TC_{2x}},\eta_{TC_{2a}},\eta_{I,C_{2x}},\eta_{I,C_{2a}})$\\
\centering
\begin{tabular}{ |c|c|c|c|c|}
\hline
Invariants & ($\vec{\eta}^{L}$,{\R $\vec{\eta}^{S}$})&dim &disp.&n.l.\\
\hline

(${\color{red}+1,-1,-1,-1,-1,+1,+1}$)
& ($0000000$,{\color{red}$0111100$}),($0001000$,{\color{red}$0110100$}),
&&& \\ 

& ($0100000$,{\color{red}$0011100$}),($0101000$,{\color{red}$0010100$}),
&&& \\

& ($0101010$,{\color{red}$0010110$})
&&& \\
\hline

(${\color{red}+1,-1,-1,+1,-1,+1,+1}$)
& ($0001000$,{\color{red}$0111100$}),($0000000$,{\color{red}$0110100$}),
&&& \\ 

& ($0101000$,{\color{red}$0011100$}),($0100000$,{\color{red}$0010100$}),
&&& \\

& ($0100010$,{\color{red}$0010110$})
&&& \\
\hline

(${\color{red}+1,-1,-1,-1,-1,-1,-1}$)
& ($0000011$,{\color{red}$0111100$}),($0001011$,{\color{red}$0110100$}),
&$4_{1,2}$ &
$k_{z}$&$k_{x},k_{x}\pm k_{y},k_{y}$ \\ 

& ($0100011$,{\color{red}$0011100$}),($0101011$,{\color{red}$0010100$}),
&$4_{3}$ & $k_{x}k_{y}k_{z},k_{z}(k_{x}^{2}-k_{y}^{2})$ &$k_{x},k_{x}\pm k_{y},k_{y},k_{z}$ \\

& ($0101001$,{\color{red}$0010110$})
&&& \\
\hline

(${\color{red}+1,-1,-1,+1,-1,-1,-1}$)
& ($0001011$,{\color{red}$0111100$}),($0000011$,{\color{red}$0110100$}),
&$4_{1}$ &
$k_{z},k_{z}(k_{x}^{2}-k_{y}^{2})$&$k_{x},k_{x}\pm k_{y},k_{y}$ \\ 

& ($0101011$,{\color{red}$0011100$}),($0100011$,{\color{red}$0010100$}),
&$4_{2,3}$ & $k_{x}k_{y}k_{z},k_{z}(k_{x}^{2}-k_{y}^{2})$ &$k_{x},k_{x}\pm k_{y},k_{y},k_{z}$ \\

& ($0100001$,{\color{red}$0010110$})
&&& \\
\hline

(${\color{red}+1,+1,-1,-1,-1,+1,+1}$)
& ($0100000$,{\color{red}$0111100$}),($0101000$,{\color{red}$0110100$}),
&&& \\ 

& ($0000000$,{\color{red}$0011100$}),($0001000$,{\color{red}$0010100$}),
&&& \\

& ($0001010$,{\color{red}$0010110$})
&&& \\
\hline

(${\color{red}+1,+1,-1,+1,-1,+1,+1}$)
& ($0101000$,{\color{red}$0111100$}),($0100000$,{\color{red}$0110100$}),
&&& \\ 

& ($0001000$,{\color{red}$0011100$}),($0000000$,{\color{red}$0010100$}),
&&& \\

& ($0000010$,{\color{red}$0010110$})
&&& \\
\hline

(${\color{red}+1,+1,-1,-1,-1,-1,-1}$)
& ($0100011$,{\color{red}$0111100$}),($0101011$,{\color{red}$0110100$}),
&$4_{1,2}$ &
$k_{z}$&$k_{x},k_{x}\pm k_{y},k_{y}$ \\ 

& ($0000011$,{\color{red}$0011100$}),($0001011$,{\color{red}$0010100$}),
&$4_{3,4}$ & $k_{x}k_{y}k_{z},k_{z}(k_{x}^{2}-k_{y}^{2})$ &$k_{x},k_{x}\pm k_{y},k_{y},k_{z}$ \\

& ($0001001$,{\color{red}$0010110$})
&&& \\
\hline

(${\color{red}+1,+1,-1,+1,-1,-1,-1}$)
& ($0101011$,{\color{red}$0111100$}),($0100011$,{\color{red}$0110100$}),
&$4_{1}$ &
$k_{z},k_{x}^{2}-k_{y}^{2}$&$k_{x}\pm k_{y}$ \\ 

& ($0001011$,{\color{red}$0011100$}),($0000011$,{\color{red}$0010100$}),
&$4_{2}$ & $k_{x}^{2}-k_{y}^{2},k_{x}k_{y}k_{z},k_{z}(k_{x}^{2}-k_{y}^{2})$ &
$k_{x}\pm k_{y},k_{z}$ \\

& ($0000001$,{\color{red}$0010110$})
&&& \\
\hline

(${\color{red}+1,-1,-1,-1,-1,+1,-1}$)
& ($0000001$,{\color{red}$0111100$}),($0001001$,{\color{red}$0110100$}),
&$4_{1,2}$ &
$k_{z},k_{x}^{2}-k_{y}^{2}$&$k_{x}\pm k_{y}$ \\ 

& ($0100001$,{\color{red}$0011100$}),($0101001$,{\color{red}$0010100$}),
&$4_{3,4}$ & $k_{x}^{2}-k_{y}^{2},k_{x}k_{y}k_{z}$ &
$k_{x}\pm k_{y},k_{z}$ \\

& ($0101011$,{\color{red}$0010110$})
&&& \\
\hline

(${\color{red}+1,-1,-1,+1,-1,+1,-1}$)
& ($0001001$,{\color{red}$0111100$}),($0000001$,{\color{red}$0110100$}),
&$4_{1}$ &
$k_{z},k_{x}^{2}-k_{y}^{2}$&$k_{x}\pm k_{y}$ \\ 

& ($0101001$,{\color{red}$0011100$}),($0100001$,{\color{red}$0010100$}),
&$4_{2}$ & $k_{x}^{2}-k_{y}^{2},k_{x}k_{y}k_{z},k_{z}(k_{x}^{2}-k_{y}^{2})$ &
$k_{x}\pm k_{y},k_{z}$ \\

& ($0100011$,{\color{red}$0010110$})
&&& \\
\hline

(${\color{red}+1,-1,-1,-1,-1,-1,+1}$)
& ($0000010$,{\color{red}$0111100$}),($0001010$,{\color{red}$0110100$}),
&$4_{1,2}$ &
$k_{z},k_{x}k_{y}$&$k_{x},k_{y}$ \\ 

& ($0100010$,{\color{red}$0011100$}),($0101010$,{\color{red}$0010100$}),
&$4_{3,4}$ & $k_{x}k_{y},k_{z}(k_{x}^{2}-k_{y}^{2})$ &$k_{x},k_{y},k_{z}$ \\

& ($0101000$,{\color{red}$0010110$})
&&& \\
\hline

(${\color{red}+1,-1,-1,+1,-1,-1,+1}$)
& ($0001010$,{\color{red}$0111100$}),($0000010$,{\color{red}$0110100$}),
&$4_{1,2}$ &
$k_{z},k_{x}k_{y}$&$k_{x},k_{y}$ \\ 

& ($0101010$,{\color{red}$0011100$}),($0100010$,{\color{red}$0010100$}),
&$4_{3,4}$ & $k_{x}k_{y},k_{z}(k_{x}^{2}-k_{y}^{2})$ &$k_{x},k_{y},k_{z}$ \\

& ($0100000$,{\color{red}$0010110$})
&&& \\
\hline

(${\color{red}+1,+1,-1,-1,-1,+1,-1}$)
& ($0100001$,{\color{red}$0111100$}),($0101001$,{\color{red}$0110100$}),
&$4_{1,2}$ &
$k_{z},k_{x}^{2}-k_{y}^{2}$&$k_{x}\pm k_{y}$ \\ 

& ($0000001$,{\color{red}$0011100$}),($0001001$,{\color{red}$0010100$}),
&$4_{3}$ & $k_{x}^{2}-k_{y}^{2},k_{x}k_{y}k_{z}$ &
$k_{x}\pm k_{y},k_{z}$ \\

& ($0001011$,{\color{red}$0010110$})
&&& \\
\hline

(${\color{red}+1,+1,-1,+1,-1,+1,-1}$)
& ($0101001$,{\color{red}$0111100$}),($0100001$,{\color{red}$0110100$}),
&$4_{1}$ &
$k_{z},k_{x}^{2}-k_{y}^{2}$&$k_{x}\pm k_{y}$ \\ 

& ($0001001$,{\color{red}$0011100$}),($0000001$,{\color{red}$0010100$}),
&$4_{2,3}$ & $k_{x}^{2}-k_{y}^{2},k_{x}k_{y}k_{z}$ &
$k_{x}\pm k_{y},k_{z}$ \\

& ($0000011$,{\color{red}$0010110$})
&&& \\
\hline

(${\color{red}+1,+1,-1,-1,-1,-1,+1}$)
& ($0100010$,{\color{red}$0111100$}),($0101010$,{\color{red}$0110100$}),
&$4_{1,2}$ &
$k_{z},k_{x}k_{y}$&$k_{x},k_{y}$ \\ 

& ($0000010$,{\color{red}$0011100$}),($0001010$,{\color{red}$0010100$}),
&$4_{3}$ & $k_{x}k_{y},k_{z}(k_{x}^{2}-k_{y}^{2})$ &$k_{x},k_{y},k_{z}$ \\

& ($0001000$,{\color{red}$0010110$})
&&& \\
\hline

(${\color{red}+1,+1,-1,+1,-1,-1,+1}$)
& ($0101010$,{\color{red}$0111100$}),($0100010$,{\color{red}$0110100$}),
&&& \\ 

& ($0001010$,{\color{red}$0011100$}),($0000010$,{\color{red}$0010100$}),
&&& \\

& ($0000000$,{\color{red}$0010110$})
&&& \\
\hline

($-1,-1,-1,-1,-1,+1,+1$)
&$\cdots$&$\cdots$&$\cdots$&$\cdots$ \\ 

($-1,-1,-1,+1,-1,+1,+1$)
&&&& \\ 

($-1,-1,-1,-1,-1,-1,-1$)
&&&& \\ 

($-1,-1,-1,+1,-1,-1,-1$)
&&&& \\ 

($-1,+1,-1,-1,-1,+1,+1$)
&&&& \\ 

($-1,+1,-1,+1,-1,+1,+1$)
&&&& \\ 

($-1,+1,-1,-1,-1,-1,-1$)
&&&& \\ 

($-1,+1,-1,+1,-1,-1,-1$)
&&&& \\ 

($-1,-1,-1,-1,-1,+1,-1$)
&&&& \\ 

($-1,-1,-1,+1,-1,+1,-1$)
&&&& \\ 

($-1,-1,-1,-1,-1,-1,+1$)
&&&& \\ 

($-1,-1,-1,+1,-1,-1,+1$)
&&&& \\ 
\hline
\end{tabular}
\end{table*}

\begin{table*}
(Extension of Supplementary Table \ref{D4hdrztinv},${\rm Invariants}\equiv (\eta_{C_{2x},C_{2y}}, \eta_T,\eta_{IT},
\eta_{TC_{2x}},\eta_{TC_{2a}},\eta_{I,C_{2x}},\eta_{I,C_{2a}})$\\
\centering
\begin{tabular}{ |c|c|c|c|c|}
\hline
Invariants & ($\vec{\eta}^{L}$,{\R $\vec{\eta}^{S}$})&dim &disp.&n.l.\\
\hline

($-1,+1,-1,-1,-1,+1,-1$)
&$\cdots$&$\cdots$&$\cdots$&$\cdots$ \\ 

($-1,+1,-1,+1,-1,+1,-1$)
&&&& \\ 

($-1,+1,-1,-1,-1,-1,+1$)
&&&& \\ 

($-1,+1,-1,+1,-1,-1,+1$)
&&&& \\ 

($-1,-1,+1,-1,-1,+1,+1$)
&&&& \\ 

($-1,-1,+1,+1,-1,+1,+1$)
&&&& \\ 

($-1,-1,+1,-1,-1,-1,-1$)
&&&& \\ 

($-1,-1,+1,+1,-1,-1,-1$)
&&&& \\ 

($-1,+1,+1,-1,-1,+1,+1$)
&&&& \\ 

($-1,+1,+1,+1,-1,+1,+1$)
&&&& \\ 

($-1,+1,+1,-1,-1,-1,-1$)
&&&& \\ 

($-1,+1,+1,+1,-1,-1,-1$)
&&&& \\ 

($-1,-1,+1,-1,-1,+1,-1$)
&&&& \\ 

($-1,-1,+1,+1,-1,+1,-1$)
&&&& \\ 

($-1,-1,+1,-1,-1,-1,+1$)
&&&& \\ 

($-1,-1,+1,+1,-1,-1,+1$)
&&&& \\ 

($-1,+1,+1,-1,-1,+1,-1$)
&&&& \\ 

($-1,+1,+1,+1,-1,+1,-1$)
&&&& \\ 

($-1,+1,+1,-1,-1,-1,+1$)
&&&& \\ 

($-1,+1,+1,+1,-1,-1,+1$)
&&&& \\ 
\hline
\end{tabular}
\end{table*}

\begin{table*}[htbp]
\caption{$G_{\mathbf K}\cong M_{\mathbf K}=D_{3}\times Z_2^T$,
${\rm Invariants}\equiv (\eta_{T},\eta_{TC'_{21}})$ for $D_{3}^{1}\times Z_2^T$,
${\rm Invariants}\equiv (\eta_{T},\eta_{TC''_{21}})$ for $D_{3}^{2}\times Z_2^T$,
we list the values of all symmetry invariants,
where type II MSG-realizable are colored in {\color{green}GREEN}, type IV MSG-realizable in {\color{blue}BLUE},
and only SSG-realizable in {\color{red}RED}.}
\label{D3drztinv}
\centering
\begin{tabular}{ |c|c|}
\hline
Invariants & ($\vec{\eta}^{L}$,{\R $\vec{\eta}^{S}$})\\
\hline
({\color{green}$+1,+1$}) &\\ 

({\color{green}$-1,+1$})  &\\ 
\hline

({\color{red}$+1,-1$})  & ($00$,{\color{red}$01$}),($10$,{\color{red}$11$})  \\ 
\hline

({\color{red}$-1,-1$})  & ($00$,{\color{red}$11$}),($10$,{\color{red}$01$}) \\ 
\hline

\end{tabular}
\end{table*}

\begin{table*}[htbp]
\caption{$G_{\mathbf K}\cong M_{\mathbf K}=D_{3d}\times Z_2^T$,
${\rm Invariants}\equiv (\eta_{I,C'_{21}},
\eta_{T},\eta_{IT},\eta_{TC'_{21}})$ for $D_{3d}^{1}\times Z_2^T$,
${\rm Invariants}\equiv (\eta_{I,C''_{21}},
\eta_{T},\eta_{IT},\eta_{TC''_{21}})$ for $D_{3d}^{2}\times Z_2^T$,
we list the values of all symmetry invariants,
where type II MSG-realizable are colored in {\color{green}GREEN}, type IV MSG-realizable in {\color{blue}BLUE}, and only SSG-realizable in {\color{red}RED}.
For only SSG-realizable ($\vec{\eta}^{L}$,{\R $\vec{\eta}^{S}$}), dim denotes the lowest dimension of irRep,
disp. are the $k$-terms which can split band degeneracy, n.l. is direction of nodal line in BZ.}
\label{D3ddrztinv}
\centering
\begin{tabular}{ |c|c|c|c|c|}
\hline
Invariants & ($\vec{\eta}^{L}$,{\R $\vec{\eta}^{S}$})&dim &disp.&n.l.\\
\hline
({\color{green}$+1,+1,+1,+1$})
&&&& \\

({\color{green}$-1,+1,+1,+1$})
&&&& \\

({\color{green}$+1,-1,-1,+1$})
&&&& \\

({\color{green}$-1,-1,-1,+1$})
&&&& \\   

({\color{blue}$+1,-1,+1,+1$})
&&&& \\

({\color{blue}$-1,-1,+1,+1$})
&&&& \\

({\color{blue}$+1,+1,-1,+1$})
&&&& \\

({\color{blue}$-1,+1,-1,+1$})
&&&& \\  
\hline

({\color{red}$+1,+1,-1,-1$})
&($0000$,{\color{red}$0011$}),($0100$,{\color{red}$0111$})

&&& \\ 
\hline

({\color{red}$+1,-1,-1,-1$})
&($0100$,{\color{red}$0011$}),($0000$,{\color{red}$0111$})

&&& \\ 
\hline

({\color{red}$-1,-1,-1,-1$})
&($1100$,{\color{red}$0011$})

&$4_{1}$
&$k_{z}$,
$k_x^3-3k_xk_y^2$,
$k_y^3-3k_yk_x^2$
&$-$ \\

&&$4_{2}$&$[k_{x},k_{y}]$&$k_{z}$  \\
\hline

({\color{red}$-1,+1,-1,-1$})
&($1100$,{\color{red}$0111$})

&$4_{1}$&$k_{z}$,
$k_x^3-3k_xk_y^2$,
$k_y^3-3k_yk_x^2$
&$-$ \\

&&$4_{2,3}$&$[k_{x},k_{y}]$&$k_{z}$  \\
\hline

({\color{red}$+1,-1,+1,-1$})
&($1100$,{\color{red}$1001$}),($0100$,{\color{red}$0001$}),
($0000$,{\color{red}$0101$})
&&& \\ 
\hline

({\color{red}$+1,+1,+1,-1$})
&($0000$,{\color{red}$0001$}),($0100$,{\color{red}$0101$})

&&& \\ 
\hline

({\color{red}$-1,+1,+1,-1$})
&($0000$,{\color{red}$1001$}),($1100$,{\color{red}$0101$})

&&& \\ 
\hline

({\color{red}$-1,-1,+1,-1$})
&($0100$,{\color{red}$1001$}),($1100$,{\color{red}$0001$})

&$4_{1}$&$k_{z}$,
$k_x^3-3k_xk_y^2$,
$k_y^3-3k_yk_x^2$
&$-$ \\

&&$4_{2,3}$&$[k_{x},k_{y}]$&$k_{z}$  \\
\hline

\end{tabular}
\end{table*}

\begin{table*}[htbp]
\caption{$G_{\mathbf K}\cong M_{\mathbf K}=C_{3h}\times Z_2^T$,
${\rm Invariants}\equiv (\eta_{T},\eta_{TM_{h}})$,
we list the values of all symmetry invariants,
where type II MSG-realizable are colored in {\color{green}GREEN}, type IV MSG-realizable in {\color{blue}BLUE},
and only SSG-realizable in {\color{red}RED}.}
\label{C3hdrztinv}
\centering
\begin{tabular}{ |c|c|}
\hline
Invariants & ($\vec{\eta}^{L}$,{\R $\vec{\eta}^{S}$})\\
\hline
({\color{green}$+1,+1$}) &\\  

({\color{green}$-1,+1$}) &\\  
\hline

({\color{red}$+1,-1$})  & ($00$,{\color{red}$01$}),($10$,{\color{red}$11$})  \\ 
\hline

({\color{red}$-1,-1$})  & ($00$,{\color{red}$11$}),($10$,{\color{red}$01$}) \\ 
\hline

\end{tabular}
\end{table*}

\begin{table*}[htbp]
\caption{$G_{\mathbf K}\cong M_{\mathbf K}=C_{6h}\times Z_2^T$,
${\rm Invariants}\equiv (\eta_{I,C_{2}},
\eta_{T},\eta_{IT},\eta_{TM_{h}})$,
we list the values of all symmetry invariants,
where type II MSG-realizable are colored in {\color{green}GREEN}, type IV MSG-realizable in {\color{blue}BLUE}, and only SSG-realizable in {\color{red}RED}.
For only SSG-realizable ($\vec{\eta}^{L}$,{\R $\vec{\eta}^{S}$}), dim denotes the lowest dimension of irRep,
disp. are the $k$-terms which can split band degeneracy, n.l. is direction of nodal line in BZ.}
\label{C6hdrztinv}
\centering
\begin{tabular}{ |c|c|c|c|c|}
\hline
Invariants & ($\vec{\eta}^{L}$,{\R $\vec{\eta}^{S}$})&dim &disp.&n.l.\\
\hline
({\color{green}$+1,+1,+1,+1$})
&$\cdots$
&$\cdots$&$\cdots$&$\cdots$\\

({\color{green}$-1,+1,+1,+1$})
&
&&&\\

({\color{green}$+1,-1,-1,+1$})
&
&&&\\

({\color{green}$-1,-1,-1,+1$})
&
&&&\\  

({\color{blue}$+1,-1,+1,+1$})
&
&&&\\

({\color{blue}$-1,-1,+1,+1$})
&
&&&\\

({\color{blue}$+1,+1,-1,+1$})
&
&&&\\

({\color{blue}$-1,+1,-1,+1$})
&
&&&\\  
\hline

({\color{red}$+1,-1,-1,-1$})
& ($0000$,{\color{red}$0111$}),($0100$,{\color{red}$0011$})
&&&\\  
\hline

({\color{red}$+1,+1,+1,-1$})
& ($0000$,{\color{red}$0001$}),($0100$,{\color{red}$0101$})
&&&\\  
\hline

({\color{red}$-1,+1,-1,-1$})
& ($1100$,{\color{red}$0111$})
&$4_{1}$&$k_{z},[k_{z}k_{x},k_{z}k_{y}],k_{x}^{3},k_{y}^{3}$&$-$\\  

&&$4_{2}$&$k_{z},k_{x}^{3},k_{y}^{3}$&$-$\\
\hline

({\color{red}$-1,-1,+1,-1$})
& ($1100$,{\color{red}$0001$}),($0100$,{\color{red}$1001$})
&$4_{1}$&$k_{z},[k_{x},k_{y}]$&$-$\\  

&&$4_{2}$&$k_{z},k_{x}^{3},k_{y}^{3}$&$-$\\
\hline

({\color{red}$+1,+1,-1,-1$})
& ($0100$,{\color{red}$0111$}),($0000$,{\color{red}$0011$})
&&&\\  
\hline

({\color{red}$+1,-1,+1,-1$})
& ($0100$,{\color{red}$0001$}),($0000$,{\color{red}$0101$}),($1100$,{\color{red}$1001$})
&&&\\  
\hline

({\color{red}$-1,-1,-1,-1$})
& ($1100$,{\color{red}$0011$})
&$4_{1}$&$k_{z},[k_{x},k_{y}]$&$-$\\  

&&$4_{2}$&$k_{z},k_{x}^{3},k_{y}^{3}$&$-$\\
\hline

({\color{red}$-1,+1,+1,-1$})
& ($1100$,{\color{red}$0101$}),($0000$,{\color{red}$1001$})
&&&\\  
\hline

\end{tabular}
\end{table*}

\begin{table*}[htbp]
\caption{$G_{\mathbf K}\cong M_{\mathbf K}=D_{6}\times Z_2^T$,
${\rm Invariants}\equiv (\eta_{C_{2},C'_{21}},
\eta_{T},\eta_{TC_{2}},\eta_{TC'_{21}})$,
we list the values of all symmetry invariants,
where type II MSG-realizable are colored in {\color{green}GREEN}, type IV MSG-realizable in {\color{blue}BLUE}, and only SSG-realizable in {\color{red}RED}.
For only SSG-realizable ($\vec{\eta}^{L}$,{\R $\vec{\eta}^{S}$}), dim denotes the lowest dimension of irRep,
disp. are the $k$-terms which can split band degeneracy.}
\label{D6drztinv}
\centering
\begin{tabular}{ |c|c|c|c|}
\hline
Invariants & ($\vec{\eta}^{L}$,{\R $\vec{\eta}^{S}$})&dim &disp.\\
\hline
({\color{green}$+1,+1,+1,+1$})
&&& \\

({\color{green}$-1,+1,-1,+1$})
&&& \\

({\color{green}$-1,-1,+1,+1$})
&&& \\

({\color{green}$+1,-1,-1,+1$})
&&& \\  
\hline

({\color{red}$+1,-1,+1,+1$})
&($1100$,{\color{red}$1000$}),($0110$,{\color{red}$0010$}),($0000$,{\color{red}$0100$})

&& \\  
\hline

({\color{red}$+1,+1,-1,-1$})
&($0110$,{\color{red}$0101$}),($0000$,{\color{red}$0011$})

&& \\  
\hline

({\color{red}$+1,-1,-1,-1$})
&($0110$,{\color{red}$0001$}),($0000$,{\color{red}$0111$})

&& \\  
\hline

({\color{red}$-1,+1,+1,+1$})
&($0000$,{\color{red}$1000$}),($1100$,{\color{red}$0100$})

&& \\  
\hline

({\color{red}$-1,-1,-1,-1$})
&($0110$,{\color{red}$1001$}),($1100$,{\color{red}$0011$})

&$4_{1}$&$[k_{x},k_{y}],k_{z}$ \\  

&& $4_{2}$ &$k_{z},k_{x}^{3}-3k_{x}k_{y}^{2},k_{y}^{3}-3k_{y}k_{x}^{2}$ \\
\hline

({\color{red}$-1,+1,-1,-1$})
& ($1100$,{\color{red}$0111$})

&$4_{1,2}$&$[k_{x},k_{y}],k_{z}$ \\  

&& $4_{3}$ &$k_{z},k_{x}^{3}-3k_{x}k_{y}^{2},k_{y}^{3}-3k_{y}k_{x}^{2}$ \\
\hline

({\color{red}$+1,+1,-1,+1$})
&($0000$,{\color{red}$0010$}),($0110$,{\color{red}$0100$})

&& \\  
\hline

({\color{red}$+1,-1,+1,-1$})
&($1100$,{\color{red}$1001$}),($0000$,{\color{red}$0101$}),($0110$,{\color{red}$0011$})

&& \\  
\hline

({\color{red}$+1,+1,+1,-1$})
&($0000$,{\color{red}$0001$}),($0110$,{\color{red}$0111$})

&& \\  
\hline

({\color{red}$-1,-1,-1,+1$})
&($0110$,{\color{red}$1000$}),($1100$,{\color{red}$0010$})

&$4_{1}$&$[k_{x},k_{y}],k_{z}$ \\  

&& $4_{2}$ &$k_{z},k_{x}^{3}-3k_{x}k_{y}^{2},k_{y}^{3}-3k_{y}k_{x}^{2}$ \\
\hline

({\color{red}$-1,+1,+1,-1$})
& ($0000$,{\color{red}$1001$}),($1100$,{\color{red}$0101$})

&& \\  
\hline

({\color{red}$-1,-1,+1,-1$})
&($1100$,{\color{red}$0001$})

&$4_{1,2}$&$[k_{x},k_{y}],k_{z}$ \\  

&& $4_{3}$ &$k_{z},k_{x}^{3}-3k_{x}k_{y}^{2},k_{y}^{3}-3k_{y}k_{x}^{2}$ \\
\hline

\end{tabular}
\end{table*}

\begin{table*}[htbp]
\caption{$G_{\mathbf K}\cong M_{\mathbf K}=D_{3h}\times Z_2^T$,
${\rm Invariants}\equiv (\eta_{M_{h},C'_{21}},
\eta_{T},\eta_{TM_{h}},\eta_{TC'_{21}})$ for $D_{3h}^{1}\times Z_2^T$,
${\rm Invariants}\equiv (\eta_{M_{h},M_{d1}},
\eta_{T},\eta_{TM_{h}},\eta_{TM_{d1}})$ for $D_{3h}^{2}\times Z_2^T$,
we list the values of all symmetry invariants,
where type II MSG-realizable are colored in {\color{green}GREEN}, type IV MSG-realizable in {\color{blue}BLUE}, and only SSG-realizable in {\color{red}RED}.
For only SSG-realizable ($\vec{\eta}^{L}$,{\R $\vec{\eta}^{S}$}), dim denotes the lowest dimension of irRep,
disp. are the $k$-terms which can split band degeneracy, n.l. is direction of nodal line in BZ.}
\label{D3hdrztinv}
\centering
\begin{tabular}{ |c|c|c|c|c|c|}
\hline
Invariants & ($\vec{\eta}^{L}$,{\R $\vec{\eta}^{S}$})&dim &disp.
& n.l.
\\
\hline
({\color{green}$+1,+1,+1,+1$})
&&&& \\

({\color{green}$-1,+1,+1,+1$})
&&&& \\

({\color{green}$-1,+1,+1,-1$})
&&&& \\

({\color{green}$-1,-1,+1,+1$})
&&&& \\

({\color{green}$+1,-1,+1,+1$})
&&&& \\

({\color{green}$+1,-1,+1,-1$})
&&&& \\   
\hline

({\color{red}$+1,+1,-1,-1$})
&($0000$,{\color{red}$0011$}),($0100$,{\color{red}$0111$}),
($0101$,{\color{red}$0110$})
&&& \\   
\hline

({\color{red}$+1,-1,-1,-1$})
&($0100$,{\color{red}$0011$}),($0000$,{\color{red}$0111$}),
($0101$,{\color{red}$0010$})
&&& \\   
\hline

({\color{red}$-1,-1,-1,-1$})
&($1100$,{\color{red}$0011$}) for $D_{3h}^{1}\times Z_2^T$

&$4_{1}$&$[k_{z}k_{x},k_{z}k_{y}],[k_{x},k_{y}]$,
&$k_z$
\\

&&&
$k_x^3-3k_xk_y^2$,
$k_y^3-3k_yk_x^2$
&
\\

&&$4_{2}$&$k_{z}$,
$k_x^3-3k_xk_y^2$,
$k_y^3-3k_yk_x^2$
&$-$\\

&($1100$,{\color{red}$0011$}),($0101$,{\color{red}$1010$}) for
$D_{3h}^{2}\times Z_2^T$
&$4_{1}$&$[k_{z}k_{x},k_{z}k_{y}],[k_{x},k_{y}],k_{z}$,
&$-$
\\

&&&
$k_x^3-3k_xk_y^2$,
$k_y^3-3k_yk_x^2$
&
\\

&&$4_{2}$&$k_{z}$,
$k_x^3-3k_xk_y^2$,
$k_y^3-3k_yk_x^2$
&$-$
\\
\hline

({\color{red}$-1,+1,-1,-1$})
&($1100$,{\color{red}$0111$}) for $D_{3h}^{1}\times Z_2^T$ and $D_{3h}^{2}\times Z_2^T$

&$4_{1}$&$k_{z}$,
$k_x^3-3k_xk_y^2$,
$k_y^3-3k_yk_x^2$
&$-$
\\   
&& $4_{2,3}$&$[k_{z}k_{x},k_{z}k_{y}]$,
$k_x^3-3k_xk_y^2$,
$k_y^3-3k_yk_x^2$
&$k_z$
\\
\hline

({\color{red}$+1,-1,-1,+1$})
&($1100$,{\color{red}$1010$}),($0000$,{\color{red}$0110$}),
($0100$,{\color{red}$0010$}),($0101$,{\color{red}$0011$})

&&& \\   
\hline

({\color{red}$+1,+1,-1,+1$})
&($0100$,{\color{red}$0110$}),($0000$,{\color{red}$0010$}),
($0101$,{\color{red}$0111$})
&&& \\   
\hline

({\color{red}$+1,+1,+1,-1$})
&($0000$,{\color{red}$0001$}),($0100$,{\color{red}$0101$}),
($0101$,{\color{red}$0100$})
&&& \\   
\hline

({\color{red}$-1,+1,-1,+1$})
& ($0000$,{\color{red}$1010$}),($1100$,{\color{red}$0110$})

&&& \\   
\hline

({\color{red}$-1,-1,-1,+1$})
&($0100$,{\color{red}$1010$}),($1100$,{\color{red}$0010$}) for $D_{3h}^{1}\times Z_2^T$
&$4_{1}$&$[k_{z}k_{x},k_{z}k_{y}],[k_{x},k_{y}],k_{z}$,
&$-$\\

&&&
$k_x^3-3k_xk_y^2$,
$k_y^3-3k_yk_x^2$
&
\\

&&$4_{2}$&$k_{z}$,
$k_x^3-3k_xk_y^2$,
$k_y^3-3k_yk_x^2$
&$-$
\\

& ($1100$,{\color{red}$0010$}) for $D_{3h}^{2}\times Z_2^T$
&$4_{1}$&$[k_{z}k_{x},k_{z}k_{y}],[k_{x},k_{y}]$,
&$k_z$\\

&&&
$k_x^3-3k_xk_y^2$,
$k_y^3-3k_yk_x^2$
&
\\

&&$4_{2}$&$k_{z}$,
$k_x^3-3k_xk_y^2$,
$k_y^3-3k_yk_x^2$
&$-$
\\
\hline

({\color{red}$-1,-1,+1,-1$})
&($1100$,{\color{red}$0001$}),($0100$,{\color{red}$1001$}) for $D_{3h}^{1}\times Z_2^T$

&$4_{1}$&$k_{z}$,
$k_x^3-3k_xk_y^2$,
$k_y^3-3k_yk_x^2$
&$-$
\\   

&($1100$,{\color{red}$0001$}),($0101$,{\color{red}$1000$}) for $D_{3h}^{2}\times Z_2^T$
& $4_{2,3}$&$[k_{z}k_{x},k_{z}k_{y}]$,
$k_x^3-3k_xk_y^2$,
$k_y^3-3k_yk_x^2$
&$k_z$
\\
\hline

\end{tabular}
\end{table*}

\begin{table*}[htbp]
\caption{$G_{\mathbf K}\cong M_{\mathbf K}=D_{6h}\times Z_2^T$,
${\rm Invariants}\equiv (\eta_{C'_{21},C''_{21}},\eta_T,\eta_{IT},
\eta_{TC'_{21}},\eta_{TC''_{21}},\eta_{I,C'_{21}},\eta_{I,C''_{21}})$,
we list the values of all symmetry invariants,
where type II MSG-realizable are colored in {\color{green}GREEN}, type IV MSG-realizable in {\color{blue}BLUE}, and only SSG-realizable in {\color{red}RED}.
For only SSG-realizable ($\vec{\eta}^{L}$,{\R $\vec{\eta}^{S}$}), dim denotes the lowest dimension of irRep,
disp. are the $k$-terms which can split band degeneracy, n.l. is direction of nodal line in BZ.}
\label{D6hdrztinv}
\centering
\begin{tabular}{ |c|c|c|c|c|}
\hline
Invariants & ($\vec{\eta}^{L}$,{\R $\vec{\eta}^{S}$})&dim &disp.&n.l.\\
\hline

({\color{green}$+1,+1,+1,+1,+1,+1,+1$})
&$\cdots$ &$\cdots$ &$\cdots$ &$\cdots$ \\

({\color{green}$+1,+1,+1,+1,+1,-1,-1$})
&&&& \\

({\color{green}$-1,+1,+1,+1,+1,+1,-1$})
&&&& \\

({\color{green}$-1,+1,+1,+1,+1,-1,+1$})
&&&& \\  

({\color{green}$-1,-1,-1,+1,+1,+1,+1$})
&&&& \\

({\color{green}$-1,-1,-1,+1,+1,-1,-1$})
&&&& \\

({\color{green}$+1,-1,-1,+1,+1,+1,-1$})
&&&& \\

({\color{green}$+1,-1,-1,+1,+1,-1,+1$})
&&&& \\  

({\color{blue}$+1,-1,+1,+1,+1,+1,+1$})
&$\cdots$ &$\cdots$ &$\cdots$ &$\cdots$ \\

({\color{blue}$+1,-1,+1,+1,+1,-1,-1$})
&&&& \\

({\color{blue}$-1,-1,+1,+1,+1,+1,-1$})
&&&& \\

({\color{blue}$-1,-1,+1,+1,+1,-1,+1$})
&&&& \\  

({\color{blue}$-1,+1,-1,+1,+1,+1,+1$})
&&&& \\

({\color{blue}$-1,+1,-1,+1,+1,-1,-1$})
&&&& \\

({\color{blue}$+1,+1,-1,+1,+1,+1,-1$})
&&&& \\

({\color{blue}$+1,+1,-1,+1,+1,-1,+1$})
&&&& \\  
\hline

({\color{red}$+1,-1,-1,+1,-1,+1,+1$})
&($0000000$,{\color{red}$0110100$}),($1100010$,{\color{red}$1010110$}),
&&& \\ 

& ($0100000$,{\color{red}$0010100$})
&&& \\
\hline

({\color{red}$+1,+1,-1,+1,-1,+1,+1$})
&($0100000$,{\color{red}$0110100$}),($0000000$,{\color{red}$0010100$})
&&& \\ 

\hline

({\color{red}$+1,+1,-1,+1,-1,-1,-1$})
&($0100011$,{\color{red}$0110100$})
&$4_{1}$&$k_{z}$& $k_{x},k_{x}\pm\sqrt{3}k_{y},$\\ 

&&&& $k_{y},\sqrt{3}k_{x}\pm k_{y}$ \\

&&$4_{2,3}$&$[k_{z}(k_{x}^{2}-k_{y}^{2}),2k_{x}k_{y}k_{z}]$ &
$k_{x},k_{x}\pm\sqrt{3}k_{y},$\\

&&&& $k_{y},\sqrt{3}k_{x}\pm k_{y},k_{z}$ \\

\hline

({\color{red}$-1,+1,-1,+1,-1,+1,-1$})
&($1100001$,{\color{red}$0110100$})
&$4|8$&$[k_{x},k_{y}],k_{z}$&$-$ \\ 

&&$4|4$& $k_{z},k_{y}^{3}-3k_{y}k_{x}^{2}$ & $k_{x},k_{x}\pm \sqrt{3}k_{y}$ \\

\hline

({\color{red}$-1,+1,-1,+1,-1,-1,+1$})
&($1100010$,{\color{red}$0110100$}),($0000000$,{\color{red}$1010110$})
&&& \\ 

\hline

({\color{red}$+1,-1,-1,+1,-1,-1,-1$})
&($1100001$,{\color{red}$1010110$}),($0100011$,{\color{red}$0010100$})
&$4_{1}$&$k_{z},k_{y}^{3}-3k_{y}k_{x}^{2}$& $k_{x},k_{x}\pm\sqrt{3}k_{y}$ \\ 

&&$4_{2,3}$ &$[k_{z}(k_{x}^{2}-k_{y}^{2}),2k_{x}k_{y}k_{z}]$, &$k_{x},k_{x}\pm\sqrt{3}k_{y},k_{z}$ \\

&&& $k_{y}^{3}-3k_{y}k_{x}^{2}$ & \\
\hline

({\color{red}$-1,-1,-1,+1,-1,+1,-1$})
&($0100011$,{\color{red}$1010110$}),($1100001$,{\color{red}$0010100$})
&$4_{1}$&$k_{z},k_{y}^{3}-3k_{y}k_{x}^{2}$& $k_{x},k_{x}\pm\sqrt{3}k_{y}$ \\ 

&&$4_{2}$ &$[k_{z}(k_{x}^{2}-k_{y}^{2}),2k_{x}k_{y}k_{z}]$, &$k_{x},k_{x}\pm\sqrt{3}k_{y},k_{z}$ \\

&&& $k_{y}^{3}-3k_{y}k_{x}^{2}$ & \\

\hline

({\color{red}$-1,-1,-1,+1,-1,-1,+1$})
&($0100000$,{\color{red}$1010110$}),($1100010$,{\color{red}$0010100$})
&$4|8$&$[k_{x},k_{y}],k_{z}$&$-$ \\ 

&&$4|4_{1,2}$& $k_{z},k_{x}^{3}-3k_{x}k_{y}^{2}$ & $k_{y},\sqrt{3}k_{x}\pm k_{y}$ \\

\hline

({\color{red}$+1,-1,-1,-1,+1,+1,+1$})
&($0000000$,{\color{red}$0111000$}),($1100001$,{\color{red}$1011001$}),
&&& \\ 

& ($0100000$,{\color{red}$0011000$})
&&& \\
\hline

({\color{red}$+1,+1,-1,-1,+1,+1,+1$})
&($0100000$,{\color{red}$0111000$}),($0000000$,{\color{red}$0011000$})
&&& \\ 

\hline

({\color{red}$+1,+1,-1,-1,+1,-1,-1$})
&($0100011$,{\color{red}$0111000$})
&$4_{1}$&$k_{z}$& $k_{x},k_{x}\pm\sqrt{3}k_{y},$\\ 

&&&& $k_{y},\sqrt{3}k_{x}\pm k_{y}$ \\

&&$4_{2,3}$&$[k_{z}(k_{x}^{2}-k_{y}^{2}),2k_{x}k_{y}k_{z}]$ &
$k_{x},k_{x}\pm\sqrt{3}k_{y},$\\

&&&& $k_{y},\sqrt{3}k_{x}\pm k_{y},k_{z}$ \\

\hline

({\color{red}$-1,+1,-1,-1,+1,+1,-1$})
&($1100001$,{\color{red}$0111000$}),($0000000$,{\color{red}$1011001$})
&&& \\ 

\hline

({\color{red}$-1,+1,-1,-1,+1,-1,+1$})
&($1100010$,{\color{red}$0111000$})
&$4|8$&$[k_{x},k_{y}],k_{z}$&$-$ \\ 

&&$4|4$& $k_{z},k_{x}^{3}-3k_{x}k_{y}^{2}$ & $k_{y},\sqrt{3}k_{x}\pm k_{y}$ \\

\hline

({\color{red}$+1,-1,-1,-1,+1,-1,-1$})
&($1100010$,{\color{red}$1011001$}),($0100011$,{\color{red}$0011000$})
&$4_{1}$&$k_{z},k_{x}^{3}-3k_{x}k_{y}^{2}$& $k_{y},\sqrt{3}k_{x}\pm k_{y}$ \\ 

&&$4_{2,3}$ &$[k_{z}(k_{x}^{2}-k_{y}^{2}),2k_{x}k_{y}k_{z}]$, &$k_{y},\sqrt{3}k_{x}\pm k_{y},k_{z}$ \\

&&& $k_{x}^{3}-3k_{x}k_{y}^{2}$ & \\

\hline

({\color{red}$-1,-1,-1,-1,+1,+1,-1$})
&($0100000$,{\color{red}$1011001$}),($1100001$,{\color{red}$0011000$})
&$4|8$&$[k_{x},k_{y}],k_{z}$&$-$ \\ 

&&$4|4_{1,2}$& $k_{z},k_{y}^{3}-3k_{y}k_{x}^{2}$ & $k_{x},k_{x}\pm\sqrt{3}k_{y}$ \\

\hline

({\color{red}$-1,-1,-1,-1,+1,-1,+1$})
&($0100011$,{\color{red}$1011001$}),($1100010$,{\color{red}$0011000$})
&$4_{1}$&$k_{z},k_{x}^{3}-3k_{x}k_{y}^{2}$& $k_{y},\sqrt{3}k_{x}\pm k_{y}$ \\ 

&&$4_{2}$ &$[k_{z}(k_{x}^{2}-k_{y}^{2}),2k_{x}k_{y}k_{z}]$, &$k_{y},\sqrt{3}k_{x}\pm k_{y},k_{z}$ \\

&&& $k_{x}^{3}-3k_{x}k_{y}^{2}$ & \\
\hline

({\color{red}$+1,-1,+1,-1,-1,+1,+1$})
&($0000000$,{\color{red}$0101100$}),($0100011$,{\color{red}$0001111$}),
&&& \\ 

& ($0100000$,{\color{red}$0001100$})
&&& \\
\hline

({\color{red}$+1,+1,+1,-1,-1,+1,+1$})
&($0100000$,{\color{red}$0101100$}),($0000000$,{\color{red}$0001100$})
&&& \\ 

\hline
\end{tabular}
\end{table*}

\clearpage

\begin{table*}
(Extension of Supplementary Table \ref{D6hdrztinv},${\rm Invariants}\equiv (\eta_{C'_{21},C''_{21}},\eta_T,\eta_{IT},
\eta_{TC'_{21}},\eta_{TC''_{21}},\eta_{I,C'_{21}},\eta_{I,C''_{21}})$ )\\
\centering
\begin{tabular}{ |c|c|c|c|c|}
\hline
Invariants & ($\vec{\eta}^{L}$,{\R $\vec{\eta}^{S}$})&dim &disp.&n.l.\\
\hline

({\color{red}$+1,+1,+1,-1,-1,-1,-1$})
&($0100011$,{\color{red}$0101100$}),($0000000$,{\color{red}$0001111$})
&&& \\ 

\hline

({\color{red}$-1,+1,+1,-1,-1,+1,-1$})
&($1100001$,{\color{red}$0101100$})
&$4|8$&$[k_{x},k_{y}],k_{z}$&$-$ \\ 

&&$4|4$& $k_{z},k_{y}^{3}-3k_{y}k_{x}^{2}$ & $k_{x},k_{x}\pm\sqrt{3}k_{y}$ \\

\hline

({\color{red}$-1,+1,+1,-1,-1,-1,+1$})
&($1100010$,{\color{red}$0101100$})
&$4|8$&$[k_{x},k_{y}],k_{z}$&$-$ \\ 

&&$4|4$& $k_{z},k_{x}^{3}-3k_{x}k_{y}^{2}$ & $k_{y},\sqrt{3}k_{x}\pm k_{y}$ \\

\hline

({\color{red}$+1,-1,+1,-1,-1,-1,-1$})
&($0100000$,{\color{red}$0001111$}),($0100011$,{\color{red}$0001100$})
&$4_{1,2}$&$k_{z}$& $k_{x},k_{x}\pm\sqrt{3}k_{y},$\\ 

&&&& $k_{y},\sqrt{3}k_{x}\pm k_{y}$ \\

&&$4_{3,4,5,6}$&$[k_{z}(k_{x}^{2}-k_{y}^{2}),2k_{x}k_{y}k_{z}]$ &
$k_{x},k_{x}\pm\sqrt{3}k_{y},$\\

&&&& $k_{y},\sqrt{3}k_{x}\pm k_{y},k_{z}$ \\

\hline

({\color{red}$-1,-1,+1,-1,-1,+1,-1$})
&($1100010$,{\color{red}$0001111$}),($1100001$,{\color{red}$0001100$})
&$4|8$&$[k_{x},k_{y}],k_{z}$&$-$ \\ 

&&$4|4$& $k_{z},k_{x}^{3}-3k_{x}k_{y}^{2}$, & $-$ \\

&&& $k_{y}^{3}-3k_{y}k_{x}^{2}$ & \\

\hline

({\color{red}$-1,-1,+1,-1,-1,-1,+1$})
&($1100001$,{\color{red}$0001111$}),($1100010$,{\color{red}$0001100$})
&$4|8$&$[k_{x},k_{y}],k_{z}$&$-$ \\ 

&&$4|4$& $k_{z},k_{x}^{3}-3k_{x}k_{y}^{2}$, & $-$ \\

&&& $k_{y}^{3}-3k_{y}k_{x}^{2}$ & \\

\hline

({\color{red}$+1,-1,-1,+1,+1,+1,+1$})
&($0000000$,{\color{red}$0110000$}),($0100011$,{\color{red}$0010011$}),
&&& \\ 

& ($0100000$,{\color{red}$0010000$})
&&& \\
\hline

({\color{red}$+1,+1,-1,+1,+1,+1,+1$})
&($0100000$,{\color{red}$0110000$}),($0000000$,{\color{red}$0010000$})
&&& \\ 

\hline

({\color{red}$+1,+1,-1,+1,+1,-1,-1$})
&($0100011$,{\color{red}$0110000$}),($0000000$,{\color{red}$0010011$})
&&& \\ 

\hline

({\color{red}$-1,+1,-1,+1,+1,+1,-1$})
&($1100001$,{\color{red}$0110000$})
&$4_{1}$&$k_{z},k_{y}^{3}-3k_{y}k_{x}^{2}$& $k_{x},k_{x}\pm\sqrt{3}k_{y}$ \\ 

&&$4_{2}$ &$k_{z},[k_{z}k_{x},k_{z}k_{y}]$, &$k_{x},k_{x}\pm\sqrt{3}k_{y}$ \\

&&& $k_{y}^{3}-3k_{y}k_{x}^{2}$ & \\

\hline

({\color{red}$-1,+1,-1,+1,+1,-1,+1$})
&($1100010$,{\color{red}$0110000$})
&$4_{1}$&$k_{z},k_{x}^{3}-3k_{x}k_{y}^{2}$& $k_{y},\sqrt{3}k_{x}\pm k_{y}$ \\ 

&&$4_{2}$ &$k_{z},[k_{z}k_{x},k_{z}k_{y}]$, &$k_{y},\sqrt{3}k_{x}\pm k_{y}$ \\

&&& $k_{x}^{3}-3k_{x}k_{y}^{2}$ & \\

\hline

({\color{red}$+1,-1,-1,+1,+1,-1,-1$})
&($0100000$,{\color{red}$0010011$}),($0100011$,{\color{red}$0010000$})
&$4_{1,2}$&$k_{z}$& $k_{x},k_{x}\pm\sqrt{3}k_{y},$  \\ 

&&&& $k_{y},\sqrt{3}k_{x}\pm k_{y}$ \\

&&$4_{3,4}$&$k_{z}$, &
$k_{x},k_{x}\pm\sqrt{3}k_{y},$\\

&&&$[k_{z}(k_{x}^{2}-k_{y}^{2}),2k_{x}k_{y}k_{z}]$& $k_{y},\sqrt{3}k_{x}\pm k_{y}$ \\

\hline

({\color{red}$-1,-1,-1,+1,+1,+1,-1$})
&($1100010$,{\color{red}$0010011$}),($1100001$,{\color{red}$0010000$})
&$4_{1}$&$[k_{x},k_{y}],k_{z}$&$-$ \\ 

&&$4_{2}$&$k_{z},k_{x}^{3}-3k_{x}k_{y}^{2}$, & $-$ \\

&&& $k_{y}^{3}-3k_{y}k_{x}^{2}$ & \\

\hline

({\color{red}$-1,-1,-1,+1,+1,-1,+1$})
&($1100001$,{\color{red}$0010011$}),($1100010$,{\color{red}$0010000$})
&$4_{1}$&$[k_{x},k_{y}],k_{z}$&$-$ \\ 

&&$4_{2}$&$k_{z},k_{x}^{3}-3k_{x}k_{y}^{2}$, & $-$ \\

&&& $k_{y}^{3}-3k_{y}k_{x}^{2}$ & \\

\hline

({\color{red}$+1,-1,+1,+1,-1,+1,+1$})
&($0000000$,{\color{red}$0100100$}),($1100001$,{\color{red}$1000101$}),
&&& \\ 

& ($0100000$,{\color{red}$0000100$})
&&& \\
\hline

({\color{red}$+1,+1,+1,+1,-1,+1,+1$})
&($0100000$,{\color{red}$0100100$}),($0000000$,{\color{red}$0000100$})
&&& \\ 

\hline

({\color{red}$+1,+1,+1,+1,-1,-1,-1$})
&($0100011$,{\color{red}$0100100$})
&$4_{1}$&$k_{z},k_{y}^{3}-3k_{y}k_{x}^{2}$& $k_{x},k_{x}\pm\sqrt{3}k_{y}$ \\ 

&&$4_{2,3}$&$[k_{z}k_{x},k_{z}k_{y}],k_{y}^{3}-3k_{y}k_{x}^{2}$, &$k_{x},k_{x}\pm\sqrt{3}k_{y}$, \\

&&& $[k_{z}(k_{x}^{2}-k_{y}^{2}),2k_{x}k_{y}k_{z}]$ & $k_{z}$ \\

\hline

({\color{red}$-1,+1,+1,+1,-1,+1,-1$})
&($1100001$,{\color{red}$0100100$}),($0000000$,{\color{red}$1000101$})
&&& \\ 

\hline

({\color{red}$-1,+1,+1,+1,-1,-1,+1$})
&($1100010$,{\color{red}$0100100$})
&$4|8$&$[k_{x},k_{y}],k_{z}$&$-$ \\ 

&&$4|4$&$k_{z},k_{x}^{3}-3k_{x}k_{y}^{2}$, & $-$ \\

&&& $k_{y}^{3}-3k_{y}k_{x}^{2}$ & \\

\hline

({\color{red}$+1,-1,+1,+1,-1,-1,-1$})
&($1100010$,{\color{red}$1000101$}),($0100011$,{\color{red}$0000100$})
&$4_{1}$& $k_{z},k_{x}^{3}-3k_{x}k_{y}^{2}$, & $-$ \\

&&& $k_{y}^{3}-3k_{y}k_{x}^{2}$ & \\ 

&&$4_{2,3}$&$[k_{x},k_{y}]$, &$k_{z}$ \\

&&& $[k_{z}(k_{x}^{2}-k_{y}^{2}),2k_{x}k_{y}k_{z}]$ & \\

\hline

({\color{red}$-1,-1,+1,+1,-1,+1,-1$})
&($0100000$,{\color{red}$1000101$}),($1100001$,{\color{red}$0000100$})
&$4|8$&$[k_{x},k_{y}],k_{z}$&$-$ \\ 

&&$4|4_{1,2}$&$k_{z},k_{y}^{3}-3k_{y}k_{x}^{2}$& $k_{x},k_{x}\pm\sqrt{3}k_{y}$ \\

\hline

({\color{red}$-1,-1,+1,+1,-1,-1,+1$})
&($0100011$,{\color{red}$1000101$}),($1100010$,{\color{red}$0000100$})
&$4_{1}$&$[k_{x},k_{y}],[k_{z}k_{x},k_{z}k_{y}]$, &$k_{z}$ \\ 

&&& $[k_{z}(k_{x}^{2}-k_{y}^{2}),2k_{x}k_{y}k_{z}]$ & \\

&&$4_{2}$&$k_{z},k_{x}^{3}-3k_{x}k_{y}^{2}$, & $-$ \\

&&& $k_{y}^{3}-3k_{y}k_{x}^{2}$ & \\

\hline

({\color{red}$+1,-1,+1,-1,+1,+1,+1$})
&($0000000$,{\color{red}$0101000$}),($1100010$,{\color{red}$1001010$}),
&&& \\ 

& ($0100000$,{\color{red}$0001000$})
&&& \\
\hline

({\color{red}$+1,+1,+1,-1,+1,+1,+1$})
&($0100000$,{\color{red}$0101000$}),($0000000$,{\color{red}$0001000$})
&&& \\ 

\hline
\end{tabular}
\end{table*}

\begin{table*}
(Extension of Supplementary Table \ref{D6hdrztinv},${\rm Invariants}\equiv (\eta_{C'_{21},C''_{21}},\eta_T,\eta_{IT},
\eta_{TC'_{21}},\eta_{TC''_{21}},\eta_{I,C'_{21}},\eta_{I,C''_{21}})$ )\\
\centering
\begin{tabular}{ |c|c|c|c|c|}
\hline
Invariants & ($\vec{\eta}^{L}$,{\R $\vec{\eta}^{S}$})&dim &disp.&n.l.\\
\hline

({\color{red}$+1,+1,+1,-1,+1,-1,-1$})
&($0100011$,{\color{red}$0101000$})
&$4_{1}$&$k_{z},k_{x}^{3}-3k_{x}k_{y}^{2}$& $k_{y},\sqrt{3}k_{x}\pm k_{y}$\\ 

&&$4_{2,3}$&$[k_{z}k_{x},k_{z}k_{y}],k_{x}^{3}-3k_{x}k_{y}^{2}$, &$k_{y},\sqrt{3}k_{x}\pm k_{y}$, \\

&&& $[k_{z}(k_{x}^{2}-k_{y}^{2}),2k_{x}k_{y}k_{z}]$ & $k_{z}$ \\

\hline

({\color{red}$-1,+1,+1,-1,+1,+1,-1$})
&($1100001$,{\color{red}$0101000$})
&$4|8$&$[k_{x},k_{y}],k_{z}$&$-$ \\ 

&&$4|4$& $k_{z},k_{x}^{3}-3k_{x}k_{y}^{2}$, & $-$ \\

&&& $k_{y}^{3}-3k_{y}k_{x}^{2}$ & \\

\hline

({\color{red}$-1,+1,+1,-1,+1,-1,+1$})
&($1100010$,{\color{red}$0101000$}),($0000000$,{\color{red}$1001010$})
&&& \\ 

\hline

({\color{red}$+1,-1,+1,-1,+1,-1,-1$})
&($1100001$,{\color{red}$1001010$}),($0100011$,{\color{red}$0001000$})
&$4_{1}$& $k_{z},k_{x}^{3}-3k_{x}k_{y}^{2}$, & $-$ \\

&&& $k_{y}^{3}-3k_{y}k_{x}^{2}$ & \\ 

&&$4_{2,3}$&$[k_{x},k_{y}]$, &$k_{z}$ \\

&&& $[k_{z}(k_{x}^{2}-k_{y}^{2}),2k_{x}k_{y}k_{z}]$ & \\

\hline

({\color{red}$-1,-1,+1,-1,+1,+1,-1$})
&($0100011$,{\color{red}$1001010$}),($1100001$,{\color{red}$0001000$})
&$4_{1}$&$[k_{x},k_{y}],[k_{z}k_{x},k_{z}k_{y}]$, &$k_{z}$ \\ 

&&& $[k_{z}(k_{x}^{2}-k_{y}^{2}),2k_{x}k_{y}k_{z}]$ & \\

&&$4_{2}$&$k_{z},k_{x}^{3}-3k_{x}k_{y}^{2}$, & $-$ \\

&&& $k_{y}^{3}-3k_{y}k_{x}^{2}$ & \\

\hline

({\color{red}$-1,-1,+1,-1,+1,-1,+1$})
&($0100000$,{\color{red}$1001010$}),($1100010$,{\color{red}$0001000$})
&$4|8$&$[k_{x},k_{y}],k_{z}$&$-$ \\ 

&&$4|4_{1,2}$ &$k_{z},k_{x}^{3}-3k_{x}k_{y}^{2}$ &$k_{y},\sqrt{3}k_{x}\pm k_{y}$ \\

\hline

({\color{red}$+1,-1,-1,-1,-1,+1,+1$})
&($0000000$,{\color{red}$0111100$}),($0100000$,{\color{red}$0011100$})
&&& \\ 

\hline

({\color{red}$+1,+1,-1,-1,-1,+1,+1$})
&($0100000$,{\color{red}$0111100$}),($0000000$,{\color{red}$0011100$})
&&& \\ 

\hline

({\color{red}$+1,+1,-1,-1,-1,-1,-1$})
&($0100011$,{\color{red}$0111100$})
&$4_{1,2}$&$k_{z}$& $k_{x},k_{x}\pm\sqrt{3}k_{y},$ \\  

&&&& $k_{y},\sqrt{3}k_{x}\pm k_{y}$ \\

&&$4_{3,4,5,6}$&$[k_{z}(k_{x}^{2}-k_{y}^{2}),2k_{x}k_{y}k_{z}]$ &
$k_{x},k_{x}\pm\sqrt{3}k_{y},$\\

&&&& $k_{y},\sqrt{3}k_{x}\pm k_{y},k_{z}$ \\

\hline

({\color{red}$-1,+1,-1,-1,-1,+1,-1$})
&($1100001$,{\color{red}$0111100$})
&$4|8$&$[k_{x},k_{y}],k_{z}$&$-$ \\ 

&&$4|4_{1,2}$&$k_{z},k_{y}^{3}-3k_{y}k_{x}^{2}$& $k_{x},k_{x}\pm\sqrt{3}k_{y}$ \\

\hline

({\color{red}$-1,+1,-1,-1,-1,-1,+1$})
&($1100010$,{\color{red}$0111100$})
&$4|8$&$[k_{x},k_{y}],k_{z}$&$-$ \\ 

&&$4|4_{1,2}$ &$k_{z},k_{x}^{3}-3k_{x}k_{y}^{2}$ &$k_{y},\sqrt{3}k_{x}\pm k_{y}$ \\

\hline

({\color{red}$+1,-1,-1,-1,-1,-1,-1$})
&($0100011$,{\color{red}$0011100$})
&$4_{1,2}$&$k_{z}$& $k_{x},k_{x}\pm\sqrt{3}k_{y},$ \\ 

&&&& $k_{y},\sqrt{3}k_{x}\pm k_{y}$ \\

&&$4_{3,4}$&$[k_{z}(k_{x}^{2}-k_{y}^{2}),2k_{x}k_{y}k_{z}]$ &
$k_{x},k_{x}\pm\sqrt{3}k_{y},$\\

&&&& $k_{y},\sqrt{3}k_{x}\pm k_{y},k_{z}$ \\

\hline

({\color{red}$-1,-1,-1,-1,-1,+1,-1$})
&($1100001$,{\color{red}$0011100$})
&$4|8$&$[k_{x},k_{y}],k_{z}$&$-$ \\ 

&&$4|4_{1,2}$&$k_{z},k_{y}^{3}-3k_{y}k_{x}^{2}$& $k_{x},k_{x}\pm\sqrt{3}k_{y}$ \\
\hline

({\color{red}$-1,-1,-1,-1,-1,-1,+1$})
&($1100010$,{\color{red}$0011100$})
&$4|8$&$[k_{x},k_{y}],k_{z}$&$-$ \\ 

&&$4|4_{1,2}$ &$k_{z},k_{x}^{3}-3k_{x}k_{y}^{2}$ &$k_{y},\sqrt{3}k_{x}\pm k_{y}$ \\
\hline

({\color{red}$-1,-1,+1,+1,+1,-1,-1$})
&($0000000$,{\color{red}$1100011$}),($0100011$,{\color{red}$1000000$}),
&&& \\ 

&($1100010$,{\color{red}$0000001$}),($1100001$,{\color{red}$0000010$})
&&& \\
\hline

({\color{red}$-1,+1,+1,+1,+1,-1,-1$})
&($0100000$,{\color{red}$1100011$})
&$4_{1,2}$&$[k_{x},k_{y}],k_{z}$&$-$ \\ 

&&$4_{3}$&$k_{z},k_{x}^{3}-3k_{x}k_{y}^{2}$, & $-$ \\

&&& $k_{y}^{3}-3k_{y}k_{x}^{2}$ & \\
\hline

({\color{red}$-1,+1,+1,+1,+1,+1,+1$})
&($0100011$,{\color{red}$1100011$}),($0000000$,{\color{red}$1000000$})
&&& \\ 

\hline

({\color{red}$+1,+1,+1,+1,+1,-1,+1$})
&($1100001$,{\color{red}$1100011$}),($0000000$,{\color{red}$0000010$})
&&& \\ 

\hline

({\color{red}$+1,+1,+1,+1,+1,+1,-1$})
&($1100010$,{\color{red}$1100011$}),($0000000$,{\color{red}$0000001$})
&&& \\ 

\hline

({\color{red}$-1,-1,+1,+1,+1,+1,+1$})
&($0100000$,{\color{red}$1000000$}),($1100001$,{\color{red}$0000001$}),
&$4_{1,2}$&$[k_{x},k_{y}],k_{z}$&$-$ \\ 

&($1100010$,{\color{red}$0000010$})
&$4_{3}$&$k_{z},k_{x}^{3}-3k_{x}k_{y}^{2}$, & $-$ \\

&&& $k_{y}^{3}-3k_{y}k_{x}^{2}$ & \\
\hline

({\color{red}$+1,-1,+1,+1,+1,-1,+1$})
&($1100010$,{\color{red}$1000000$}),($0100011$,{\color{red}$0000001$}),
&$4_{1}$&$[k_{x},k_{y}],k_{z}$&$-$ \\ 

&($0100000$,{\color{red}$0000010$})
&$4_{2}$&$k_{z},k_{x}^{3}-3k_{x}k_{y}^{2}$, & $-$ \\

&&& $k_{y}^{3}-3k_{y}k_{x}^{2}$ & \\
\hline

({\color{red}$+1,-1,+1,+1,+1,+1,-1$})
&($1100001$,{\color{red}$1000000$}),($0100000$,{\color{red}$0000001$}),
&$4_{1}$&$[k_{x},k_{y}],k_{z}$&$-$ \\ 

&($0100011$,{\color{red}$0000010$})
&$4_{2}$&$k_{z},k_{x}^{3}-3k_{x}k_{y}^{2}$, & $-$ \\

&&& $k_{y}^{3}-3k_{y}k_{x}^{2}$ & \\
\hline

({\color{red}$+1,-1,+1,-1,+1,+1,-1$})
&($0000000$,{\color{red}$0101001$}),($1100001$,{\color{red}$1001000$}),
&&& \\ 

&($1100010$,{\color{red}$1001011$}),($0100011$,{\color{red}$0001010$})
&&& \\
\hline

({\color{red}$+1,+1,+1,-1,+1,+1,-1$})
&($0100000$,{\color{red}$0101001$})
&$4|8$&$[k_{x},k_{y}],k_{z}$&$-$ \\ 

&&$4|4$ &$k_{z},k_{x}^{3}-3k_{x}k_{y}^{2}$ &$k_{y},\sqrt{3}k_{x}\pm k_{y}$ \\
\hline

({\color{red}$+1,+1,+1,-1,+1,-1,+1$})
&($0100011$,{\color{red}$0101001$}),($0000000$,{\color{red}$0001010$})
&&& \\ 

\hline

({\color{red}$-1,+1,+1,-1,+1,+1,+1$})
&($1100001$,{\color{red}$0101001$}),($0000000$,{\color{red}$1001000$})
&&& \\ 

\hline

({\color{red}$-1,+1,+1,-1,+1,-1,-1$})
&($1100010$,{\color{red}$0101001$}),($0000000$,{\color{red}$1001011$})
&&& \\ 

\hline
\end{tabular}
\end{table*}

\begin{table*}
(Extension of Supplementary Table \ref{D6hdrztinv},${\rm Invariants}\equiv (\eta_{C'_{21},C''_{21}},\eta_T,\eta_{IT},
\eta_{TC'_{21}},\eta_{TC''_{21}},\eta_{I,C'_{21}},\eta_{I,C''_{21}})$ )\\
\centering
\begin{tabular}{ |c|c|c|c|c|}
\hline
Invariants & ($\vec{\eta}^{L}$,{\R $\vec{\eta}^{S}$})&dim &disp.&n.l.\\
\hline

({\color{red}$+1,-1,+1,-1,+1,-1,+1$})
&($1100010$,{\color{red}$1001000$}),($1100001$,{\color{red}$1001011$}),
&$4|8$&$[k_{x},k_{y}],k_{z}$&$-$ \\ 

&($0100000$,{\color{red}$0001010$})
&$4|4$& $k_{z},k_{x}^{3}-3k_{x}k_{y}^{2}$, & $-$ \\

&&& $k_{y}^{3}-3k_{y}k_{x}^{2}$ & \\
\hline

({\color{red}$-1,-1,+1,-1,+1,+1,+1$})
&($0100000$,{\color{red}$1001000$}),($0100011$,{\color{red}$1001011$}),
&$4_{1}$&$k_{z},k_{x}^{3}-3k_{x}k_{y}^{2}$& $k_{y},\sqrt{3}k_{x}\pm k_{y}$ \\ 

&($1100010$,{\color{red}$0001010$})
&$4_{2,3}$&$[k_{z}k_{x},k_{z}k_{y}],k_{x}^{3}-3k_{x}k_{y}^{2}$, &$k_{y},\sqrt{3}k_{x}\pm k_{y}$, \\

&&& $[k_{z}(k_{x}^{2}-k_{y}^{2}),2k_{x}k_{y}k_{z}]$ & $k_{z}$ \\
\hline

({\color{red}$-1,-1,+1,-1,+1,-1,-1$})
&($0100011$,{\color{red}$1001000$}),($0100000$,{\color{red}$1001011$}),
&$4_{1}$& $k_{z},k_{x}^{3}-3k_{x}k_{y}^{2}$, & $-$ \\ 

&($1100001$,{\color{red}$0001010$})
&& $k_{y}^{3}-3k_{y}k_{x}^{2}$ & \\

&&$4_{2,3}$&$[k_{x},k_{y}]$, &$k_{z}$ \\

&&& $[k_{z}(k_{x}^{2}-k_{y}^{2}),2k_{x}k_{y}k_{z}]$ & \\
\hline

({\color{red}$+1,-1,+1,+1,-1,-1,+1$})
&($0000000$,{\color{red}$0100110$}),($1100010$,{\color{red}$1000100$}),
&&& \\ 

&($1100001$,{\color{red}$1000111$}),($0100011$,{\color{red}$0000101$})
&&& \\
\hline

({\color{red}$+1,+1,+1,+1,-1,-1,+1$})
&($0100000$,{\color{red}$0100110$})
&$4|8$&$[k_{x},k_{y}],k_{z}$&$-$ \\ 

&&$4|4$& $k_{z},k_{y}^{3}-3k_{y}k_{x}^{2}$ & $k_{x},k_{x}\pm\sqrt{3}k_{y}$ \\
\hline

({\color{red}$+1,+1,+1,+1,-1,+1,-1$})
&($0100011$,{\color{red}$0100110$}),($0000000$,{\color{red}$0000101$})
&&& \\ 

\hline

({\color{red}$-1,+1,+1,+1,-1,-1,-1$})
&($1100001$,{\color{red}$0100110$}),($0000000$,{\color{red}$1000111$})
&&& \\ 

\hline

({\color{red}$-1,+1,+1,+1,-1,+1,+1$})
&($1100010$,{\color{red}$0100110$}),($0000000$,{\color{red}$1000100$})
&&& \\ 

\hline

({\color{red}$+1,-1,+1,+1,-1,+1,-1$})
&($1100001$,{\color{red}$1000100$}),($1100010$,{\color{red}$1000111$}),
&$4|8$&$[k_{x},k_{y}],k_{z}$&$-$ \\ 

&($0100000$,{\color{red}$0000101$})
&$4|4$& $k_{z},k_{x}^{3}-3k_{x}k_{y}^{2}$, & $-$ \\

&&& $k_{y}^{3}-3k_{y}k_{x}^{2}$ & \\
\hline

({\color{red}$-1,-1,+1,+1,-1,-1,-1$})
&($0100011$,{\color{red}$1000100$}),($0100000$,{\color{red}$1000111$}),
&$4_{1}$& $k_{z},k_{x}^{3}-3k_{x}k_{y}^{2}$, & $-$ \\ 

&($1100010$,{\color{red}$0000101$})
&& $k_{y}^{3}-3k_{y}k_{x}^{2}$ & \\

&&$4_{2,3}$&$[k_{x},k_{y}]$, &$k_{z}$ \\

&&& $[k_{z}(k_{x}^{2}-k_{y}^{2}),2k_{x}k_{y}k_{z}]$ & \\
\hline

({\color{red}$-1,-1,+1,+1,-1,+1,+1$})
&($0100000$,{\color{red}$1000100$}),($0100011$,{\color{red}$1000111$}),
&$4_{1}$&$k_{z},k_{y}^{3}-3k_{y}k_{x}^{2}$& $k_{x},k_{x}\pm\sqrt{3}k_{y}$ \\ 

&($1100001$,{\color{red}$0000101$})
&$4_{2,3}$&$[k_{z}k_{x},k_{z}k_{y}],k_{y}^{3}-3k_{y}k_{x}^{2}$, &$k_{x},k_{x}\pm\sqrt{3}k_{y}$, \\

&&& $[k_{z}(k_{x}^{2}-k_{y}^{2}),2k_{x}k_{y}k_{z}]$ & $k_{z}$ \\
\hline

($-1,-1,-1,-1,-1,+1,+1$)
&$\cdots$ &$\cdots$ &$\cdots$ &$\cdots$ \\  

($-1,+1,-1,-1,-1,+1,+1$)
&&&& \\  

($-1,+1,-1,-1,-1,-1,-1$)
&&&& \\  

($+1,+1,-1,-1,-1,+1,-1$)
&&&& \\  

($+1,+1,-1,-1,-1,-1,+1$)
&&&& \\  

($-1,-1,-1,-1,-1,-1,-1$)
&&&& \\  

($+1,-1,-1,-1,-1,+1,-1$)
&&&& \\  

($+1,-1,-1,-1,-1,-1,+1$)
&&&& \\  

($-1,-1,-1,+1,-1,+1,+1$)
&&&& \\  

($-1,+1,-1,+1,-1,+1,+1$)
&&&& \\  

($-1,+1,-1,+1,-1,-1,-1$)
&&&& \\  

($+1,+1,-1,+1,-1,+1,-1$)
&&&& \\  

($+1,+1,-1,+1,-1,-1,+1$)
&&&& \\  

($-1,-1,-1,+1,-1,-1,-1$)
&&&& \\  

($+1,-1,-1,+1,-1,+1,-1$)
&&&& \\  

($+1,-1,-1,+1,-1,-1,+1$)
&&&& \\  

($-1,-1,-1,-1,+1,+1,+1$)
&&&& \\  

($-1,+1,-1,-1,+1,+1,+1$)
&&&& \\  

($-1,+1,-1,-1,+1,-1,-1$)
&&&& \\  

($+1,+1,-1,-1,+1,+1,-1$)
&&&& \\  

($+1,+1,-1,-1,+1,-1,+1$)
&$\cdots$ &$\cdots$ &$\cdots$ &$\cdots$ \\  

($-1,-1,-1,-1,+1,-1,-1$)
&&&& \\  

($+1,-1,-1,-1,+1,+1,-1$)
&&&& \\  

($+1,-1,-1,-1,+1,-1,+1$)
&&&& \\  

($-1,-1,+1,-1,-1,+1,+1$)
&&&& \\  

($-1,+1,+1,-1,-1,+1,+1$)
&&&& \\  

($-1,+1,+1,-1,-1,-1,-1$)
&&&& \\  

($+1,+1,+1,-1,-1,+1,-1$)
&&&& \\  

($+1,+1,+1,-1,-1,-1,+1$)
&&&& \\  

($-1,-1,+1,-1,-1,-1,-1$)
&&&& \\  

($+1,-1,+1,-1,-1,+1,-1$)
&&&& \\  

($+1,-1,+1,-1,-1,-1,+1$)
&&&& \\  
\hline
\end{tabular}
\end{table*}

\begin{table*}[htbp]
\caption{$G_{\mathbf K}\cong M_{\mathbf K}=O\times Z_2^T$,
${\rm Invariants}\equiv (\eta_{T},\eta_{TC_{2z}},\eta_{TC_{2f}})$,
we list the values of all symmetry invariants,
where type II MSG-realizable are colored in {\color{green}GREEN}, type IV MSG-realizable in {\color{blue}BLUE}, and only SSG-realizable in {\color{red}RED}.
For only SSG-realizable ($\vec{\eta}^{L}$,{\R $\vec{\eta}^{S}$}), dim denotes the lowest dimension of irRep,
disp. are the $k$-terms which can split band degeneracy.}
\label{Odrztinv}
\centering
\begin{tabular}{ |c|c|c|c|}
\hline
Invariants & ($\vec{\eta}^{L}$,{\R $\vec{\eta}^{S}$})&dim &disp.\\
\hline
({\color{green}$+1,+1,+1$})
&&&\\

({\color{green}$+1,-1,+1$})
&&&\\

({\color{green}$-1,+1,+1$})
&&&\\

({\color{green}$-1,-1,+1$})
&&&\\    
\hline

({\color{red}$+1,+1,-1$})
& ($000$,{\color{red}$001$}),($110$,{\color{red}$111$})
&&\\
\hline

({\color{red}$+1,-1,-1$})
&($100$,{\color{red}$111$})
&$4|8$&$[k_{x},k_{y},k_{z}]$\\

&&$4|4_{1,2}$&$[k_{x},k_{y},k_{z}]$\\
\hline

({\color{red}$-1,+1,-1$})
&($100$,{\color{red}$001$})
&$4|8$&$[k_{x},k_{y},k_{z}]$\\

&&$4|4$&$[k_{x},k_{y},k_{z}]$\\
\hline

({\color{red}$-1,-1,-1$})
&($000$,{\color{red}$111$}),($110$,{\color{red}$001$})
&&\\  
\hline
\end{tabular}
\end{table*}

\begin{table*}[htbp]
\caption{$G_{\mathbf K}\cong M_{\mathbf K}=O_{h}\times Z_2^T$,
${\rm Invariants}\equiv (\eta_{C_{2x},C_{2y}},\eta_T,\eta_{IT},\eta_{TC_{2a}},\eta_{I,C_{2a}})$,
we list the values of all symmetry invariants,
where type II MSG-realizable are colored in {\color{green}GREEN}, type IV MSG-realizable in {\color{blue}BLUE}, and only SSG-realizable in {\color{red}RED}.
For only SSG-realizable ($\vec{\eta}^{L}$,{\R $\vec{\eta}^{S}$}), dim denotes the lowest dimension of irRep,
disp. are the $k$-terms which can split band degeneracy, n.l. is direction of nodal line in BZ.}
\label{Ohdrztinv}
\centering
\begin{tabular}{ |c|c|c|c|c|}
\hline
Invariants & ($\vec{\eta}^{L}$,{\R $\vec{\eta}^{S}$})&dim &disp.&n.l.\\
\hline

({\color{green}$+1,+1,+1,+1,+1$})
&$\cdots$&$\cdots$&$\cdots$&$\cdots$ \\

({\color{green}$+1,+1,+1,+1,-1$})
&&&& \\

({\color{green}$-1,-1,-1,+1,+1$})
&&&& \\

({\color{green}$-1,-1,-1,+1,-1$})
&&&& \\  

({\color{blue}$+1,-1,+1,+1,+1$})
&&&& \\

({\color{blue}$+1,-1,+1,+1,-1$})
&&&& \\

({\color{blue}$-1,+1,-1,+1,+1$})
&&&& \\

({\color{blue}$-1,+1,-1,+1,-1$})
&&&& \\ 
\hline

({\color{red}$+1,-1,+1,-1,+1$})
&($00000$,{\color{red}$01010$}),($01001$,{\color{red}$00011$}),

&&& \\ 

& ($01000$,{\color{red}$00010$})
&&& \\
\hline

({\color{red}$+1,+1,+1,-1,+1$})
&($01000$,{\color{red}$01010$}),($00000$,{\color{red}$00010$})

&&& \\ 
\hline

({\color{red}$+1,+1,+1,-1,-1$})
&($01001$,{\color{red}$01010$}),($00000$,{\color{red}$00011$})

&&& \\ 
\hline

({\color{red}$+1,-1,+1,-1,-1$})
&($01000$,{\color{red}$00011$}),($01001$,{\color{red}$00010$})

&$4|12$ & $[k_{x},k_{y},k_{z}]$&$-$\\ 

&
&$4|4_{1}$ & $k_{x}k_{y}k_{z}$,&
$k_{x},k_{y},k_{z}$,\\

&&& $(k_{x}^{2}-k_{y}^{2})(k_{x}^{2}-k_{z}^{2})(k_{y}^{2}-k_{z}^{2})$
& $k_{x}\pm k_{y},k_{x}\pm k_{z}$,\\

&&&& $k_{y}\pm k_{z}$ \\

&& $4|4_{2,3}$ & $k_{x}k_{y}k_{z}[k_{x}^{2}+k_{y}^{2}-2k_{z}^{2},\sqrt{3}(k_{x}^{2}-k_{y}^{2})]$,&
$k_{x},k_{y},k_{z}$,\\

&&& $(k_{x}^{2}-k_{y}^{2})(k_{x}^{2}-k_{z}^{2})(k_{y}^{2}-k_{z}^{2})$
& $k_{x}\pm k_{y},k_{x}\pm k_{z}$,\\

&&&& $k_{y}\pm k_{z}$,\\
&&&& $k_{x}+k_{y}+k_{z}$,\\
&&&& $k_{x}+k_{y}-k_{z}$,\\
&&&& $k_{x}-k_{y}+ k_{z}$,\\
&&&& $k_{x}-k_{y}- k_{z}$ \\
\hline

({\color{red}$+1,-1,-1,+1,+1$})
&($00000$,{\color{red}$01100$}),($01001$,{\color{red}$00101$}),

&&& \\ 

& ($01000$,{\color{red}$00100$})
&&& \\
\hline

({\color{red}$+1,+1,-1,+1,+1$})
&($01000$,{\color{red}$01100$}),($00000$,{\color{red}$00100$})

&&& \\ 
\hline

({\color{red}$+1,+1,-1,+1,-1$})
&($01001$,{\color{red}$01100$}),($00000$,{\color{red}$00101$})

&&& \\ 
\hline

({\color{red}$+1,-1,-1,+1,-1$})
&($01000$,{\color{red}$00101$}),($01001$,{\color{red}$00100$})

&$4|12$ & $[k_{x},k_{y},k_{z}]$&$-$\\ 

&
&$4|4_{1}$ & $k_{x}k_{y}k_{z}$,&
$k_{x},k_{y},k_{z}$,\\

&&& $(k_{x}^{2}-k_{y}^{2})(k_{x}^{2}-k_{z}^{2})(k_{y}^{2}-k_{z}^{2})$
& $k_{x}\pm k_{y},k_{x}\pm k_{z}$,\\

&&&& $k_{y}\pm k_{z}$ \\

&& $4|4_{2}$ & $k_{x}k_{y}k_{z}$, & $k_{x},k_{y},k_{z}$, \\

&&& $k_{x}k_{y}k_{z}[k_{x}^{2}+k_{y}^{2}-2k_{z}^{2},\sqrt{3}(k_{x}^{2}-k_{y}^{2})]$,&
$k_{x}\pm k_{y},k_{x}\pm k_{z}$, \\

&&& $(k_{x}^{2}-k_{y}^{2})(k_{x}^{2}-k_{z}^{2})(k_{y}^{2}-k_{z}^{2})$
& $k_{y}\pm k_{z}$\\
\hline
\end{tabular}
\end{table*}

\begin{table*}
(Extension of Supplementary Table \ref{Ohdrztinv},
${\rm Invariants}\equiv (\eta_{C_{2x},C_{2y}},\eta_T,\eta_{IT},\eta_{TC_{2a}},\eta_{I,C_{2a}})$ )\\
\centering
\begin{tabular}{ |c|c|c|c|c|}
\hline
Invariants & ($\vec{\eta}^{L}$,{\R $\vec{\eta}^{S}$})&dim &disp.&n.l.\\
\hline

({\color{red}$+1,-1,-1,-1,+1$})
&($00000$,{\color{red}$01110$}),($01000$,{\color{red}$00110$})

&&& \\ 
\hline

({\color{red}$+1,+1,-1,-1,+1$})
&($01000$,{\color{red}$01110$}),($00000$,{\color{red}$00110$})

&&& \\ 
\hline

(${\color{red}+1,+1,-1,-1,-1}$)
&($01001$,{\color{red}$01110$})

&$4|12$ & $[k_{x},k_{y},k_{z}]$&$-$\\ 

&
&$4|4_{1}$ & $k_{x}k_{y}k_{z}$,&
$k_{x},k_{y},k_{z}$,\\

&&& $(k_{x}^{2}-k_{y}^{2})(k_{x}^{2}-k_{z}^{2})(k_{y}^{2}-k_{z}^{2})$
& $k_{x}\pm k_{y},k_{x}\pm k_{z}$,\\

&&&& $k_{y}\pm k_{z}$ \\

&& $4|4_{2,3}$ & $k_{x}k_{y}k_{z}[k_{x}^{2}+k_{y}^{2}-2k_{z}^{2},\sqrt{3}(k_{x}^{2}-k_{y}^{2})]$,&
$k_{x},k_{y},k_{z}$,\\

&&& $(k_{x}^{2}-k_{y}^{2})(k_{x}^{2}-k_{z}^{2})(k_{y}^{2}-k_{z}^{2})$
& $k_{x}\pm k_{y},k_{x}\pm k_{z}$,\\

&&&& $k_{y}\pm k_{z}$,\\
&&&& $k_{x}+k_{y}+k_{z}$,\\
&&&& $k_{x}+k_{y}-k_{z}$,\\
&&&& $k_{x}-k_{y}+ k_{z}$,\\
&&&& $k_{x}-k_{y}- k_{z}$ \\
\hline

({\color{red}$+1,-1,-1,-1,-1$})
&($01001$,{\color{red}$00110$})

&$4|12$ & $[k_{x},k_{y},k_{z}]$&$-$\\ 

&
&$4|4_{1}$ & $k_{x}k_{y}k_{z}$,&
$k_{x},k_{y},k_{z}$,\\

&&& $(k_{x}^{2}-k_{y}^{2})(k_{x}^{2}-k_{z}^{2})(k_{y}^{2}-k_{z}^{2})$
& $k_{x}\pm k_{y},k_{x}\pm k_{z}$,\\

&&&& $k_{y}\pm k_{z}$ \\

&& $4|4_{2}$ & $k_{x}k_{y}k_{z}[k_{x}^{2}+k_{y}^{2}-2k_{z}^{2},\sqrt{3}(k_{x}^{2}-k_{y}^{2})]$,&
$k_{x},k_{y},k_{z}$,\\

&&& $(k_{x}^{2}-k_{y}^{2})(k_{x}^{2}-k_{z}^{2})(k_{y}^{2}-k_{z}^{2})$
& $k_{x}\pm k_{y},k_{x}\pm k_{z}$,\\

&&&& $k_{y}\pm k_{z}$,\\
&&&& $k_{x}+k_{y}+k_{z}$,\\
&&&& $k_{x}+k_{y}-k_{z}$,\\
&&&& $k_{x}-k_{y}+ k_{z}$,\\
&&&& $k_{x}-k_{y}- k_{z}$ \\
\hline

($-1,-1,-1,-1,+1$)
&$\cdots$&$\cdots$&$\cdots$&$\cdots$ \\ 

($-1,+1,-1,-1,+1$)
&&&& \\ 

($-1,+1,-1,-1,-1$)
&&&& \\ 

($-1,-1,-1,-1,-1$)
&&&& \\ 

($-1,-1,+1,+1,+1$)
&&&& \\ 

($-1,+1,+1,+1,+1$)
&&&& \\ 

($-1,+1,+1,+1,-1$)
&&&& \\ 

($-1,-1,+1,+1,-1$)
&&&& \\ 

($-1,-1,+1,-1,+1$)
&&&& \\ 

($-1,+1,+1,-1,+1$)
&&&& \\ 

($-1,+1,+1,-1,-1$)
&&&& \\ 

($-1,-1,+1,-1,-1$)
&&&& \\ 

\hline
\end{tabular}
\end{table*}

\begin{table*}[htbp]
\caption{The definitions of symmetry invariants of 58 type III magnetic point groups
are given in terms of factor system $\omega$ or irRep $\rho$,
where
$I={\rm spacial\ inversion}$, $M={\rm mirror\ reflection\ plane}$, $T={\rm time\ reversal}$, $Z_{2}^{IT}=\{E, IT\}$, $Z_{2}^{T_{c}}=\{E, TC_{2}\}$,$Z_{2}^{T_{m}}=\{E, TM_{}\}$.
The invariants are interpreted as the following:
$\eta_{IT}\equiv\omega_{2}(IT,IT)$,
$\eta_{TC_{2}}\equiv\omega_{2}(TC_{2},TC_{2})$,
$\eta_{TM}\equiv\omega_{2}(TM,TM)$,
$\eta_{I,C_{2z}}\equiv\frac{\omega_{2}(I,C_{2z})}{\omega_{2}(C_{2z},I)}$,
$\eta_{I,C_{2x}}\equiv\frac{\omega_{2}(I,C_{2x})}{\omega_{2}(C_{2x},I)}$,
$\eta_{I,C^{+}_{4z}}\equiv\frac{\omega_{2}(I,C^{+}_{4z})}{\omega_{2}(C^{+}_{4z},I)}$,
$\eta_{I,C^{}_{2}}\equiv\frac{\omega_{2}(I,C^{}_{2})}{\omega_{2}(C^{}_{2},I)}$,
$\eta_{C_{2x},C_{2y}}\equiv\frac{\omega_{2}(C_{2x},C_{2y})}{\omega_{2}(C_{2y},C_{2x})}$,
$\eta_{M_{x},M_{y}}\equiv\frac{\omega_{2}(M_{x},M_{y})}{\omega_{2}(M_{y},M_{x})}$,
$\eta_{C_{2},C'_{21}}\equiv\frac{\omega_{2}(C_{2},C'_{21})}{\omega_{2}(C'_{21},C_{2})}$,
$\eta_{C_{2},M_{d1}}\equiv\frac{\omega_{2}(C_{2},M_{d1})}{\omega_{2}(M_{d1},C_{2})}$,
$\eta_{M_{h},C'_{21}}\equiv\frac{\omega_{2}(M_{h},C'_{21})}{\omega_{2}(C'_{21},M_{h})}$,
$\eta_{M_{h},M_{d1}}\equiv\frac{\omega_{2}(M_{h},M_{d1})}{\omega_{2}(M_{d1},M_{h})}$, $\eta_{I,C^{'}_{21}}\equiv\frac{\omega_{2}(I,C^{'}_{21})}{\omega_{2}(C^{'}_{21},I)}$, $\eta_{I,C^{''}_{21}}\equiv\frac{\omega_{2}(I,C^{''}_{21})}{\omega_{2}(C^{''}_{21},I)}$. The Number of $G$ is according to Table 7.1 of book~\cite{Bradley2010}
For a type III magnetic point group, if not all the sets of symmetry invariants can be realized at high-symmetry points in BZ of MSGs,
we list the values of all symmetry invariants,
where MSG-realizable in {\color{blue}BLUE}, and only SSG-realizable in {\color{red}RED}. The BLACK sets of invariants cannot be realized in MSG, or even SSG. For only SSG-realizable invariants, dim denotes the lowest dimension of irRep,
disp. are the $k$-terms which can split band degeneracy, n.l. is direction of nodal line in BZ.
} \label{tab:invrtstypeIII}
\centering
\begin{tabular}{ |c|c|c|c|c|c|c| }
\hline
Number &
 $G$ & Generators & Invariants &
 dim & disp. & n.l.\\
\hline

1
& $C_{1}\times Z_{2}^{IT}$
& $IT$
& $\eta_{IT}$
&&&\\
\hline

2
&$C_{1}\times Z_{2}^{T_{c}}$
& $TC_{2z}$
& $\eta_{TC_{2z}}$
&&&\\
\hline

3
& $C_{1}\times Z_{2}^{T_m}$
& $TM_z$
& $\eta_{TM_z}$
&&&\\
\hline

4
& $\mathscr{C}_{2}\times Z_{2}^{IT}$
& $C_{2z}$, $IT$
& ($\eta_{IT}$,$\eta_{TM_z}$)
&&&\\
\hline

5
& $C_{1h}\times Z_2^{IT}$
& $M_z$, $IT$
& ($\eta_{IT}$,$\eta_{TC_{2z}}$)
&&&\\
\hline

6
& $C_{i}\times Z_{2}^{T_c}$
& $I$,$TC_{2z}$
& ($\eta_{TC_{2z}}$,$\eta_{TM_z}$)
&&&\\
\hline

7
& $\mathscr{C}_{2}\times Z_2^{T_c}$
& $C_{2z}$,$TC_{2x}$
& ($\eta_{TC_{2x}}$,$\eta_{TC_{2y}}$)
&&&\\
\hline

\end{tabular}
\end{table*}

\begin{table*}
(Extension of Supplementary Table \ref{tab:invrtstypeIII} )\\
\centering
\begin{tabular}{ |c|c|c|c|c|c|c|}
\hline

Number &
 $G$ & Generators & Invariants &
  dim & disp. & n.l.\\
\hline

26
&$C_{4v}\times Z_2^{IT}$
&$C_{4z}^+$,$M_{x}$,$IT$
&($\eta_{M_{x},M_{y}}$,$\eta_{IT}$,
$\eta_{TC_{2x}}$,$\eta_{TC_{2a}}$)
&&&\\

&&& ({\color{blue}$+1,+1,+1,+1$})
&&&\\

&&& ({\color{blue}$+1,+1,-1,+1$})
&&&\\

&&& ({\color{blue}$-1,-1,+1,+1$})
&&&\\

&&& ({\color{blue}$-1,-1,-1,+1$})
&&&\\
\hline

&&& ({\color{red}$+1,+1,+1,-1$})
&&&\\

\hline

&&& ({\color{red}$+1,+1,-1,-1$})
&&&\\
\hline

\end{tabular}
\end{table*}

\begin{table*}
(Extension of Supplementary Table \ref{tab:invrtstypeIII} )\\
\centering
\begin{tabular}{ |c|c|c|c|c|c|c|}
\hline

Number &
 $G$ & Generators & Invariants &
 dim & disp. & n.l.\\
\hline

32
& $C_{3}\times Z_2^{T_m}$
& $TS_3^-$
& $\eta_{TM_h}$
&&&\\

&&& ({\color{blue}$+1$})
&&&\\

\hline

&&& ({\color{red}$-1$})
&&&\\
\hline

33
& $C_{3h}\rtimes Z_2^{T_c}$
& $S_3^-$,$TC'_{21}$
& ($\eta_{TC'_{21}}$,$\eta_{TM_{v1}}$)
&&&\\

&
& $S_3^-$,$TC''_{21}$
& ($\eta_{TC''_{21}}$,$\eta_{TM_{d1}}$)
&&&\\

&&& ({\color{blue}$+1,+1$})
&&&\\

&&& ({\color{blue}$+1,-1$})
&&&\\
\hline

\end{tabular}
\end{table*}

\begin{table*}
(Extension of Supplementary Table \ref{tab:invrtstypeIII} )\\
\centering
\begin{tabular}{ |c|c|c|c|c|c|c|}
\hline

Number &
 $G$ & Generators & Invariants &
 dim & disp. & n.l. \\
\hline

47
& $C_{3v}^1\times Z_2^{T_c}$
& $M_{d1}$,$TC_6^+$
& ($\eta_{TC_2}$,$\eta_{TM_{v1}}$)
&&&\\

& $C_{3v}^2\times Z_2^{T_c}$
& $M_{v1}$,$TC_6^+$
& ($\eta_{TC_2}$,$\eta_{TM_{d1}}$)
&&&\\

\hline

48
& $D_{3h}^1\times Z_2^{IT}$
& $S_3^-$,$C'_{21}$,$IT$
& ($\eta_{M_{h},C'_{21}}$,$\eta_{IT}$,
$\eta_{TC_2}$,$\eta_{TM_{d1}}$)
&&&\\

& $D_{3h}^2\times Z_2^{IT}$
& $S_3^-$,$M_{d1}$,$IT$
& ($\eta_{M_{h},M_{d1}}$,$\eta_{IT}$,
$\eta_{TC_2}$,$\eta_{TC'_{21}}$)
&&&\\

&&& ({\color{blue}$+1,+1,+1,+1$})
&&&\\

&&& ({\color{blue}$+1,+1,-1,+1$})
&&&\\

&&& ({\color{blue}$+1,+1,-1,-1$})
&&&\\
\hline

\end{tabular}
\end{table*}

\begin{table*}
(Extension of Supplementary Table \ref{tab:invrtstypeIII} )\\
\centering
\begin{tabular}{ |c|c|c|c|c|c|c|}
\hline

Number &
 $G$ & Generators & Invariants &
 dim & disp. & n.l. \\
\hline

50
& $D_6\times Z_2^{IT}$
& $C_6^+$,$C'_{21}$,$IT$
& ($\eta_{C_{2},C'_{21}}$,$\eta_{IT}$,
$\eta_{TM_h}$,$\eta_{TM_{d1}}$)
&&&\\

&&& ({\color{blue}$+1,+1,+1,+1$})
&&&\\

&&& ({\color{blue}$+1,+1,+1,-1$})
&&&\\

&&& ({\color{blue}$-1,+1,+1,+1$})
&&&\\

&&& ({\color{blue}$-1,+1,+1,-1$})
&&&\\

&&& ({\color{blue}$-1,-1,+1,+1$})
&&&\\

&&& ({\color{blue}$-1,-1,+1,-1$})
&&&\\
\hline

\end{tabular}
\end{table*}

\begin{table*}
(Extension of Supplementary Table \ref{tab:invrtstypeIII} )\\
\centering
\begin{tabular}{ |c|c|c|c|c|c|c|}
\hline

Number &
 $G$ & Generators & Invariants &
 dim & disp. & n.l. \\
\hline

56
& $O\times Z_2^{IT}$
& $C_{2x}$,$C_{2y}$,$C_{2a}$,$C_{2f}$,$IT$
& ($\eta_{IT}$,$\eta_{TM_z}$,$\eta_{TM_{da}}$)
&&&\\

&&& ({\color{blue}$+1,+1,+1$})
&&&\\

&&& ({\color{blue}$-1,+1,+1$})
&&&\\

&&& ({\color{blue}$+1,+1,-1$})
&&&\\

&&& ({\color{blue}$-1,+1,-1$})
&&&\\

\hline

&&& ({\color{red}$-1,-1,+1$})
&&&\\

\hline

&&& ({\color{red}$-1,-1,-1$})
&&&\\

\hline

&&& ($+1,-1,+1$)
&&&\\

&&& ($+1,-1,-1$)
&&&\\

\hline

57
& $T_d\times Z_2^{IT}$
& $C_{2x}$,$C_{2y}$,$M_{da}$,$M_{df}$,$IT$
& ($\eta_{IT}$,$\eta_{TM_z}$,$\eta_{TC_{2a}}$)
&&&\\

&&& ({\color{blue}$+1,+1,+1$})
&&&\\

&&& ({\color{blue}$-1,+1,+1$})
&&&\\

&&& ({\color{blue}$+1,-1,+1$})
&&&\\

&&& ({\color{blue}$-1,-1,+1$})
&&&\\

\hline

&&& ({\color{red}$+1,+1,-1$})
&&&\\

\hline

&&& ({\color{red}$+1,-1,-1$})
& $4|8$
& $[k_x,k_y,k_z]$
& $-$
\\

&&&
& $4|4_{1,2}$
&$[k_x(k_y^2-k_z^2),
k_y(k_z^2-k_x^2),
k_z(k_x^2-k_y^2)]$
& $k_x$, $k_y$, $k_z$,
\\

&&&&&
& $k_x+k_y+k_z$,
\\

&&&&&
& $k_x+k_y-k_z$,
\\

&&&&&
& $k_x-k_y+k_z$,
\\

&&&&&
& $k_x-k_y-k_z$
\\

\hline

&&& ({\color{red}$-1,-1,-1$})
&&&\\

\hline

&&& ({\color{red}$-1,+1,-1$})
& $4|8$
& $[k_x,k_y,k_z]$
& $-$
\\

&&&
& $4|4$
&$[k_x(k_y^2-k_z^2),
k_y(k_z^2-k_x^2),
k_z(k_x^2-k_y^2)]$,
& $k_x$, $k_y$, $k_z$,
\\

&&&&
& $(k_x^2-k_y^2)
(k_x^2-k_z^2)
(k_y^2-k_z^2)$
& $k_x+k_y+k_z$,
\\

&&&&&
& $k_x+k_y-k_z$,
\\

&&&&&
& $k_x-k_y+k_z$,
\\

&&&&&
& $k_x-k_y-k_z$
\\

\hline

58
& $T_h\rtimes Z_2^{T_c}$
& $C_{31}^{-},C_{33}^{+},I,TC_{2a}$
& ($\eta_{C_{2x},C_{2y}}$,$\eta_{TC_{2a}}$,$\eta_{TM_{da}}$)
&&&\\

&&& ({\color{blue}$+1,+1,+1$})
&&&\\

&&& ({\color{blue}$-1,+1,+1$})
&&&\\

&&& ({\color{blue}$+1,+1,-1$})
&&&\\

&&& ({\color{blue}$-1,+1,-1$})
&&&\\

\hline

&&& ({\color{red}$+1,-1,+1$})
&&&\\

\hline

&&& ({\color{red}$+1,-1,-1$})
&&&\\

\hline

&&& ($-1,-1,+1$)
&&&\\

&&& ($-1,-1,-1$)
&&&\\

\hline

\end{tabular}
\end{table*}

\end{document}